\documentclass[preprint]{aastex}
\pdfoutput=1
\usepackage{isotope}
\usepackage{capt-of}
\usepackage{longtable}
\usepackage{multirow}
\usepackage{lscape}
\usepackage{color}
\usepackage[section] {placeins}
\def\newasrev{New~Astron.~Rev.}%
\def\nar{New~Astron.~Rev.}%
\def\rpp{Rep.~Prog.~Phys.}%
\def\nar{New~Astron.~Rev.}%
\def\pasa{Publ.~Astronom.~Soc.~Aus.}%
\newcommand{\sect}[1]{Section\,\ref{#1}}

\newcommand{\fig}[1]{Fig.\,\ref{#1}}

\newcommand{\tab}[1]{Table\,\ref{#1}}
\newcommand{\tabs}[1]{Tables\,\ref{#1}}

\newcommand{\pdcz}{pulse-driven convective zone}

\newcommand{\setone}{Set\,1}
\newcommand{\setopo}{Set\,1.1}
\newcommand{\setopt}{Set\,1.2}

\newcommand{\spr}{\mbox{$s$-process}}
\newcommand{\sprn}{\mbox{$s$ process}}
\newcommand{\ipr}{\mbox{$i$-process}}
\newcommand{\iprn}{\mbox{$i$ process}}
\newcommand{\npr}{\mbox{$n$-process}}
\newcommand{\nprn}{\mbox{$n$ process}}
\newcommand{\rpr}{\mbox{$r$-process}}
\newcommand{\rprn}{\mbox{$r$ process}}
\newcommand{\ppr}{\mbox{$p$-process}}
\newcommand{\pprn}{\mbox{$p$ process}}
\newcommand{\apr}{\mbox{$\alpha$-process}}
\newcommand{\aprn}{\mbox{$\alpha$ process}}

\newcommand{\p}{\ensuremath{\mem{p}}}

\newcommand{\hevi}{\ensuremath{^{4}\mem{He}}}

\newcommand{\cdr}{\isotope[13]{C}}
\newcommand{\czw}{\isotope[12]{C}}

\newcommand{\nvi}{\ensuremath{^{14}\mem{N}}}

\newcommand{\ofu}{\ensuremath{^{15}\mem{O}}}
\newcommand{\ose}{\isotope[16]{O}}

\newcommand{\nezw}{\ensuremath{^{22}\mem{Ne}}}

\newcommand{\msun}{\ensuremath{\, M_\odot}}
\newcommand{\lsun}{\ensuremath{\, L_\odot}}

\newcommand{\teff}{\ensuremath{T_{\rm eff}}}
\newcommand{\kelv}{\ensuremath{\,\mathrm K}}
\newcommand{\mzams}{\ensuremath{M_{\rm ZAMS}}}
\newcommand{\mppnp}{\textsf{mppnp}}
\newcommand{\sppn}{\textsf{sppn}}
\newcommand{\se}{\textsf{se}}
\newcommand{\MESA}{\texttt{MESA}}

\newcommand{\beq}{\begin{equation}}
\newcommand{\beqa}{\begin{eqnarray}}
\newcommand{\eeq}{\end{equation}}
\newcommand{\eeqa}{\end{eqnarray}}
\newcommand{\bedis}{\begin{displaymath}}
\newcommand{\edis}{\end{displaymath}}

\newcommand{\mem}[1]{\ensuremath{\mathrm{ #1}}}

\newcommand{\apgeq}{\ensuremath{\stackrel{>}{_\sim}}}

\newcommand{\hp}{\ensuremath{H_\mem{p}}}

\newcommand{\Ye}{\mathrm{Y_e}}

\newcommand{\cc}{\ensuremath{\,\rm cm^{-3}}}

\graphicspath{{figures/}}

\slugcomment{DRAFT: \today}

\shorttitle{NuGrid stellar data Set I}
\shortauthors{}

\begin{document}

\title{NuGrid stellar data set I.
Stellar yields from H to Bi for stars with metallicities $Z = 0.02$ and $Z = 0.01$}
\author{M. Pignatari\altaffilmark{1,2,3,14},
F.~Herwig\altaffilmark{4,5,14},
R.~Hirschi\altaffilmark{6,7,14},
M.~Bennett\altaffilmark{6,14},
G.~Rockefeller\altaffilmark{8,14},
C.~Fryer\altaffilmark{8,14},
F.~X.~Timmes\altaffilmark{9,5,14},
C.~Ritter\altaffilmark{4,14},
A.~Heger\altaffilmark{10,5,14},
S.~Jones\altaffilmark{4,5,14},
U.~Battino\altaffilmark{3,14},
%C.~Ritter\altaffilmark{9,14},
A.~Dotter\altaffilmark{11,14},
R.~Trappitsch\altaffilmark{12,14},
S.~Diehl\altaffilmark{8},
U.~Frischknecht\altaffilmark{3},
A.~Hungerford\altaffilmark{8,14},
G.~Magkotsios\altaffilmark{5,14},
C.~Travaglio\altaffilmark{13,14},
P.~Young\altaffilmark{9}
}

\altaffiltext{1}{E.A. Milne Centre for Astrophysics, Dept of Physics \& Mathematics, University of Hull, HU6 7RX, United Kingdom}
\altaffiltext{2}{Konkoly Observatory, Research Centre for Astronomy and Earth Sciences, Hungarian Academy of Sciences, Konkoly Thege Miklos ut 15-17, H-1121 Budapest, Hungary}
\altaffiltext{3}{Department of Physics, University of Basel, Klingelbergstrasse 82, CH-4056 Basel, Switzerland}
\altaffiltext{4}{Department of Physics \& Astronomy, University of Victoria, Victoria, BC, V8P5C2 Canada.}
\altaffiltext{5}{The Joint Institute for Nuclear Astrophysics, Notre Dame, IN 46556, USA}
\altaffiltext{6}{Keele University, Keele, Staffordshire ST5 5BG, United Kingdom.}
\altaffiltext{7}{Institute for the Physics and Mathematics of the Universe (WPI), University of Tokyo, 5-1-5 Kashiwanoha, Kashiwa 277-8583, Japan}
\altaffiltext{8}{Computational Physics and Methods (CCS-2), LANL, Los Alamos, NM, 87545, USA.}
\altaffiltext{9}{Arizona State University (ASU), School of Earth and Space Exploration (SESE), PO Box 871404, Tempe, AZ, 85287-1404, USA.}
\altaffiltext{10}{Monash Centre for Astrophysics, School of Mathematical Sciences, Monash University, Vic 3800, Australia.}
%\altaffiltext{9}{Goethe-Universität Frankfurt, Senckenberganlage 31, 60325 Frankfurt am Main, Germany.}
\altaffiltext{11}{Research School of Astronomy and Astrophysics, Australian National University, Weston Creek, ACT 2611, Australia.}
\altaffiltext{12}{Department of the Geophysical Sciences and Chicago Center for Cosmochemistry, Chicago, IL 60637, USA.}
\altaffiltext{13}{Osservatorio Astronomico di Torino, Torino, Italy.}
\altaffiltext{14}{NuGrid collaboration, \url{http://www.nugridstars.org}}

%altaffiltext{1}{Keele University, Keele, Staffordshire ST5 5BG, United Kingdom.}
%altaffiltext{2}{Joint Institute for Nuclear Astrophysics, University of Notre Dame, Notre Dame, IN 46556, United States.}
%altaffiltext{3}{Universita' di Torino, Torino, Via Pietro Giuria 1, 10126 Italy.}
%altaffiltext{4}{GSI, Darmstadt, Germany.}
%altaffiltext{5}{Forschungszentrum Karlsruhe, Institut für Kernphysik, Postfach 3640, 76021 Karlsruhe, Germany.}
%\affil{University of Torino, Italy}
%email{marco@astro.keele.ac.uk}
%\affil{University of Victoria, B.C., Canada}
%\email{fherwig@uvastro.phys.uvic.ca}

\newcommand{\X}[1]{\textbf{\color{red}{#1}}}

\begin{abstract}
We provide a set of stellar evolution and nucleosynthesis calculations
that applies established physics assumptions simultaneously to low-
and intermediate-mass and massive star models. Our goal is to provide
an internally consistent and comprehensive nuclear production and
yield data base for applications in areas such as pre-solar grain
studies.  Our non-rotating models assume convective boundary mixing
where it has been adopted before. We include 8 (12) initial masses for
$Z = 0.01$ ($0.02$).  Models are followed either until the end of the
asymptotic giant branch phase or the end of Si burning, complemented
by a simple analytic core-collapse supernova models with two options
for fallback and shock velocities. The explosions show 
which pre-supernova yields will most strongly be effected
by the explosive nucleosynthesis. We discuss how these two explosion
parameters impacts the light elements and the $s$ and $p$ process. For
low- and intermediate-mass models our stellar yields from H to Bi
include the effect of convective boundary mixing at the He-intershell
boundaries and the stellar evolution feedback of the mixing process
that produces the \cdr\ pocket. All post-processing nucleosynthesis
calculations use the same nuclear reaction rate network and nuclear
physics input.  We provide a discussion of  the nuclear
production across the entire mass range organized by element
group. All our stellar nucleosynthesis profile and time evolution
output is available electronically, and tools to explore the data on
the NuGrid VOspace hosted by the Canadian Astronomical Data Centre are
introduced.
\end{abstract}


%\tableofcontents

\section{Introduction}
\label{intro}

%Stars and their outbursts can produce all the elements beside H. 
%Elements heavier than Li form in stars and their outbursts.
All the elements heavier than H can be formed 
in stars and their outbursts.
Understanding the processes that have lead to the abundance
distribution in the solar system is one of the fundamental goals of
stellar nucleosynthesis and galactic astronomy.  The solar system
abundance distribution has been formed through nucleosynthesis in
several generations of different stars. Despite significant
progress, details regarding the chemical evolution of the Galaxy
remain poorly understood
\citep[e.g.,][]{tinsley:80,timmes:95,goswami:00,travaglio:04,gibson:06,kobayashi:06}.
%This makes  the analysis of the solar abundances challenging, since
%not only stellar yields need to be known for different metallicities,
%but the respective contribution from different stellar sources
%depends on the dynamical evolution of the Galaxy.
This makes understanding the origin of the solar abundances challenging.
Complete, metallicity-dependent stellar yields would provide part of the answer,
but the respective contribution from different stellar sources
depends on the dynamical evolution of the Galaxy.
The analysis of spectroscopic observations of unevolved stars in the local
disk of the Galaxy carries a similar degeneracy to the analysis of
stellar nucleosynthesis.
The observation of evolved low- and
intermediate mass stars
\citep[e.g.,][]{busso:01,garcia-hernandez:06,hernandez:08,abia:10,abia:12}
and of the ejecta of core-collapse supernova (CCSN)
\citep[e.g.,][]{kjaer:10,isensee:10,isensee:12,hwang:12} can provide
information about the intrinsic nucleosynthesis of these objects and constrain some of the modelling uncertainties.

A closer source of information about stellar nucleosynthesis
processes is hidden in primitive meteorites.  Small dust
grains of presolar origin---which were produced in ancient stars whose lives ended before the formation of our solar system---can be
found on Earth preserved in meteorites 
\citep[][]{lewis:87,bernatowicz:87,amari:90,bernatowicz:91,huss:94,nittler:95,choi:99}. 
These are assumed to
carry a relatively unmodified
nucleosynthesis signature from the environment of their parent stars
\citep[e.g.,][]{zinner:03,clayton:04}.

Stars with different initial masses and metallicities contribute in different ways to the production of
elements. Low- and intermediate-mass stars contribute to the chemical evolution of the interstellar medium over longer time scales than
massive stars, firstly during the advanced hydrostatic phases via a
stellar wind and (predominantly) late in their lives during the asymptotic giant branch phase
\cite[AGB e.g.,][]{iben:83,busso:99,herwig:05}. These stars also have the possibility to contribute to element production much later in time as Type Ia supernovae \citep[SNIa,
  e.g.,][]{nomoto:84,timmes:95,hillebrandt:00,dominguez:01,thielemann:04,travaglio:11,pakmor:12,seitenzahl:13,hillebrandt:13}.
During the AGB phase, light elements like carbon, nitrogen and
fluorine can be significantly produced, depending on the initial stellar mass, in addition to heavy \spr\ elements
\citep[e.g.,][]{herwig:04a,karakas:10,cristallo:11,bisterzo:11}.  In
particular, low-mass AGB stars are responsible for the production of
the main \spr\ component in the solar system explaining the
\spr\ abundances between strontium and lead; they are also responsible for the strong
\spr\ component, which mainly contributes to the solar lead inventory
\citep[e.g.,][]{gallino:98,travaglio:01a,sneden:08}.

Massive stars ($M \gtrsim 8\msun$) provide the first contribution to the
elemental chemical evolution owing to their short lifetimes.  They
produce metals both during their evolution and in the core-collapse supernova
explosions (CCSN) marking their deaths. During their evolution, massive stars contribute to the
chemical enrichment of the interstellar medium via winds; in these winds
it is predominantly light elements up to silicon that are released
\citep[for instance carbon and nitrogen, which are H- and
  He-burning products; see,
  e.g.,][]{meynet:06}.
Most $\alpha$-elements up to the iron group
are produced during the advanced evolutionary stages
\citep[e.g.,][]{thielemann:85} and/or by the final CCSN
\citep[e.g.,][]{woosley:95,thielemann:96,rauscher:02}.  Massive
stars are also the main site for the weak \sprn \citep[e.g.,][]{kaeppeler:11}.  
The weak \spr\ component
\citep[forming most of the \spr\ abundances in the solar system
  between iron and strontium, e.g.,][]{travaglio:04} is produced
during convective core He-burning and convective shell C-burning
stages \citep[e.g.,][]{raiteri:91a, raiteri:91b,the:07,pignatari:10}.
Since the \spr\ yields from massive stars are mostly ejected during
the CCSN explosion, partial or more extreme modifications triggered by
explosive nucleosynthesis need to be considered for these elements
\citep[e.g.,][]{thielemann:96,rauscher:02}. One example is the
classical \pprn\ (also known as the $\gamma$ process) which forms
proton-rich nuclei due to the photo-disintegration of \spr\ products
in deep \spr-rich layers \citep[][]{arnould:03}.

The \sprn\ is responsible for about half of the abundances
of trans-iron elements in the solar system. The \rprn\ is responsible
for the production of the majority of the remaining abundances, however there are some distinct discrepancies between the predictions from the \rpr\ residual
method \citep[e.g.][]{arlandini:99} and direct observations of
elemental abundances of metal-poor, \rpr-rich stars
\citep[][]{sneden:08,roederer:10}. The astrophysical source of the \rprn\ has been associated to neutrino-driven winds during CCSN
events, merging of their remnants or in jets from magnetorotationally-driven SNe \citep[e.g.,][]{kratz:08,thielemann:11,winteler:12,perego:14}.
The scenarios in which the conditions for \rprn\ nucleosynthesis are postulated to arise are the neutrino-induced winds from the CCSNe either 
before the formation of the reverse shock 
\citep[e.g.,][]{woosley:94,wanajo:01,farouqi:10}
or after fallback has begun 
\citep[e.g.,][]{fryer:06,arcones:07},
polar jets exuding from rotating 
magneto-hydrodynamical explosions 
of CCSNe \citep[][]{nishimura:06}, and
neutron-rich matter ejected from merging neutron stars
\citep[][]{freiburghaus:99} and neutron-star-black
hole mergers \citep[][]{surman:08}.
For a review of the different scenarios and recent $r$-process 
results see \citet{thielemann:11}, \citet{winteler:12} and \cite{korobkin:12}.

Many applications in astronomy and meteoritics require stellar yield
and nuclear production data. Presently, for AGB stars one may use the
yields of \citet{Karakas:2010et} which are available for a suitable range of
metallicities and initial masses but are limited to providing only the light
elements. Heavy element predictions for elemental compositions based
on the parameterized post-processing method are available from
\citet{bisterzo:10}. \spr\ yields from stellar evolution models are
available for a wide range of metallicities from the FRUITY database
\citep{cristallo:11,cristallo:14}. These yields are limited to low-mass stars ($M\leq
3\msun$), except for low metallicities where models up to M=$6\msun$ are included \citep[][]{straniero:14}.
For super-AGB stars, there is a much more limited amount of choice and while one may use the models of
\citet{Siess:2010hj,doherty:14}, the yields for heavy elements are not provided. Several
choices are available for massive star yields
\citep[e.g.][]{woosley:95,chieffi:04,nomoto:06}. These different
investigators have used different assumptions for the stellar micro-physics (e.g. opacities and nuclear reaction rates) and macro-physics (e.g. mixing assumptions and mass loss); the method with which the numerical solution to the equations of stellar evolution are found is also a factor that one can not ignore. Thus, yield tables stitched together from a range of
sources such as these do not only suffer from the inevitable uncertainties in many of
the ingredients required for such calculations \citep[see, e.g.,][]{romano:10,few:14,molla:15}, but also from a significant 
internal inconsistency. 
%This heavily limits the efficiency of feedback from confrontating stellar models with observations into the modeling process.
This introduces an additional degree of degeneracy in the feedback obtained from galactical chemical evolution studies about the physics and the assumptions implemented in stellar models.

The NuGrid research platform aims to address this issue by providing
different sets of stellar yields to be used for Galactic Chemical
Evolution (GCE), nuclear sensitivity and uncertainty studies, and
direct comparison with stellar observations. Each set will represent a
%comprehensive 
adequate
coverage of low-mass, intermediate-mass and massive star
models for a given set of physics assumptions and using the same
modeling codes for all masses. 
In this work, we present our first step toward achieving these goals. 
The first set of stellar models and their yields in the NuGrid production flow (\setone)
includes a grid of stellar masses 
%from $1.65$ to $60\msun$ at 2 metallicities: $Z$ = 0.02 and 0.01. 
from $1.65$ to $60\msun$ at metallicity $Z$ = 0.02, and from $1.65$ to $25\msun$ at metallicity $Z$ = 0.01.
%Two different codes are still used for massive stars and for low- and intermediate-mass stars. On the other hand, the same initial abundances, nuclear reaction rates, opacity tables... are used.
 Even though two different codes are still used in this study for massive stars and for low- and intermediate-mass stars, the same initial abundances, nuclear reaction rates, and opacity tables are used (see \sect{stars} for more details). Most importantly, the stellar models are post-processed with the same nucleosynthesis post-processing code. This allows to compare the nucleosynthesis results from different stellar codes, disentangle nuclear physics uncertainties from stellar uncertainties, and infer about the impact of a number of approximations that have to be made in 1D stellar codes \citep[e.g.,][]{jones:15,lattanzio:15}.

%Models for additional masses and
%three additional lower metallicities for both solar-scaled and
%$\alpha$-enhanced initial composition are presently computed and will
%be released as the \setone\ extension.

Our massive star simulations include one-dimensional, simplified CCSN
models which are used in order to qualitatively study explosive nucleosynthesis. While other
studies may have adopted a more realistic approach to the problem of explosive nucleosynthesis, the
uncertainties and limits of simulation capabilities of CCSN nucleosynthesis in 1D remain a
significant obstacle \citep[see, e.g.\ discussion in][]{roberts:10,perego:15,ertl:15}.
Our goal is to provide an estimate of the explosive contribution to stellar yields, including a
general understanding on how pre-explosive abundances are modified by
the explosion \citep[e.g.,][]{woosley:95,limongi:00,rauscher:02,nomoto:06}.
Therefore, the explosive SN yields presented in this work can be used 
for GCE calculations and for direct comparison with observations \citep[e.g.,][]{pignatari:15}, 
but keeping in mind their intrinsic limitations.

Together, the stellar models represent the Stellar Evolution and
Explosion (SEE) library and all of these models are then
post-processed using \mppnp\ 
%(see Appendix~\ref{ap:codes}) 
to calculate the nucleosynthesis during the evolution of each model, 
which comprises the post-processing data (PPD) library.  The SEE 
and PPD libraries associated with \setone\ and with this work are 
available online (see Appendix~\ref{ap:codes}).

Simulations for super-AGB stars
\citep[e.g.,][]{siess:07,poelarends:08,doherty:10,ventura:11,doherty:14},
electron-capture SNe \citep[e.g.,][]{nomoto:84,hoffman:08,wanajo:09},
SNIa \citep[e.g.,][]{hillebrandt:00,seitenzahl:13} and
\rprn\ \citep[][]{thielemann:11,winteler:12,kratz:14,nishimura:15} are not included in this work.
%, but at least super-AGB star models will be added in the \setone\ extension. 
The paper is organized as
follows: in \sect{stars} the stellar evolution codes and CCSN
models are described and in \sect{nucleo} we present the
post-processing calculations and the stellar yields of \setone.
Finally, in \sect{sec:concl}, we summarize the main conclusions of
this work and discuss future prospects.  Details regarding the physics
assumption and published data can be found in Appendices
\ref{sub_ap:physics_package} and \ref{database}.

\section{Stellar Evolution Calculations}
\label{stars}
The stellar evolution models for \setone\ were calculated with two
stellar evolution codes, \MESA\ and GENEC.  \MESA\ \citep[described in
detail in][]{paxton:11}, revision 3372, was used for low- and
intermediate-mass stars while GENEC 
\citep[][]{2008ApSS.316...43E,bennett:12,pignatari:13} 
was used for massive stars.  
GENEC is a well established research 
and production code for simulating the evolution of stars (massive stars in particular), 
but is not designed to simulate in detail 
the complex thermal pulse (TP) evolution and nucleosynthesis during the AGB phase.
On the other hand, 
%when \setone\ features 
%were designed 
\MESA\ calculations have been shown to produce results 
that are quantitatively consistent with established stellar evolution codes that are designed specifically to simulate the evolution of AGB stars
%to simulate the evolution of specifically low- and intermediate-mass stars including the TP phase.
\citep[e.g., EVOL,][]{herwig:04a,paxton:11}.
%while when \setone\ features 
%were designed \MESA\ massive star evolution 
%and nucleosynthesis was not fully 
%tested until advanced Si-burning stages.
Models of non-rotating massive stars calculated using the \MESA\ code do
provide results that are overall consistent with other stellar evolution codes 
including GENEC \citep{paxton:11,paxton:13}.
A detailed analysis comparing different stellar codes is provided by \cite{sukhbold:14} and \cite{jones:15};
\cite{jones:15} also explored the impact of those differences on the nucleosynthesis until the end of central He burning.
\setone\ includes models at two metallicities: $Z = 0.02$ (\setopt) and $Z
= 0.01$ (\setopo). \setopt\ includes models with initial masses, $M = 
1.65, 2, 3, 4, 5, 15, 20, 25, 32, 60 \msun$ and \setopo\ includes models
with initial masses, $M = 1.65, 2, 3, 4, 5, 15, 20, 25 \msun$.  
In particular, the $M = 1.65 \msun$ stars are low-mass stars, the $M = 2, 3, 4~\mathrm{and}~5 \msun$ stars are intermediate-mass stars, and the $M = 15, 20, 25, 32, 60 \msun$ stars are massive stars \citep[][]{herwig:05}.
The main
input physics used in the models is described below. Note that the
models do not include the effects of rotation and magnetic fields.

\subsection{Input Physics}\label{st_ev_co}
The massive star models computed using GENEC were calculated with the same input physics
as the \MESA\ low- and intermediate-mass models wherever possible. The main
differences in the input physics between the two codes are concerned with the treatment of convective boundary mixing and the prescriptions for mass loss; the differences are described in the corresponding sections below.
Improvements in input physics such as updated solar composition from \cite{asplund:09}, low-temperature opacities from \cite{marigo:09} and rotation \citep[][]{ekstroem:12}and magnetic fields \citep[e.g.,][]{heger:05} were not included in these calculations for two main reasons. The first is to be able to compare to past results \citep[e.g.,][]{schaller:92,woosley:95}. The second is to provide a basic set of yield that will provide a standard of comparison for future grid of yields including these improvements in input physics.

\subsubsection{Initial Composition and Opacities}
%The initial elemental abundances are scaled from the solar distribution given by \citet{grevesse:93}, 
In this work the initial element abundances are scaled
to $Z = 0.01$ and $Z = 0.02$ from \citet{grevesse:93} 
and the isotopic percentage for each element is given by \citet{lodders:03}.
The initial composition corresponds
directly to the OPAL Type 2 opacity tables that were used in both \MESA\ and GENEC for the present work \citep{1996ApJ...456..902R}. For low temperatures outside of the OPAL domain, the opacities from \citet{2005ApJ...623..585F} are used.
%The initial abundances for the \setopt\ (Z=0.02) and \setopo\ (Z=0.01) models
% are displayed in Table \ref{tab:iniabund}.
% MP: here was located table with initial abundances.

\subsubsection{Nuclear Reaction Network and Rates}
In \MESA, the \emph{agb.net} nuclear reaction network was used, which includes the p-p chains, the CNO cycles, the triple-$\alpha$ reaction and the following $\alpha$-capture reactions:
\isotope[12]{C}($\alpha$, $\gamma$)\isotope[16]{O}, 
\isotope[14]{N}($\alpha$,$\gamma$)\isotope[18]{F}($e^+,\nu$)\isotope[18]{O}, 
\isotope[18]{O}($\alpha$,$\gamma$)\isotope[22]{Ne}, 
\isotope[13]{C}($\alpha$, n)\isotope[16]{O}, and
\isotope[19]{F}($\alpha$, p)\isotope[22]{Ne}. 
In particular, we assume that the He-shell flash convection is dominated by the triple-$\alpha$ reaction, and we did not consider the \isotope[22]{Ne}$+\alpha$ reactions.

GENEC also includes the main reactions for the hydrogen and helium-burning phases and in addition accounts for the fusion of carbon, the fusion of oxygen and an $\alpha$-chain network 
for the neon-, oxygen- and silicon-burning phases. The following isotopes are 
included in the network explicitly: \isotope[1]{H}, \isotope[3]{He}, \isotope[4]{He}, 
\isotope[12]{C}, \isotope[13]{C}, \isotope[14]{N}, \isotope[15]{N}, \isotope[16]{O}, 
\isotope[17]{O}, \isotope[18]{O}, \isotope[20]{Ne}, \isotope[22]{Ne}, 
\isotope[24]{Mg}, \isotope[25]{Mg}, \isotope[26]{Mg}, \isotope[28]{Si}, 
\isotope[32]{S}, \isotope[36]{Ar}, \isotope[40]{Ca}, \isotope[44]{Ti}, 
\isotope[48]{Cr}, \isotope[52]{Fe}, \isotope[56]{Ni}. 
Note that additional isotopes are included implicitly to follow the p-p chains, 
CNO tri-cycles and the combined ($\alpha$,p)-(p,$\gamma$) 
reactions in the advanced stages.

%In both codes, most reaction rates were taken from the NACRE
%\citep{angulo:99} reaction rate compilation for the experimental rates
%and from their
%website\footnote{http://pntpm3.ulb.ac.be/Nacre/nacre.htm} for the
%theoretical rates.
In both codes, most of the reaction rates were provided by the NACRE compilation 
\citep[][]{angulo:99}.
%; the experimental rates were taken from \citet{angulo:99} and the theoretical rates from the NACRE website\footnote{http://pntpm3.ulb.ac.be/Nacre/nacre.htm}. [SJ: this sentence was really awkward - does it still have the same meaning as you intended?]
There are, however, a few exceptions that should be clarified. In GENEC, the
rate of \citet{2003PhRvC..67f5804M} was used for \isotope[14]{N}(p
,$\gamma$)\isotope[15]{O} below $0.1$ GK and the lower limit NACRE
rate was used for temperatures above $0.1$ GK. This combined rate is
very similar to the more recent LUNA rate \citep{imbriani:04} at
relevant temperatures, which was used in \MESA. In both codes, the
\citet{fynbo:05} rate was used for the triple-$\alpha$ reaction and
the \citet{kunz:02} rate was used for
\isotope[12]{C}($\alpha,\gamma$)\isotope[16]{O}.  In GENEC, the
\isotope[22]{Ne}($\alpha$, n)\isotope[25]{Mg} rate was taken from
\citet{jaeger:01} and used for $T \leq 1$ GK; the NACRE rate was used
for higher temperatures.  The \isotope[22]{Ne}($\alpha$,
n)\isotope[25]{Mg} rate competes with \isotope[22]{Ne}($\alpha,
\gamma$)\isotope[26]{Mg}, where the NACRE rate was used 
%[SJ: this small sentence is really awkward, too - not sure what to do about that, any ideas?].  
The key reaction rates responsible for the energy generation are the
same for the high- (GENEC) and intermediate- and low-mass (\MESA)
stellar models.

\subsubsection{Mass Loss}
For the low- and intermediate-mass stellar models, we adopted in \MESA\ the Reimers 
mass loss formula \citep{reimers:75}
with $\eta_\mem{R}=0.5$ for the RGB phase. For the AGB phase we used
the mass loss formula from \citet{bloecker:95a} with $\eta_\mem{B}=0.01$
for the O-rich phase. During the TP phase carbon is recurrently mixed into the stellar envelope from the helium inter-shell by the third dredge-up. Once the surface C/O ratio exceeds about 1.15 we increased the
mass loss parameters to $\eta_\mem{B}=0.04$ for the 1.65 and $2\msun$ tracks
and to $\eta_\mem{B}=0.08$ for the $3\msun$ tracks. This choice is
motivated by observational constraints on the maximum
level of C enhancement seen in C-rich stars and planetary nebulae 
\citep{herwig:04c} as well as by %recent 
hydrodynamics simulations 
investigating mass loss rates in C-rich giants \citep[e.g.,][]{mattsson:10,mattsson:11}. 
%and preliminary evaluations of how they would influence AGB stellar evolution (Mattsson \etal, private communication).
In order to explore the influence of the Mattson mass loss rate for C-stars we have calculated some preliminary stellar evolution tracks, and the mass loss parameters were chosen to reflect findings from these tests (Mattsson et al., in prep).
The choice to enhance the mass loss rate is also motivated by %stellar observations, including
considering counts of C- and O-rich stars in
the Magellanic Clouds \citep[e.g.][]{Marigo:2007cw}, which together indicate
that the C-rich phase cannot last for more than at most a dozen thermal
pulses. While the Magellanic Clouds are more metal poor than the AGB models considered here, theoretical hydrodynamics calculations by \cite{mattsson:08} and observations of AGB stars in the galactic halo \citep[e.g.,][]{lagadec:12} and in metal poor galaxies \citep[e.g.,][]{sloan:09} including the Magellanic Clouds \citep[][]{groenewegen:09} indicate that mass-loss rates in the final C-rich AGB phase should not significantly change with metallicity. 
We refer to \cite{nanni:13}, \cite{karakas:14} and \cite{straniero:14} for more details.

The $5\msun$ tracks are dominated by hot-bottom burning and do
not become C-rich.
% for most of their TP-AGB evolution. 
We adopt $\eta_\mem{B}=0.05$ from the beginning of the AGB phase for the models
tracks with this mass. 
%[SJ: You say here 'also motivated by stellar observations', but earlier in this paragraph you already put a motivation from stellar observations - we should probably combine them, otherwise it looks like you just added the last point later.]

For massive star models, several mass loss rates are used depending on the effective
temperature \teff~and the evolutionary stage of the star in GENEC.
For main sequence massive stars where $\log \teff > 3.9$, mass
loss rates are taken from \citet{2001A&A...369..574V}.  Otherwise the
rates are taken from \citet{1988A&AS...72..259D}. For lower
temperatures ($\log \teff < 3.7$) however, a scaling law of the form
\begin{equation}
\dot{M} = - 1.479 \times 10^{-14} \times \left(\frac{L}{\lsun}\right)^{1.7}
\end{equation}
is used, where $\dot{M}$ is the mass loss rate in $\msun$ yr$^{-1}$, 
$L$ is the stellar luminosity. 
%[SJ: This is confusing - so de Jager (1988) is only used for log Teff between 3.7 and 3.9?! I don't think that's right... I'll check. I also think we should put where that scaling law comes from, but I don't know the answer...MP: I understand your concern SJ, for now I do not change it, but better check with Raphael.]
During the Wolf-Rayet (W-R) phase,
mass loss rates by \citet{2000A&A...360..227N} are used.

\subsubsection{Convective boundary mixing}
\label{sec:CBM}
The Schwarzschild criterion was used in all models (\MESA\ \& GENEC) for the placement of the convective boundary.
The \MESA\ code allows for the exponential
diffusive convective boundary mixing (CBM) or overshooting introduced by 
\citet{herwig:97} based on
hydrodynamic simulations by \citet{freytag:96}. More recent
hydrodynamic simulations of He-shell flash convection zone also
show convection-induced mixing at convective boundaries
\citep{herwig:07a,herwig:06a}. The nature of the instabilities
observed in the deep interior, however, is different then the buoyancy-driven
overshooting situation found in shallow surface convection
studies by \citet{freytag:96}. We therefore refer to our
exponentially decaying mixing model at the convective boundary rather as CBM which may represent a variety of physical processes causing mixing across the Schwarzschild boundary.
Treating the convective boundary mixing as a diffusive processes may be justified in the case of the formation of \isotope[13]{C} pocket if the physics processes of internal gravity waves \citep{denissenkov:03t} applies. If the mixing process is more hydrodynamic in nature an advection scheme may be more appropriate.

In \MESA\ models, a CBM efficiency of $f_\mem{ov}=0.014$ was used at all boundaries, except during 
the dredge-up, when  $f_\mem{DUP}=0.126$ was used to generate a
 \cdr-pocket for the \spr\ according to \citet{herwig:02a}, and
$f_\mem{PDCZ}=0.008$ (where PDCZ stands for \pdcz) 
was used at the bottom of the He-shell flash convection zone. Because of the latter choice our models reproduce the observational constraints, especially the O mass fraction of $\approx 0.1 -- 0.15$, from H-deficient post-AGB stars  \citep{werner:06}. This approach was followed as well by \cite{miller-bertolami:06}. AGB simulations without CBM at the bottom of the PDCZ have so far not been able to reproduce the abundance of H-deficient post-AGB stars which show the exposed intershell of the former AGB star. Detailed AGB models adopting this CBM treatment have been presented by \citet{weiss:09} and their models show generally good agreement with our models (\sect{sub:st_ev_tr}). \citet{Kamath:2012jk} find that it is possible to explain the observed C/O and C isotopic ratios for AGB stars when adopting intershell abundances of models with CBM at the bottom of the PDCZ, for at least one globular cluster of the Magellanic Cloud. CBM at the bottom of the He-shell flash convection zone is supported by hydrodynamic simulations \citep[][]{herwig:07a}.

The core overshooting value for the $1.65\msun$ case is $1/2$
of the value appropriate for higher masses, as motivated by the
investigation of \citet{vandenberg:06} using star cluster data on
low-mass stars.

In GENEC, convective mixing is treated as instantaneous from hydrogen up to neon
burning. From oxygen burning onwards (since the
evolutionary timescale is becoming too small to justify the
instantaneous mixing assumption), convective mixing in GENEC is treated as a
diffusive process as is the case at all times in the \MESA\ calculations. 
In GENEC overshooting is only included for hydrogen- and
helium-burning cores, where an overshooting parameter of $\alpha_\mem{ov} = 0.2
H_P$ is used as in previous non-rotating grids of models \citep{schaller:92}.

A recent comparison between \MESA\  and GENEC can be found in \citet{jones:15} where $f_\mem{ov}=0.022$ was used in MESA to match the $\alpha_\mem{ov} = 0.2 H_P$ in GENEC. For this study we initially planned to use the EVOL code \citep{herwig:00} for the low-mass models. We compared convective cores with overshooting in $9\msun$ stellar models from the GENEC code and the EVOL code to ensure that convective core sizes are matching at the transition mass. For the EVOL code $f_\mem{ov}=0.016$ matched approximately the GENEC model with $\alpha_\mem{ov} = 0.2 H_P$. For stars around $2\msun$ it was determined by \citet{paxton:11} that  $f_\mem{ov}=0.014$ matches observational constraints of the main-sequence width in \MESA\ models, and we have adopted this value for main-sequence core convection in our MESA low- and intermediate mass models. CBM and its dependence on initial mass is still uncertain but there is support for an overshooting efficiency that broadly increases with initial mass \citep{deupree:00}. The overshooting efficiencies adopted here for AGB and massive stars are well within the range of values used in the literature, see e.\,g. \citet{martins:13}.

\subsubsection{Additional \MESA\ Code Information}
The low- and intermediate-mass models (1.65, 2, 3, 4 and $5\msun$) have
been calculated with the \MESA\ code (rev.\,3372), for which a comprehensive code
description and comparison (including GENEC for massive stars)
is provided by \citet{paxton:11}. Concerning stellar evolution before and during the AGB phase, results from \MESA\ have been compared in detail to results obtained with the EVOL stellar evolution code \citep[e.g.][]{bloecker:95a,herwig:99a,herwig:04a}. 
In particular, the $2\msun$, $Z = 0.01$ \MESA\ stellar model has been compared to the corresponding track of \citet{herwig:04b} from the pre-main sequence to the tip of the AGB 
by \citet{paxton:11}. The two stellar models share a similar evolution in the HR diagram, and key properties such as main-sequence lifetime and age at first thermal pulse, H-free core mass at the end of He-core burning and core mass at first thermal pulse differ by less than 5\%.
During the AGB, similar occurrence and efficiency of third dredge-up, interpulse periods and evolution of C/O ratio in the AGB envelope as well as subsequent C-star
formation are obtained \citep[][]{paxton:11}.
%Concerning AGB stellar evolution, results from \MESA\ have been compared in detail to results obtained with the EVOL stellar evolution code \citep[e.g.][]{bloecker:95a,herwig:99a,herwig:04a}. 
%The $2\msun$, $Z = 0.01$ \MESA\ AGB stellar model has in fact  been compared in detail to the corresponding track of \citet{herwig:04b}, which shares extremely similar features in terms of occurrence and efficiency of third dredge-up as well as subsequent C-star formation. 
%Intermediate-mass \MESA\ models of correspondingly lower metallicity have been confirmed to be very similar to the EVOL stellar evolution tracks published by \citet{herwig:04a}.

The following settings were used in \MESA:
\begin{itemize}
\item structure, nuclear burning and time-dependent mixing operators were
always solved together using a joint operator method;
\item in addition to the default \MESA\ mesh refinement, enhanced resolution
was applied in regions with gradients in H, \hevi,\,\cdr\ and \nvi\ in order to resolve the \cdr\ pocket during the
entire interpulse time. This is needed to accurately follow
$s$-process nucleosynthesis;
\item the mixing-length parameter used is $1.73\hp$, as calibrated
for a solar model;
\item additional time step controls are used to
allow for sufficient resolution of the He-shell flashes as well as the
evolution of the thin H-burning shell during the interpulse evolution;
\item OPAL Type 2 opacity tables \citep[][]{1996ApJ...456..902R},
 and 
\item the atmosphere option \emph{simple\_photosphere}.
\end{itemize}

\subsection{Stellar evolution tracks}
\label{sub:st_ev_tr}

\begin{figure}
\includegraphics[width=\textwidth]{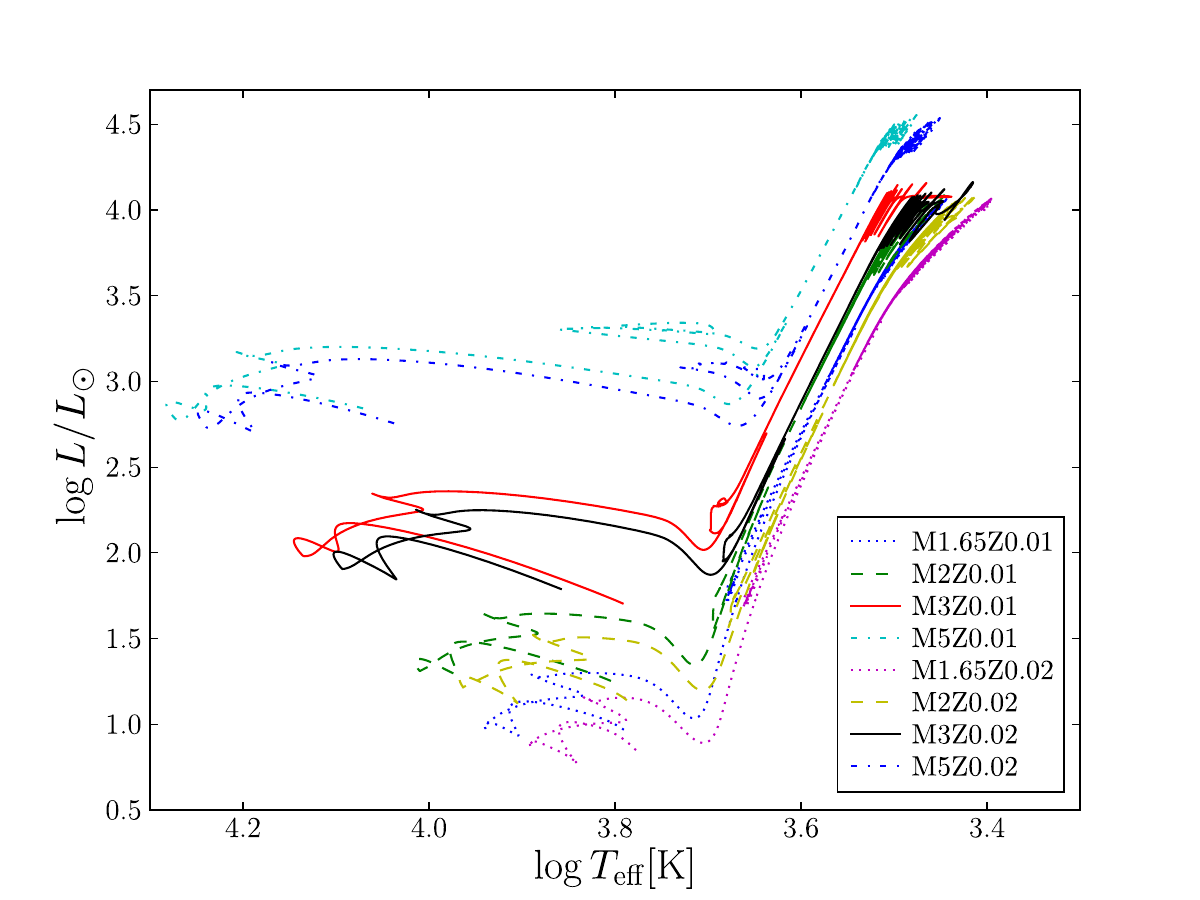}
% figure created in /astro/fherwig/rpod2/M/Set_1.2_simple_at
\caption{H-R diagram for low- and intermediate-mass models. Labels give
  the initial stellar mass followed by 'S1' for \setopo\ models ($Z=0.01$)
  and 'S2' indicates accordingly \setopt\ models ($Z=0.02$).
  Toward the end of the sequence the tracks show
  wide loops indicating an instability towards the end of the
  evolution that has been omitted from the plot for clarity (see text
  for details).}
\label{fig:low-mass-HRD}
\end{figure}

\begin{figure}
\includegraphics[width=\textwidth]{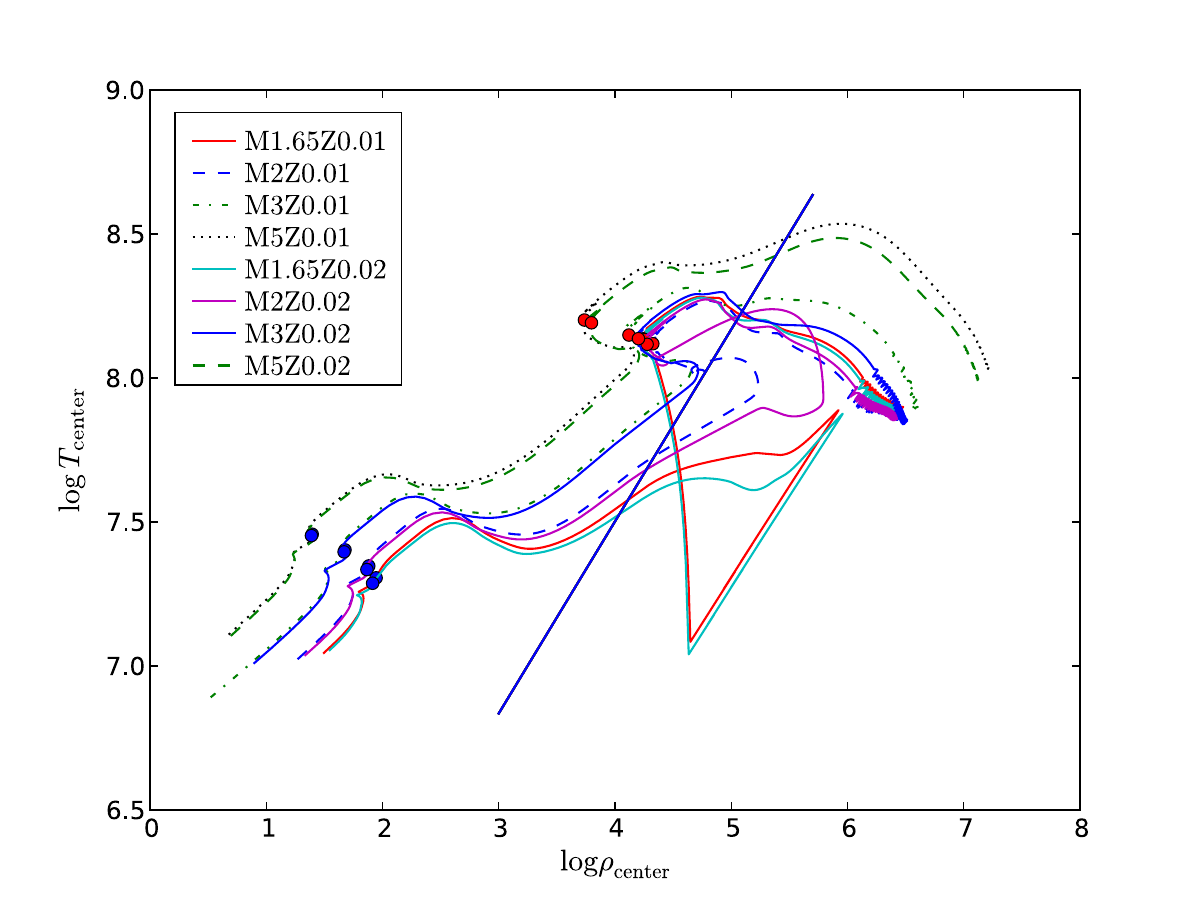}
% figure created with SEE/work/vis_lowmass.py
\caption{Central temperature, $T_{\rm c}$, as a function of central density,
  $\rho_{\rm c}$, for low mass and intermediate mass models from \setone. The 
  labels indicate initial mass and metallicity as in \fig{fig:low-mass-HRD}.
  H- and He-ignition points for the core burning stages are indicated by 
  blue and red colored points, which are determined at the point when the 
  principal fuel is depleted by 1\% from its maximum value
  %[RH: Tcenter AND RHOcenter BETWEEN BRACKETS].
}
\label{fig:low-mass-tcrhoc}
\end{figure}

\begin{figure}
\includegraphics[width=0.7\textwidth]{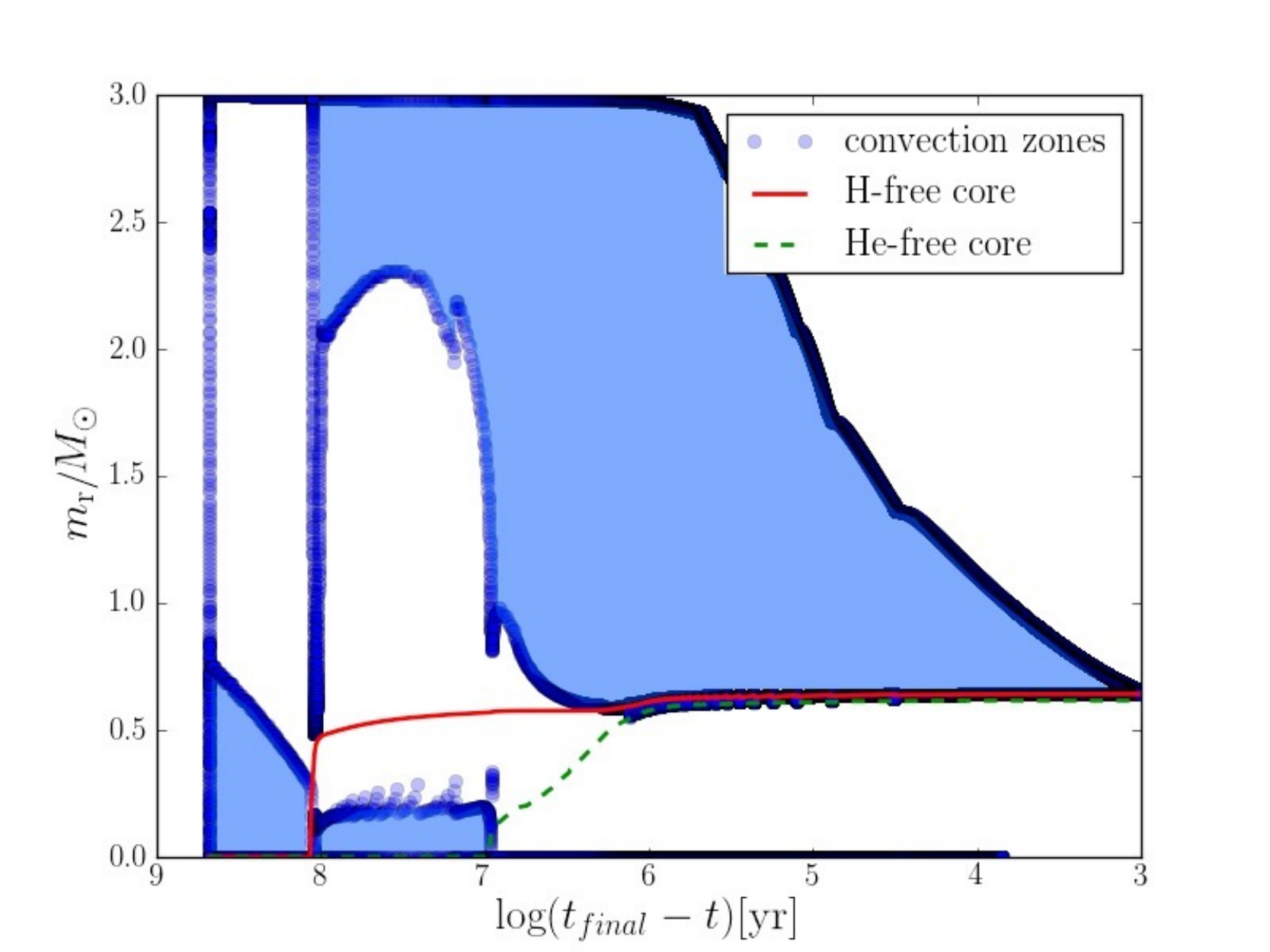}
\includegraphics[width=0.7\textwidth]{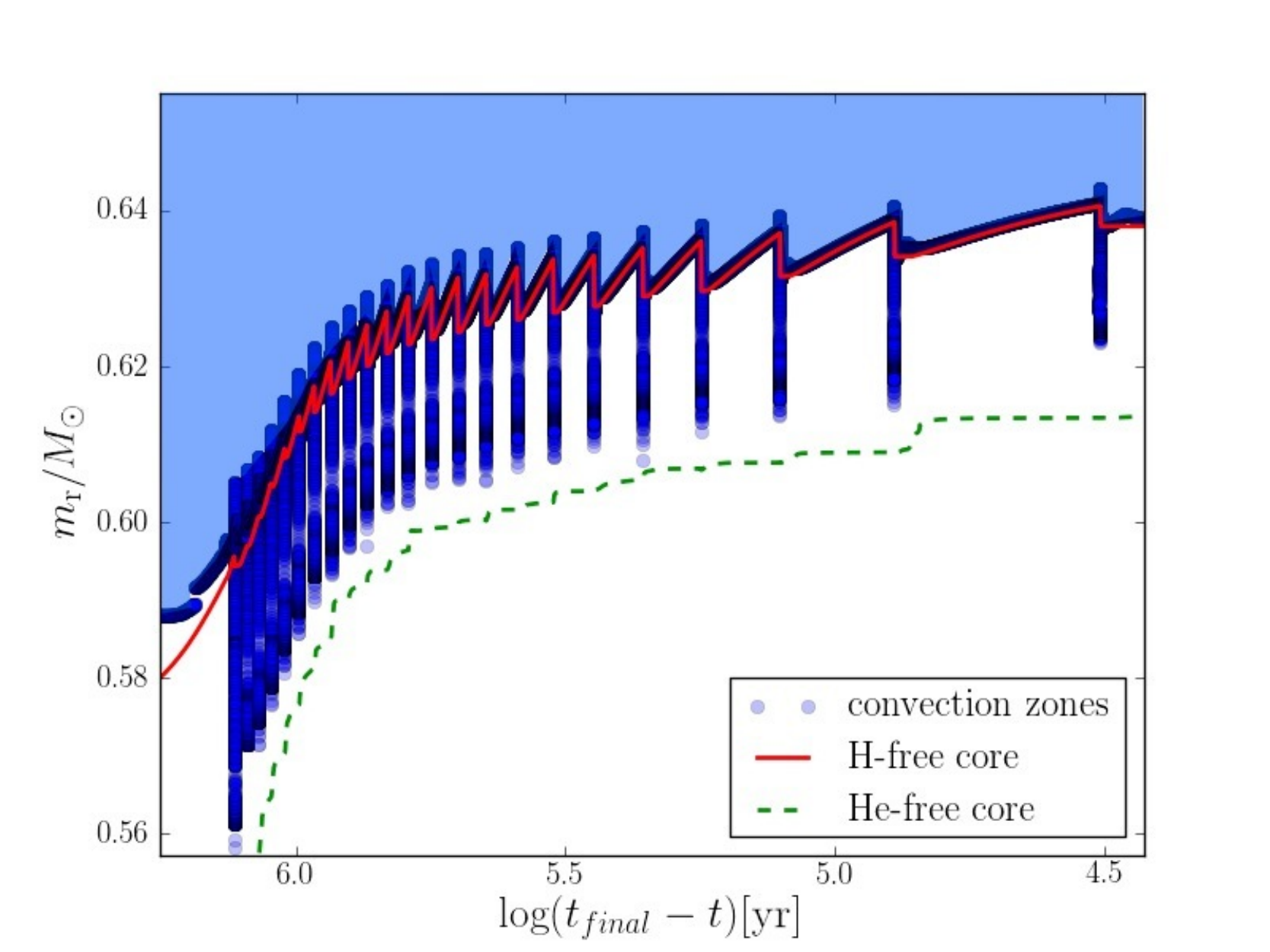}
% figure created in /astro/fherwig/rpod2/M/Set_1.2_simple_atm and
% modified for /Users/fherwig/ARTUR/L/PAPER/SpringerChapter/manuscript/Figures
\caption{Top panel: Kippenhahn diagram of $\mzams=3\msun$ stellar
  evolution calculation with $Z=0.02$ from the pre-main sequence to
  the end of the TP-AGB evolution. The position of convection zones
  and mass coordinates of the H- and He-free cores are shown as a
  function of the logarithm of the time left until the end of the
  TP-AGB. Bottom panel: Zoom-in of top panel, showing the sequence of
  thermal pulses.}
\label{fig:kippenhahn_3Msun}
\end{figure}

\begin{figure}
\includegraphics[width=0.54\textwidth]{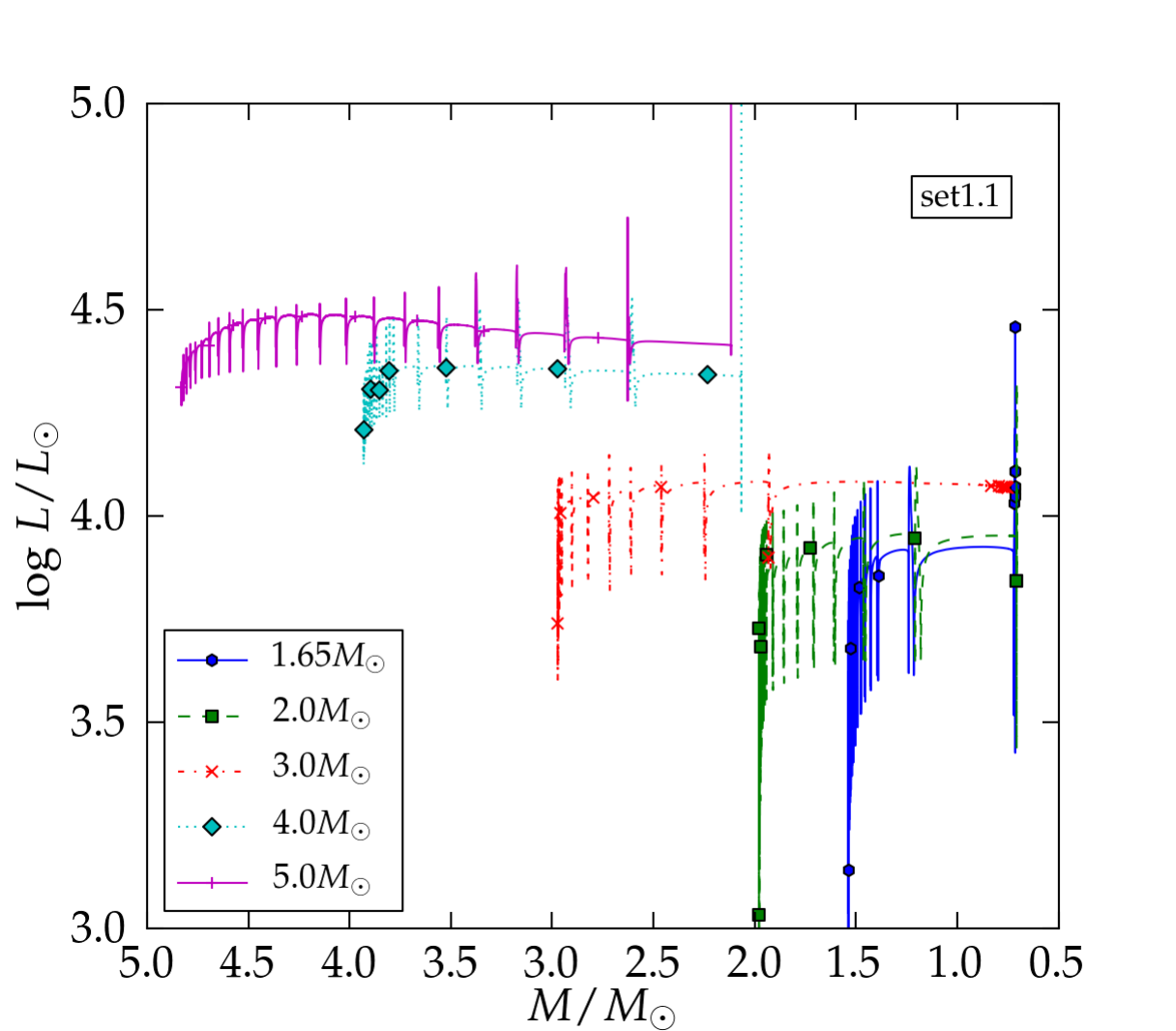}
\includegraphics[width=0.54\textwidth]{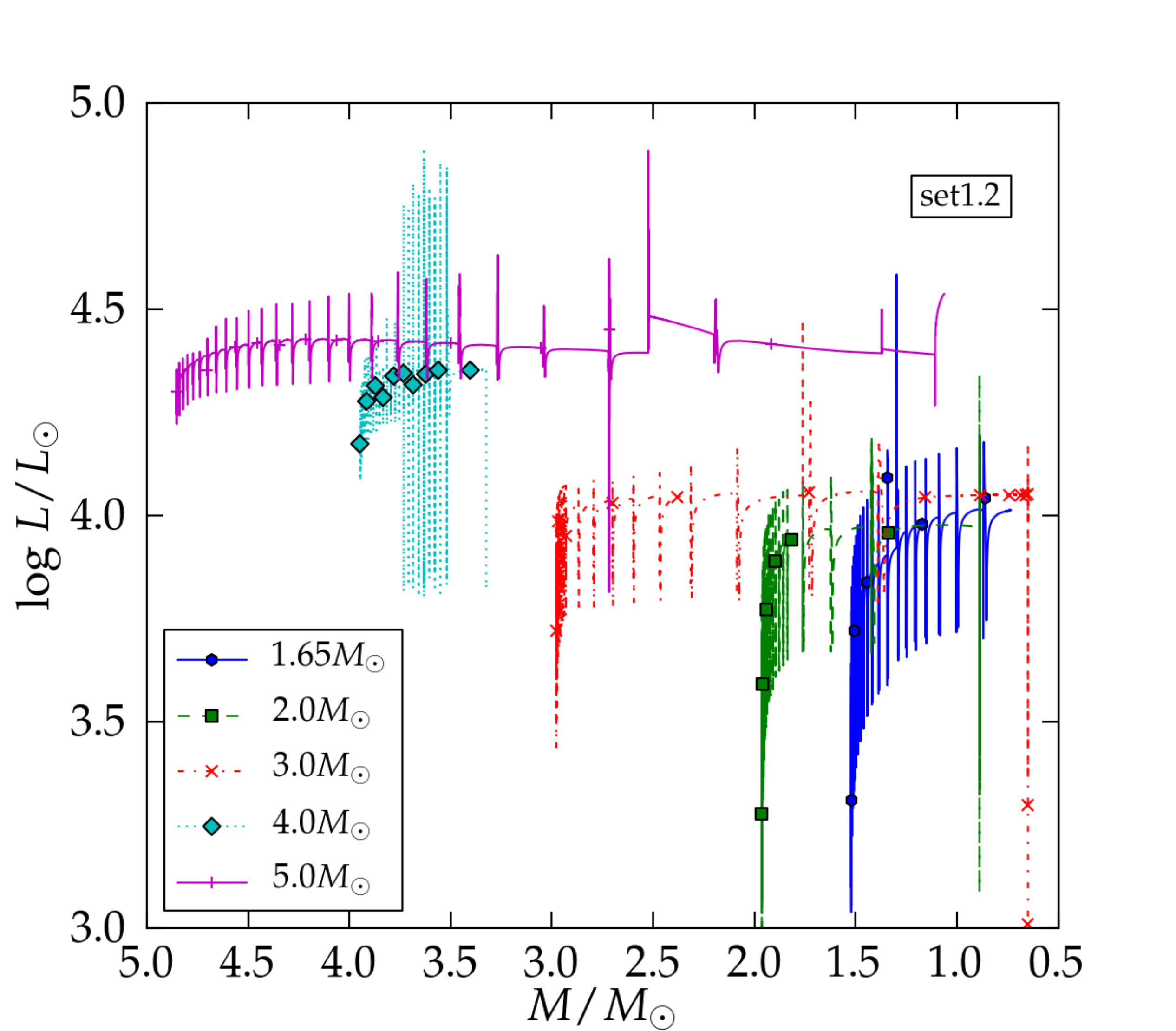}
% figure created in /nfs/rpod2/home/critter/set1extension/AGB_stars/model_properties/paper1_tex
\caption{AGB luminosities of \setopo\ and \setopt.}
\label{fig:agb_lum}
\end{figure}

\begin{figure}
\includegraphics[width=0.58\textwidth]{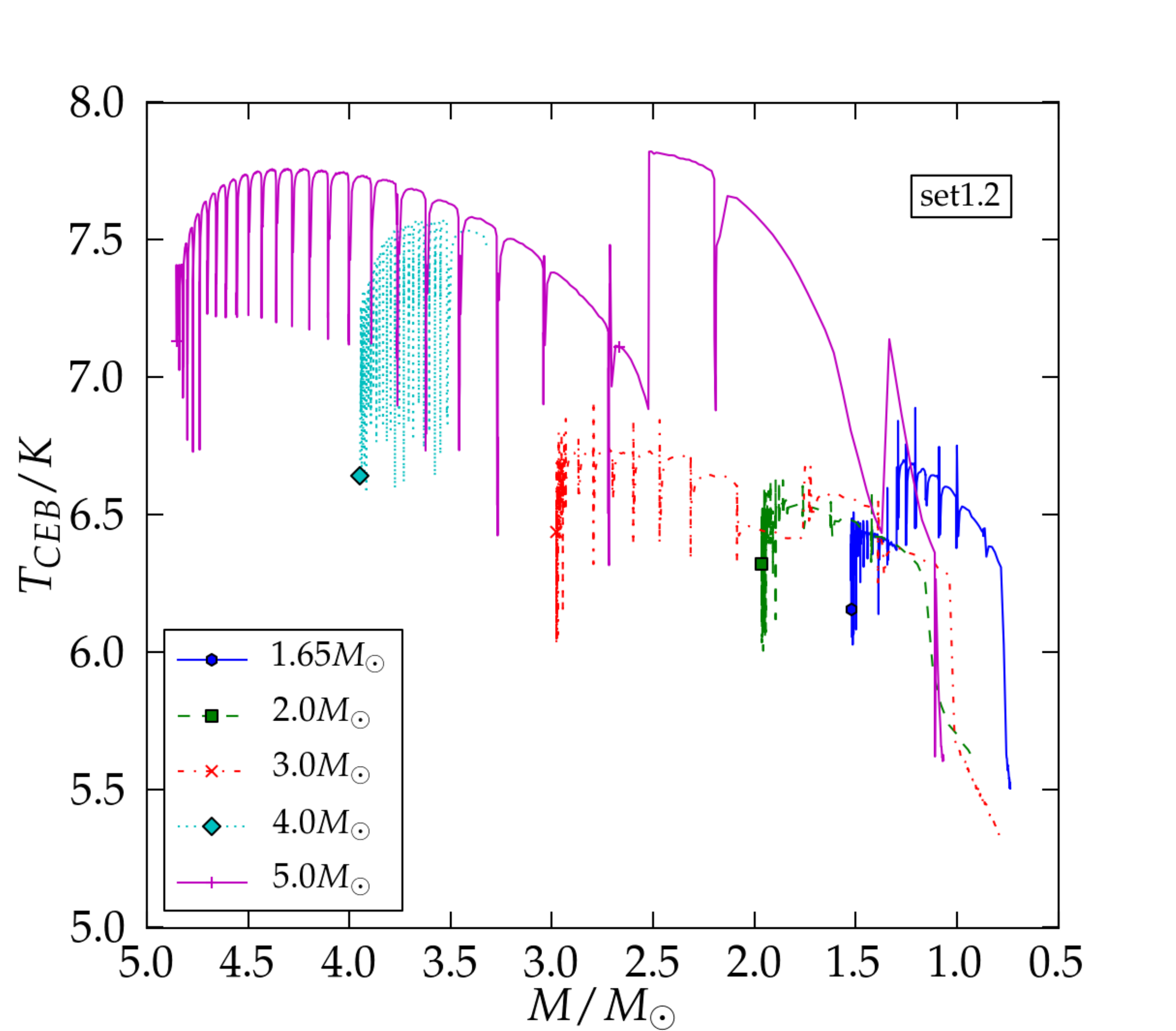}
\includegraphics[width=0.58\textwidth]{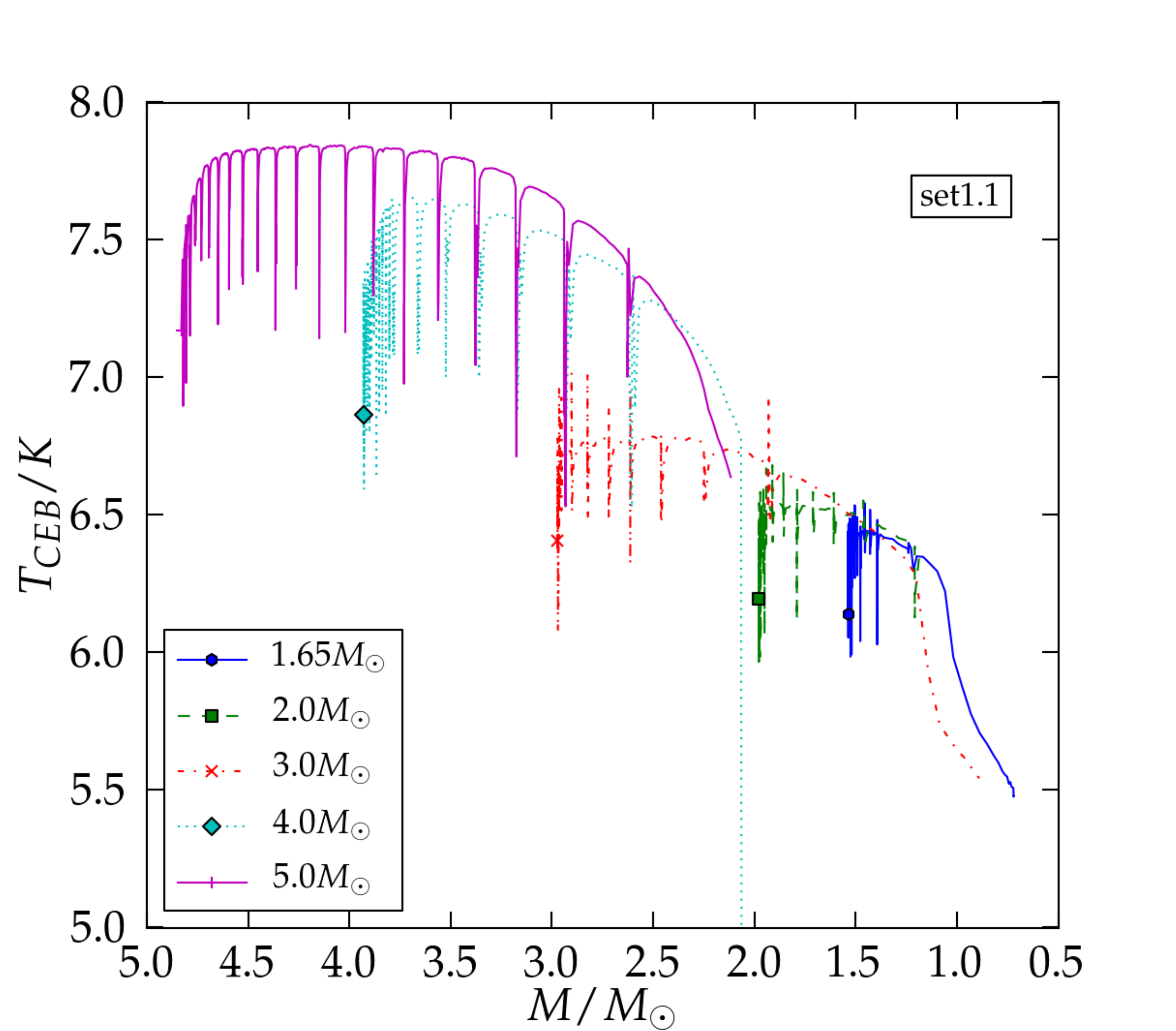}
% figure created in /nfs/rpod2/home/critter/set1extension/AGB_stars/model_properties/paper1_tex
\caption{Temperature at the bottom of the convective envelope for \setopt\ and \setopo.}
\label{fig:agb_TCEB}
\end{figure}

The H-R diagram for low mass and intermediate mass stellar models is shown in
\fig{fig:low-mass-HRD}, and the evolution of central temperature and
density in \fig{fig:low-mass-tcrhoc}.  In \fig{fig:kippenhahn_3Msun} we
also show, as an example, the Kippenhahn diagram for the $3\msun, Z=0.02$ model.
%evolution for $Z=0.02$. 
The final core masses and lifetimes calculated for all low mass and intermediate mass stellar models are shown in
\tab{tab:cores_agb_set1}. The main features during the AGB evolution are summarized in \tabs{agbprop_1p2}, \ref{agbprop_1p1}, \ref{agb_model1p2prop2} and \ref{agb_model1p1prop2}. The AGB surface luminosity and temperatures at the bottom of the convective envelope are given in \fig{fig:agb_lum} and \fig{fig:agb_TCEB}. The $3$ and $4\msun$ models with $Z=0.02$ have average luminosities of $11,000\lsun$ and $20,000\lsun$.
% when the thermal pulses are fully evolved. 
This is in good agreement with the results of \citet{herwig:98b} obtained with the EVOL code.

Our $2\msun$, $Z=0.02$ calculation compares well to that of \citet{weiss:09}, except the core mass at the first thermal pulse. It is $0.510\msun$ for our model and $0.478\msun$ ($0.518\msun$) for the $M_\mathrm{ini}=2\msun$ ($2.6\msun$) \citet{weiss:09} models. Their and our $2\msun$ simulations have $13$ and $12$ thermal pulses with 3DUP. The final C/O ratio is in our model $1.476$ and $1.204$ ($1.426$) in the \citet{weiss:09} $M_\mathrm{ini}=2\msun$ ($2.6\msun$) models. The average luminosity in our model is $\log L \approx 3.95$ while that of \citet{weiss:09} is a bit lower ($\log L \approx 3.80$) consistent with the lower core mass of their model.

Our $5\msun$ stellar model with  $Z=0.02$ has a
final core mass of $M = 0.8747\msun$. The highest temperature
obtained at the bottom of the AGB envelope is 6.56$\times$10$^7$ K.
%Concerning the massive AGB stars, our $5\msun$ models have as a final core
%mass of M = 0.8747 and 0.9098 $\msun$ at Z=0.02 and 0.01, respectively. The
%highest temperature obtained at the bottom of the AGB envelope are
%6.56$\times$10$^7$ K and 6.86$\times$10$^7$ K for solar and half-solar
%metallicity, indicating efficient HBB. 
For the same mass and metallicity, \cite{cristallo:15} obtained $M =
0.8462~\msun$ and about 8$\times$10$^6$~K, and \cite{karakas:12} $M =
0.8726~\msun$ and 5.74$\times$10$^7$~K.
%For the same mass and metallicity Z=0.0007, \cite{straniero:14} obtained M =
%0.947 $\msun$ and about 8$\times$10$^6$ K, and \cite{karakas:10} at Z=0.0001 M
%= 0.929 $\msun$ and 9.25$\times$10$^7$ K.  The lower metallicities in those
%models result in larger cores %and higher HBB temperatures compared to our
%models, but the fundamental processes of TP AGB evolution are similar at
%different Z, apart from an initial mass offset.
The total number of TPs is 25 with TDUP after each pulse except the
first one.  \cite{cristallo:15} and \cite{karakas:12} models experience 10 and
25 thermal pulses respectively, while our model has been followed for 25 thermal pulses when the total mass has decreased to $2.198\msun$. Our TP-AGB life time is $1.38 \times 10^5\mathrm{yr}$ while that of \cite{cristallo:15} is $1.04 \times 10^5\mathrm{yr}$. Our lifetime after 10 thermal pulses is $0.483 \times 10^5\mathrm{yr}$, about one half of the lifetime of the model of \cite{cristallo:15} after the same number of thermal pulses. This implies that their interpulse lifetime is about twice that of our model for these first 10 thermal pulses. The interpulse time at the last TP in our $5\msun$ model is $0.7\times10^4\mathrm{yr}$ while \cite{karakas:12} report $1.3\times 10^4\mathrm{yr}$. 
%The total number of TPs at Z=0.02 and 0.01 is 25 and 22 with TDUP after each
%pulse except the first one.  \cite{straniero:14} and \cite{karakas:10} models
%experience 35 and 56 thermal pulses respectively.  %The difference in number
%may be related to the start of the C-rich phase and the choice of its enhanced
%mass loss.
The total lifetime of our model of 1.17$\times$10$^8$ yrs is in
agreement with the lifetime of 1.19$\times$10$^8$ yrs and 1.06$\times$10$^8$
yrs, found by \cite{cristallo:15} and \cite{karakas:12}.
%The total lifetime of our models of 1.17$\times$10$^8$ yrs and
%1.08$\times$10$^8$ yrs for $Z=0.02$ and $Z=0.01$ is longer then the lifetime
%of 9$\times$10$^7$ yrs found in \cite{straniero:14}.%  and by
%\cite{karakas:10}.  The efficient DUP with $\lambda_{DUP} = 0.8$ and higher
%leads to a 
For the total mass dredged up we obtain 3.72$\times$10$^{-2}\msun$.
This value is about a factor of two lower than the 6.47$\times$10$^{-2}\msun$
obtained in \cite{karakas:12}, but much larger than the
4.06$\times$10$^{-3}\msun$ in \cite{cristallo:15}.
%The total mass dredged up 3.72$\times$10$^{-2}\msun$ at $Z=0.02$ and
%2.56$\times$10$^{-2}\msun$ at $Z=0.01$.  These values are comparable with the
%3.5$\times$10$^{-2}\msun$ in \cite{straniero:14}, and lower than the
%$0.122\msun$ obtained in \cite{karakas:10}.  %The reason lies in the higher
%number of TPs reached in those models.  %Our $\lambda_{DUP}$ is similar to the
%values reached in \cite{karakas:10} of up to 0.96.
The maximum temperature in the PDCZ is found to be 3.43$\times$10$^8$
K. The value is consistent with \cite{karakas:12} model which gives
3.44$\times$10$^8$ K, and is about 10\% larger than the 3.12$\times$10$^8$ K by
\cite{cristallo:15}. This difference might be due to their smaller core mass. Overall the three models agree with each other although significant difference between either pair of models can be identified.
%The maximum temperature in the PDCZ is found to be 3.43$\times$10$^8$ K and
%3.46$\times$10$^8$ K at Z=0.02 and 0.01 respectively. The values by
%\cite{straniero:14} of 3.71$\times$10$^8$ K and \cite{karakas:10} of
%3.6$\times$10$^8$ K might owe their higher value to the larger core mass.
%\textbf{[MP@FH,CR: sergio has a much smaller number of pulses, and as a
%consequence a smaller amount of material DUP. Is this because of the mass loss?
%Can you check this and add a sentence pointing to this difference?]}

\begin{figure}
\includegraphics[width=0.54\textwidth]{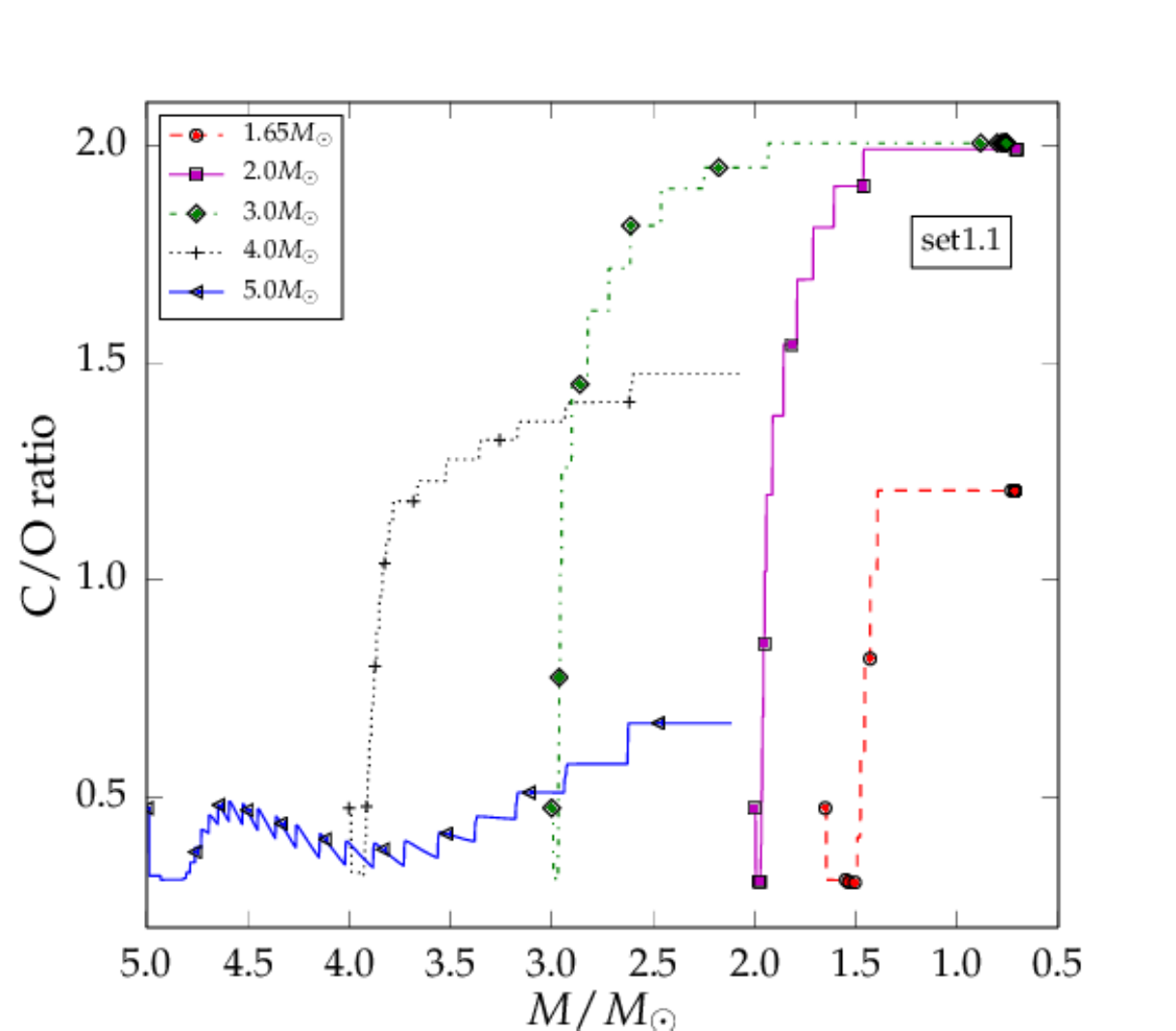}
\includegraphics[width=0.54\textwidth]{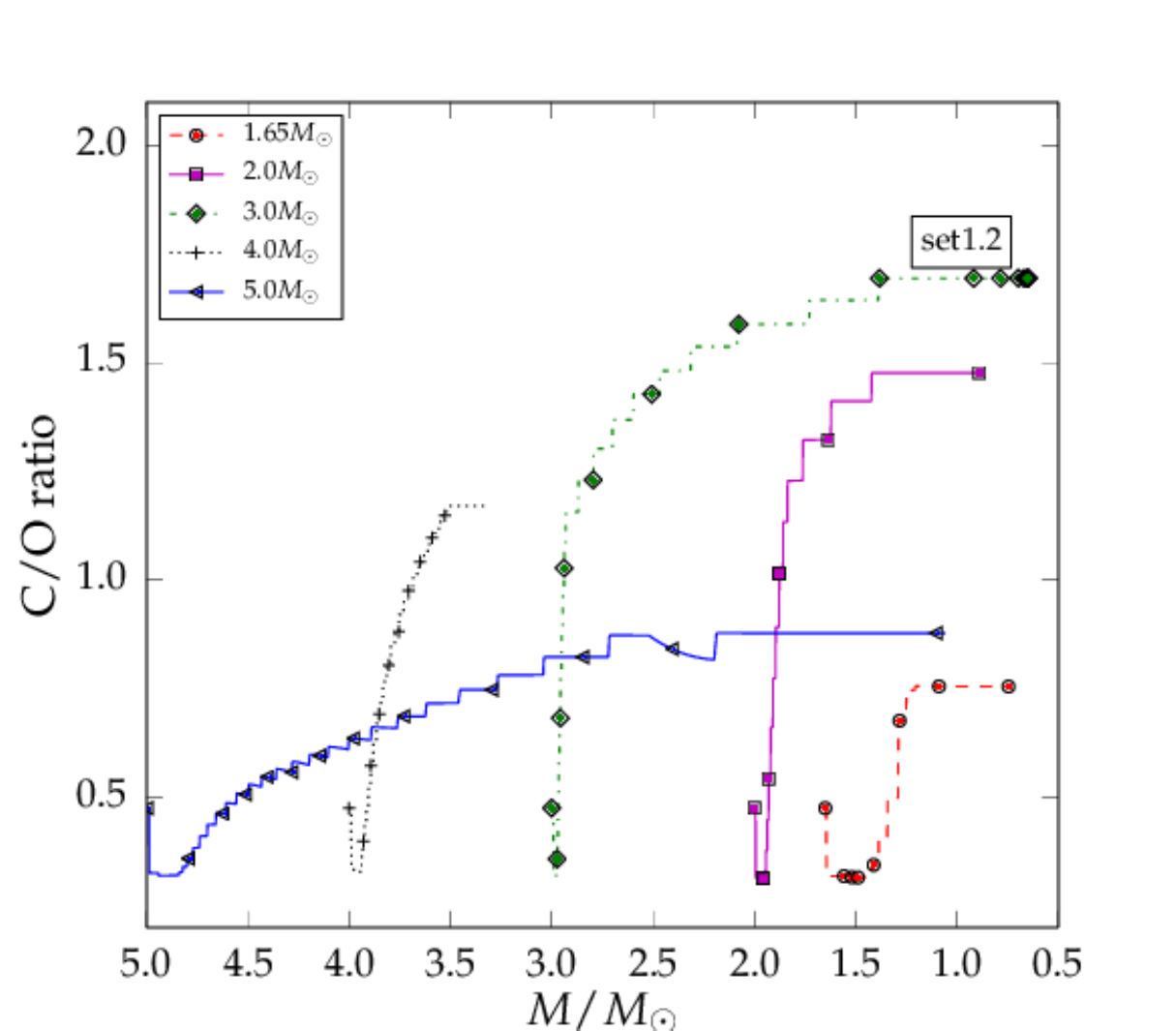}
% figure created in /astro/fherwig/rpod2/M/Set_1.2_simple_at
% figures created in /astro/critter/set1extension/AGB_stars/CO_ratio
\caption{Evolution of the C/O number ratio as a function of
  stellar mass. Since the stellar mass decreases with time the C/O
  evolution corresponds to a time sequence from left to right.
  Labels are the same as in Fig.\,\ref{fig:low-mass-HRD}.}
\label{fig:papI_CO-Mstar}
\end{figure}

Convective boundary mixing during the thermal
pulse phase is important for nucleosynthesis in two locations: the
bottom of the He-shell flash convection zone during the TP
and the bottom of the convective envelope during 
%and, in particular, toward the end of 
the third dredge-up phase. It also influences the
efficiency of the third dredge-up which is responsible for mixing C
and O from the intershell to the surface, which eventually is
responsible for the formation of C-stars (\fig{fig:papI_CO-Mstar}). 

\begin{figure}
\centering
\includegraphics[width=0.58\textwidth]{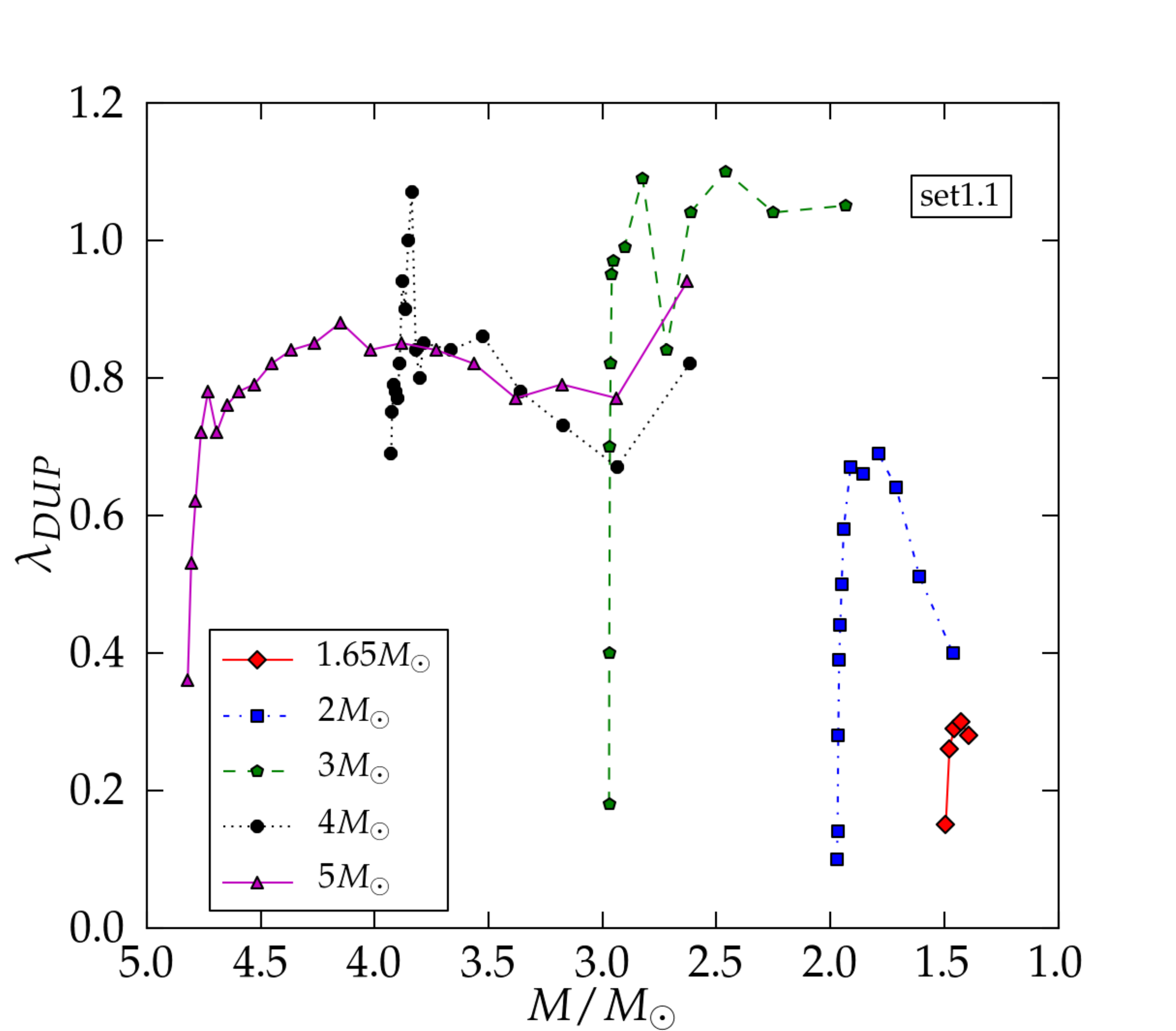} \\
\includegraphics[width=0.58\textwidth]{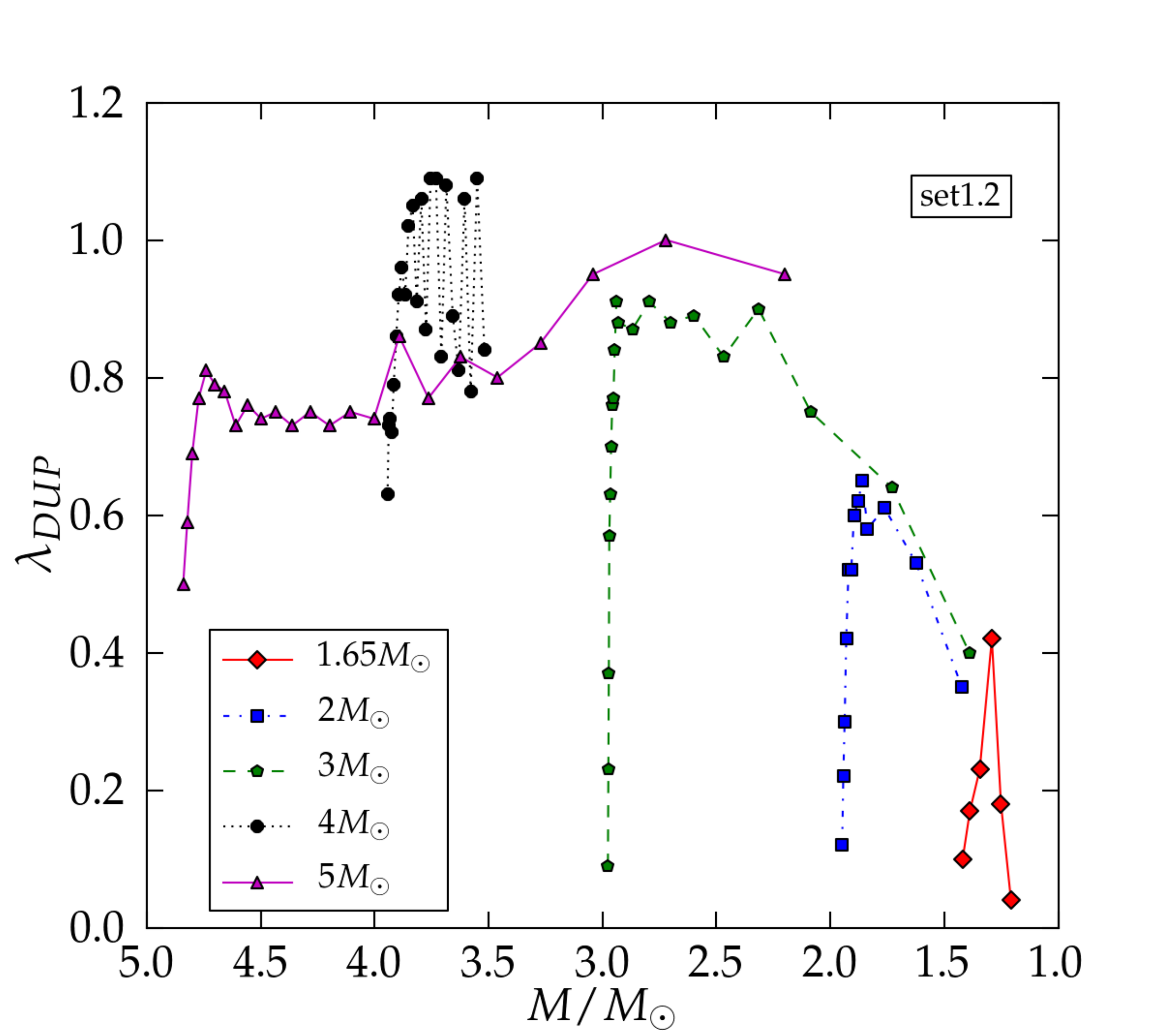}
\caption{Evolution of dredge-up parameter $\lambda_{DUP}$ at each TP,
starting with the second pulse. Top panel: \setopo; bottom panel: \setopt\ models.}
\label{fig:agb_lambda_starmass_winds}
\end{figure}

The efficiency of mixing processed material from the core to the envelope is expressed with the dredge-up parameter
\begin{equation}
\lambda = \frac{\Delta M_{DUP}}{\Delta M_{H}}
\end{equation}
where $\Delta M_{DUP}$ is the dredged up mass and $\Delta M_{H}$ is
the hydrogen free core growth during the last interpulse phase. The evolution of the
dredge-up parameter as calculated in our models is shown in 
\fig{fig:agb_lambda_starmass_winds}.  The parameter reflects the evolutionary
behavior of the core and envelope mass. In our models the dredge-up efficiency is decreasing with increasing Z, decreasing core mass and decreasing envelope mass as expected \citep{lattanzio:89a}. For the $3\msun$, $Z=0.02$ model $\lambda \approx 0.8 \dots 0.9$ which compares to  $\lambda \approx 0.6 \dots 0.7$ for models with the same initial parameters by \citep{karakas:14}. These differences are consistent with the different assumptions of convective boundary mixing in the two sets of calculations.
The evolution of $\lambda$ appears to be discontinuous for some of the AGB models when the maximum $\lambda$ values are reached in the evolution, with variations up to 30\% from one TDU to the next. This is due to the CBM feedback to the stellar behavior before and during the TDU, both at the bottom of the convective TP \citep[e.g.,][]{mowlavi:99,herwig:00} and at the bottom of the TDU itself \citep[e.g.,][]{herwig:04a}.
In particular, the $4\msun$ model at $Z=0.02$ shows a peculiar zig-zag pattern with variations of $\lambda$ on the order of 30\%. The same extreme pattern is not obtained in the other models. This is due to the CBM activation during the TDU, where some minor H burning remains and may switch the CBM at the base of the convective envelope between $f_\mem{DUP}=0.126$ and $f_\mem{ov}=0.014$. 
%[SJ: if this is true, then I suggest we say that this affects the 4 Mo model because it is on just on the edge of being a HBB/HDUP star (whichever one...)?]}

The most obvious consequence of the third dredge-up is the transformation of an initially O-rich star into a C star (\fig{fig:papI_CO-Mstar}). The C/O ratio in the intershell is due to primary He burning and therefore nearly the same for the two metallicities, and the dredge-up efficiency is similar as well. The larger C/O ratio reached in the  $Z=0.01$ \setopo\ is simply due to the fact that the initial amount of O in the envelope is only half compared to the $Z=0.02$ case. For the $5\msun$ case however the $Z=0.02$ case reaches a higher final C/O ratio because hot-bottom burning \citep[HBB, ][]{bloecker:91,lattanzio:92b} is activated already in the $Z=0.01$ case and this reduces the C/O ratio. Toward the end of the $5\msun$, $Z=0.01$ simulation dredge-up becomes again more important than HBB and the C/O ratio increases again \citep[][]{Frost:1998vn}.

\begin{figure}
\centering
\resizebox{11.3cm}{!}{\rotatebox{0}{\includegraphics{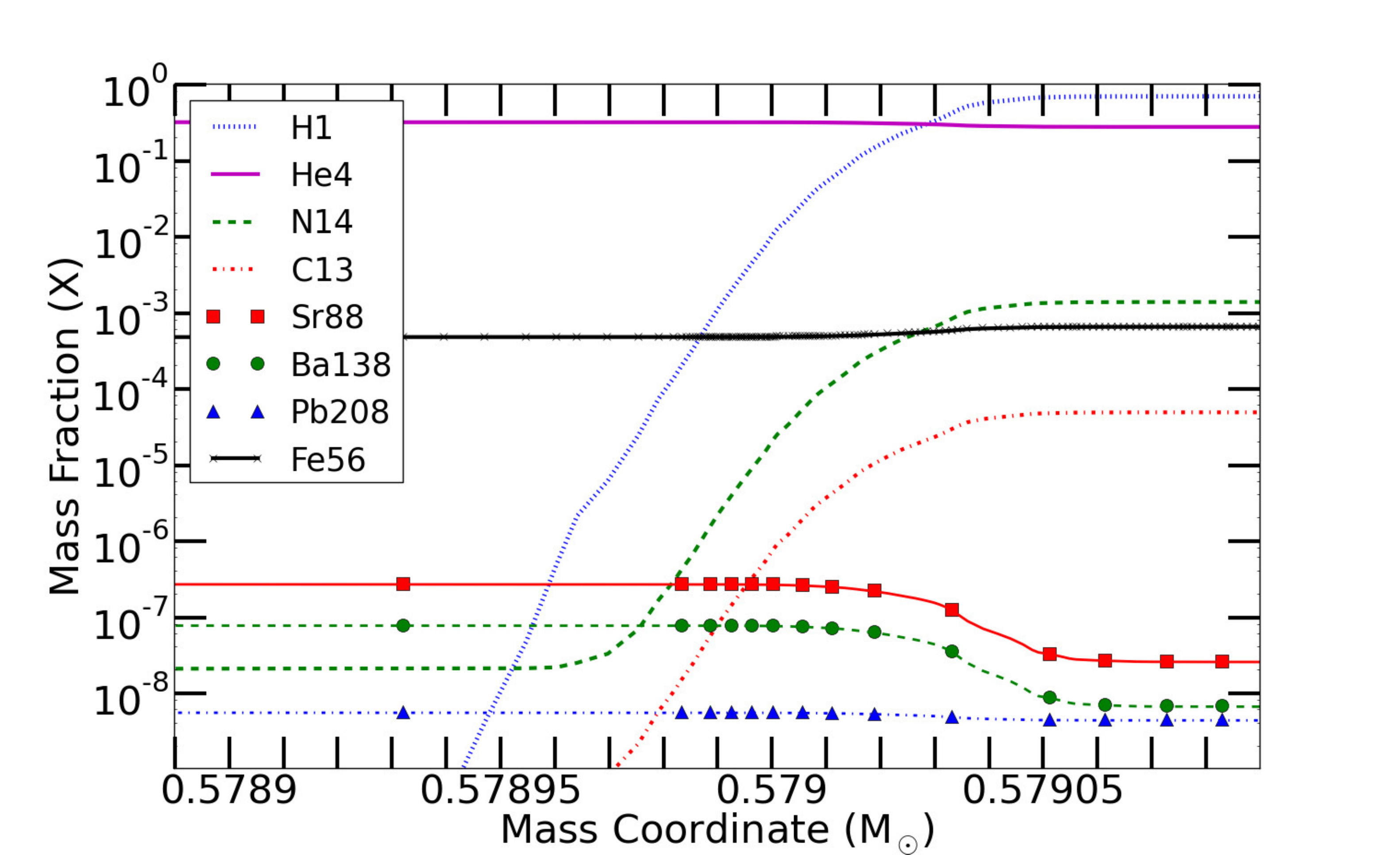}}}
\resizebox{11.3cm}{!}{\rotatebox{0}{\includegraphics{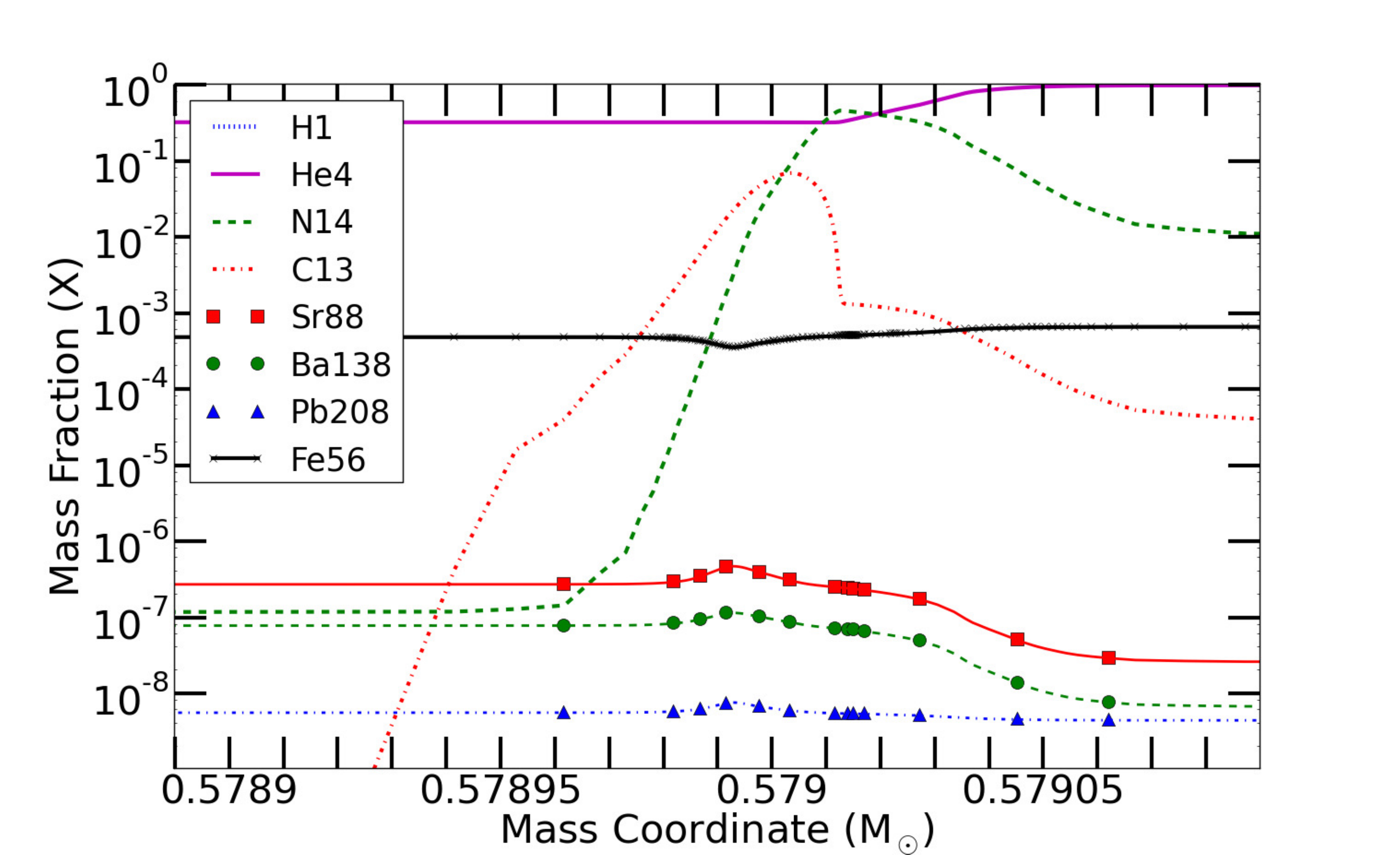}}}
\resizebox{11.3cm}{!}{\rotatebox{0}{\includegraphics{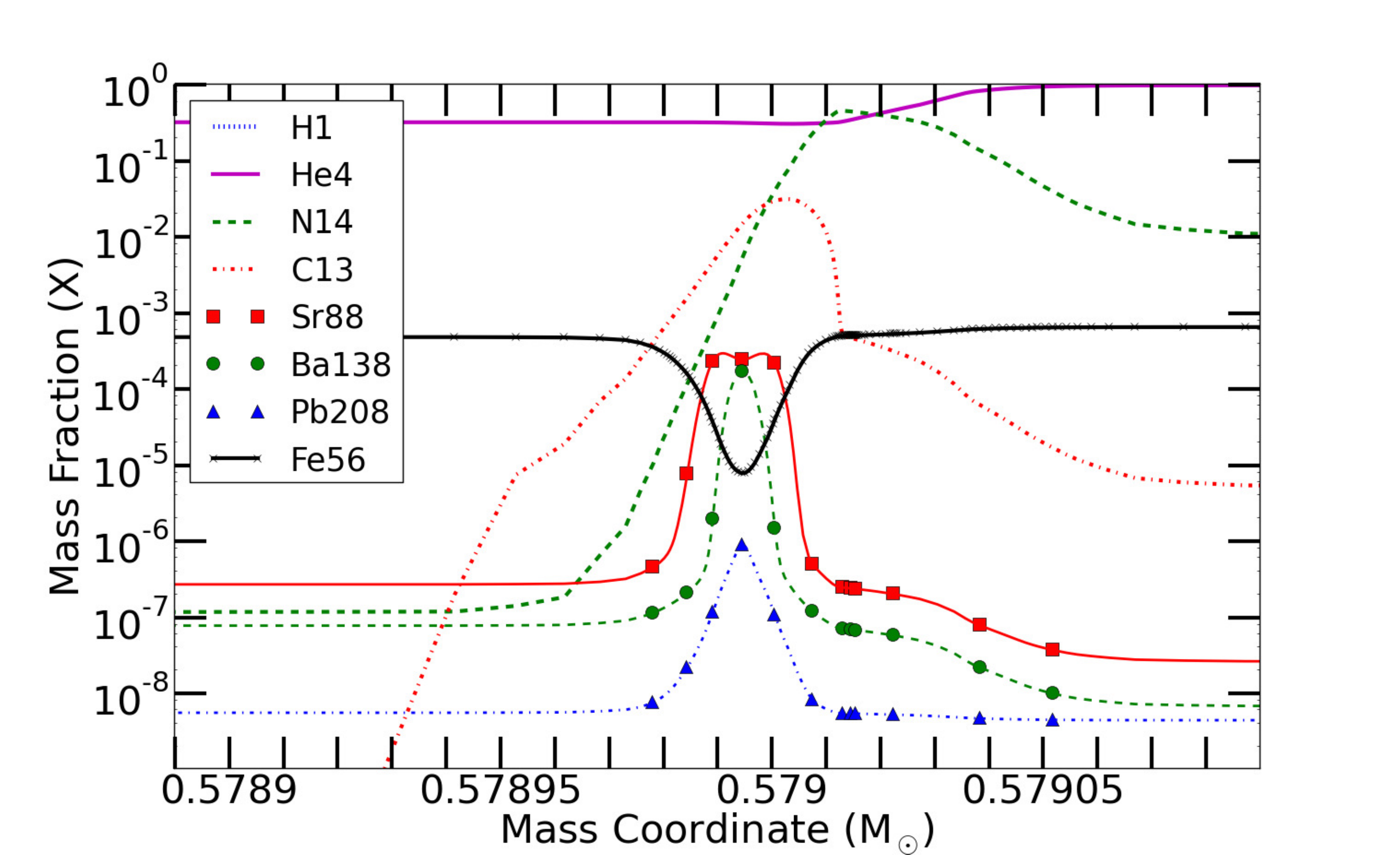}}}
%\resizebox{7.5cm}{!}{\rotatebox{0}{\includegraphics{20_post_pre_comparison_set1p2_d.pdf}}}
%\resizebox{7.5cm}{!}{\rotatebox{0}{\includegraphics{summary_exp_25_set1p2_r.pdf}}}
\caption{\isotope[13]{C}-pocket and neutron magic $s$-nuclei formation. The top panel refers to the moment just after the maximum penetration of the hydrogen-shell during TDU event, which is followed by the radiative burning of the \isotope[13]{C}-pocket with the consequent neutron release and $s$-nuclei synthesis (middle and bottom panel). Also \isotope[56]{Fe} seeds are plotted. The simulations are from the $2\msun$ star, \setopo.
}
\label{fig:c13poc_snuc_form}
\end{figure}

AGB stellar models often show a good agreement with many \spr\ heavy-element abundance observables \citep[e.g.,][]{gallino:97b,goriely:00,busso:99,cristallo:11,bisterzo:11,lugaro:12}, while in other cases are less successful: e.g., see e.g., \cite{vaneck:03} for Pb in CEMP stars, \cite{desmedt:12} and \cite{desmedt:14} for post-AGB stars, and the S, Y, Zr region for many CEMP-s stars \citep[][]{lugaro:12,bisterzo:12}.
Nevertheless, the present established scenario to produce the \spr\ in AGB stars is that at the end of the third dredge-up a partially mixed zone of H and \czw\ leaves behind the conditions for the formation of a \cdr-enriched layer (\fig{fig:c13poc_snuc_form}). Such a layer can subsequently release neutrons under (mostly) radiative conditions during
the interpulse phase.
%Many investigations of the \spr\ \citep[e.g.,][]{gallino:97b,goriely:00,busso:99,cristallo:11,bisterzo:11,lugaro:12} have shown that there is a good agreement between the abundance evolution predicted by stellar models and many heavy-element abundance observables if at the end of the third dredge-up a partially mixed zone of H and \czw\ 
%%[SJ: I don't quite understand this - 'partial mixing zone' is not a clear term and 'of H and \czw\ ' is also ill-defined] 
%leaves behind the conditions for the formation of a \cdr-enriched layer (\fig{fig:c13poc_snuc_form}). Such a layer can subsequently release neutrons under (mostly) radiative conditions during the interpulse phase.  
In our low-mass AGB stellar models we achieve this partial mixing zone through the exponential CBM algorithm (cf.\ \sect{st_ev_co}).
 
The 
%intermediate-mass 
massive AGB stellar models with $5\msun$ encounter just over 20 TPs with third dredge-up. After the initial transient phase the dredge-up parameter is $\lambda \approx 0.8$ (\fig{fig:agb_lambda_starmass_winds}). The temperature at the bottom of the convective envelope $T_\mathrm{CEB}$ in our $ M_\mathrm{ini}=5\msun$, $Z=0.02$ calculation peaks close to $5\cdot10^7\kelv$ (\fig{fig:agb_TCEB}), in good agreement with the results presented by \citet{karakas:12}. 
In the last two pulses of our $5\msun$ sequence $T_\mathrm{CEB}$ is enhanced because of the modified convection and opacity assumptions that we make trying to overcome the well-known modelling problems for higher-mass and higher-Z TP-AGB models \citep{lau:12}. 
Therefore, this final jump in $T_\mathrm{CEB}$ is an artifact of this approximation introduced to simulate more thermal pulses. 
Also concerning the model with $ M_\mathrm{ini}=1.65\msun$, $Z=0.02$, the $T_\mathrm{CEB}$ discontinuity is due to the same opacity modification introduce to aid convergence.
 
\begin{figure}
\centering
\resizebox{11.3cm}{!}{\rotatebox{0}{\includegraphics{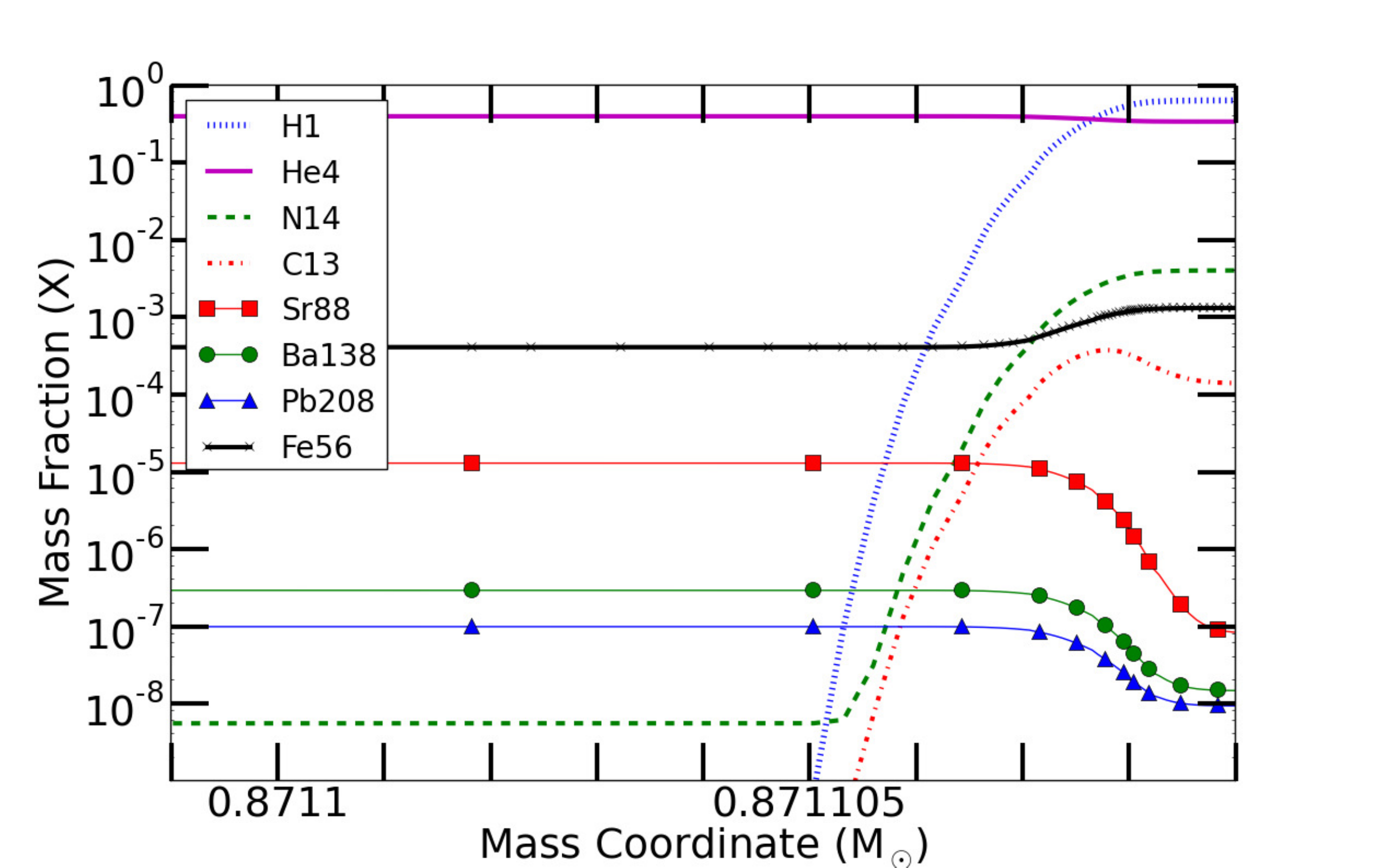}}}
\resizebox{11.3cm}{!}{\rotatebox{0}{\includegraphics{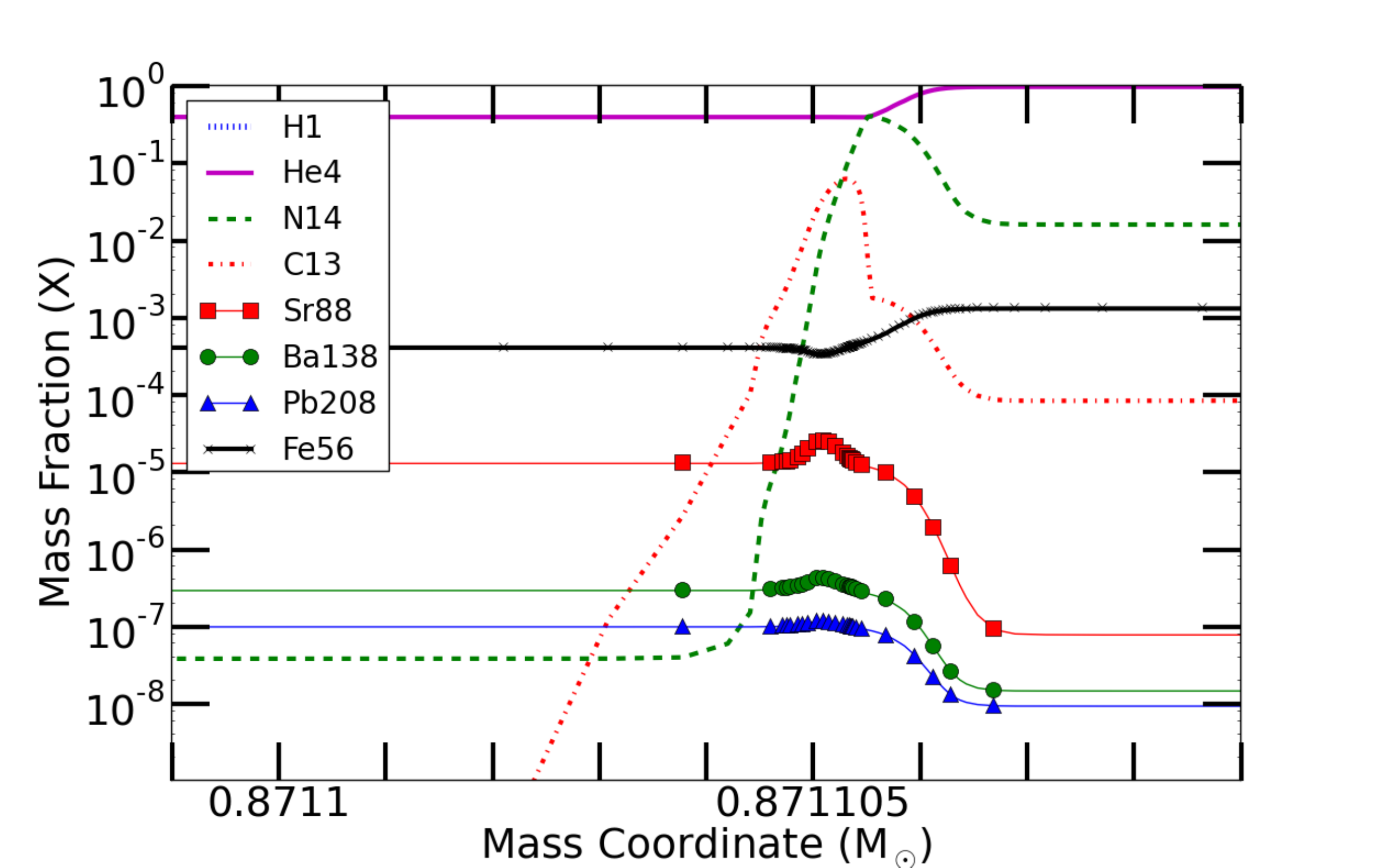}}}
\resizebox{11.3cm}{!}{\rotatebox{0}{\includegraphics{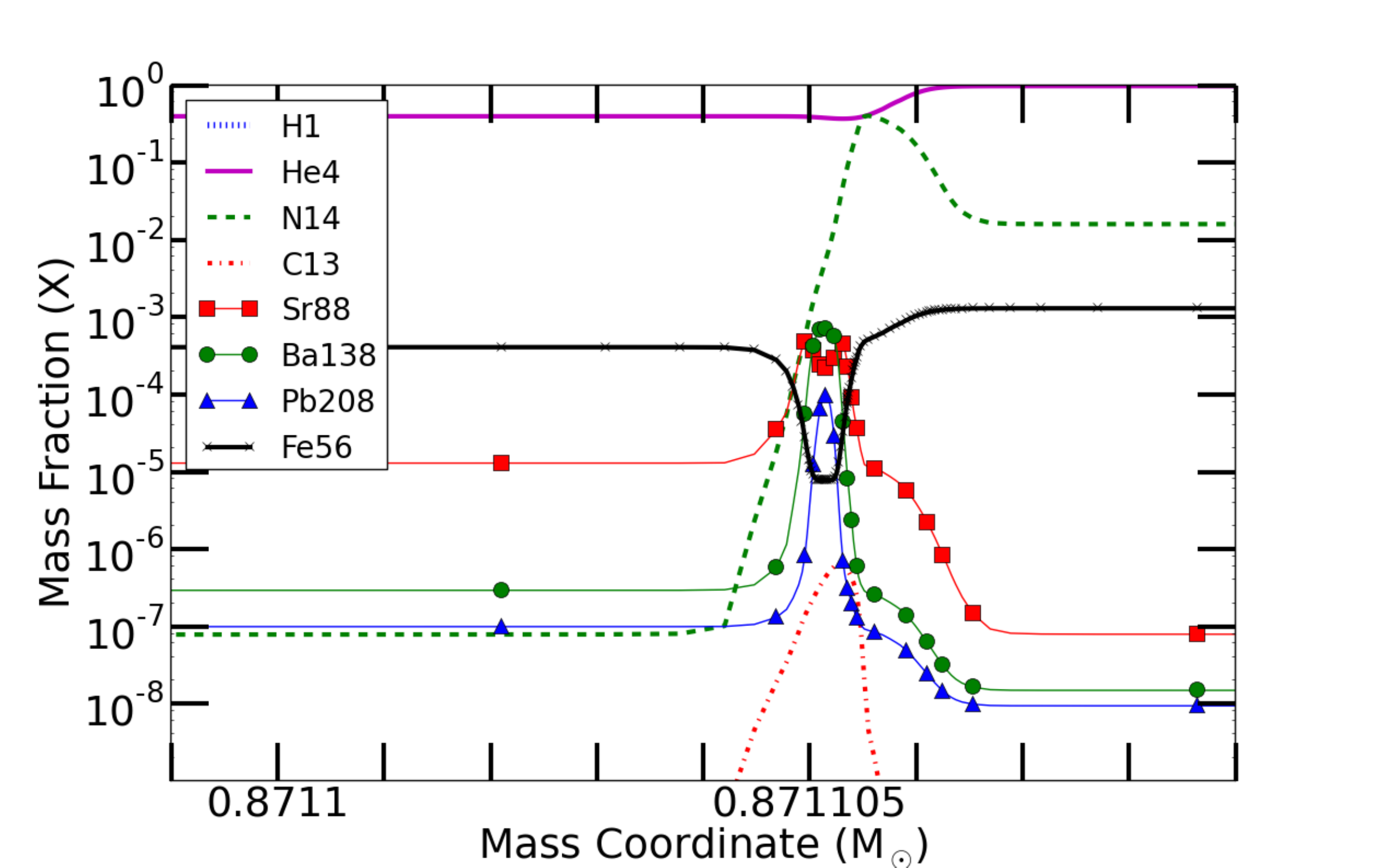}}}
\caption{Same as \fig{fig:c13poc_snuc_form}, but for the $5\msun$ star, \setopt.
}
\label{fig:c13poc_snuc_form_5msunset1p2}
\end{figure}

Inspection of the H-burning luminosity shows that at these high metallicities the models do not show the hot dredge-up reported for lower-Z models \citep[e.g.,][]{herwig:04a}. The \isotope[13]{C}-pocket forms just as in the lower-mass cases, but it contains only about $10^{-6}\msun$. It is post-processed and well resolved, as shown in \fig{fig:c13poc_snuc_form_5msunset1p2}.

\begin{figure}
\includegraphics[width=\textwidth]{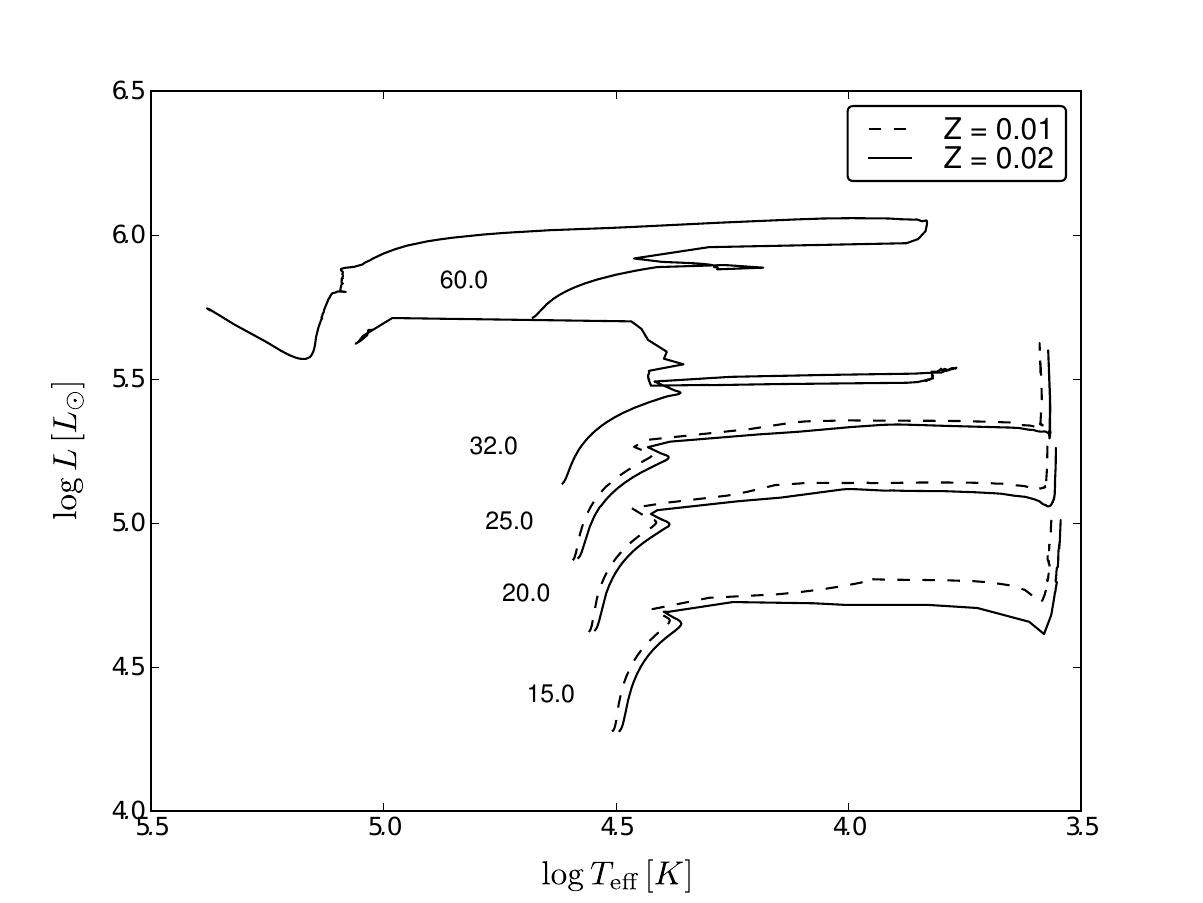}
\caption{H-R diagram for massive star models in \setopo\ ($Z=0.01$)
  and \setopt\ ($Z=0.02$). The evolution of the \setopt\ models in 
  the H-R diagram is shown also in Fig.\ 3 of \citet{bennett:12}.}
\label{fig:hrd_ms_set1}
\end{figure}

\begin{figure}
\includegraphics[width=\textwidth]{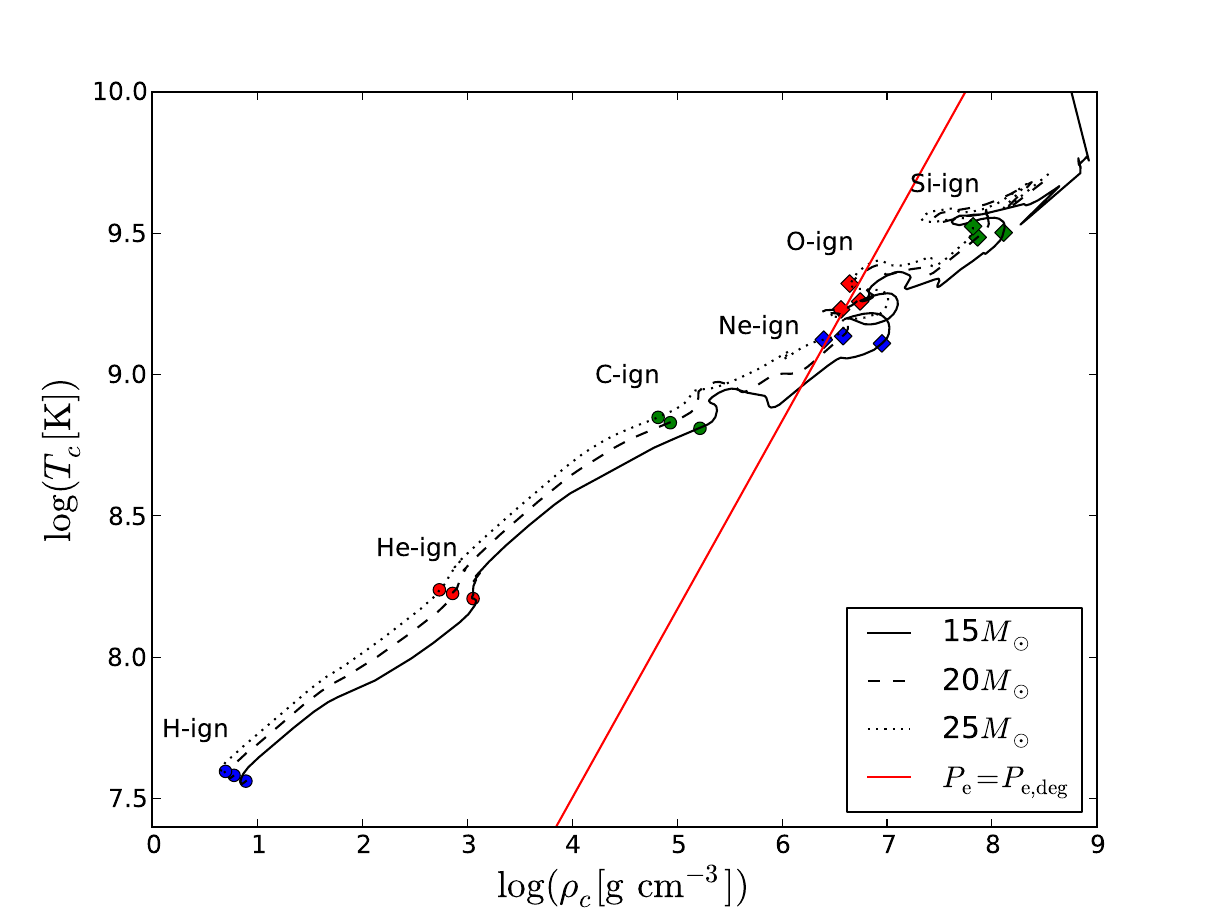}
\caption{Central temperature, $T_{\rm c}$, as a function of central density,
  $\rho_{\rm c}$, for the 15, 20 and $25\msun$ massive star models of
  \setopo.  Ignition points for the core burning stages are indicated
  by the colored points, which are determined at the point when the
  principal fuel is depleted by 0.3\% from its maximum value.}
\label{fig:tcrhoc1_1}
\end{figure}

\begin{figure}
\includegraphics[width=\textwidth]{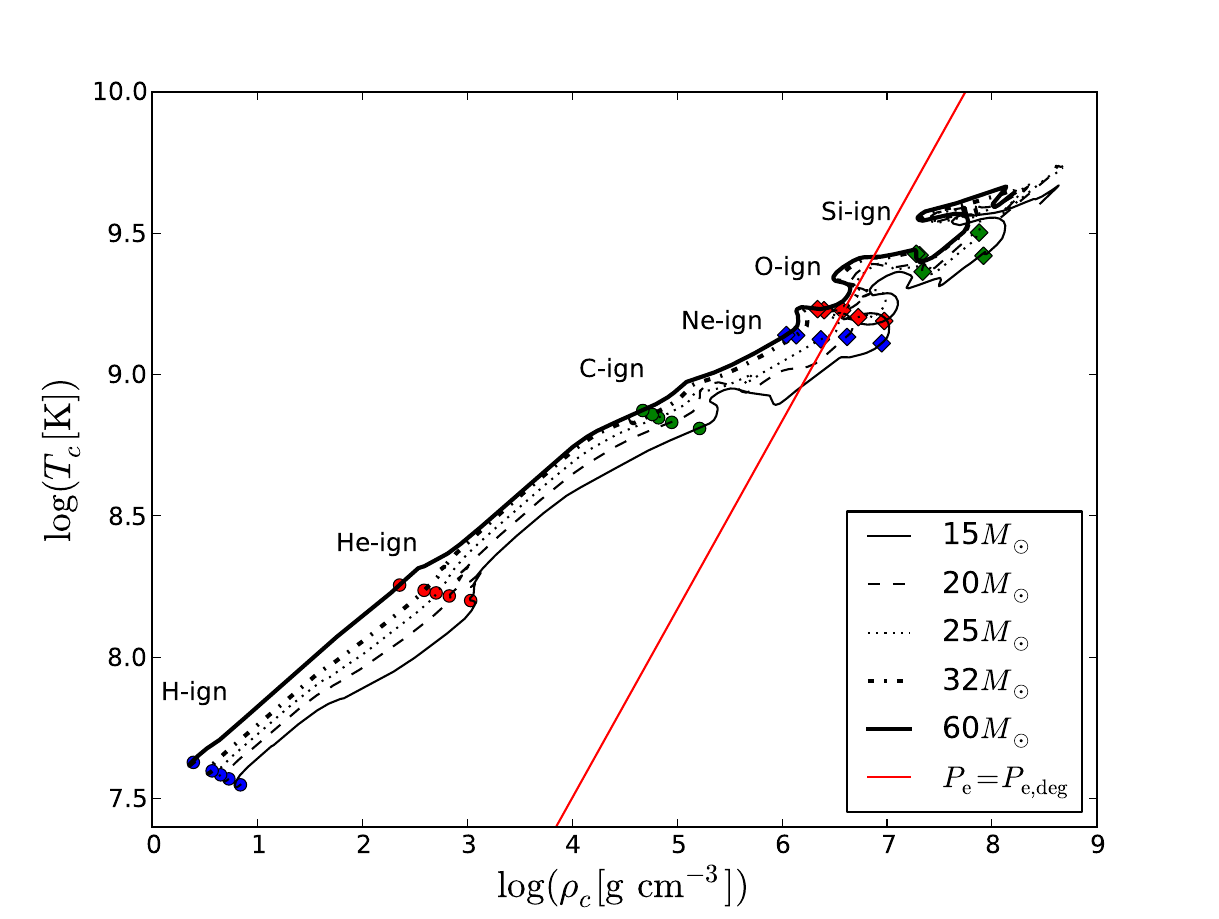}
\caption{Central temperature, $T_{\rm c}$, as a function of central density,
  $\rho_{\rm c}$, for the 15, 20, 25, 32 and $60\msun$ massive star
  models of \setopt.  Ignition points for the core burning stages are
  indicated by the colored points, which are determined at the point
  when the principal fuel is depleted by 0.3\% from its maximum value.
}
\label{fig:tcrhoc1_2}
\end{figure}

\begin{figure}
\centering
%\resizebox{7.5cm}{!}{\rotatebox{0}{\includegraphics{kipm015z10S002.pdf}}}
%\resizebox{7.5cm}{!}{\rotatebox{0}{\includegraphics{kipm020z10S002.pdf}}}
%\resizebox{7.5cm}{!}{\rotatebox{0}{\includegraphics{kipm025z10S002.pdf}}}
\resizebox{7.5cm}{!}{\rotatebox{0}{\includegraphics{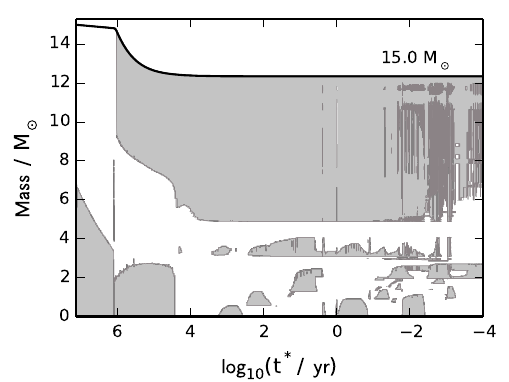}}}
\resizebox{7.5cm}{!}{\rotatebox{0}{\includegraphics{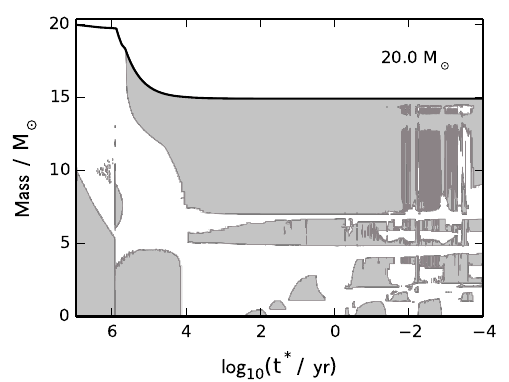}}}
\resizebox{7.5cm}{!}{\rotatebox{0}{\includegraphics{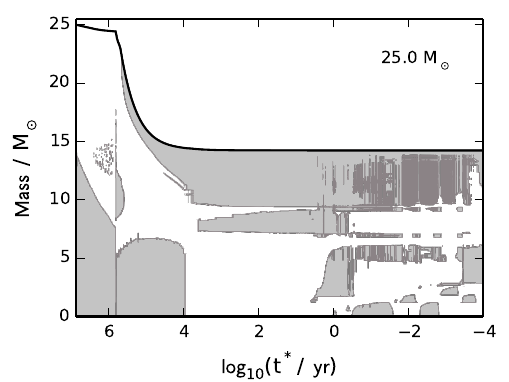}}}
\caption{Kippenhahn diagrams for the 15, 20, and $25\msun$ models from \setopo.
}
\label{fig: kipp_set1p1_ms}
\end{figure}

\begin{figure}
\centering
%\resizebox{7.5cm}{!}{\rotatebox{0}{\includegraphics{kipm015z20S002.pdf}}}
%\resizebox{7.5cm}{!}{\rotatebox{0}{\includegraphics{kipm020z20S002.pdf}}}
%\resizebox{7.5cm}{!}{\rotatebox{0}{\includegraphics{kipm025z20S002.pdf}}}
%\resizebox{7.5cm}{!}{\rotatebox{0}{\includegraphics{kipm032z20S002.pdf}}}
%\resizebox{7.5cm}{!}{\rotatebox{0}{\includegraphics{kipm060z20S002.pdf}}}
\resizebox{7.5cm}{!}{\rotatebox{0}{\includegraphics{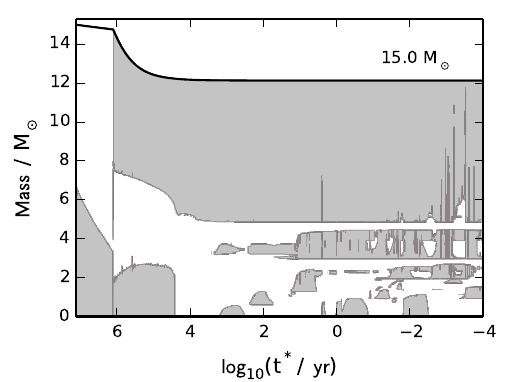}}}
\resizebox{7.5cm}{!}{\rotatebox{0}{\includegraphics{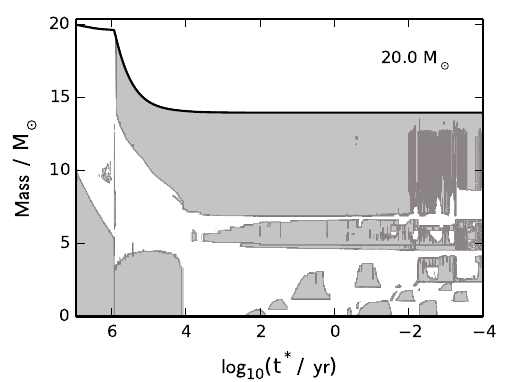}}}
\resizebox{7.5cm}{!}{\rotatebox{0}{\includegraphics{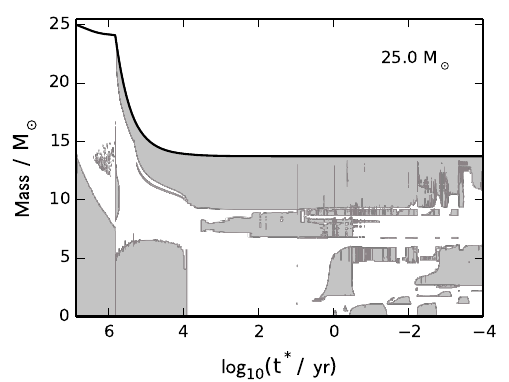}}}
\resizebox{7.5cm}{!}{\rotatebox{0}{\includegraphics{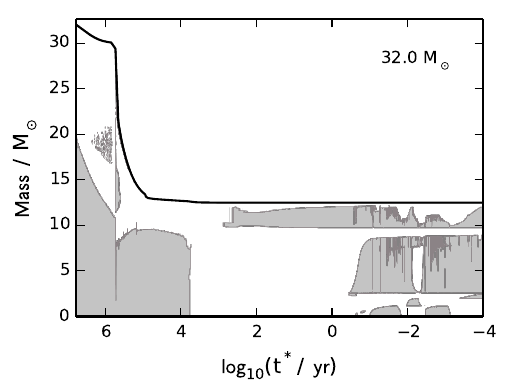}}}
\resizebox{7.5cm}{!}{\rotatebox{0}{\includegraphics{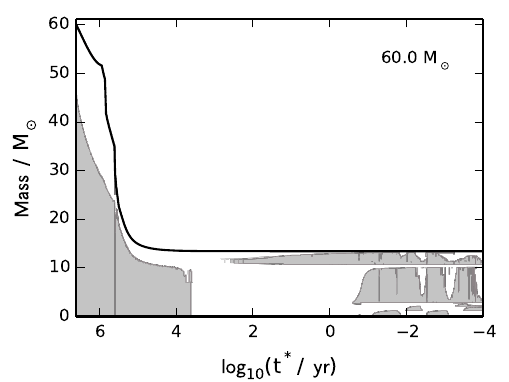}}}
\caption{Kippenhahn diagrams for the 15, 20, 25, 32 and $60\msun$ models from \setopt.
}
\label{fig: kipp_set1p2_ms}
\end{figure}

Full details regarding the \setopt\ massive stars can be found in
\citet{bennett:12}.  In this work the stellar evolution data is extended to
include \setopo\ models. The Hertzsprung$-$Russell diagram for all models in \setone\ are shown 
in \fig{fig:hrd_ms_set1} and the evolutionary tracks in the 
$T_{\rm c}$-$\rho$$_{\rm c}$ plane are shown in Figs. \ref{fig:tcrhoc1_1} and
\ref{fig:tcrhoc1_2}, 
%The usual paths of massive stars are obtained
%\citep[see e.g.,][]{hirschi:04}. 
both of which are consistent with previous results \citep[see e.g.,][]{hirschi:04}.
In particular, models with masses M
$\leq  25 \msun$ end up as red super giants (RSGs), and the \setopt~32~and~$60\msun$~models
end as Wolf-Rayet stars. In \fig{fig: kipp_set1p1_ms}
and \ref{fig: kipp_set1p2_ms} Kippenhahn diagrams of the massive stars
are shown. 
%The final core masses of these models are comparable to other grids of models calculated with GENEC \citep[][]{hirschi:04,ekstrom:12}. The choice of $0.2\hp$ for the extend of overshooting during the core H- and He-burning phases implies that core masses are slightly larger than in the other grids using $0.1\hp$ for core overshooting. 
The final core masses of these models are comparable to other grids of models calculated with GENEC with the same overshooting \citep{schaller:92}.
The choice of $0.2\hp$ for the extent of overshooting during
the core H- and He-burning phases implies that core masses are
slightly larger than in other GENEC grids using $0.1\hp$ for core
overshooting \citep[][]{hirschi:04,ekstrom:12}.
%The final stellar masses at both $Z = 0.01$ and 0.02 are typically lower than the models obtained using other stellar evolution codes, due to the mass loss prescriptions used in the RSG phase, which is based on observational constrains (see \S \ref{st_ev_co}).
The final stellar masses at both Z = 0.01 and 0.02 are typically lower than the models obtained using other stellar evolution codes. This is due to the different mass loss prescriptions used for RSG (see \S \ref{st_ev_co} for the mass loss rates used in GENEC) in different codes, which are empirical and still uncertain.
%RH additional sentences discussing why 32 and 60Mo models were only calculated at Z=0.02
Although the fate of massive stars is still not well understood \citep[see e.\,g.][]{UJMA12,smartt:15}, the probable fate of stars above 30\,$M_\odot$ at metallicities lower than solar is a collapse without explosion (although the dependence of mass loss rates on metallicity is also uncertain). Furthermore the winds of massive stars only enrich the ISM in light elements (up to aluminium). Based on our $Z=0.02$ simulation we expect their contribution to heavy elements will be small and therefore did not compute 32 and 60\,$M_\odot$ models at $Z=0.01$.

The core masses for all of the massive star models are shown in 
\tab{tab:cores1_set1}.  The core masses are determined at the end of
silicon burning and are defined as the mass coordinate
where a criterion for the core mass is satisfied. The helium-core
mass, $M^{75\%}_{\mathrm{\alpha}}$, is defined by the mass coordinate
where \isotope[4]{He} abundance becomes lower than 0.75 in mass (note that
the 32 and $60\msun$ stars become W-R stars and have
lost their entire H-rich envelope). For the CO-core mass,
$M_{\mathrm{CO}}$, the position corresponds to the mass coordinate
where the \isotope[4]{He} abundance falls below 0.001 toward the center of
the star.  For the silicon-core mass, $M_{\mathrm{Si}}$, the position
corresponds to a mass coordinate where the sum of Si, S, Ar, Ca and Ti
mass fraction abundances, for all isotopes, is 0.5.  
% - we don't calculate the Fe core mass
%The iron core
%mass, $M_{\mathrm{Fe}}$, is determined in a similar way, using in this
%case the sum of Cr, Fe and Ni isotopes. 
The core-burning lifetimes for
hydrostatic-burning stages are presented in Table
\ref{tab:lifetimes_set1} for the \setopt\ and \setopo\ massive star
models. The lifetimes are defined for each stage as the difference in
age from the point where the principal fuel for that stage (\isotope[1]{H} 
for hydrogen burning, \isotope[4]{He} for helium burning, etc.)  is depleted 
by 0.3\% from its maximum value to the age where the abundance of
that fuel is depleted below a mass fraction of $10^{-5}$. There are
exceptions, however, for carbon burning and neon burning where this value is
$10^{-3}$, and oxygen burning where it is $10^{-2}$. These
criteria are necessary to ensure that a lifetime is calculated in
those cases where residual fuel is unburnt and to ensure that the
burning stages are correctly separated (for example, the mass fraction
abundance of \isotope[12]{C} at neon ignition for the \setopt\ $60\msun$
model is $4.123 \times 10^{-5}$). The lifetime of the advanced stages
is quite sensitive to the mass fractions of isotopes defining the
lifetime, particularly for stages following carbon burning.

\subsection{The approximations of CCSN explosion}
\label{subsec:exp}

Stellar winds play a role dispersing nuclides into the circumstellar medium, particularly for the light elements carbon and nitrogen. The bulk
of the nucleosynthetic yields from massive stars, however, are ejected by the
supernova explosion. In the deeper layers (most importantly the silicon
and oxygen layers, although potentially also in the neon and carbon layers), the supernova shock drives further nuclear burning. Determining the ultimate yield including this explosive burning is a complex problem
\citep[e.g.,][]{woosley:95,chieffi:98,limongi:00,woosley:02,nomoto:06,tominaga:07,thielemann:11} 
and specific discussions are needed for different species \citep[see for
  example][]{rauscher:02,tur:09}.  In this paper, our stellar
models follow the evolution of the star through silicon burning,
but not to collapse.  Instead of forcing a collapse, we model the
explosive nucleosynthesis using a semi-analytic description for the shock
heating and subsequent evolution of the matter to produce a qualitative
picture of explosive nuclear burning.

The first step in our semi-analytic prescription is the determination
of the mass-cut defining the line between matter ejected and matter
falling back onto the compact remnant \citep{fryer:12}. We use the
prescription outlined in \citet{fryer:12} for the final compact remnant mass as
a function of the initial stellar mass and metallicity (\tab{tab: coo_fallback}). 
Under the convective-engine paradigm, the explosion energy is a function of the ram pressure of the infalling stellar material, and hence depends upon the time of the explosion.  The mass of the final compact remnant depends both on this time and on the amount of material that falls back after the launch of the explosion. This fallback depends strongly on the explosion energy.
In accordance with \citet{fryer:12}, two explosion models are considered
for each massive star model, labeled as \emph{delayed} and \emph{rapid}.
%Based on the convective engine, the explosion time is related to the
%explosion energy and different explosion energies produce different
%remnant masses.
%Considering that the (neutrino-driven) convective engine is the driving mechanism behind the supernova explosion, the explosion time is related to the explosion energy; different explosion energies thus produce different remnant masses due to varying amounts of accretion onto the compact remnant. 
We include the two models here to give a range of
remnant masses. In general, the rapid explosion produces smaller
remnant masses than the delayed explosion.  For more massive stars,
the rapid explosion model fails, producing large remnants.  
%Neglecting
%mass fallback, \citet{timmes:96} set the remnant mass to the iron core
%mass.  
Comparing our remnant masses to the core masses in Table
\ref{tab:cores1_set1}, we note that a direct correspondence between
core mass and remnant mass does not exist with the new remnant-mass
prescription in \citet{fryer:12} that includes both supernova engine
and fallback effects.  Beyond the mass cut, our stellar structure is
in agreement with pre-collapse stellar models \citep{limongi:06,woosley:02,young:07}. 
In particular, the stellar structure outside of the final mass cut 
is not expected to vary much between the end of core Si-burning and the collapse stage so the results presented here are not affected by the fact that we did not follow 
the pre-collapse phase \citep[see e.g., comparison in][]{paxton:11}.
Hence, our semi-analytic prescription for the shock will produce the same yield
with a pre-collapse star as it does with our end-of-silicon-burning
models.

We determine the shock velocity in the analytical explosion model using the Sedov
blastwave solution~\citep{sedov:46} throughout the stellar structure.
The density and temperature of each zone are assumed to spike 
suddenly following the shock jump conditions in the strong shock
limit \citep{chevalier:89}.  The pressure ($P$) is given by
\begin{equation}
P = (\gamma+1)/2 \rho v_{\rm shock}^2,
\end{equation}
where $\gamma$ is the pre-shock adiabatic index determined from our
stellar models, $\rho$ is the pre-shock density, and $v_{\rm shock}$ 
is the shock velocity.  After being shocked, the pressure is radiation
dominated, allowing us to calculate the post-shock temperature
($T_{\rm shock}$),
\begin{equation}
T_{\rm shock} = (3P/a)^{1/4},
\end{equation}
where $a$ is the radiation constant.  The post-shock density
($\rho_{\rm shock}$) is given by
\begin{equation}
\rho_{\rm shock}/\rho = (\gamma+1)/(\gamma-1).
\end{equation}

After the material is shocked to its peak explosive temperature
and density, it cools.  For these models, we use a variant of
the adiabatic exponential decay \citep{Hoyle:64,Fowler:64},
\begin{equation}
T(t)=T_{\rm shock} e^{-t/(3\tau)}
\end{equation}
and
\begin{equation}
\rho(t)=\rho_{\rm shock} e^{-t/(\tau)},
\end{equation}
where $t$ is the time after the the material is shocked,
$\tau=446/\rho_{\rm shock}^{1/2}$\,s, and $\rho_{\rm shock}$
is the post-shock density in g$\cc$.

\begin{figure}
\centering
\resizebox{7.5cm}{!}{\rotatebox{0}{\includegraphics{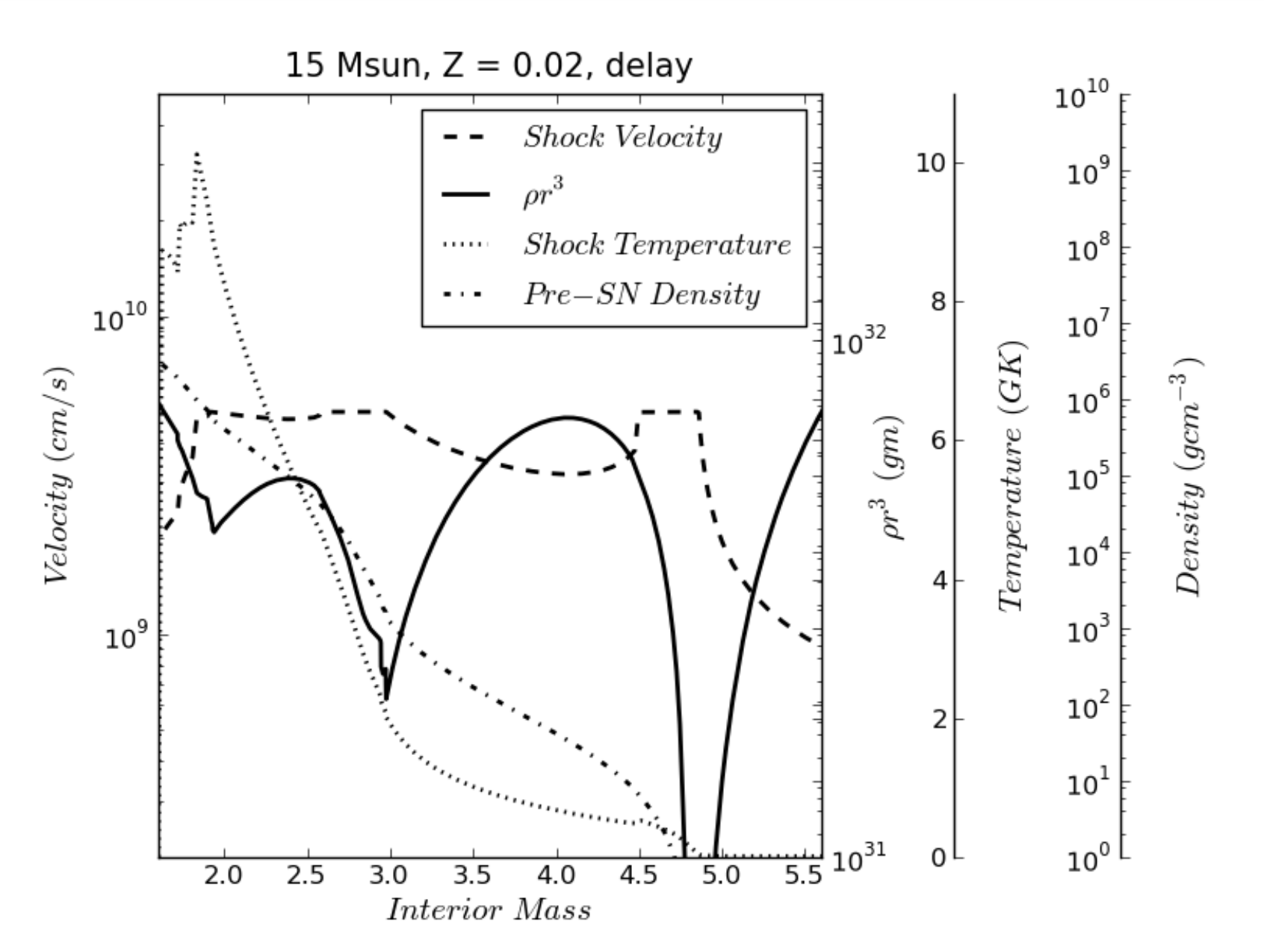}}}
\resizebox{7.5cm}{!}{\rotatebox{0}{\includegraphics{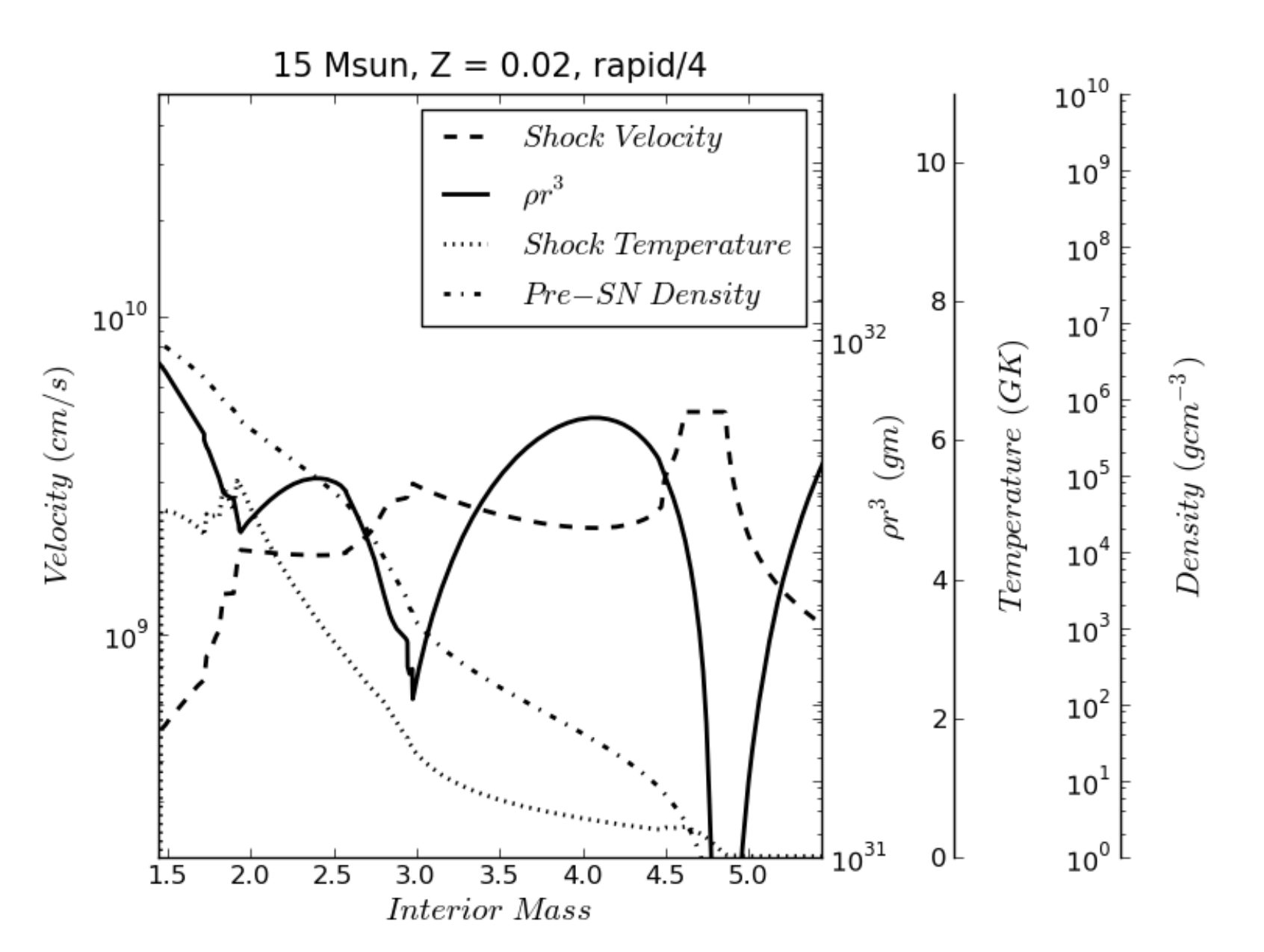}}}
\resizebox{7.5cm}{!}{\rotatebox{0}{\includegraphics{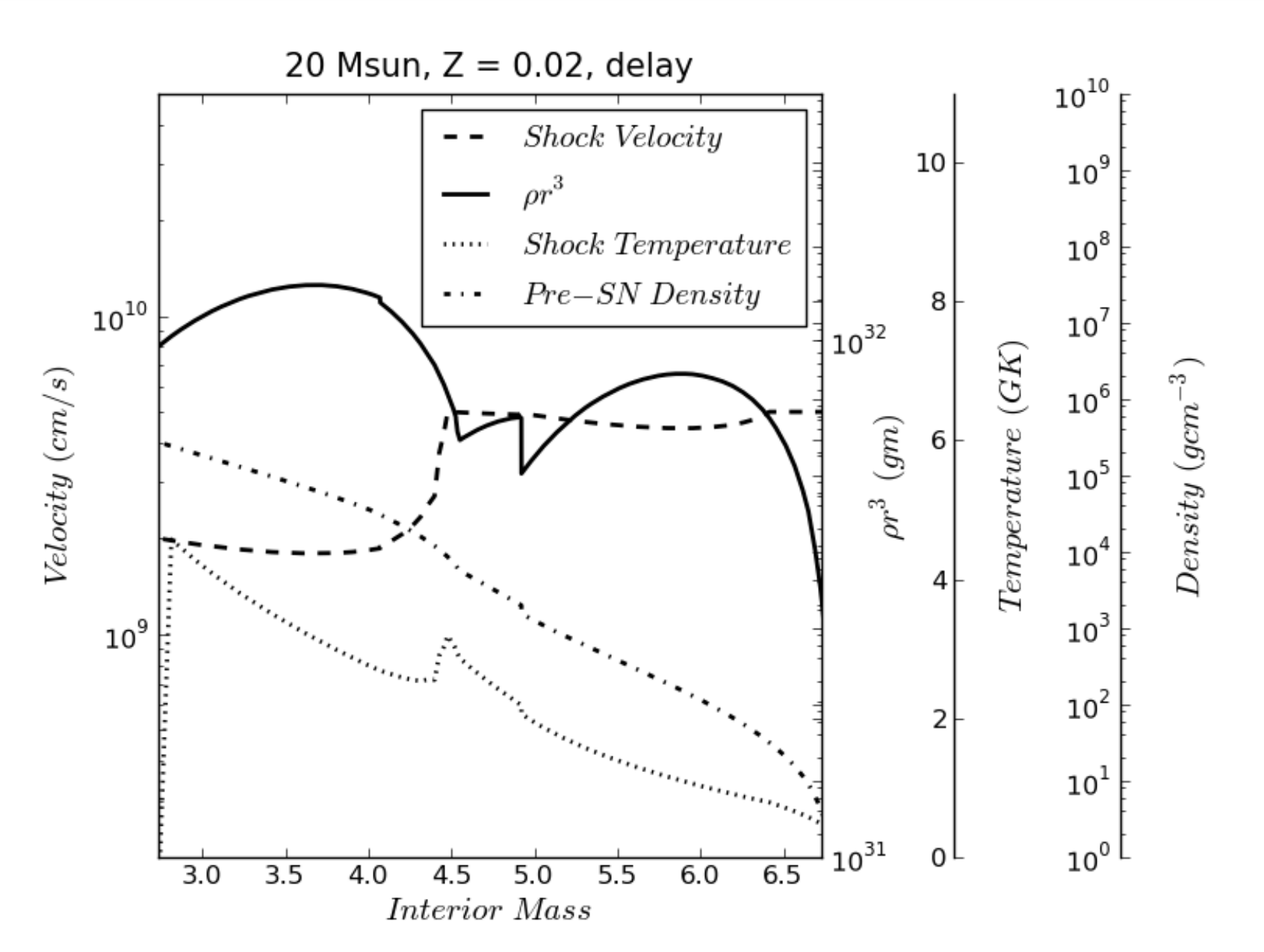}}}
\resizebox{7.5cm}{!}{\rotatebox{0}{\includegraphics{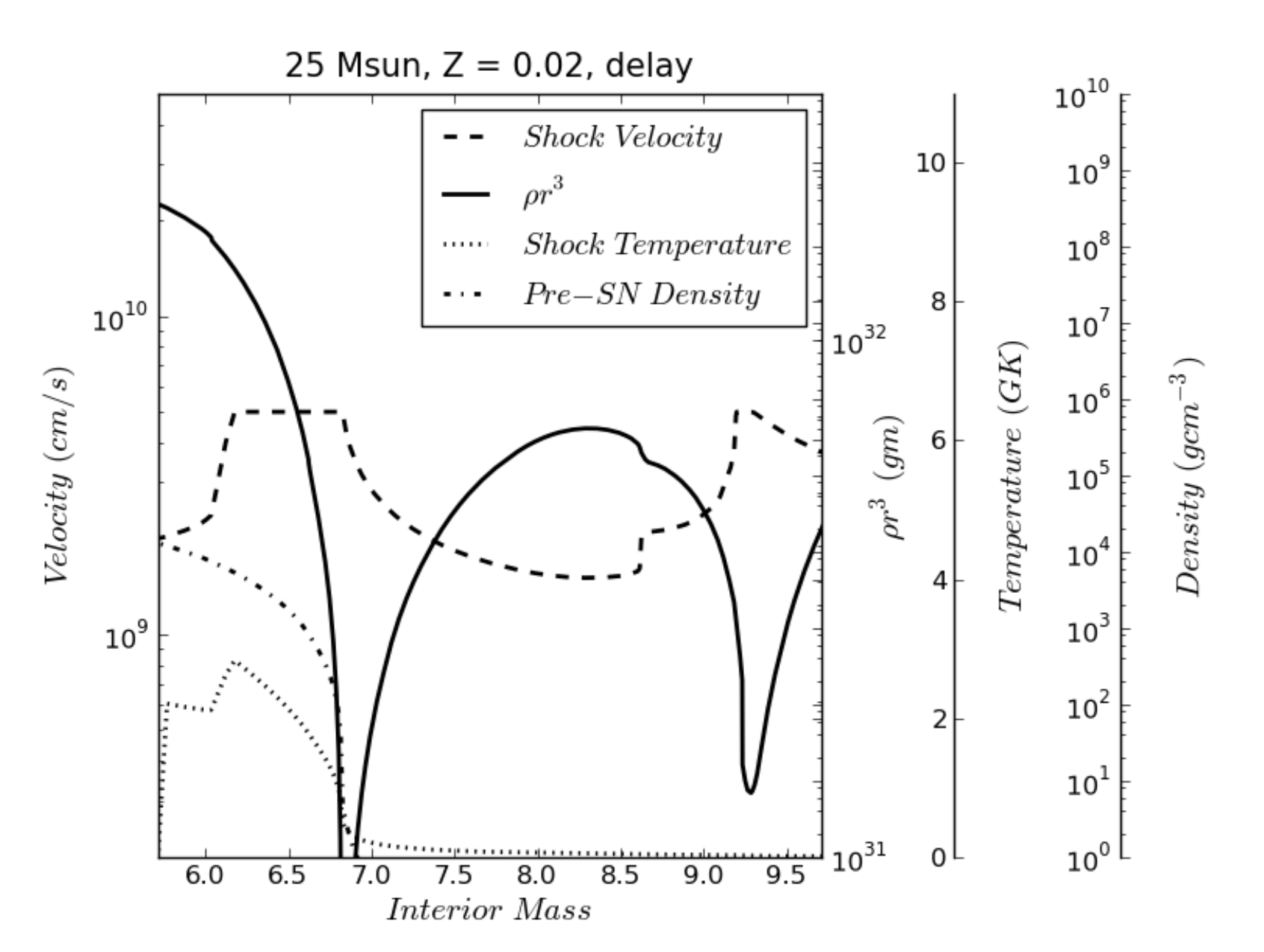}}}
\caption{Details for a selection of four CCSN models of \setopt.
Basic pre-supernova structure information (density and $\rho~r^3$ profiles),
and shock velocity and temperature at the first cycle
of SN simulations. Initial mass and metallicity of the model is given
on top of each plot.  In the same location, also the type of
fall-back prescription is also indicated, namely with
$rapid$ and $delay$. See the text for explanation. The models selected
are two $15\msun$ models with $delayed$ SN explosion and
$rapid/4$ (where the shock velocity from explosion $rapid$
is reduced by a factor 4), a $20\msun$ and a $25\msun$ models
with $delayed$ SN explosion.
}
\label{fig:summary_exp_set1p2}
\end{figure}

The details of the explosion for our \setopt\ model with the
delayed explosion model are shown in \fig{fig:summary_exp_set1p2}. The lower mass
cut is determined using the prescription in \citet{fryer:12}.  Aside from the mass cut, there is no difference between our implementation of
the rapid and delayed explosions (we implement the same shock
velocities).  In this manner, our delayed/rapid comparisons highlight
the effect of the mass cut on the yield.  We use an initial velocity
of $2\times10^{9} \mathrm{cm~s^{-1}}$, and we define this as the setup for our standard model
(on par with reasonably strong velocities at the launch of a shock in core-collapse
calculations).  
We added two additional $15\msun$ models to \setopt\ using the $rapid$ explosion model, in which the initial shock velocity is reduced by a factor of two and four (i.e., assuming an initial $v_{\rm shock}$ = $1\times10^{9} \mathrm{cm~s^{-1}}$ and $5\times10^{8} \mathrm{cm~s^{-1}}$, respectively).
For comparison, the explosion characteristics for the model with $v_{\rm shock}$ = $5\times10^{8} \mathrm{cm~s^{-1}}$ is shown in \fig{fig:summary_exp_set1p2}.

The strong shocks in our standard model produce at similar densities higher shock temperatures
than common one-dimensional models of CCSN \citep[e.g.,][]{woosley:95}, affecting the explosive nucleosynthesis. In particular, 
%because of the higher temperatures
the present nucleosynthesis calculations may show many similarities with
hypernovae or the high energetic components of asymmetric supernovae
\citep[e.g.,][]{nomoto:09}.  At the elemental boundary layers,
the shock can accelerate a small amount of material to high velocities
as it travels down the density gradient.  In most explosion calculations
\citep{young:07}, viscous forces limit this acceleration and we
artificially cap our maximum velocity to $v_{\rm shock}=
5 \times 10^{9} \mathrm{cm~s^{-1}}$.

With these analytic explosion models, we are able to understand the
trends in explosive burning.  To compare in detail post-explosive and
pre-explosive abundances, we refer to the production factors presented
in \sect{nucleo}, and to the complete yields tables provided online.

\section{Post-processing nucleosynthesis calculations}
\label{nucleo}

In this section, first we present the stellar yields obtained for the models
described in \sect{stars}, and the tools adopted for the nucleosynthesis
simulations.  In the second part of the section we discuss the production of
the elements by the nucleosynthesis processes considered in our models. 
%Each element lighter than Bi is formed by one or more stable isotopes, besides
%Tc and Pm.
In order to understand the production of elements, we first need to disentangle
the different nucleosynthesis processes contributing to their isotopes.  More
than one process might potentially contribute to the isotope inventory, and
this combination might change with the galactic evolution time.  For instance,
about 92\% of the neutron-magic isotope \isotope[138]{Ba} observed in the Solar
System is produced by the \sprn, with a smaller contribution from the
\rpr~\citep[][]{bisterzo:14}, while its production in old metal-poor \rpr\
stars was only due to the \rprn\ \citep[e.g.,][]{sneden:08,roederer:14a}.
Furthermore, the same nucleosynthesis process can be activated in different
types of stars, eventually overlapping their respective contribution to the
interstellar medium. The isotope \isotope[12]{C} is a main product of He
burning in stars, and its abundance in the Solar System was made by the He
burning activated in both AGB stars and massive stars
\citep[e.g.,][]{kobayashi:11}. 
%\isotope[14]{N} in the early Galaxy is only made by proton captures in massive
%stars, while the contribution from AGB stars becomes dominant at later stages
%\citep[e.g.,][]{tosi:07}.
Based on these considerations, a comprehensive nucleosynthesis analysis often
requires to consider different types of stars.

%The interpretation of observations can be easier for e.g., galactic
%archaeology studies, where the contribution from massive stars dominates the
%production of light elements \citep[e.g.,][]{nomoto:13}, and for all other
%stellar observations where the signature of intrinsic processes can be
%isolated and/or changes the pristine abundance concentrations due to
%galactical chemical evolution \citep[e.g., the \spr\ signature in galactic AGB
%stars,][]{zamora:09}. 
%\textbf{The interpretation of observations might be easier for e.g., galactic archaeology studies, where the contribution from massive stars dominates the production of light elements \citep[e.g.,][]{nomoto:13}. More in general, the comparison between theoretical stellar models and observations becomes more powerful when the signature of intrinsic processes changes the pristine abundance concentrations due to galactical chemical evolution and it can be isolated \citep[e.g., the \spr\ signature in galactic AGB stars,][]{zamora:09}.} 
The interpretation of observations can be easier for, e.g.,
galactic archaeology studies, where the contribution from massive stars
dominates the production of light elements \citep[e.g.][]{nomoto:13}. More
generally, comparing theoretical stellar models with observations is more
instructive when a single nucleosynthesis process modifies the abundance of an
element. This makes it much easier to trace and isolate the origin of the
process using galactic chemical evolution simulations
\citep[e.g.][]{zamora:09}.

Therefore, we decided to briefly describe the production of the elements
dividing them by small groups (C N and O in \sect{c-n-o}; F, Ne and Na in
\sect{f-ne-na}; Mg, Al and Si in \sect{mg-al-si}), and by mass regions
(intermediate elements between P and Sc in \sect{intermediate_elements};
iron-group elements in \sect{iron_elements}, heavy elements between Ni and Zr
in \sect{ni-zr} and beyond Zr in \sect{zr-pb}). A similar approach has been
separately used in the past to describe the nucleosynthesis in massive stars
\citep[e.g.,][]{woosley:95} and in AGB stars \citep[e.g.,][]{ventura:13}. Here
%, for the first time, 
we apply the same methodology but discussing together the nucleosynthesis in
our models for low-mass, intermediate-mass and massive stars.

In general, charged particle reactions in the different stellar evolutionary
stages are responsible for the chemical inventory of light elements, up to the
iron group \citep[e.g.,][]{woosley:02,karakas:14}. Neutron captures are
responsible for the majority of the element production beyond Fe
\citep[][]{kaeppeler:11,thielemann:11}, but they have to be included when
considering the production for the production of a number of light isotopes.
For instance, the neutron capture on \isotope[22]{Ne} is relevant for the
production of Na at solar metallicities in massive stars, while it is less
important for the production of Na in AGB stars (see \cite{mowlavi:99a} and
\sect{f-ne-na}). 

The neutron-rich isotope \isotope[36]{S} has a different origin compared to the other S stable isotopes, and it is fully produced by neutron captures, in both AGB stars and massive stars (\sect{intermediate_elements}).\\
Even if a specific nucleosynthesis process is not efficiently contributing for the galactic chemical evolution of an element, nevertheless it may be possible to observe the abundance signature associated to that process in other stellar associations or in single stars.
%, providing valuable information to compare with theoretical stellar predictions.
For instance, AGB stars are not relevant for the chemical inventory of Ti, but the Ti isotopic ratios can be measured in presolar carbon-rich grains carrying the \spr\ signature from their parent AGB stars \citep[e.g.,][]{zinner:14}. 
In metal-poor globular clusters (GCs), the second generation of stars are Na-rich and O-poor compared to the older pristine population \citep[e.g.,][]{gratton:12}. In GCs, the Na enrichment is due to proton captures in fast rotating massive stars and/or in massive AGB stars. On the other hand, in the Milky Way for the typical metallicity range of GCs Na is mainly made by C burning in massive stars, before the CCSNe explosion \citep[][]{thielemann:96}.

%In summary, a comprehensive analysis of how the elements are made in all types of stars is the final goal of nuclear astrophysics. 
%Aligned with this final goal, here we start to present stellar yields for AGB stars and massive stars for two metallicities, and we summarize the nucleosynthesis results for different group of elements.
%A comprehensive analysis of how the elements are made in stars requires the challenging task to consider their production by different nucleosynthesis processes. 
Here we present stellar yields for AGB stars and massive stars for two metallicities, and we summarize our nucleosynthesis results for different group of elements.

\subsection{Nucleosynthesis code and calculated data}
\label{sec:nuc-code}

The nucleosynthesis simulations 
%presented in this paper 
are calculated using the multizone frame \mppnp\ of the NuGrid post-processing code
\citep[e.g.,][]{herwig:08a,Pignatari:2012dw}.
A detailed description of the code and the post-processing method is
available in Appendix \ref{ap:codes}.

Thermodynamic and structural information regarding the stellar models and
CCSN explosion simulations is described in \sect{stars} and provides the input
for the nucleosynthesis calculations. The size of the nuclear
network increases dynamically as needed, up to a limit of 5234
isotopes during the CCSN explosion with 74313 reactions. The NuGrid
physics package uses nuclear data from a wide range of sources,
including the major nuclear physics compilations and many other
individual rates \citep[\sect{sub_ap:physics_package}, ][]{herwig:08}.
As explained in \sect{ap:codes} the post-processing code must adopt the
same rates as the underlying stellar evolution calculations for charged
particle reactions relevant for energy generation (\sect{stars}). These
include triple-$\alpha$ and \isotope[12]{C}($\alpha$,$\gamma$)\isotope[16]{O}
reactions from \citet{fynbo:05} and \citet{kunz:02}, respectively, as
well as the \isotope[14]{N}(p,$\gamma$)\isotope[15]{O} reaction \citep{imbriani:05}.
The neutron source reaction 
\isotope[13]{C}($\alpha$,n)\isotope[16]{O} 
is taken from
\citet{heil:08} and the competing 
\isotope[22]{Ne}($\alpha$,n)\isotope[25]{Mg}
and
\isotope[22]{Ne}($\alpha$,$\gamma$)\isotope[26]{Mg}
reactions are taken from
\citet{jaeger:01} and \citet{angulo:99}, respectively. Experimental
neutron capture reaction rates are taken, when available, from the KADoNIS
compilation \citep{dillmann:06}. For neutron capture rates not
included in KADoNIS, we adopt data from the Basel REACLIB database,
revision 20090121 \citep{rauscher:00}. The $\beta-$decay rates are
from \citet{oda:94} or \citet{fuller:85} for light species and from
\citet{langanke:00} and \citet{aikawa:05} for the iron group and for
species heavier than iron; exceptions are the isomers of \isotope[26]{Al},
\isotope[85]{Kr}, 
\isotope[115]{Cd}, 
\isotope[176]{Lu}, 
and \isotope[180]{Ta}. 
For isomers below the
thermalization temperature the isomeric state and the ground state are
considered as separate species and terrestrial $\beta-$decay rates are
used \citep[e.g.,][]{ward:76}.

In \tab{tab:isotopic_prodfac_set1.2_winds} the isotopic overproduction
factors---the final products normalized to their initial abundances---are
given for stellar winds in \setopt. In \tabs{tab:isotopic_prodfac_set1.2_pre} and
\ref{tab:isotopic_prodfac_set1.2_exp} the pre-explosive and explosive
overproduction factors are given for massive stars at the same metallicity.
Radioactive isotopes have been assumed to have decayed.

The overproduction factors, 
$OP_{im}$, for a given model of initial mass, $M$, for element/isotope $i$ is given by
%\begin{equation}
% OP_{im} = \frac{EM_{im}}{M_{\rm ini} X^0_i},
%\end{equation}
\begin{equation}
 OP_{im} = \frac{EM_{im}}{M_{\rm ej} X^0_i},
\end{equation}
where $EM_{im}$ is the total ejected mass of element/isotope $i$,
$M_{\rm ej}$ is the ejected mass of the model, and $X^0_i$ is the
initial mass fraction of element/isotope $i$.

The total ejected masses include the contributions from both stellar
winds and the SN explosion for massive stars and solely from the wind for low-
and intermediate-mass stars.
The wind contribution is given by:
\begin{equation}
% stellar yields: mp^{\rm wind}_{im} = \int^{\tau(m)}_0 \dot{M}(m,t)[X^S_i(m,t) - X^0_i] dt 
% ejected masses: no  ``- X^0_i''
EM^{\rm wind}_{im} = \int^{\tau(m)}_0 \dot{M}(m,t)\,X^S_i(m,t) dt \label{eqn:wind}
\end{equation}
where $\tau(m)$ is the final age of the star, $\dot{M}(m,t)$ is the mass loss rate, $X^S_i$ is the surface mass-fraction abundance;
%, $X^0_i$ is the initial mass-fraction abundance.  
the SN contribution is given by:
\begin{equation}
% stellar yields: mp^{\rm preSN}_{im} = \int^{m_{\tau}}_{M_{{\rm rem},m}} [X_i(m_r) - X^0_i] dm_r 
% ejected masses: no `` - X^0_i'' term
EM^{\rm SN}_{im} = \int^{m_{\tau}}_{M_{{\rm rem},m}}\, X_i(m_r) dm_r \label{eqn:sn}
\end{equation}
where $m_{\tau}$ is the total mass of the star at $\tau(m)$, $M_{{\rm rem},m}$ is the compact remnant mass and $X_i(m_r)$ is the mass fraction abundance of element/isotope $i$ at mass coordinate $m_r$. 
%RH: IF YIELDS ARE GIVEN IN THE ONLINE VERSION, ONE NEEDS TO BE CLEAR
%AND NOT CONFUSE EJECTED MASSES AND STELLAR YIELDS. TO BE CHECKED WHEN
%CONTENT IS FINALISED.
The same data are given in \tabs{tab:element_prodfac_set1.2_winds},
\ref{tab:element_prodfac_set1.2_pre}, and
\ref{tab:element_prodfac_set1.2_exp} for the elemental abundances. As mentioned before,
the radiogenic contribution is included.
Similar information is provided for \setopo\ in \tabs{tab:isotopic_prodfac_set1.1_winds},
\ref{tab:isotopic_prodfac_set1.1_pre}
and \ref{tab:isotopic_prodfac_set1.1_exp} for isotopes, and
in \tabs{tab:element_prodfac_set1.1_winds},
\ref{tab:element_prodfac_set1.1_pre}, and
\ref{tab:element_prodfac_set1.1_exp} for elements, respectively.
Complete tables are provided online together with the analogous
production factors, stellar yields in form of ejected masses 
\citep[given in solar masses; for details, see][for example]{bennett:12} and net yields \citep[see definition in, e.g.,][]{hirschi:05a}. 
The same tables are also provided online for two additional $15\msun$ models of \setopt, $rapid$ explosion, where the initial shock velocity is assumed to be lower by a factor of two and four (\sect{subsec:exp} for more details).

\begin{figure}
\centering
\resizebox{7.5cm}{!}{\rotatebox{0}{\includegraphics{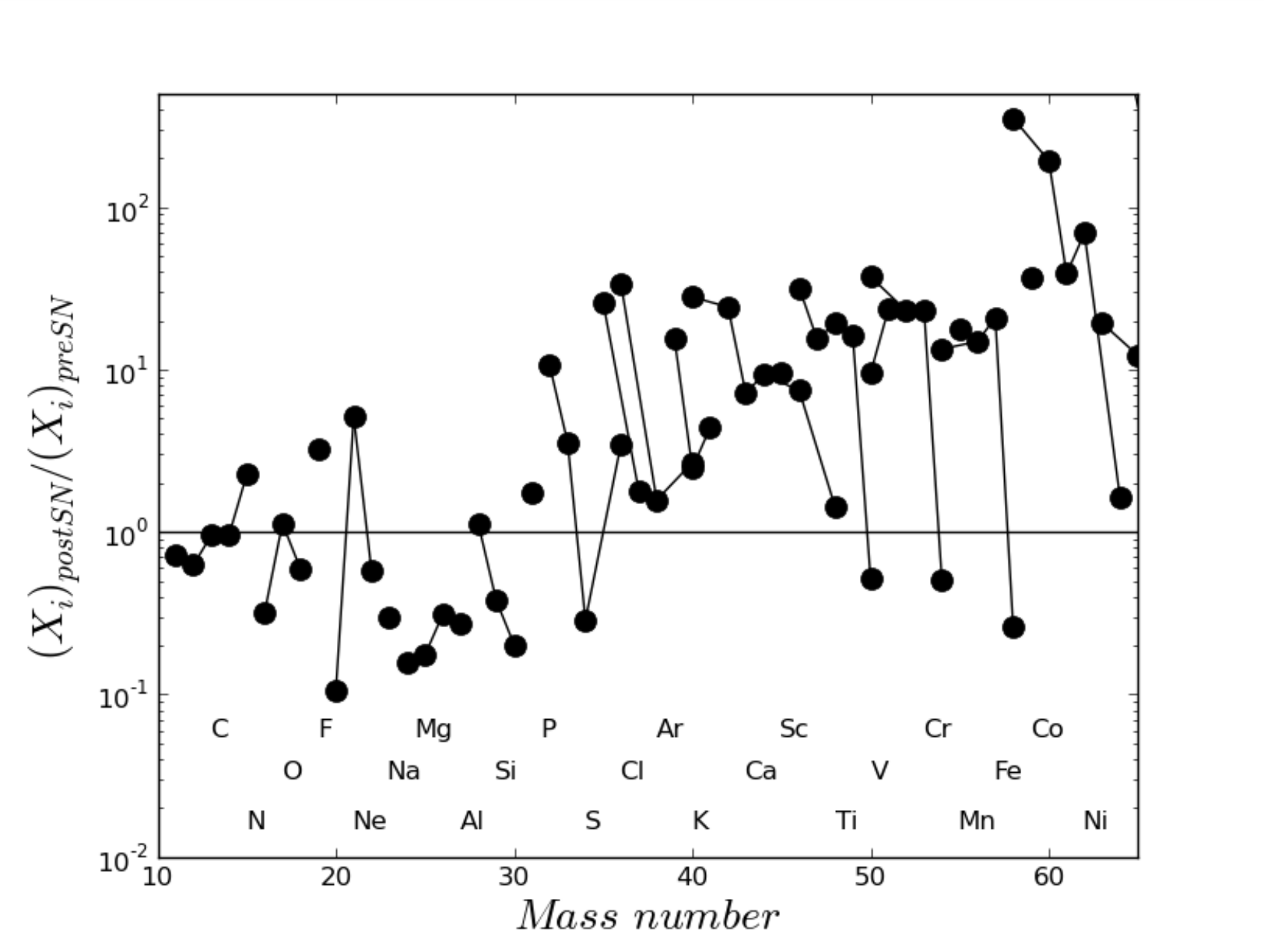}}}
\resizebox{7.5cm}{!}{\rotatebox{0}{\includegraphics{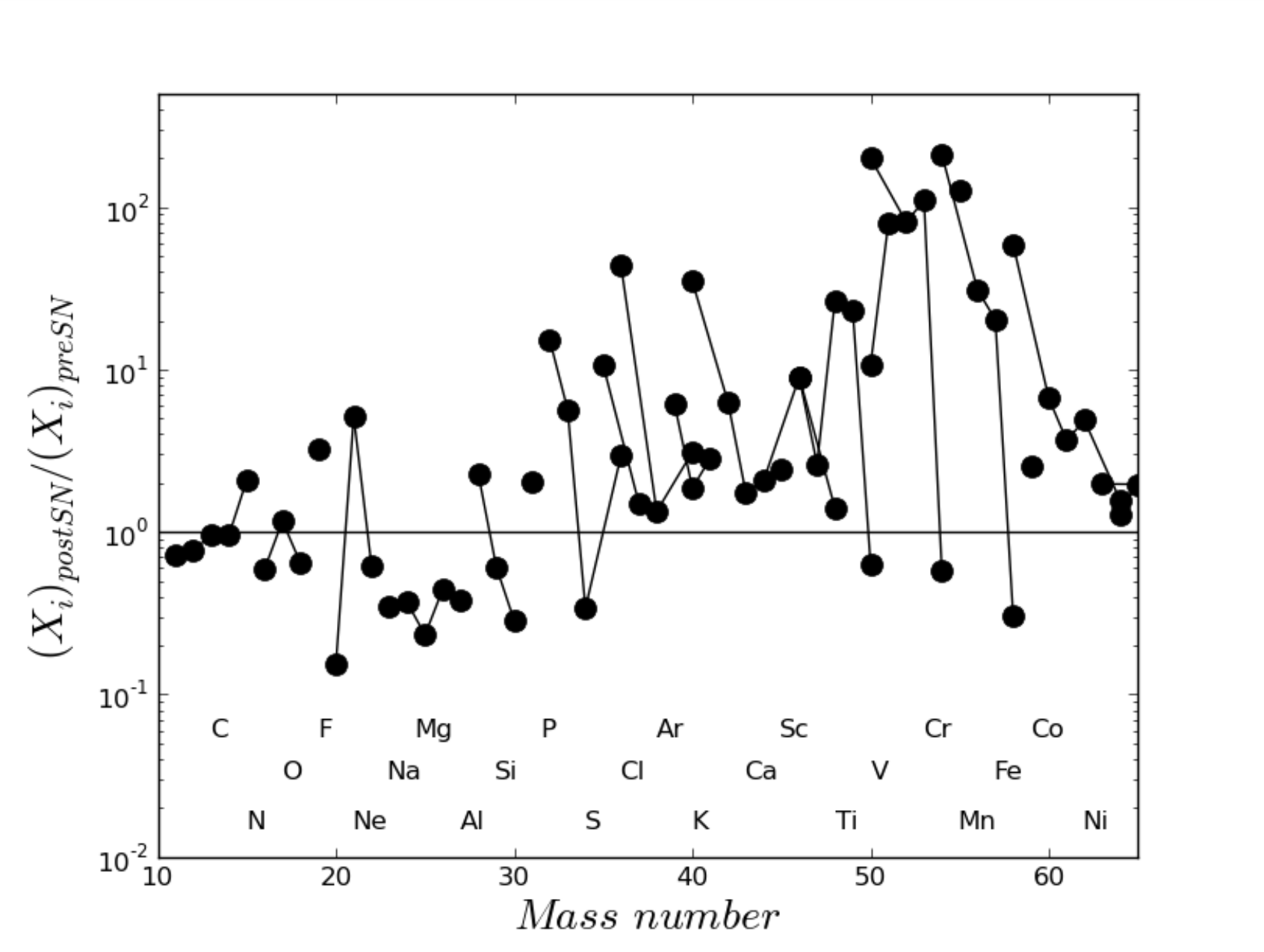}}}
\resizebox{7.5cm}{!}{\rotatebox{0}{\includegraphics{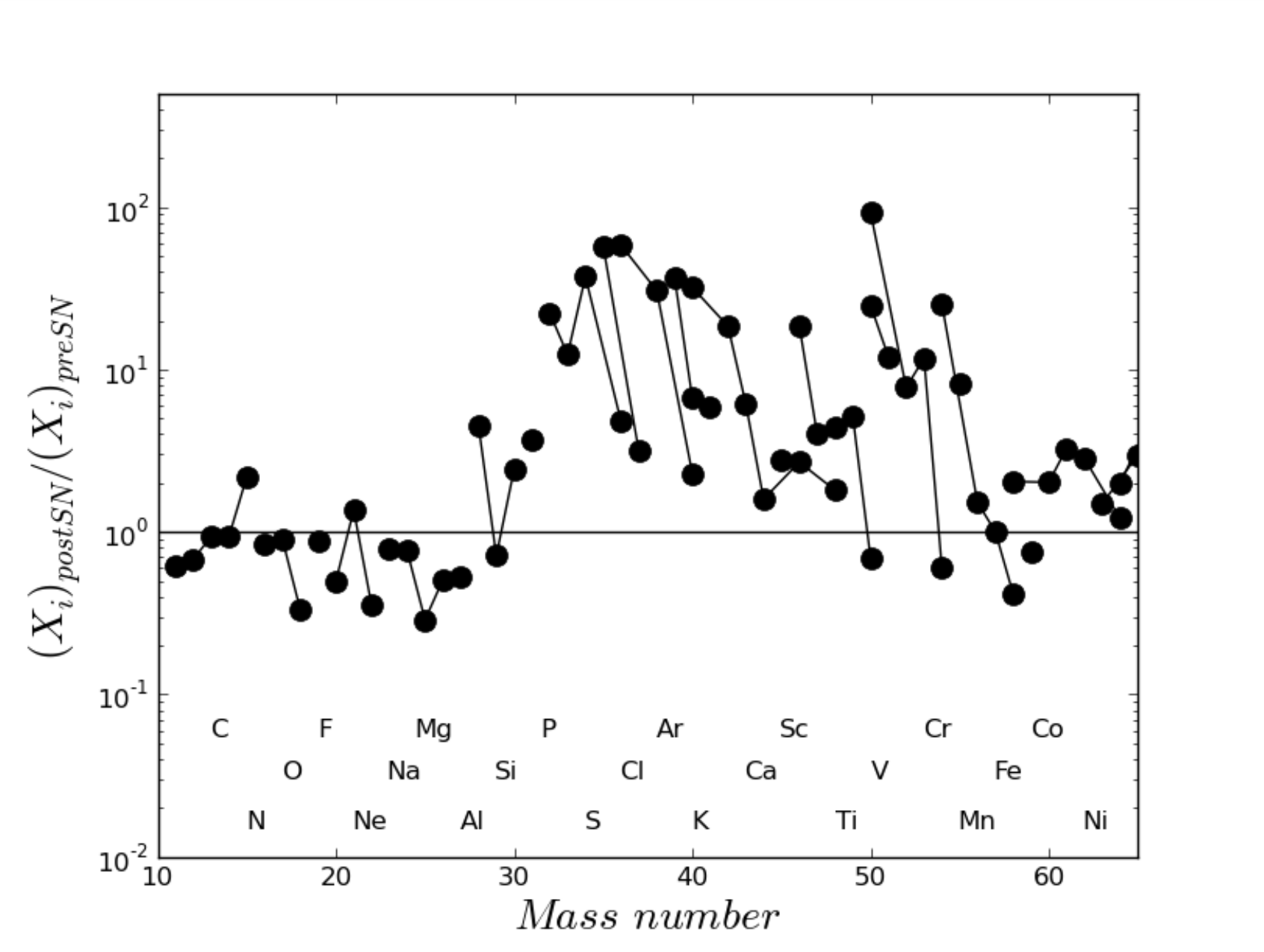}}}
%\resizebox{7.5cm}{!}{\rotatebox{0}{\includegraphics{20_post_pre_comparison_set1p2_d.pdf}}}
%\resizebox{7.5cm}{!}{\rotatebox{0}{\includegraphics{summary_exp_25_set1p2_r.pdf}}}
\resizebox{7.5cm}{!}{\rotatebox{0}{\includegraphics{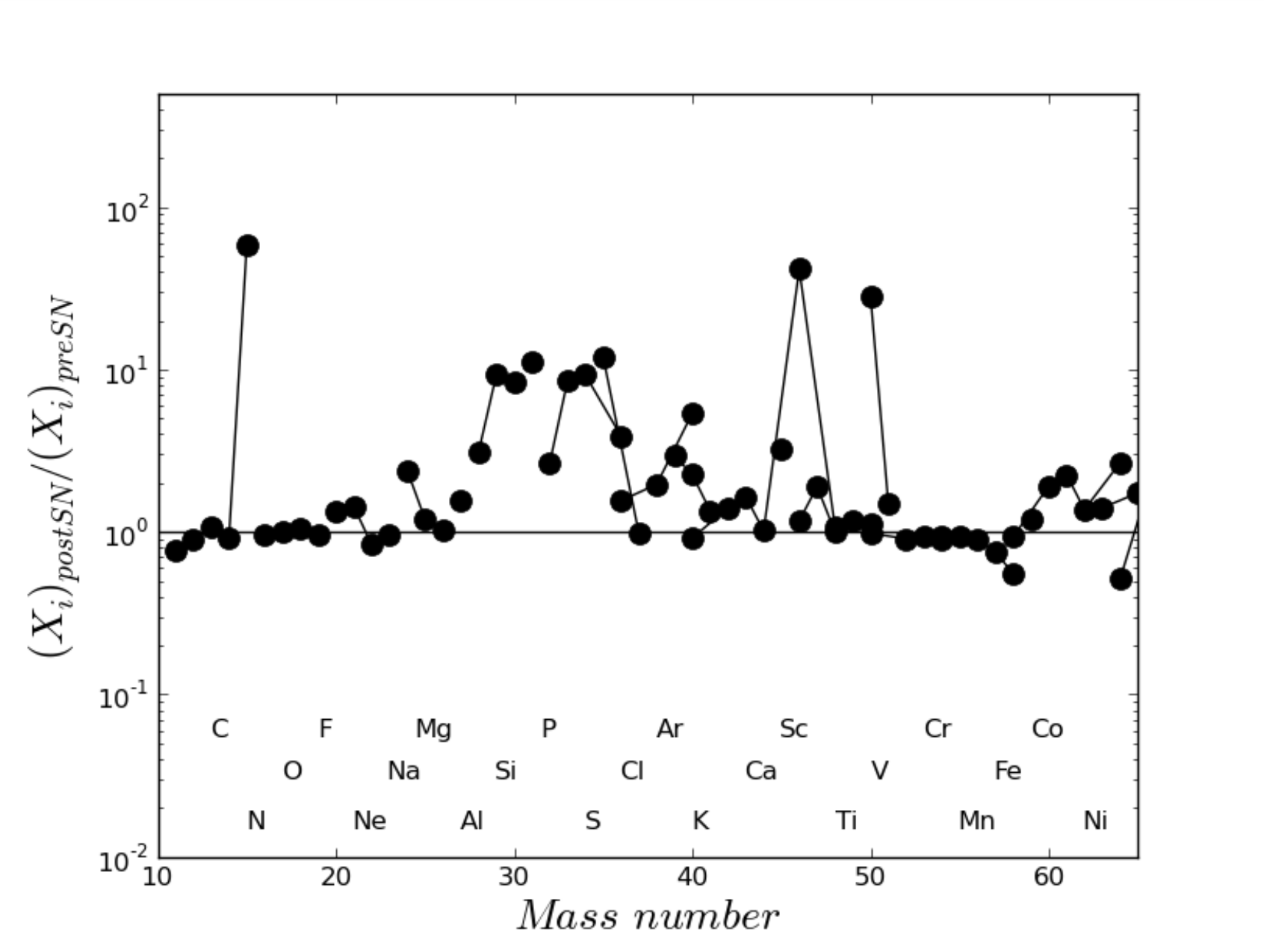}}}
\caption{
  Final isotopic distribution between
  C and Ni after the explosion are compared to
  pre-explosive abundances for the models in
  \fig{fig:summary_exp_set1p2}:
  two $15\msun$ models with $delayed$
  SN explosion and $rapid$/4
  (where the shock velocity from explosion $rapid$
  is reduced by a factor 4), a $20\msun$ and a
  $25\msun$ models with $delayed$ SN explosion.
  For a detailed comparison for all the species and for
  all the models we refer to the complete online
  tables \ref{tab:isotopic_prodfac_set1.2_exp}
  and \ref{tab:isotopic_prodfac_set1.2_pre} for \setopt,
  and tables \ref{tab:isotopic_prodfac_set1.1_exp}
  and \ref{tab:isotopic_prodfac_set1.1_pre} for \setopo.
}
\label{fig:post_versus_pre_set1p2}
\end{figure}

The analysis of nucleosynthesis in one-dimensional explosion simulations provides
fundamental information that is required to understand how species are formed or modified
under these extreme conditions \citep[e.g.,][]{woosley:95}. 
%We note again that the SN explosions are treated in a simplistic way (see \sect{subsec:exp}). 
%[SJ: I'm not sure you need to mention again the simplicity of our explosions---I'm also not convinced that what we do is really *that* much simpler than what is done in KEPLER and other models, though I agree that it is very simple compared to ``real'' simulations of supernovae : ) ].
The primary goal of our SN yield calculations is to estimate which
elements and isotopes would be strongly affected by explosive
nucleosynthesis in the CCSN. An overview of this information is available in
\fig{fig:post_versus_pre_set1p2} for a selection of models. At a given shock density our explosions feature
shock temperatures larger than usual 1-D CCSN simulations \citep[e.g.,][]{woosley:95}, and our models therefore
give some insight into the yields of such explosions. Complete tables
with pre-explosive and post-explosive abundances, overproduction factors,
production factors, yields in solar masses and net yields, as well as the thermodynamic histories
from these models, are available online (Appendix \ref{database}).
%Because of the intrinsic limitations of the present explosive yields we recommend taking caution when using them in, e.g., galactic chemical evolution simulations. While these data can be representative for a number of elemental and/or isotopic ratios, they generally depend on the initial choices made for the explosion simulations (\sect{stars}).
Despite the intrinsic limitations of 1D SN yields, these data can provide already important insights for a number of elemental and isotopic ratios. On the other hand, they should also be used as diagnostic tools to derive constraints for more realistic multi-dimensional hydrodynamics CCSN simulations, and study %in order to derive constraints about 
e.g., the CCSN engine and the SN-shock propagation producing these yields \citep[e.g.][]{hix:14,wongwathanarat:15}.

Based on our calculations we present in the following a
discussion of the different element groups and their production in
different mass regimes and evolution phases. 
%We refer to, e.g., \cite{woosley:73,arnett:85,thielemann:85,woosley:95,thielemann:96,chieffi:98,limongi:00,rauscher:02,woosley:02,nomoto:13}for similar analyses and discussions; while these discussions focus only on massive stars, we consider also the nucleosynthesis in low-mass and intermediate-mass stars \citep[e.g.,][]{bisterzo:10,cristallo:11,ventura:13,karakas:14}. 
There is a comprehensive literature for the nucleosynthesis in massive
stars \citep[][]{woosley:73,arnett:85,thielemann:85,woosley:95,thielemann:96,chieffi:98,limongi:00,rauscher:02,woosley:02,nomoto:13}
as well as for low and intermediate mass stars \citep[e.g.,][]{bisterzo:10,cristallo:11,ventura:13,karakas:14,cristallo:15}.
The Solar System abundances are comprised of contributions from different stellar sources. In our analysis we compare the production of the same isotope in different types of stars.

The discussion will
follow the yield plots (Figs. \ref{fig:CNONaAl_set1p2} to
\ref{fig:fe_set1p2_sum}) for \setopt. Similar plots are available
online for all stable isotopes and elements for both
metallicities. The yield plots show the weighted stellar yields in the
following sense. For each initial mass the ejected amount (during the
wind as well as during the final SN or wind ejection as appropriate)
in solar masses is weighted by a Salpeter IMF 
($\alpha$ exponent = 2.35) sampled by non-uniform initial mass intervals, 
normalized to $1\msun$, and represented by a
dashed black line in the yield plots.
%The initial mass intervals are
%chosen to represent initial masses with similar nucleosynthetic
%production mechanisms compared to the available stellar evolution
%model representing it.
The initial mass intervals are chosen in such a way that initial masses in the same interval
are considered to possess similar nucleosynthetic production mechanisms that are represented
by one of the stellar models in our set.
The dashed line corresponds to 
%zero yields, i.e. to 
the return of the same amount of material that was present in the star 
from the initial abundance distribution. 
A yield line above or below the dashed line thus corresponds
to production and destruction, respectively. These plots therefore
allow us to compare the contribution from stars with different initial masses through their
production factors (the ratio of the yield line with the IMF line) as
well as the relative importance of the contributing mass range (via
the difference of the yield line and the IMF line) under the
assumption that stars of all masses have enough time to return their
winds and ejecta.  While low- and intermediate mass stars eject all
their yields during the wind phase (into which even a rapid superwind
phase at the end is included), we distinguish for the massive stars
between contributions from different processes; the wind yields are
the ejecta returned during the pre-SN stellar evolution mass
loss; the pre-SN contribution is an imaginary component that
represents the ejecta that the SN would mechanically expel
without any explosive nucleosynthesis. 
It is basically the integral of the to be ejected layers just before the explosion. 
For the SN contribution 
%including the explosive nucleosynthesis 
different options are shown, reflecting some of the uncertainties in modeling
the explosions. 
Notice here that the explosive contribution is separated from the wind contribution, as in \tabs{tab:isotopic_prodfac_set1.2_exp}, \ref{tab:element_prodfac_set1.2_exp}, \ref{tab:isotopic_prodfac_set1.1_exp} and \ref{tab:element_prodfac_set1.1_exp}.
In other words, these figures show the wind yields and the explosive yields weighted over the Salpeter initial mass function, providing the stellar yields representative of each mass range.
In this work we do not include models representative
for the mass range $7 - 11 \msun$.
In such a range there are super-AGB stars, electron-capture supernovae
and the lowest mass iron-core collapse supernovae \citep[][]{jones:13}.
Therefore, in \fig{fig:CNONaAl_set1p2} to 
\ref{fig:fe_set1p2_sum} this mass range is shaded.  
%and \ref{fig:summary_exp_set1.2_15_energytest}.

The production of Li, Be, and B is not 
fully available in this release, since
our stellar models miss some important physics processes that contribute
to their their nucleosynthesis. Li production from intermediate mass
stars through Hot Bottom Burning (HBB) during the AGB phase
\citep[initial mass higher than $\sim4\msun$,
e.g.,][]{lattanzio:99} is present in the 4 and $5\msun$ models
%\footnote{Model predictions for Li have to be taken from the \MESA\ profile output which was computed with coupled mixing and nuclear burning operators. The \mppnp\ post-processing output employs an operator split which does not accurately resolve the Cameron-Fowler transport mechanism with the present time stepping algorithm. 
%%This will be addressed in future releases.}. 
Model predictions for Li have to be taken from the \MESA\ profile output which was computed with coupled mixing and nuclear burning operators. The \mppnp\ post-processing output employs an operator split which does not accurately resolve the Cameron-Fowler transport mechanism with the present time stepping algorithm.
A finer mass grid is required, however, for a thorough characterization of HBB Li yields. 
Li may also be produced as a result of extra-mixing (the so-called cool bottom process) in AGB and RGB stars with lower initial masses \citep{sackmann:99,nollett:03,denissenkov:11,palmerini:11}. Such non-standard mixing processes are not included in this model generation.
Furthermore, in these stars Li predictions are also quite uncertain, as shown by \cite{lattanzio:15}. Indeed, by comparing the results from different codes (including \MESA) \cite{lattanzio:15} show that Li is drastically affected by e.g., the time-step criterion and spatial mesh refinement, and that a preliminary convergence analysis need to be done before safely using Li stellar yields.

Production of Be and B in stars is mostly due to neutrino irradiation
on \isotope[4]{He} and \isotope[12]{C} respectively, during CCSN
\citep[e.g.,][]{woosley:02,nakamura:10,banerjee:13} and hypernovae 
\citep[][]{fields:02}. 
In the present models we do not
include neutrino nucleosynthesis.

\subsection{C, N, and O}
\label{c-n-o}

C is efficiently produced by both low-mass and massive stars
\citep[e.g.,][]{goswami:00,woosley:02} in He shell burning. In massive
stars \czw\ can originate from the portion of He-core ashes which is
ejected by the SN explosion;
% \citep[e.g.,][]{thielemann:85}. 
a non-negligible contribution from Wolf-Rayet stars with masses larger
than 25-$30\msun$ has been suggested in order to reproduce carbon
abundances in the Galactic disk \citep[e.g.,][]{gustafsson:99}. In low
mass stars, \czw\ comes from the triple-$\alpha$ reaction in the He-shell
flash and is brought to the surface in the third dredge-up mixing
following the thermal pulse \citep[e.g.,][and references
therein]{herwig:05}.

\begin{figure} \centering
 \includegraphics[width=0.46\textwidth]{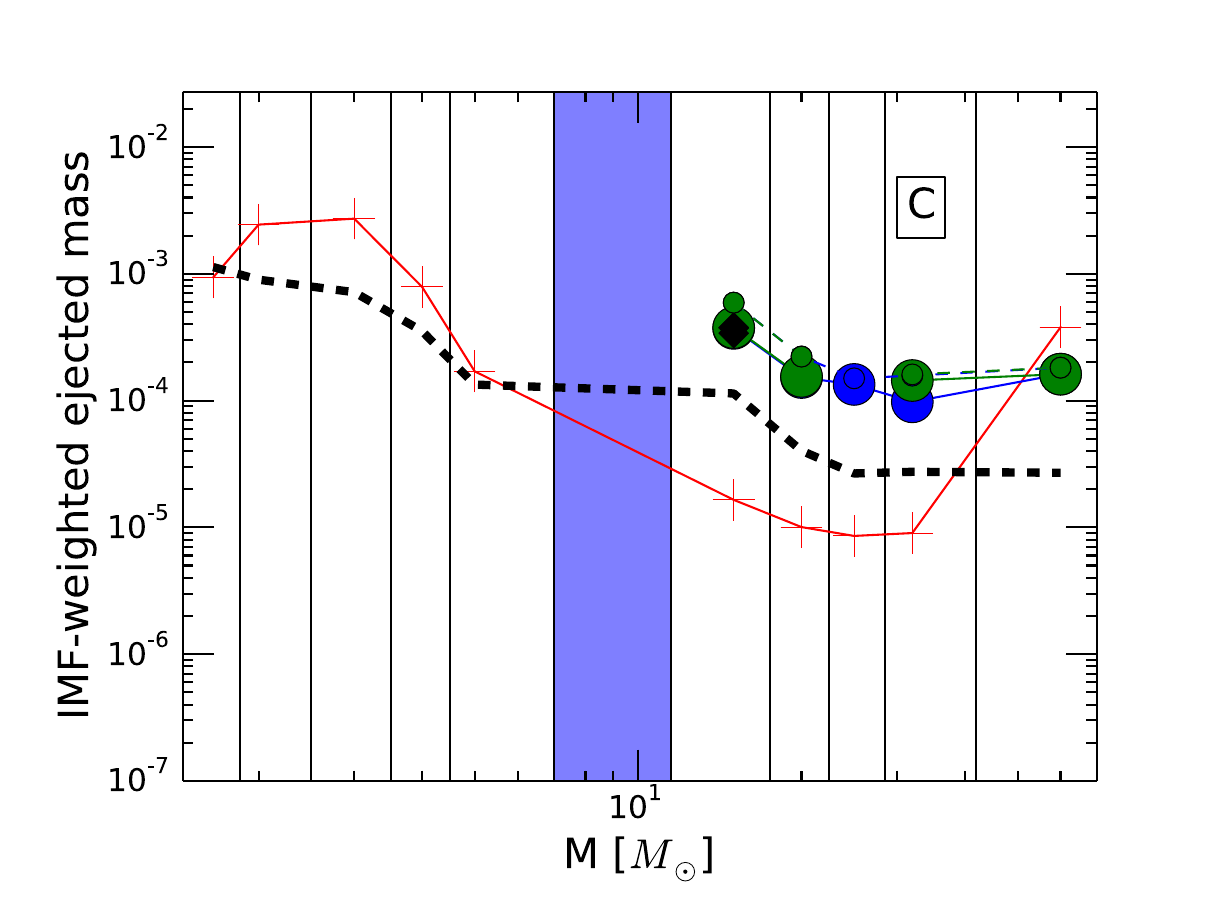}
 \includegraphics[width=0.46\textwidth]{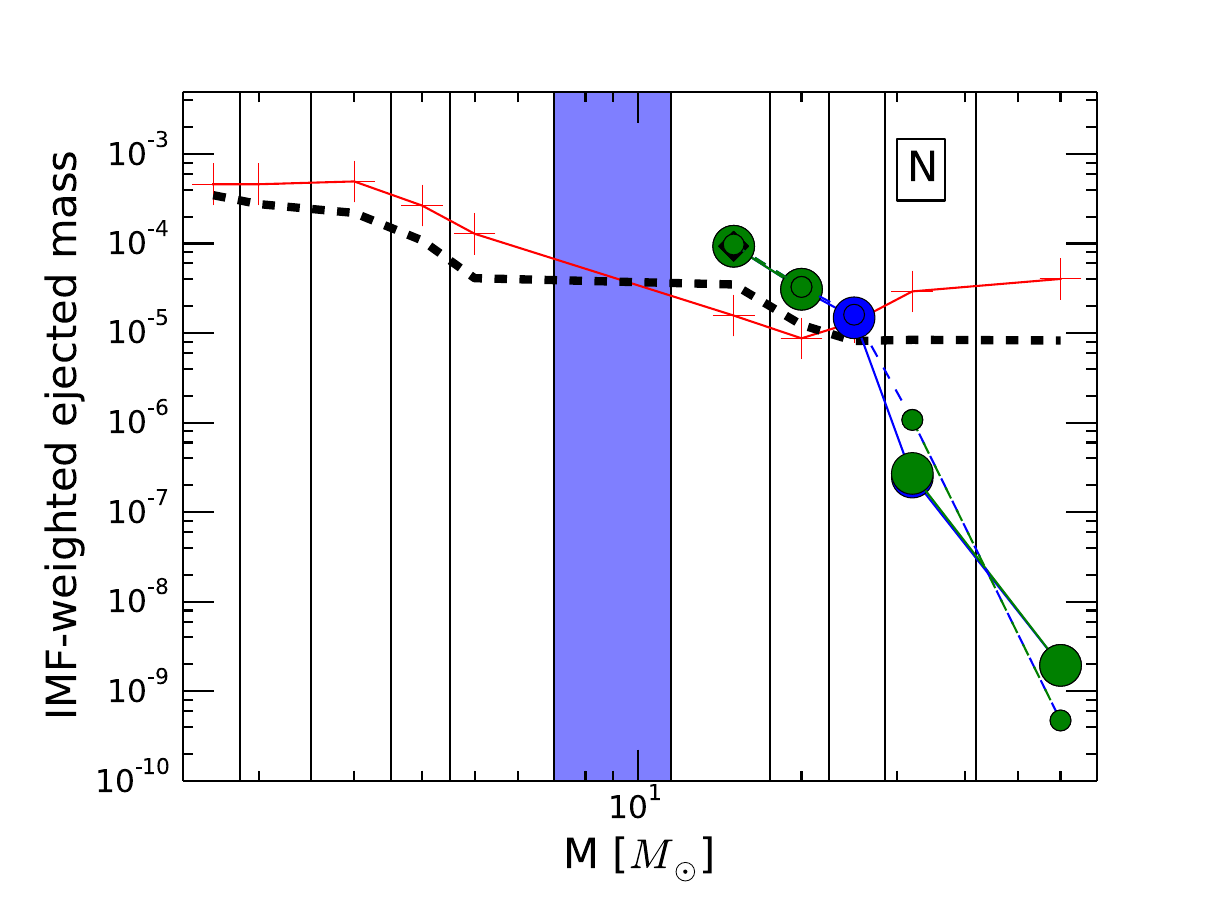}
 \includegraphics[width=0.46\textwidth]{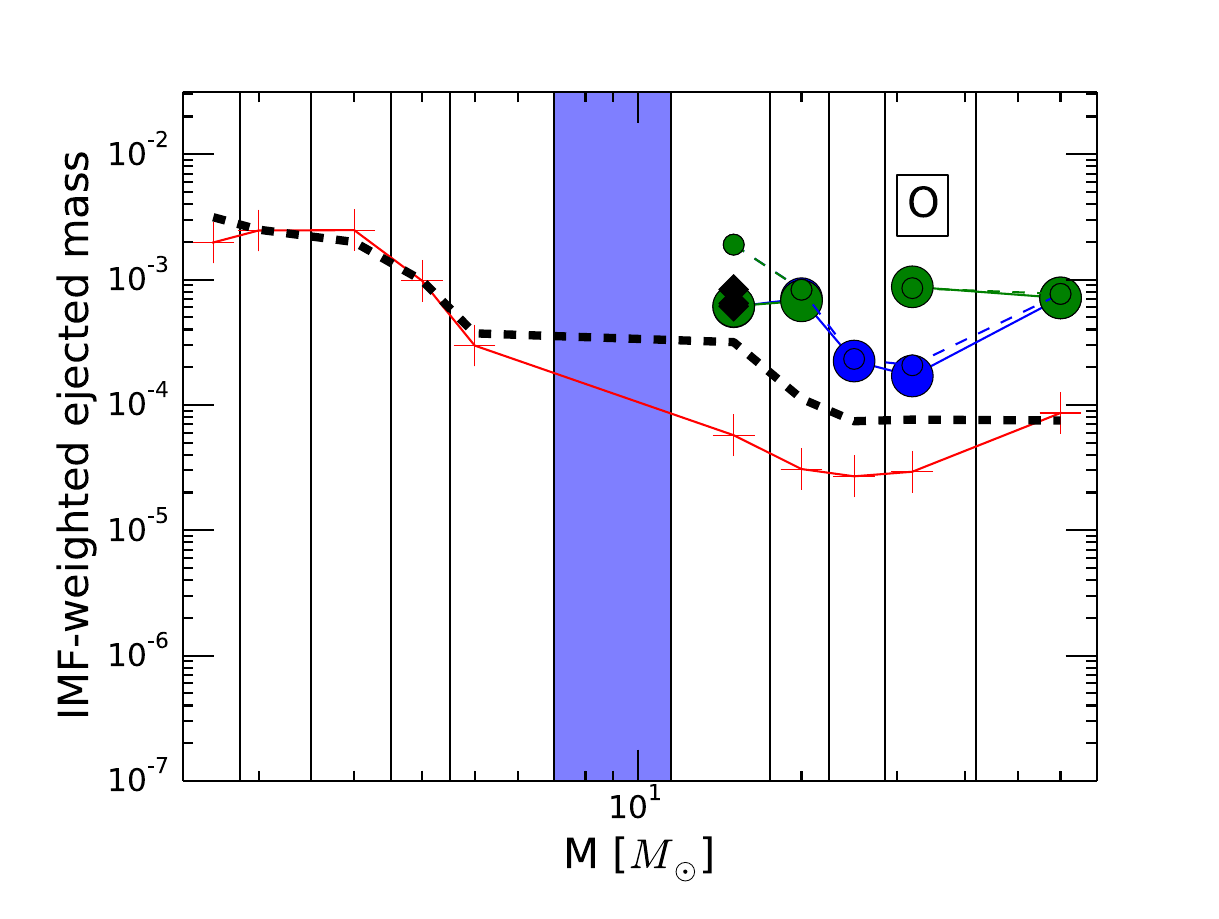}
 \includegraphics[width=0.46\textwidth]{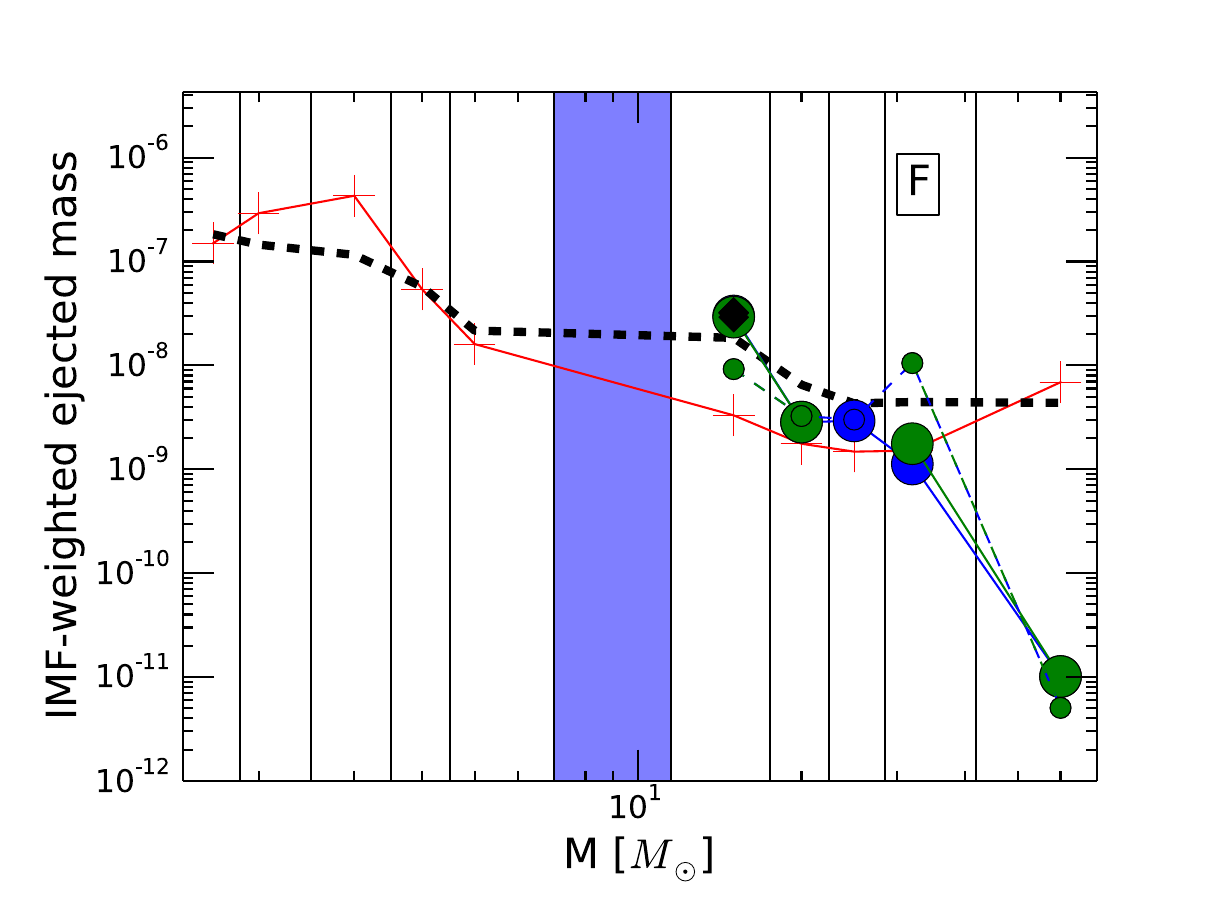}
 \includegraphics[width=0.46\textwidth]{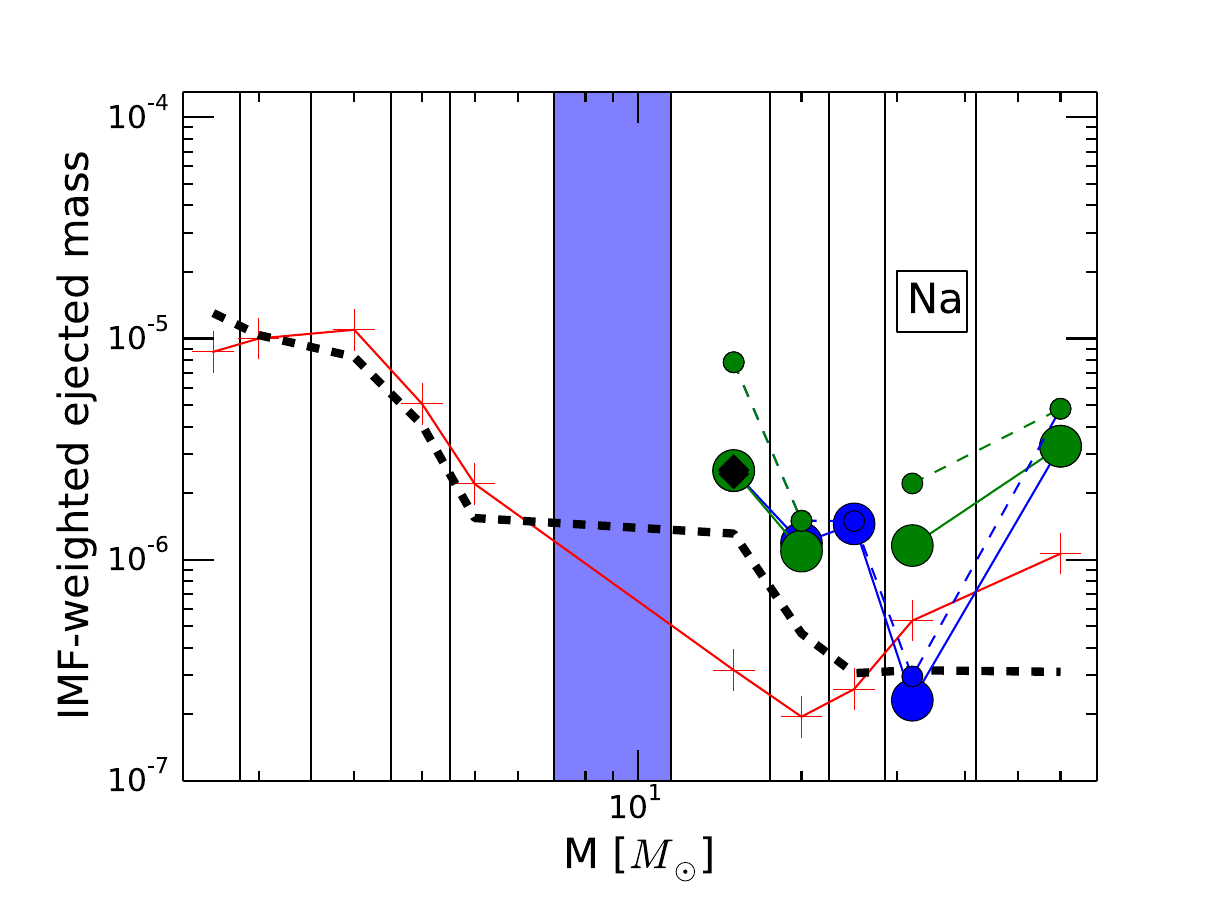}
 \includegraphics[width=0.46\textwidth]{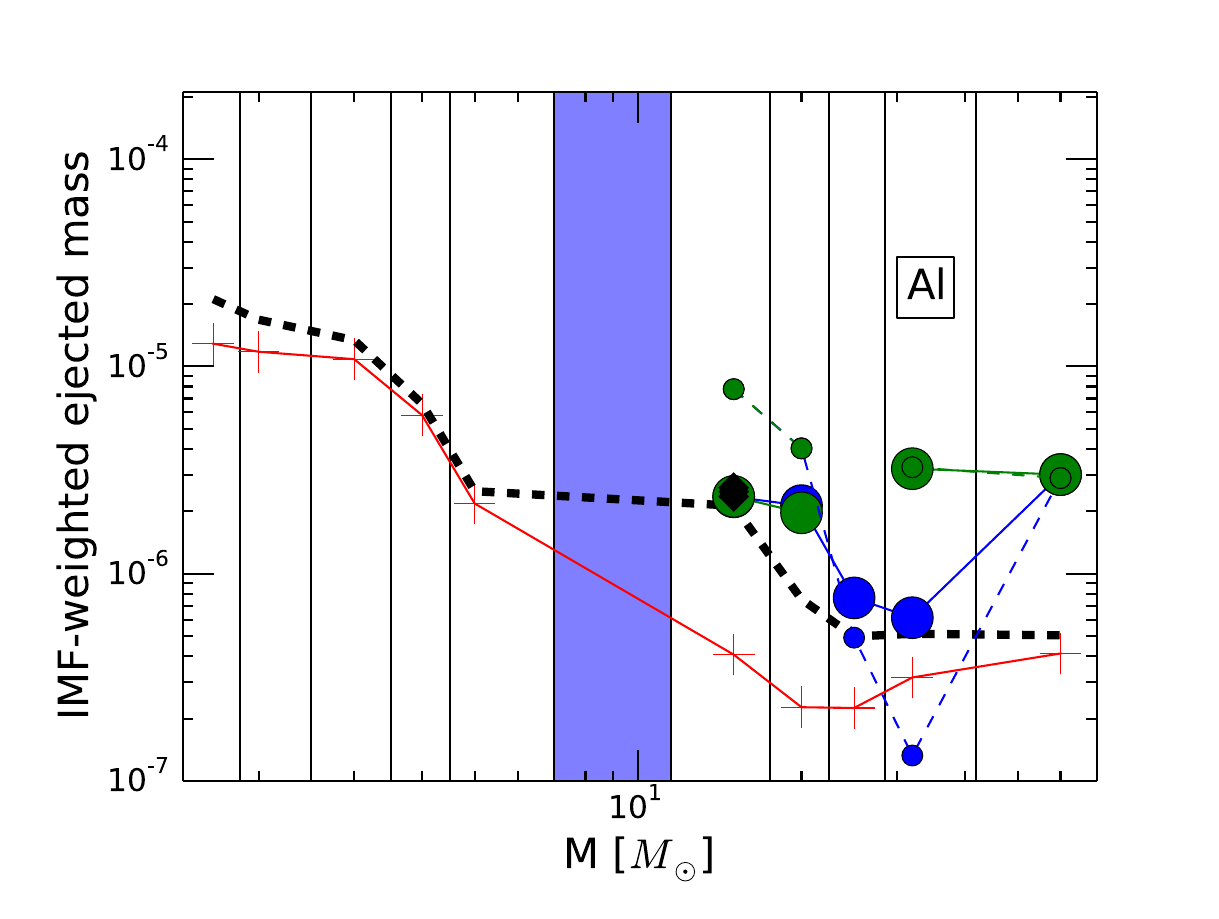}
\caption{ \label{fig:CNONaAl_set1p2}
IMF-weighted ejected masses in solar mass unit for different 
stellar masses from \setopt.  
The $\Delta$M limits 
used for the IMF weight are shown. No models representative of the mass range
$M = 7-11 \msun$ are considered (see text \sect{sec:nuc-code}). 
Big red crosses are the contribution by stellar winds. 
Small green and blue circles are the pre-supernova abundances, 
between the remnant mass and the surface of
the star when core collapse starts, associated to the SN fallback
prescription \emph{delay} and \emph{rapid} (cf.\ \sect{subsec:exp}). 
Green and blue large circles are the abundances including the explosive contributions
according to the two fallback assumptions. 
Black diamonds show the yields including \emph{rapid} SN with reduced 
explosion energy. In order to clarify if a
model has a positive contribution to the chemical enrichment of e.g., carbon,
we report the initial content for comparison (black-dashed line).
A positive production requires that the yield is larger than
the value given by the dashed line.
}
 \end{figure}

\begin{figure}
\centering
\includegraphics[width=0.46\textwidth]{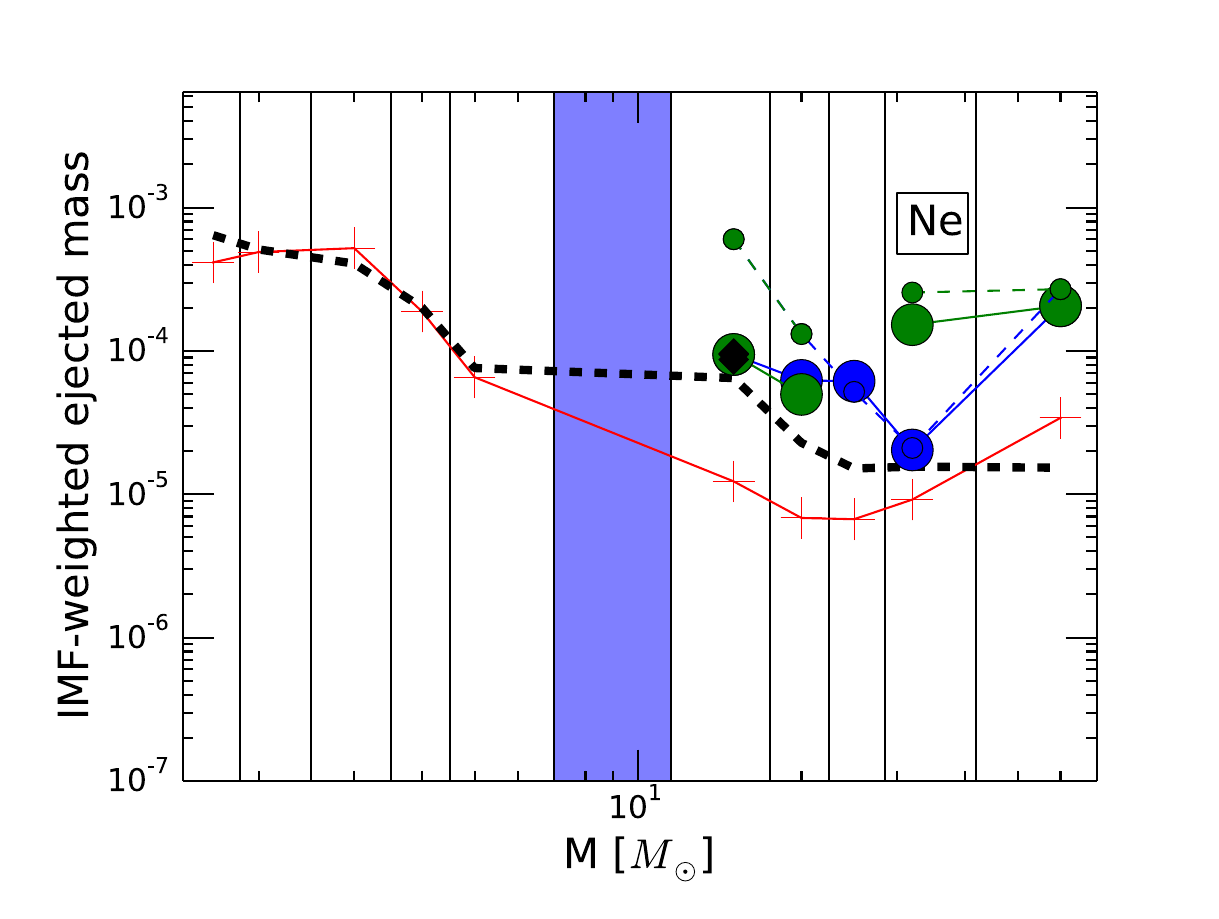}
\includegraphics[width=0.46\textwidth]{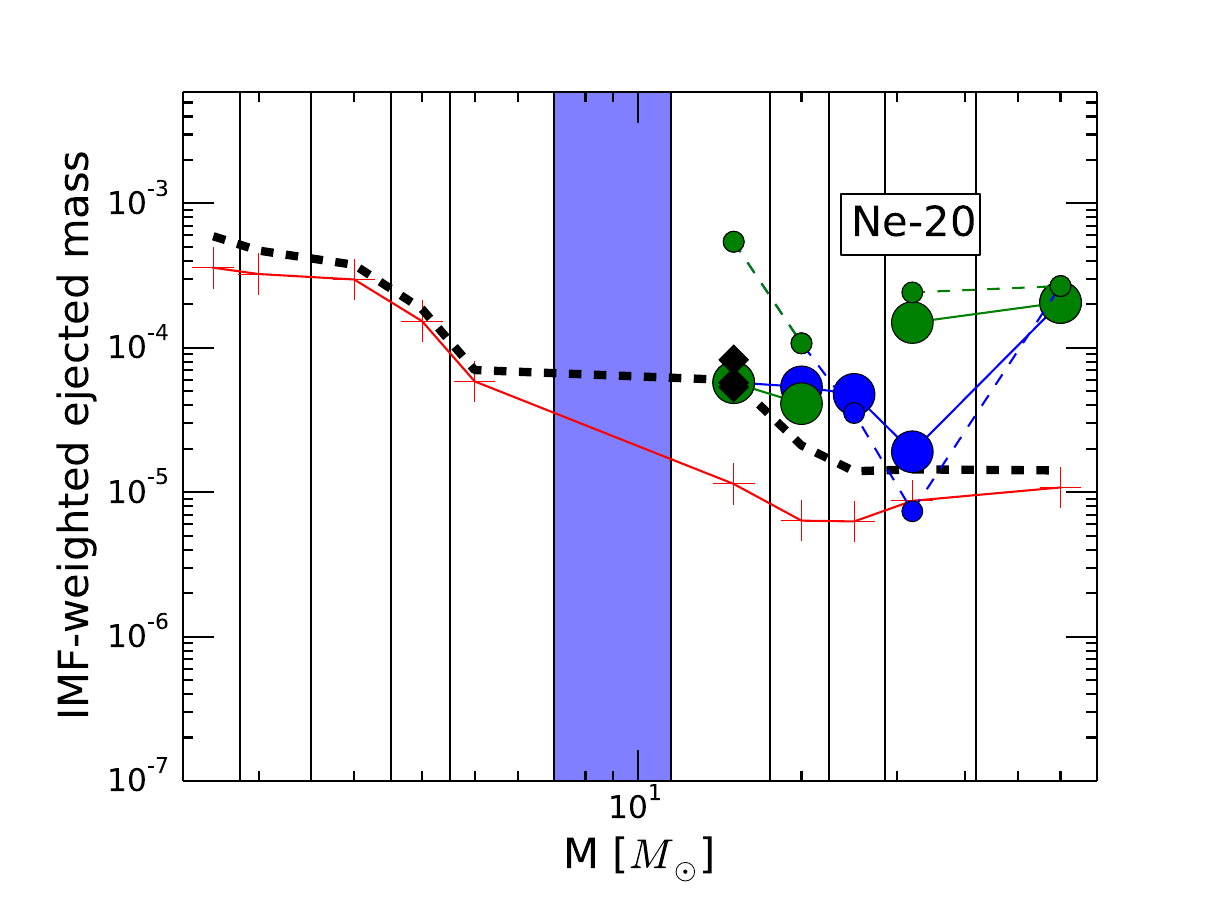}
\includegraphics[width=0.46\textwidth]{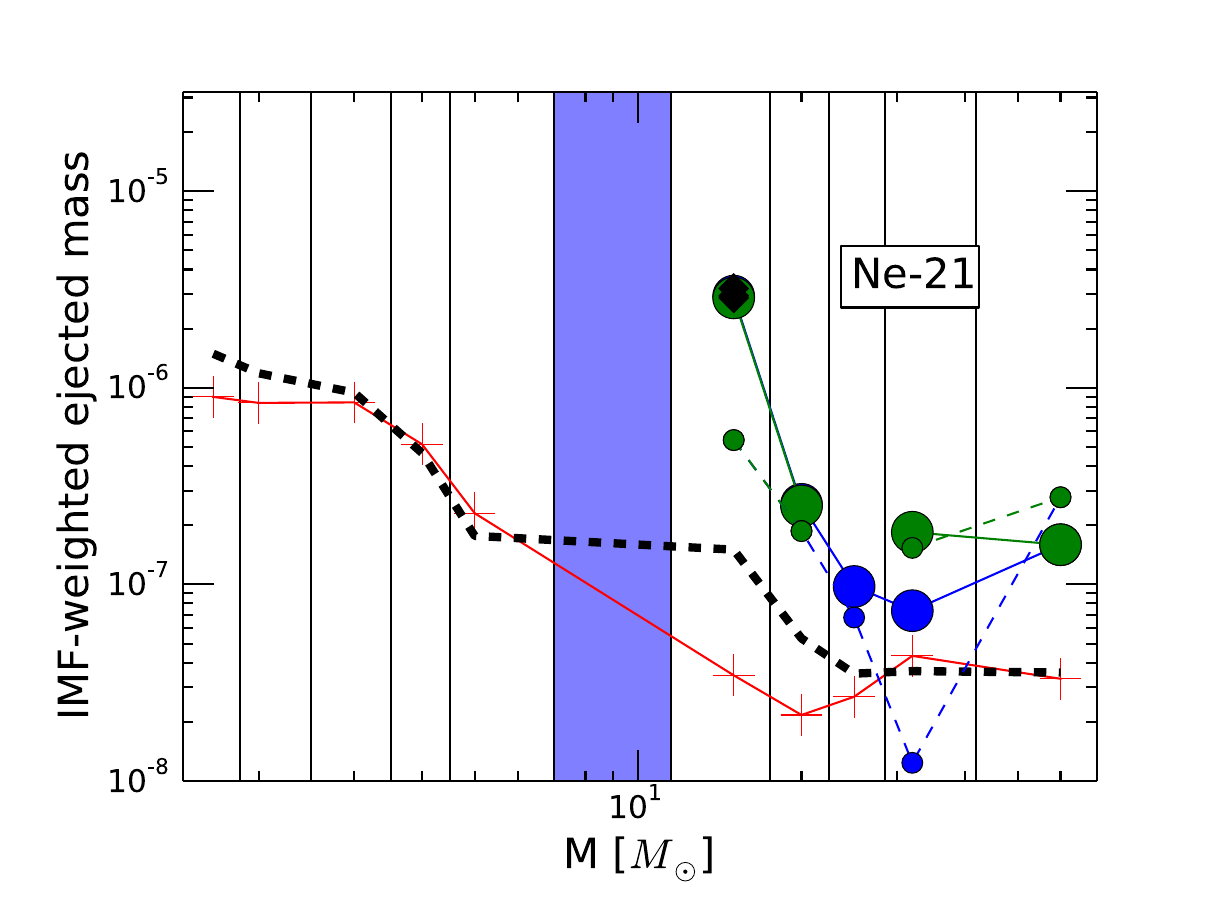}
\includegraphics[width=0.46\textwidth]{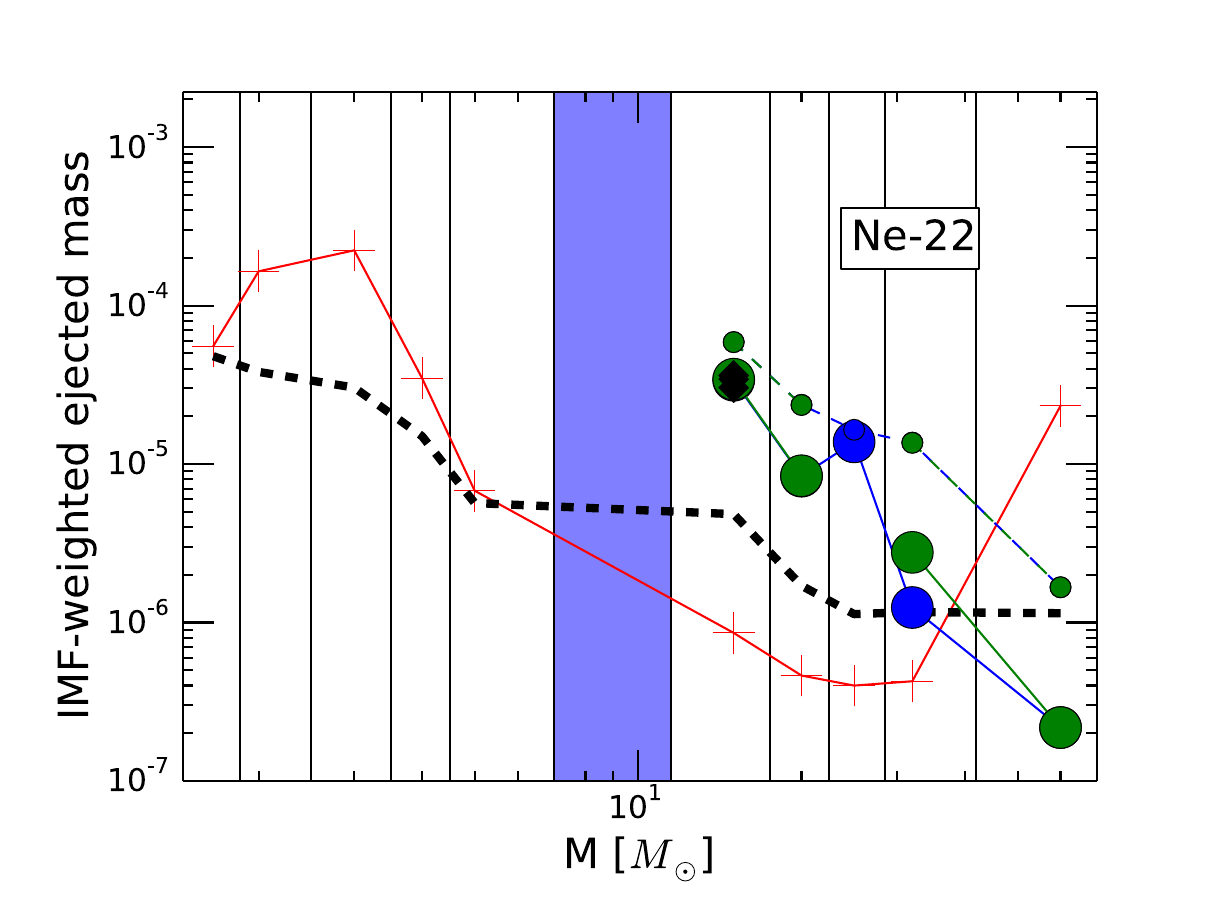}
\includegraphics[width=0.46\textwidth]{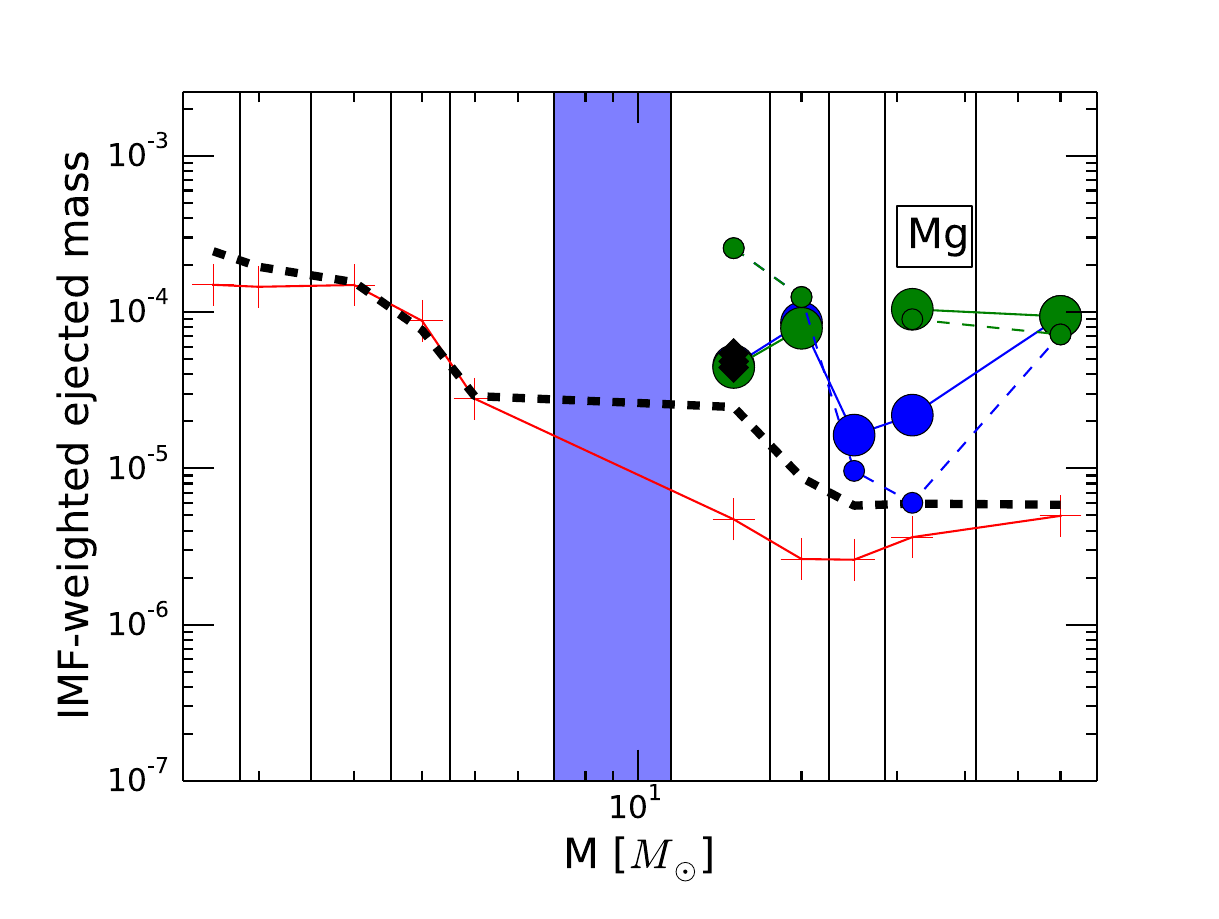}
\includegraphics[width=0.46\textwidth]{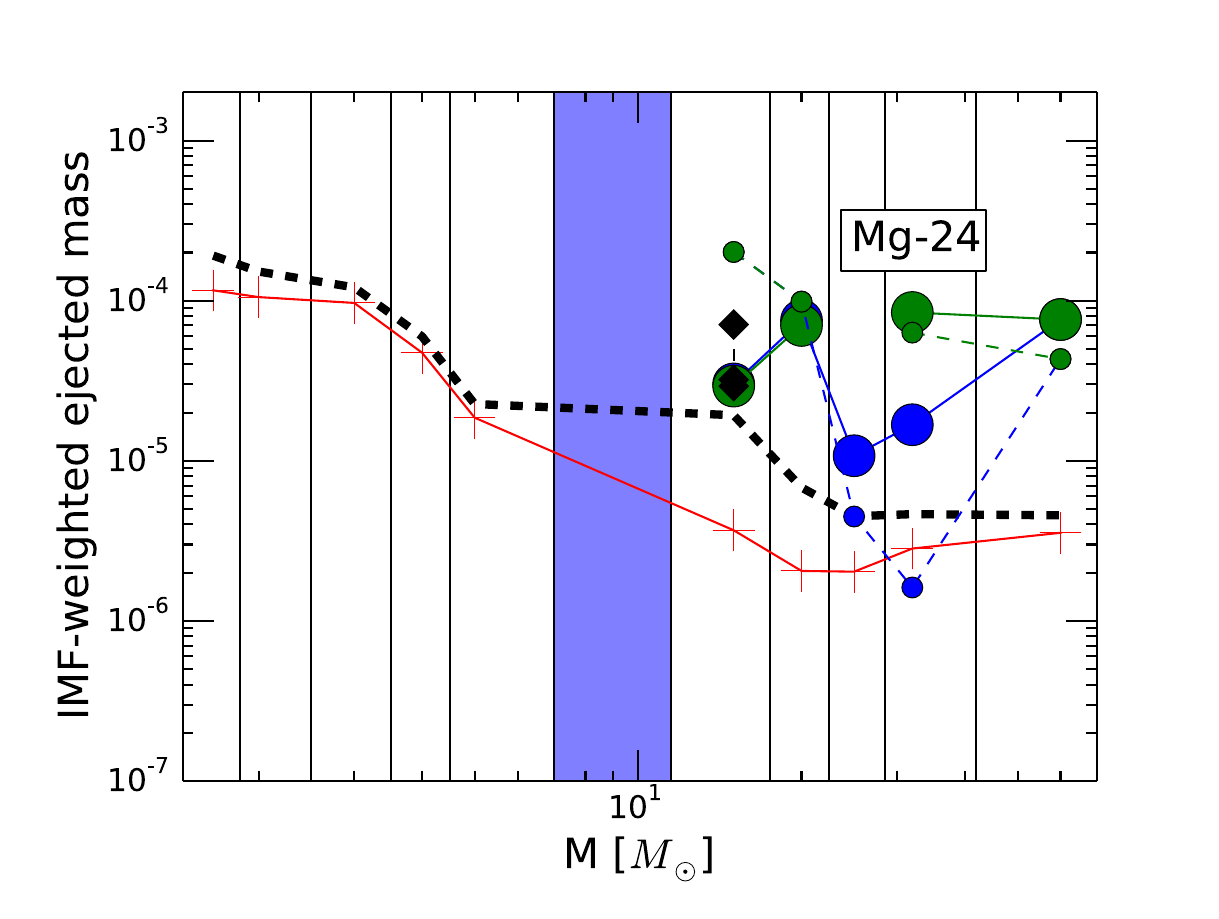}
\includegraphics[width=0.46\textwidth]{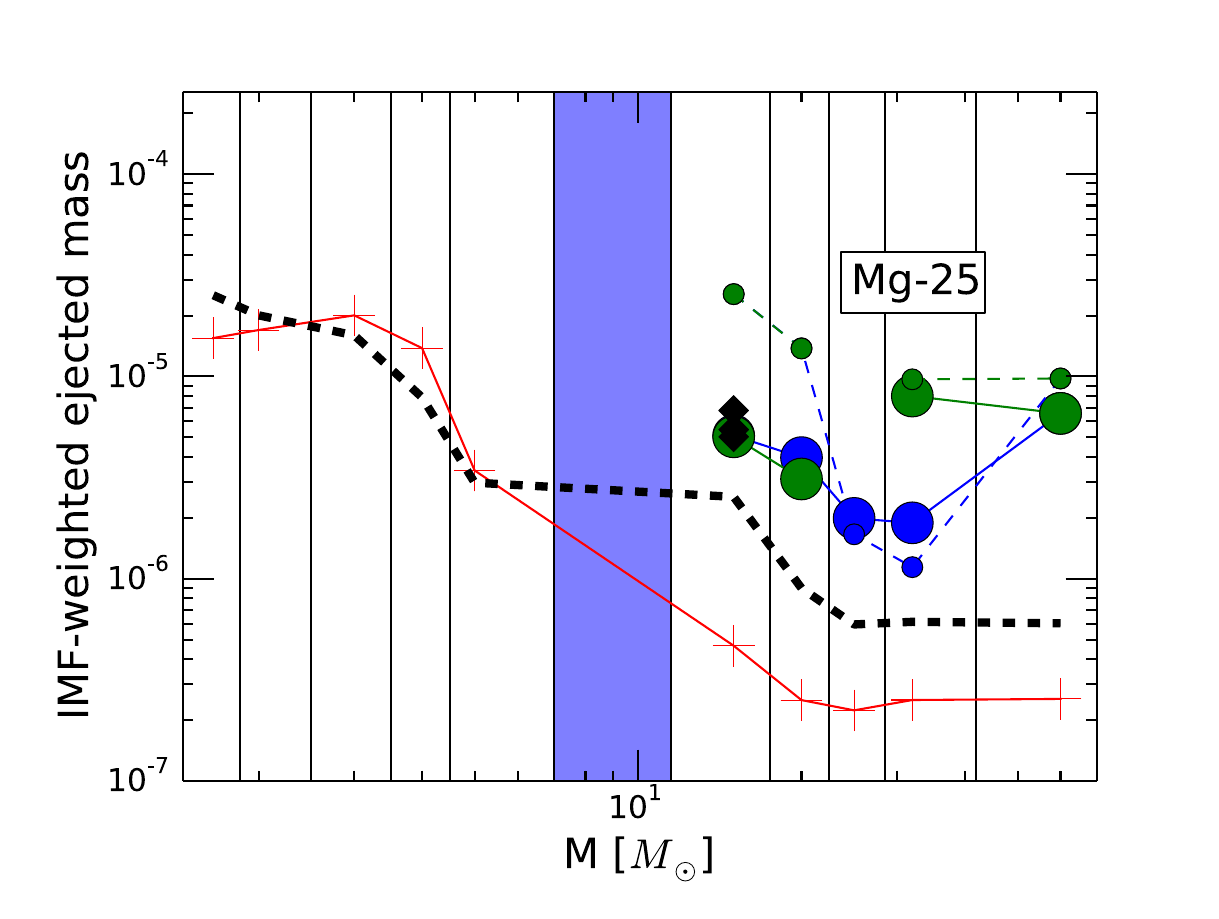}
\includegraphics[width=0.46\textwidth]{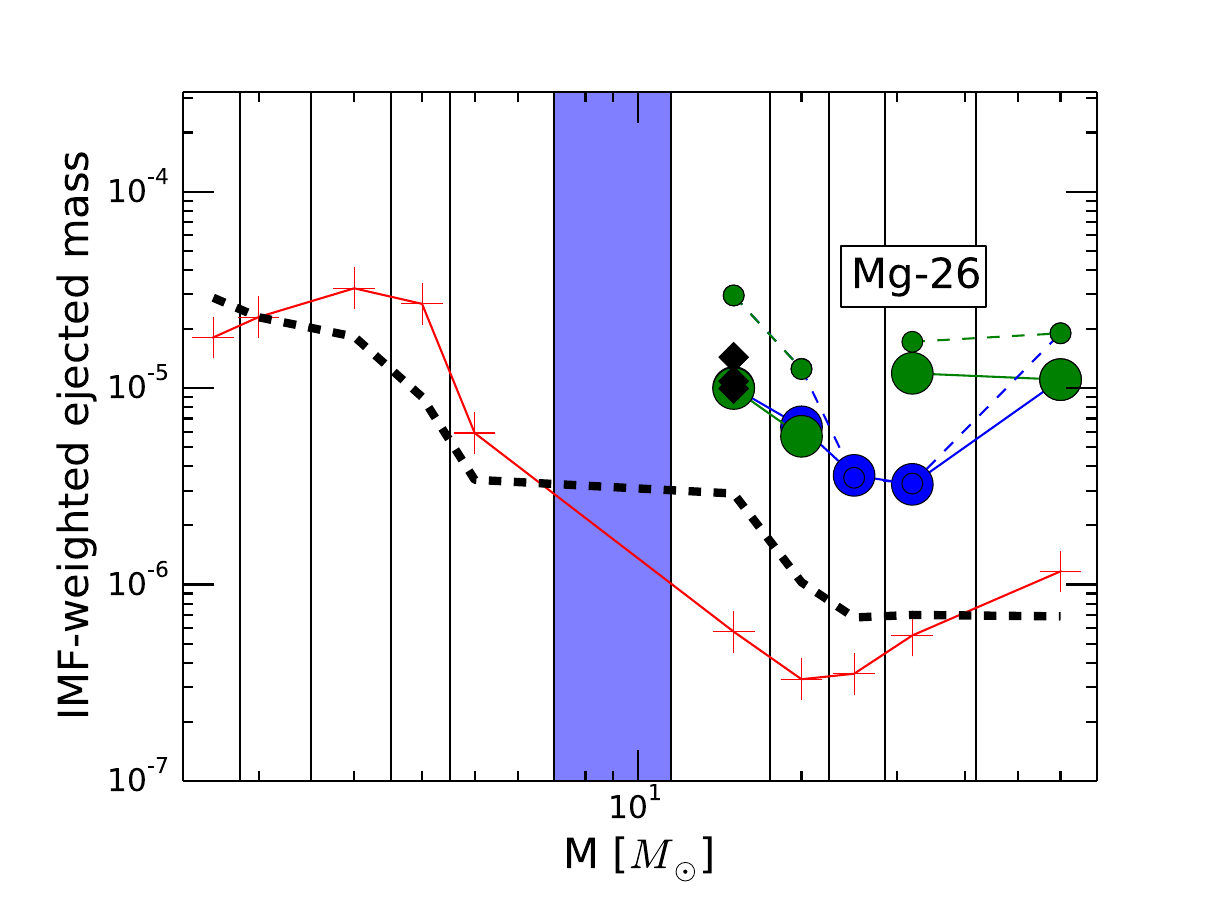}
\caption{Same as \fig{fig:CNONaAl_set1p2} for Ne and Mg and their stable isotopes.}
\label{fig:NeMg_set1p2_sum}
\end{figure}

\begin{figure}
\centering
\includegraphics[width=0.46\textwidth]{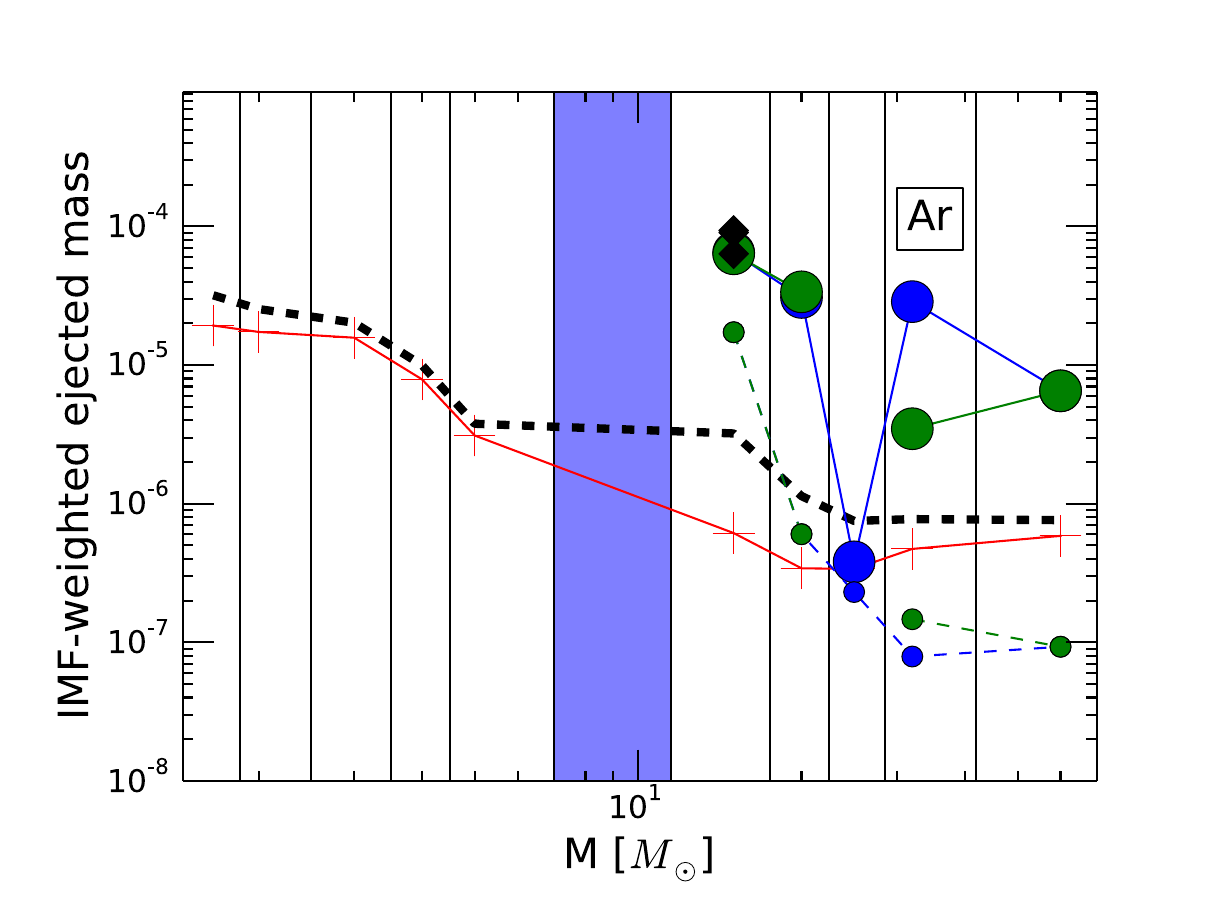}
\includegraphics[width=0.46\textwidth]{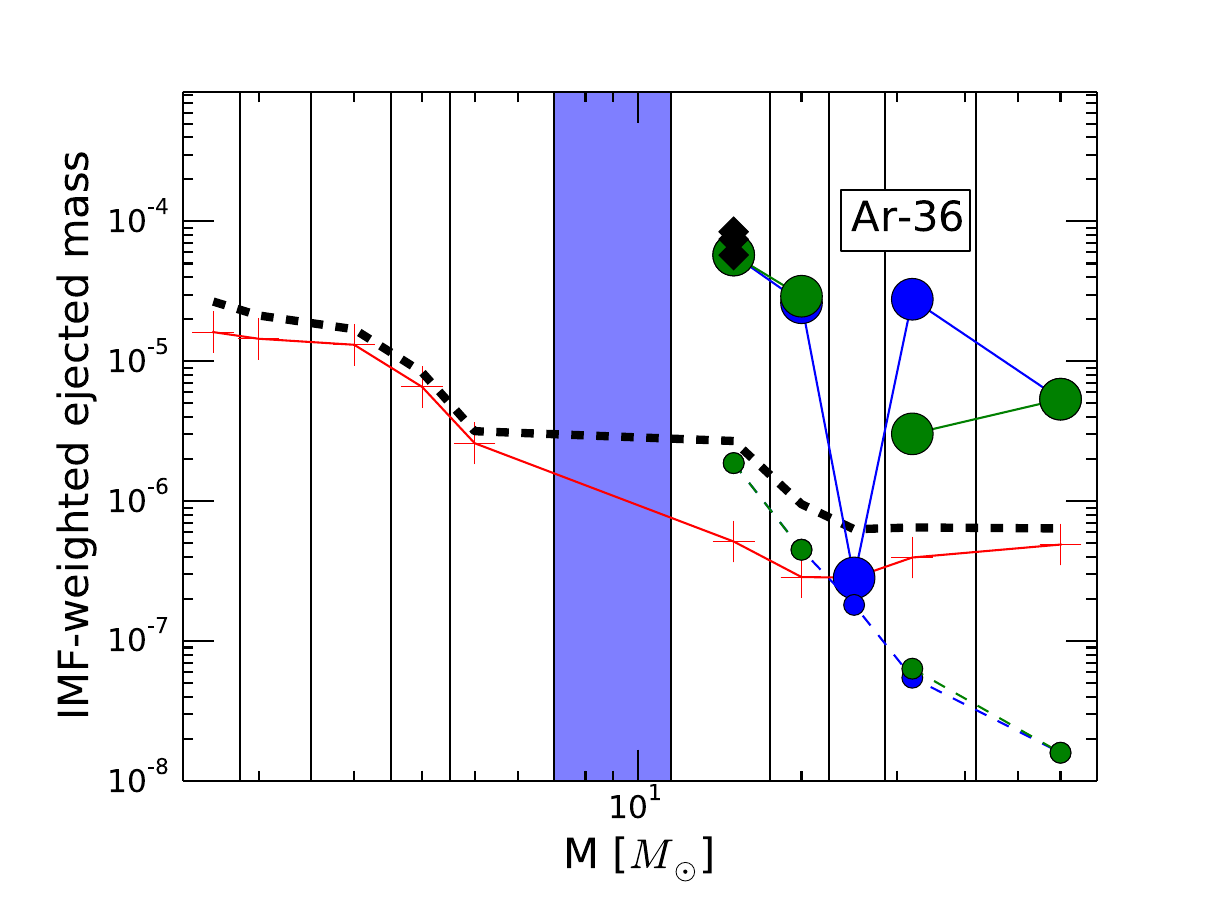}
\includegraphics[width=0.46\textwidth]{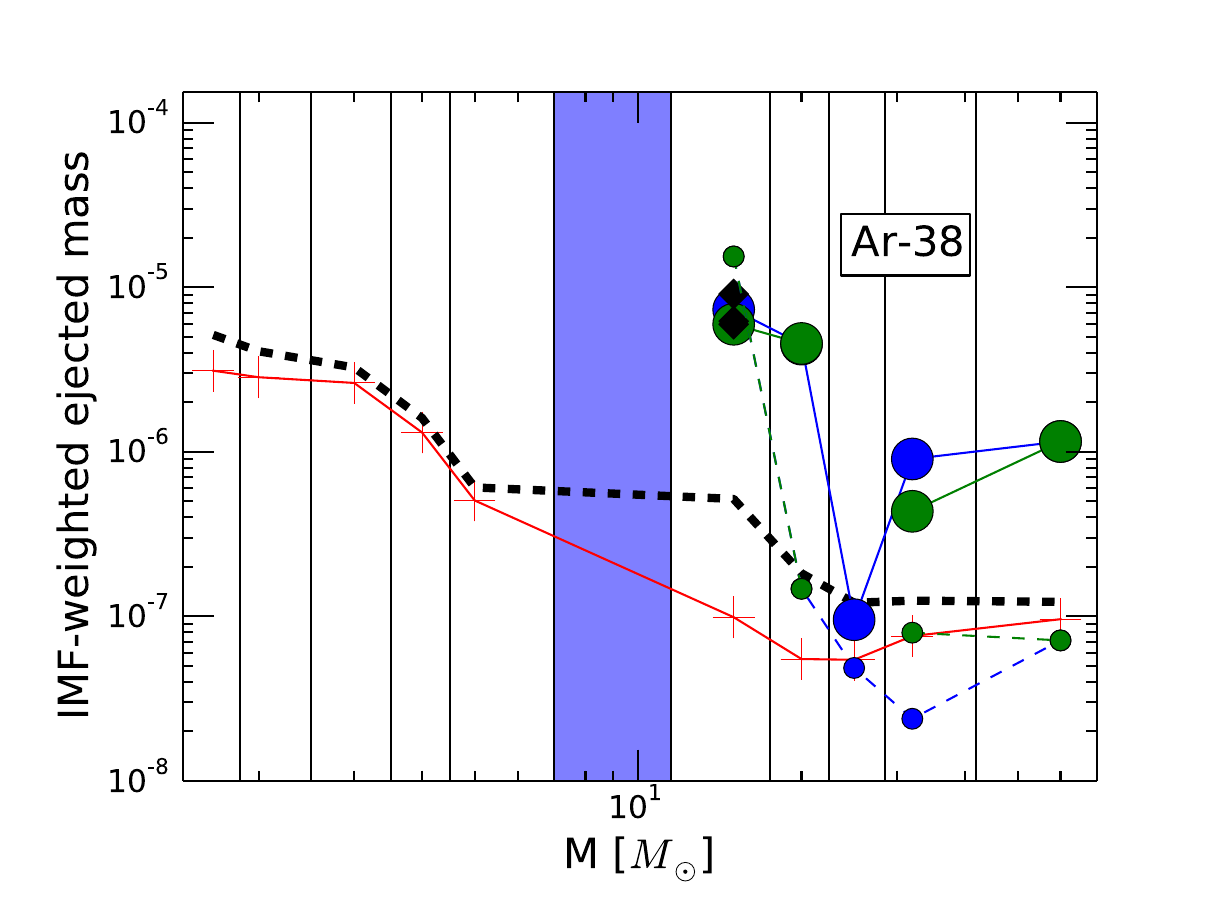}
\includegraphics[width=0.46\textwidth]{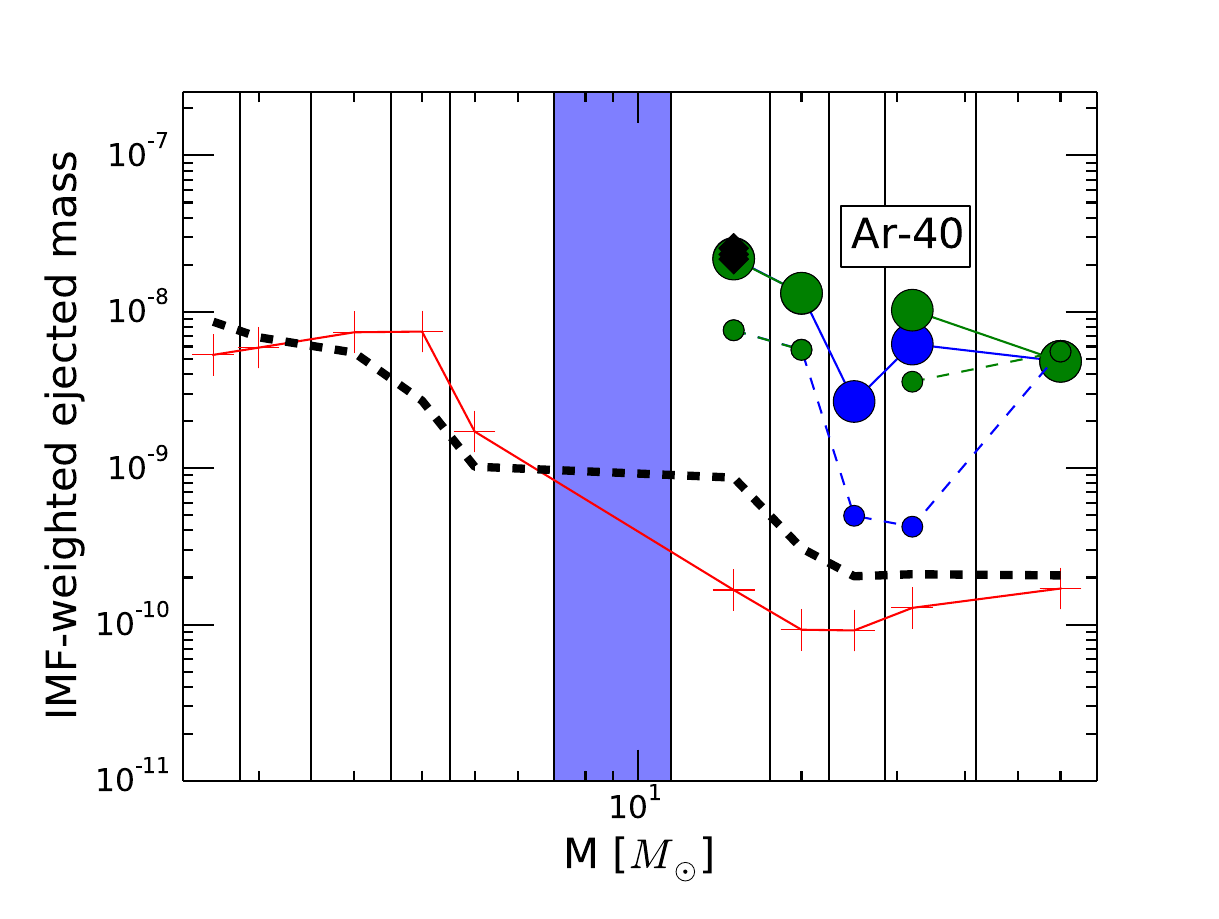}
\includegraphics[width=0.46\textwidth]{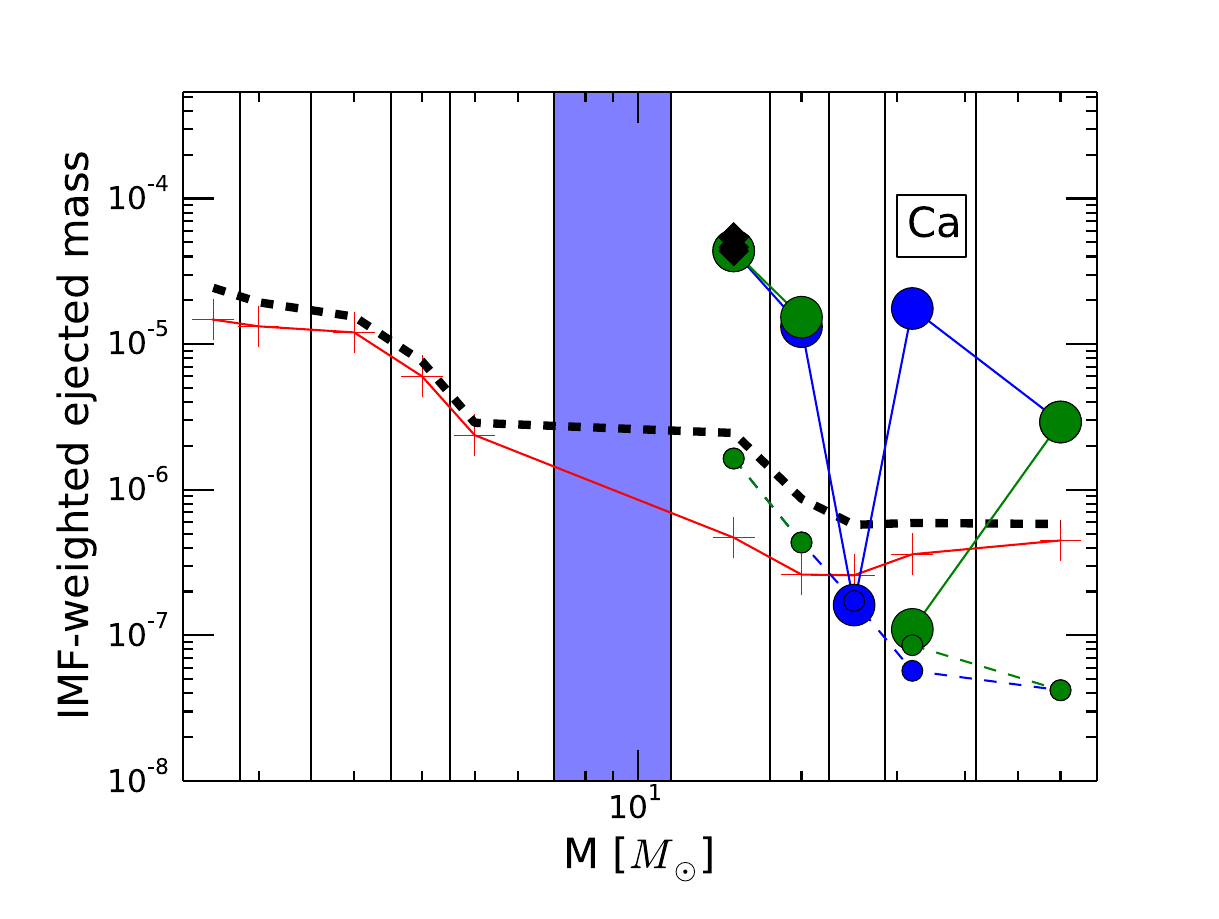}
\includegraphics[width=0.46\textwidth]{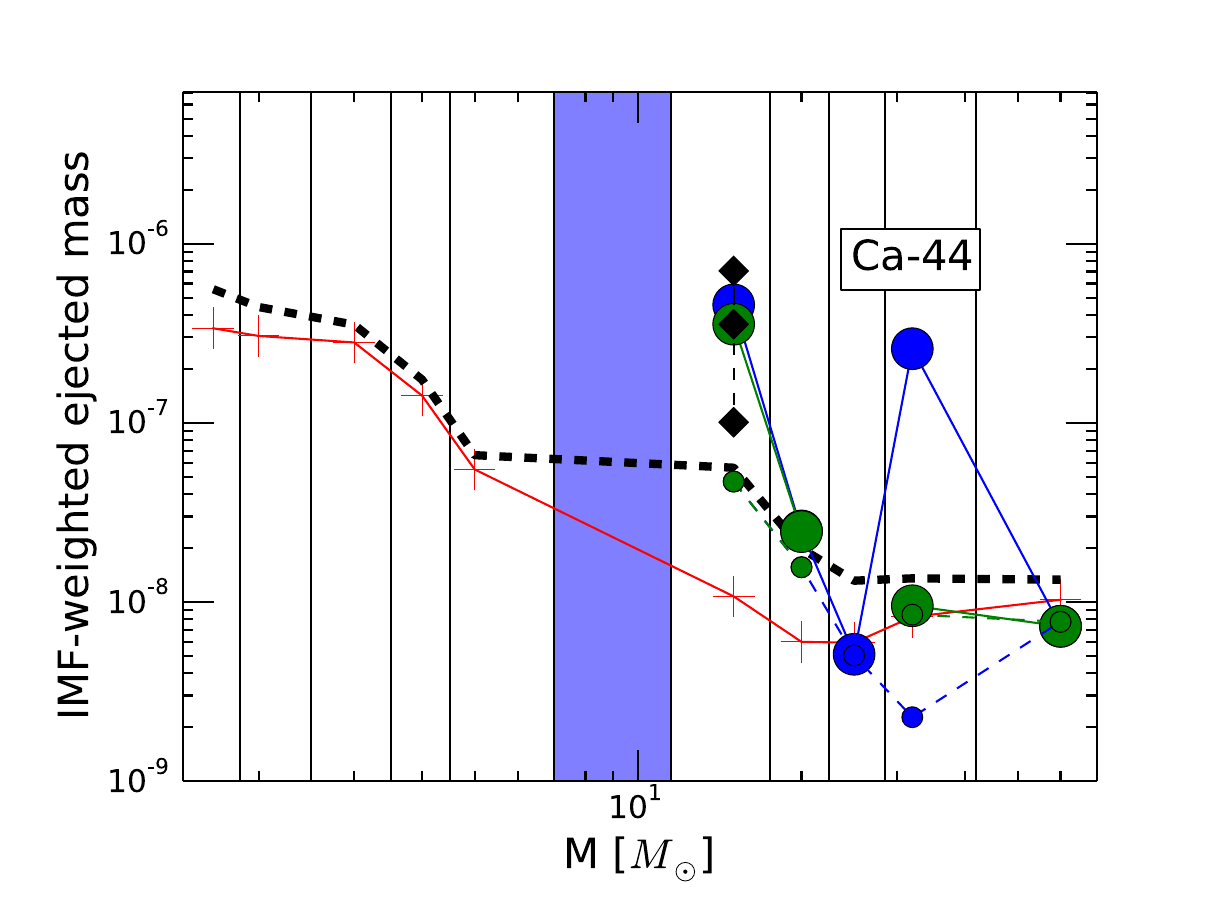}
\includegraphics[width=0.46\textwidth]{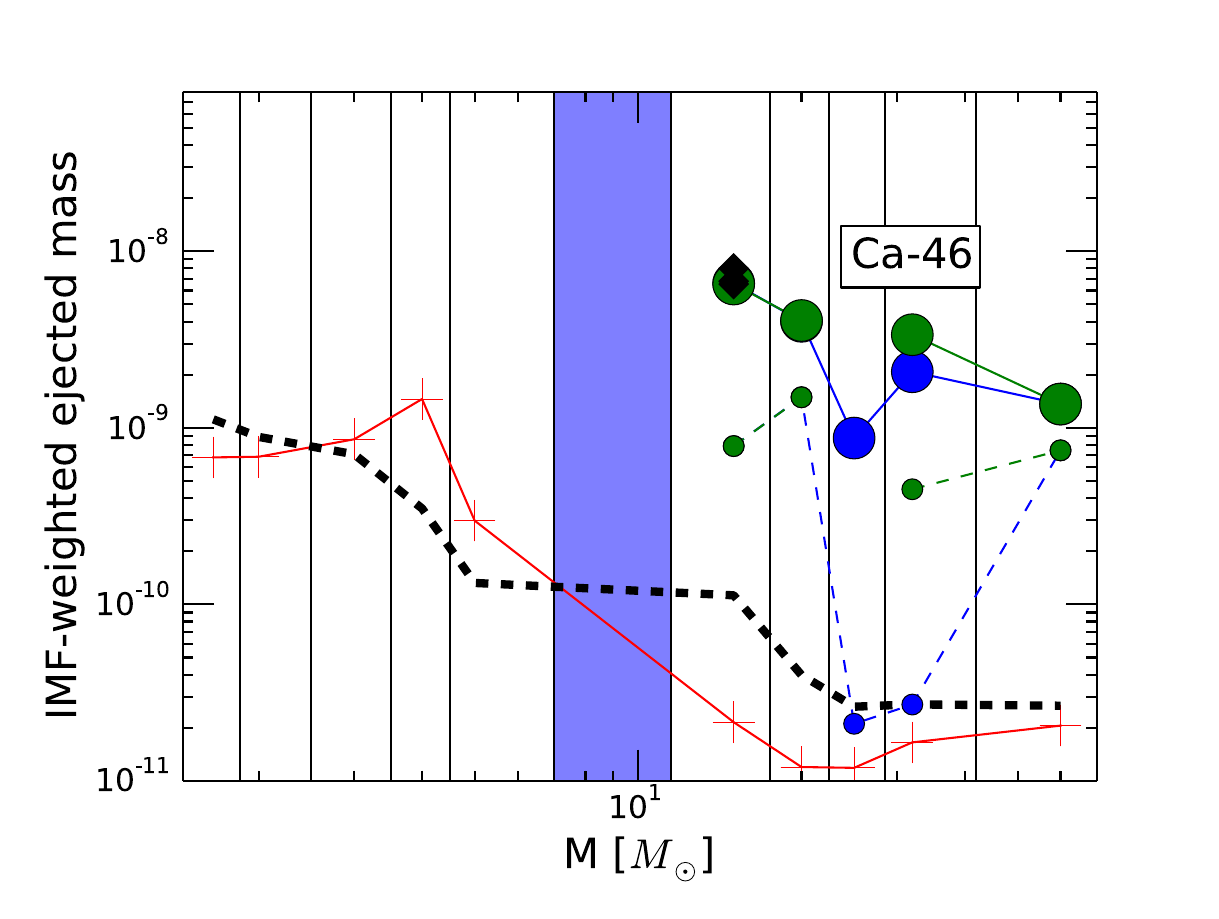}
\includegraphics[width=0.46\textwidth]{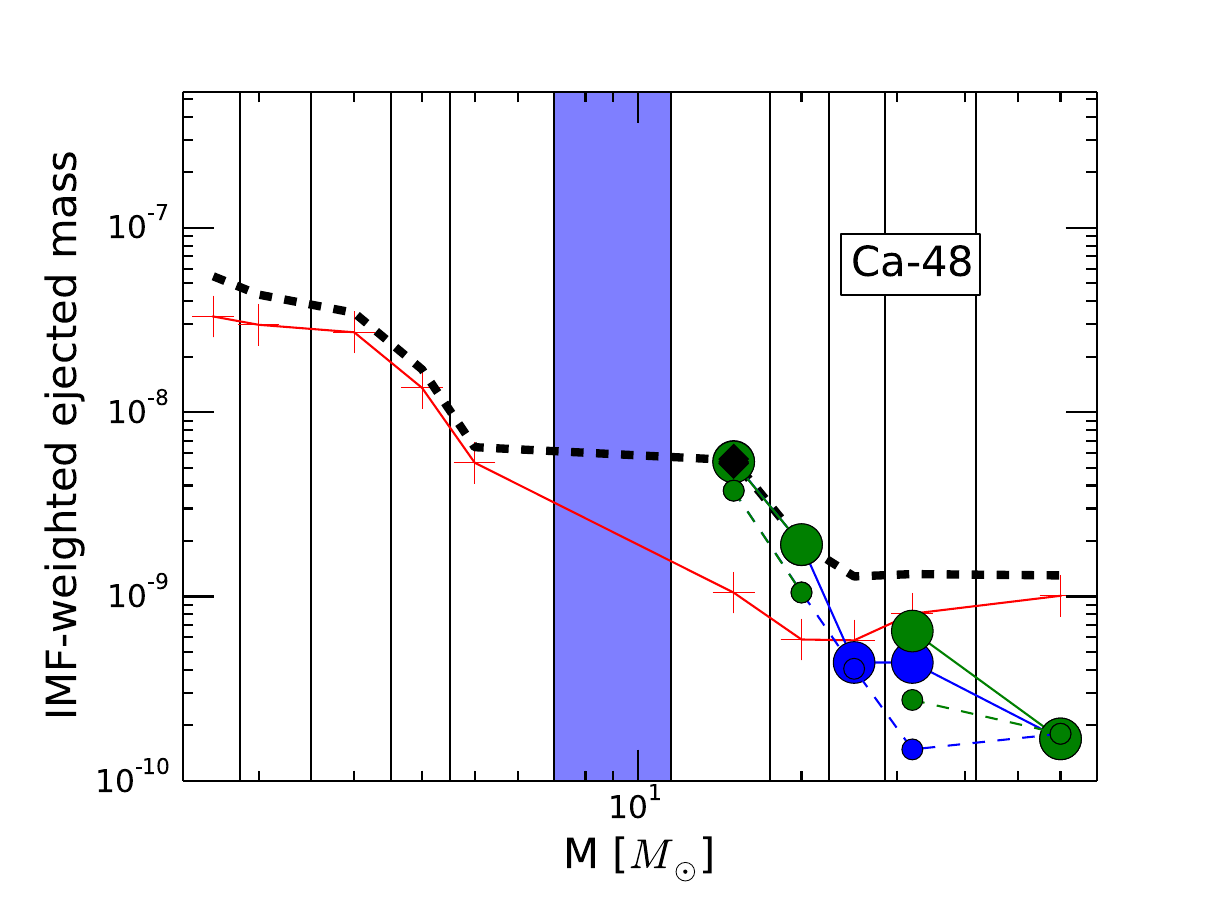}
\caption{Same as \fig{fig:CNONaAl_set1p2} for Ar and Ca and some of their stable isotopes.}
\label{fig:ca_set1p2_sum}
\end{figure}

In our calculations (\fig{fig:CNONaAl_set1p2}, \tabs{tab:element_prodfac_set1.2_winds} and
\ref{tab:element_prodfac_set1.1_winds} for wind contributions, Tables
\ref{tab:element_prodfac_set1.1_exp} and
\ref{tab:element_prodfac_set1.2_exp} for explosive contributions) the
production factors of low-mass stars and massive stars are similar 
\citep[see also][]{dray:03}. 
The \czw\ yields are similar for both metallicities corresponding to the primary
nature of C production; the weighted yield from massive stars is a
factor of about 5--10 lower than from the low-mass star regime, and
comes mostly from (pre-)SN ejecta. Only the $60\msun$ model has a
dominant wind contribution, while the massive star models with lower initial masses are
dominated by C formed during the pre-SN evolution and ejected in the
explosion. An exception is the $25\msun$, $Z=0.01$ case with rapid
explosion, where the fall-back mass is larger compared to other models
of the same mass and the amount of carbon ejected is insignificant. In
general, our models confirm previous results that the
production factor of carbon tends to increase with the initial stellar
mass. %\citep[see also][]{dray:03}.

For low-mass stars the C production increases with the initial mass,
peaking at the $3\msun$ models and then decreasing again for the 4 and $5\msun$
models by a factor of approximately 2 due to HBB
\citep[e.g.,][]{lattanzio:99,herwig:03c}. We do not include possible
effects due to binary evolution, which may reduce the C
contribution from AGB stars \citep[by about 15 \%, according to
e.g.,][]{tout:99}.

N in the solar system is mostly produced by 
%intermediate-mass 
AGB stars 
\citep[e.g.,][and \fig{fig:CNONaAl_set1p2}]{spite:05}.
In more massive stars, the amount of N from winds is similar to the SN explosion
ejecta for the $25\msun$ model (\tabs{tab:element_prodfac_set1.2_winds} and
\ref{tab:element_prodfac_set1.2_exp}) due to the enhanced mass loss
efficiency; while in the $32\msun$ and the $60\msun$ models the contribution from winds dominates. 
The N production only weakly depends on the SN explosion 
%or the initial mass 
and is mostly located in the more external He-rich layers of the star that have not yet been
processed by He burning;
%\citep[e.g.,][]{woosley:95}. 
the isotope \nvi\ is converted
to \nezw\ under helium burning conditions \citep[e.g.,][]{peters:68}. In
%low- and intermediate-mass 
AGB stars the amount of N lost by stellar winds increases with
initial mass (\tab{tab:element_prodfac_set1.2_winds}). 
%The higher nitrogen production in intermediate mass stars (i.e., $5\msun$) is due to HBB \citep[e.g,][]{lattanzio:99}. 
In particular, in the 
%4 and 
$5\msun$ models the production of \isotope[14]{N} increases while \isotope[12]{C} decreases, due to HBB \citep[e.g,][]{lattanzio:99}.
Again, as with C, production factors of N for low-, intermediate-, and high-mass stars are
similar but, in terms of weighted yields, 
%intermediate-mass 
AGB stars
dominate N production for both metallicities (\fig{fig:CNONaAl_set1p2}).

After H and He, O is the most abundant element in the Solar System. 
Most of it %of the solar O
is considered to be produced in massive stars, and possibly from low-mass 
AGB stars due to the O enrichment in the He intershell \citep{herwig:99a}.
Most of the O from massive stars is ejected by the SN explosion, but is of 
pre-SN origin. %\citep[e.g.,][]{thielemann:85,chieffi:98}. 
Thus, according to
standard one-dimensional SN models, the amount of ejected oxygen
increases with initial mass \citep[see e.g.,][]{thielemann:96}. Our models take 
into account fallback and, as a result, the $20\msun$ model ejects more
\ose\ than both the $15\msun$ and $25\msun$ models (\tab{tab:element_prodfac_set1.2_exp}). 
The amount of ejected O increases again in the $32\msun$ and $60\msun$ models 
because of the correspondingly smaller compact remnant masses.
The high temperature in the $15\msun$ case (see \sect{subsec:exp}) leads to the 
destruction of a large fraction of O made during the pre-SN phase (cf.,
 \tabs{tab:element_prodfac_set1.2_exp} and
\ref{tab:element_prodfac_set1.2_pre}; \fig{fig:CNONaAl_set1p2}).

Our AGB models produce O due to the CBM applied at the bottom of the
He-shell flash convection zone (see \sect{stars}).  O is then brought
to the envelope along with C during the third dredge-up.
\isotope[16]{O} is a primary product of the He burning
  reaction \isotope[12]{C}($\alpha$,$\gamma$)\isotope[16]{O}
  following the triple-$\alpha$ reaction in the He intershell region.
  For instance, from \tabs{tab:isotopic_prodfac_set1.2_winds} and
  \ref{tab:isotopic_prodfac_set1.1_winds} the overproduction factors
  for the M=$2\msun$ star at Z=0.02 and Z=0.01 corresponds to the same
  increase of $\Delta X(\isotope[16]{O}) \approx 0.005$, independent
  of the initial abundance. 
%Adopting the solar distribution by \cite{asplund:09} with an O abundance between our two initial compositions, would not change the prediction of a primary production of O in our AGB models. Specifically, using the \cite{asplund:09} initial O abundance we would have obtained an \isotope[16]{O} overproduction of 1.66.  
This source of O may be relevant to the total O inventory in the Galaxy (see \fig{fig:CNONaAl_set1p2}, and \tab{tab:element_prodfac_set1.2_winds} and \ref{tab:element_prodfac_set1.1_winds}, and discussion in \cite{delgado-inglada:15}), but galactic chemical evolution simulations are needed to verify this possibility. For a comparison with O yields provided by other groups, we refer to \sect{comparison}.

\subsection{F, Ne, and Na}
\label{f-ne-na}

F is produced in massive stars during the CCSN---predominantly via neutrino
spallation on \isotope[20]{Ne} \citep[e.g.,][]{woosley:88,kobayashi:11a}, the Wolf-Rayet (WR) wind
phase \citep[][]{meynet:00}---and low-mass AGB stars
\citep[e.g.,][]{jorissen:92,lugaro:04,cristallo:07,stancliffe:07,karakas:08}. 
No relevant contribution is expected from massive AGB stars, since
HBB in the envelope destroys \isotope[19]{F} via proton capture
\citep[][]{smith:05,karakas:07}. F enhancement has been confirmed 
spectroscopically only in AGB stars \citep[][]{abia:10,lucatello:11},
but chemical evolution studies seem to indicate that all the sources
above are required in order to explain the abundance evolution of this
element in the galaxy \citep[][]{renda:04,kobayashi:11a}. Our simulations have no
contributions from neutrino spallation during SNe or rotationally
induced mixing and identify AGB stars with $\mzams\leq 3\msun$ as the
most productive source of F. 
Contributions from WR stars or from CCSNe are, however, considered. 
In particular, in \setopt\ only for the $60\msun$ model is the wind contribution positive, and only for the $15\msun$ star is the explosive contribution positive (\fig{fig:CNONaAl_set1p2}). In the massive star models at \setopo\ metallicity, all of the wind contributions are negative and it is only the $15\msun$ explosion that leads a small positive net massive star production factor.
%However, contributions from WR stars (see the $60\msun$ star) or from CCSN
%($15\msun$ case). All wind contributions for massive star \setopo\
%models are negative, and only the $15\msun$ explosion leads overall to a
%small positive massive star production factor.

Our models (\fig{fig:NeMg_set1p2_sum}) confirm that Ne is produced as
\isotope[20]{Ne} in massive stars. 
%\citep[e.g.,][]{thielemann:85,woosley:02}.
\isotope[20]{Ne} is efficiently produced already during the
pre-explosive evolution of massive stars in the C-burning layers.
During the CCSN, \isotope[20]{Ne} in the deeper layers of the ejecta is processed and destroyed 
by the SN shock wave, whereas more external parts of C-burning Ne-rich
layers are ejected almost unchanged. %\citep[e.g.,][]{woosley:95}. 
Notice
that some production of Ne is obtained at the bottom of the explosive He
shell, depending on the SN shock temperatures. A similar effect can be 
observed for the $\alpha$-elements Mg, Si, S, Ar, and Ca. Due to
similarly high explosion temperatures, hypernova models or the high energy
component of asymmetric CCSN explosion models show such a production for
\isotope[28]{Si} \citep[e.g.,][]{nomoto:09}. Those specific signatures identify
a stellar region at the bottom of the He shell called C/Si zone, which
provide a suitable location for carbide grains condensation in the ejecta. 
%[SJ: This sentence is a little confusing - maybe when you mention explosive He shell before you should say where that is, and then re-work this sentence]. 
Furthermore, the existence of the C/Si zone may be consistent
with observations of CasA and SN1987A objects \citep[][]{pignatari:13}.

\isotope[21]{Ne} shows a small overproduction compared to its initial abundance in massive AGB stars. The isotope is made by neutron capture on \isotope[20]{Ne} and via the reaction \isotope[18]{F}($\alpha$,p)\isotope[21]{Ne} in the He intershell \citep[for the impact of this last reaction channel and its uncertainty, see][]{karakas:08}, but it is depleted by HBB \citep[e.g.,][]{doherty:14}. 
%via the proton capture on \isotope[20]{Ne} during HBB, and 
On the other hand, \isotope[21]{Ne} is efficiently produced in massive stars 
(\fig{fig:NeMg_set1p2_sum}). Finally, \isotope[22]{Ne} is mostly produced in
low-mass AGB stars; some of it may be primary depending upon the third
dredge-up, where of \czw\ can be returned as \nvi\ to the next thermal pulse
He-shell flash convection zone. 
%[SJ: Please check that this is still true]. 
\isotope[22]{Ne} has an additional contribution from CCSN and from the stellar winds of more massive WR stars 
(the $60\msun$ star in our stellar set).

\isotope[23]{Na} is efficiently made during hydrostatic carbon burning in massive stars, like \isotope[20]{Ne}.
%\citep[e.g.,][]{thielemann:96}. 
Its pre-SN abundance is partially destroyed by CCSN (\fig{fig:CNONaAl_set1p2}).
Similarly to \isotope[20]{Ne}, \isotope[23]{Na} is directly made by C-fusion reaction. On the other hand, it receives a relevant additional contribution by proton capture and neutron capture on \isotope[22]{Ne}. 
Due to the secondary nature of this isotope, the final massive star yields of Na decrease with the decreasing of the initial metallicity , causing the known odd-even effect with the neighbor elements Ne and Mg \citep[e.g.,][]{woosley:95,limongi:00}.
Na may be ejected during the WR phase of more massive stars (e.g., the 32 and $60\msun$ models) via proton
capture on \isotope[22]{Ne}. The same nucleosynthesis path is responsible for most of the Na produced in low-mass 
AGB and massive AGB stars \citep[e.g.,][]{cristallo:06,lucatello:11}.
According to these simulations AGB stars are efficient producers of Na compared to massive stars at the same metallicity, with the strong contribution of the $3\msun$ and $5\msun$ stars (\fig{fig:CNONaAl_set1p2}).

\subsection{Mg, Al, and Si}
\label{mg-al-si}
Mg is mostly produced in massive stars, % \citep[e.g.,][]{thielemann:85,woosley:02}.
however the individual Mg isotopes show a more complex behavior (\fig{fig:NeMg_set1p2_sum}).
The isotope \isotope[24]{Mg} is only produced in massive stars; in the 15 and $20\msun$, $Z= 0.02$ models 
\isotope[24]{Mg} is produced during the pre-explosive
phase, with a partial depletion due to nucleosynthesis during CCSN.
%\citep[see also][]{thielemann:96}. 
On the other hand, for larger masses
explosive nucleosynthesis provides an additional contribution to
\isotope[24]{Mg}. The dependence on the initial mass is due to the large amount
of material falling back on the SN remnant in the 25 and $32\msun$
models, where most of the pre-explosive \isotope[24]{Mg} will not
be ejected and the explosive He shell component dominates the final
abundance. \isotope[25]{Mg} and \isotope[26]{Mg} are produced also by the AGB stars, more specifically in the He-shell flash convection zones of more massive AGB stars due to 
$\alpha$-capture by \nezw\ \citep[e.g.,][]{karakas:07}.

Al is efficiently produced in massive stars---mainly
in C-burning zones---with no contribution from AGB stars (\fig{fig:CNONaAl_set1p2}).
\isotope[27]{Al} shares nuclear production conditions with 
\isotope[25]{Mg} and \isotope[26]{Mg} in the 15 and $20\msun$ stars,
%\citep[see also discussion in][]{woosley:95}
and has the same dependence as
\isotope[24]{Mg} on the amount of material falling back after the SN explosion.

Si is efficiently produced in massive stars (\fig{fig:p_set1p2}). 
The origin from explosive nucleosynthesis is
always larger than the pre-explosive contribution. 
%\citep[e.g.,][]{woosley:95,thielemann:96}.
The $15\msun$ case shows an increase of the Si yield with decreasing explosion
energy. In order to account for all of the Si inventory observed in the
Solar System, a contribution from SNIa (not considered here) is needed
%\citep[e.g.,][]{thielemann:86,travaglio:11,kusakabe:11}. 
\citep[e.g.,][]{seitenzahl:13}. 
The neutron-rich isotopes \isotope[29,30]{Si} are mostly made by neutron captures on \isotope[28]{Si} during both pre-SN and explosive C burning \citep[][]{rauscher:02}.

\subsection{From P to Sc}
\label{intermediate_elements}

\begin{figure}
\centering
\includegraphics[width=0.46\textwidth]{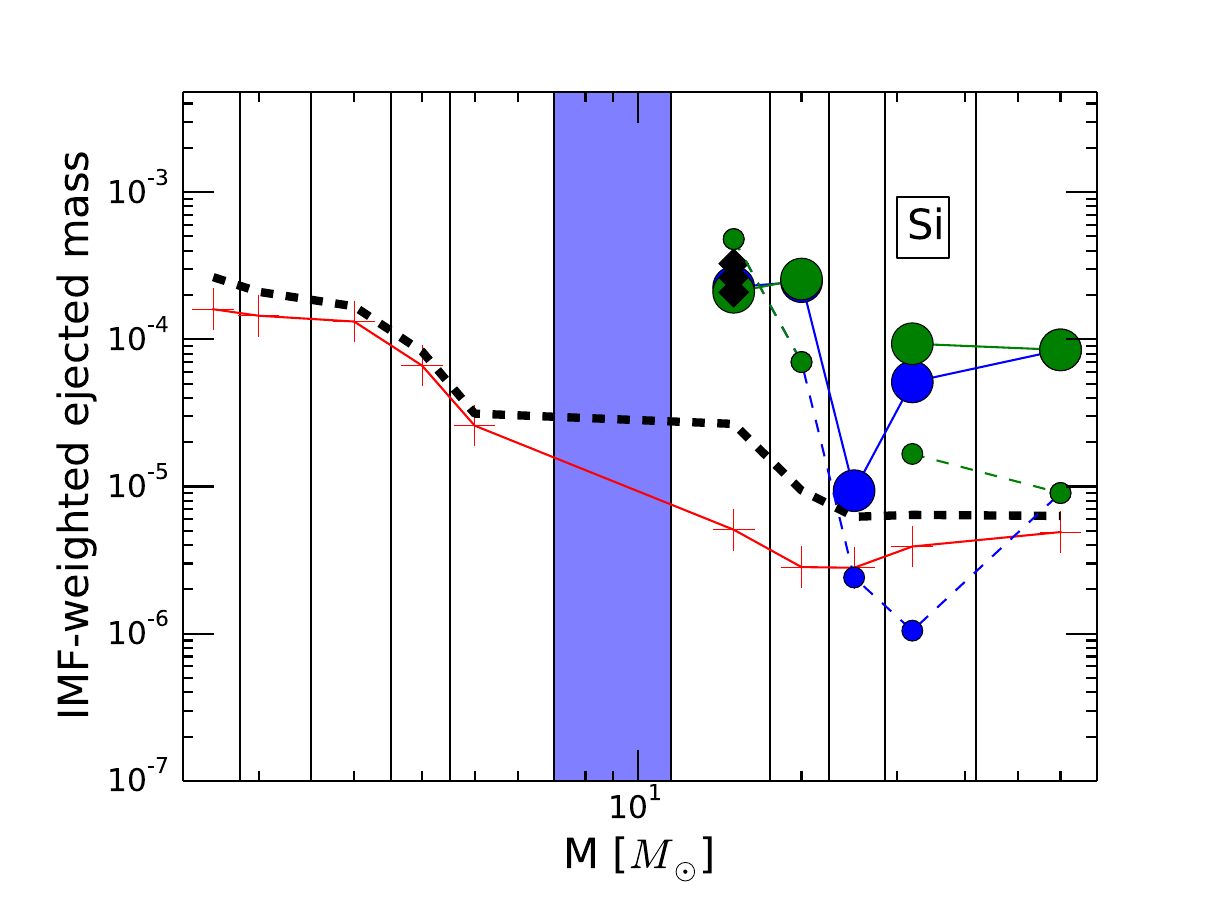}
\includegraphics[width=0.46\textwidth]{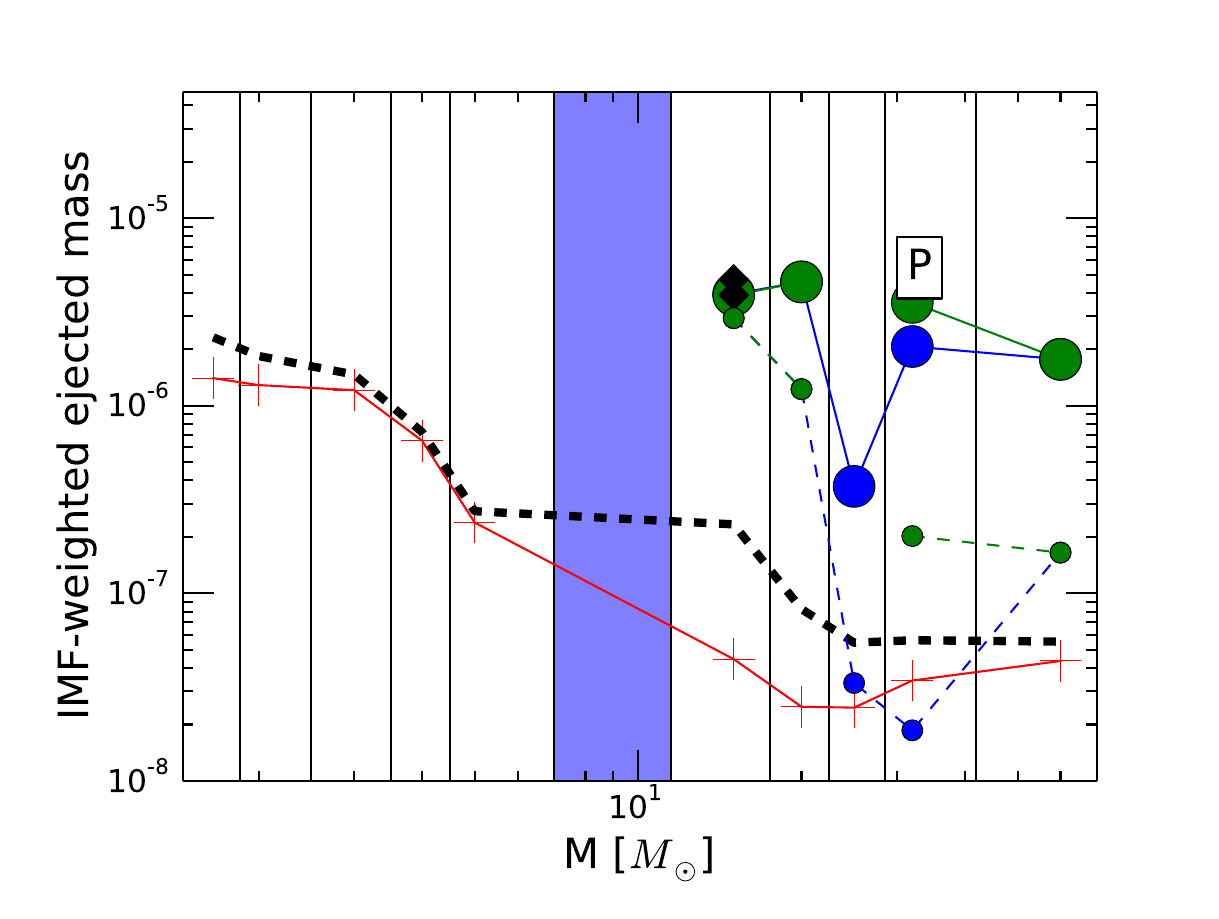}
\includegraphics[width=0.46\textwidth]{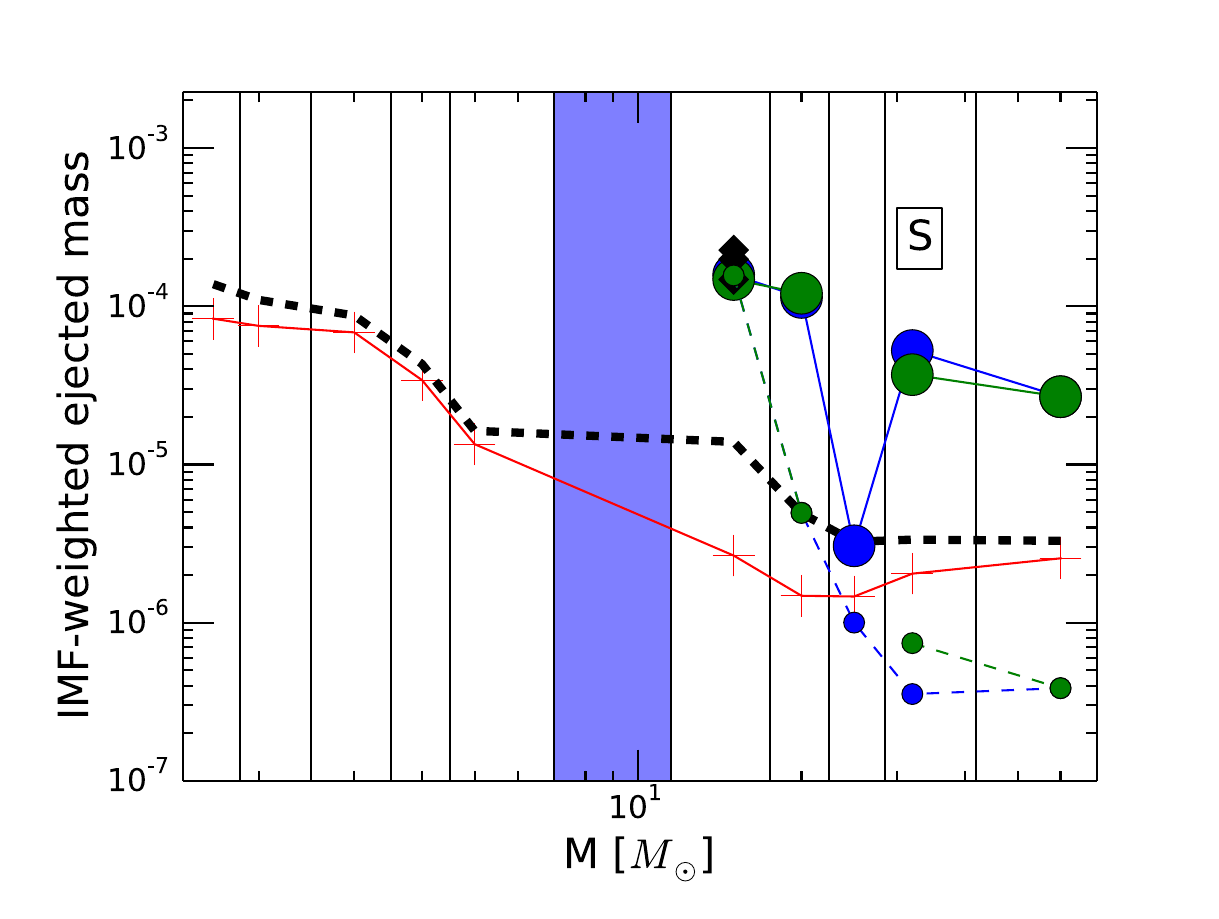}
\includegraphics[width=0.46\textwidth]{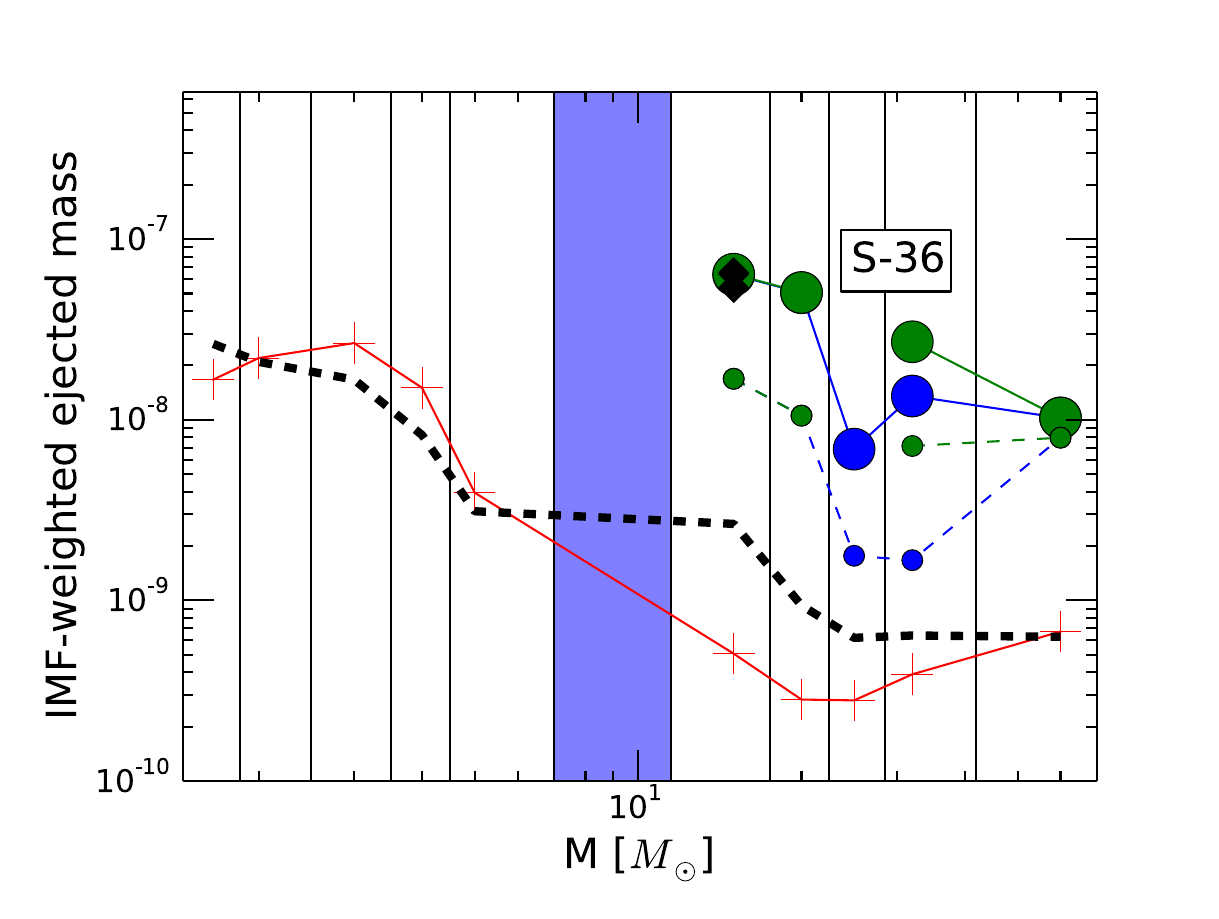}
\includegraphics[width=0.46\textwidth]{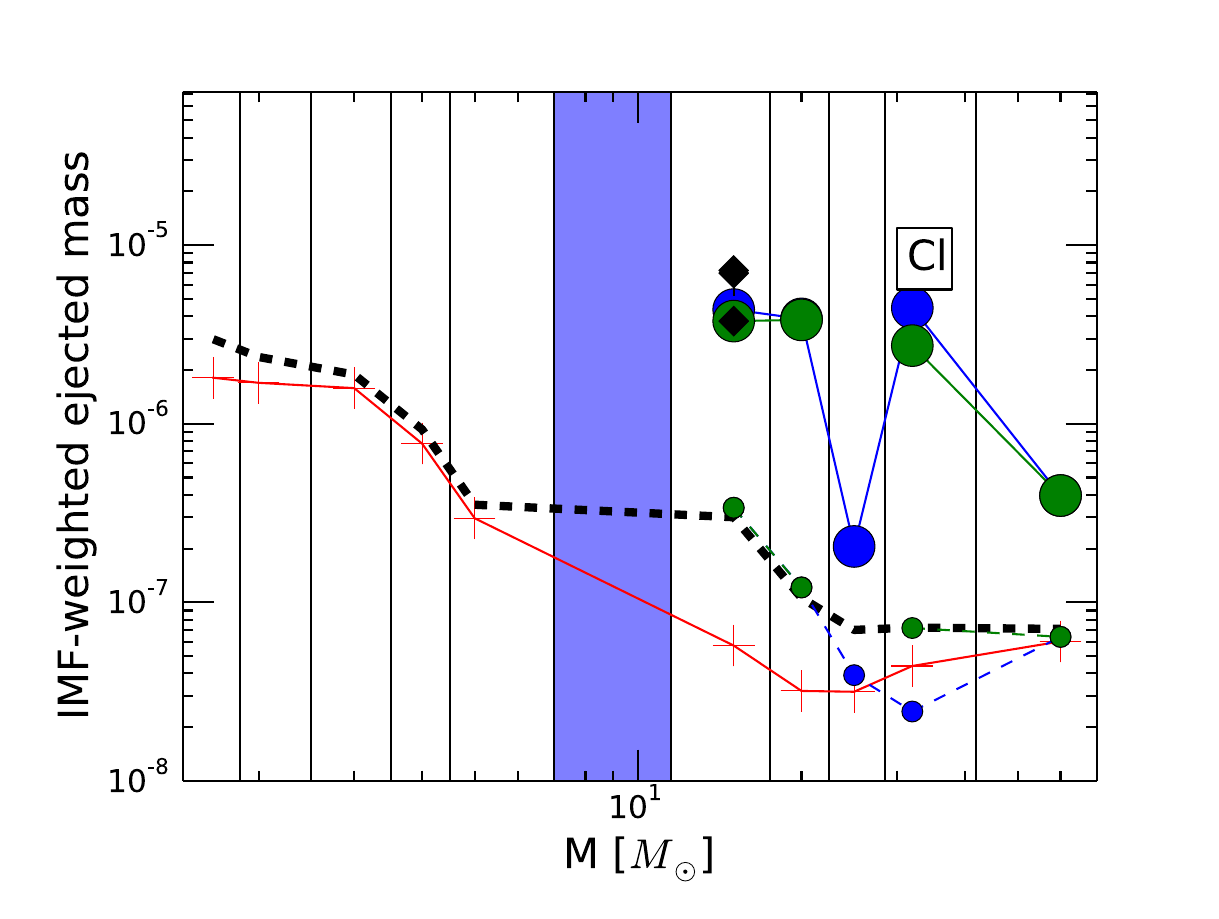}
\includegraphics[width=0.46\textwidth]{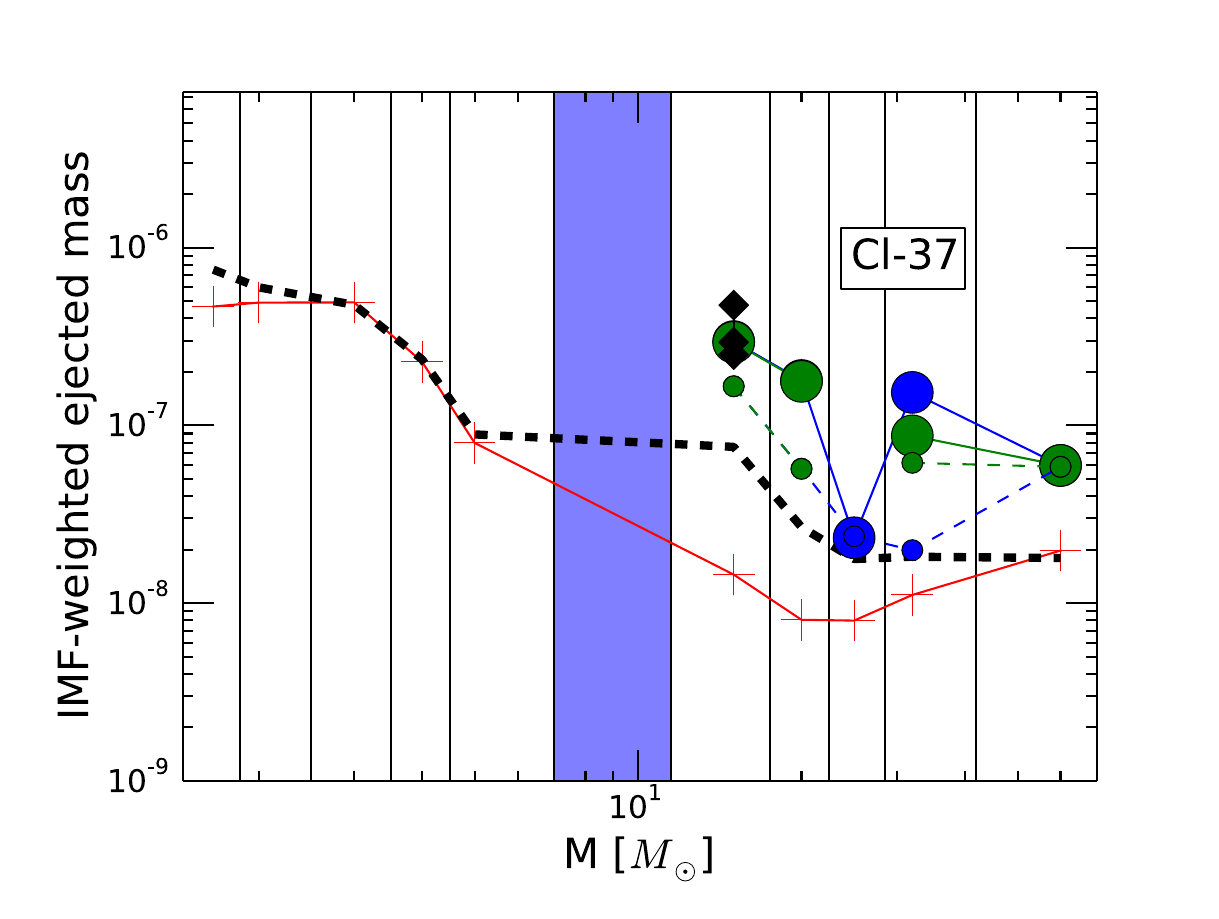}
\includegraphics[width=0.46\textwidth]{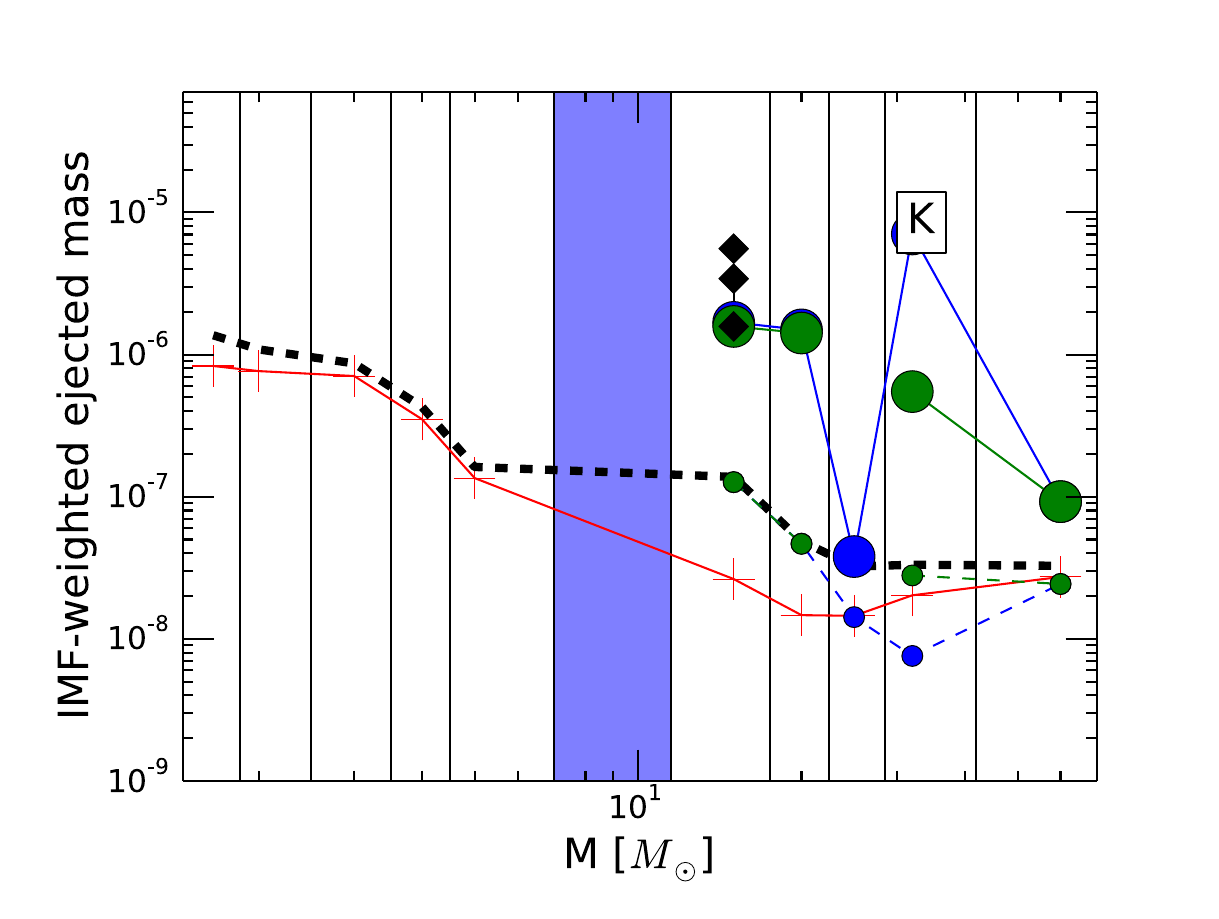}
\includegraphics[width=0.46\textwidth]{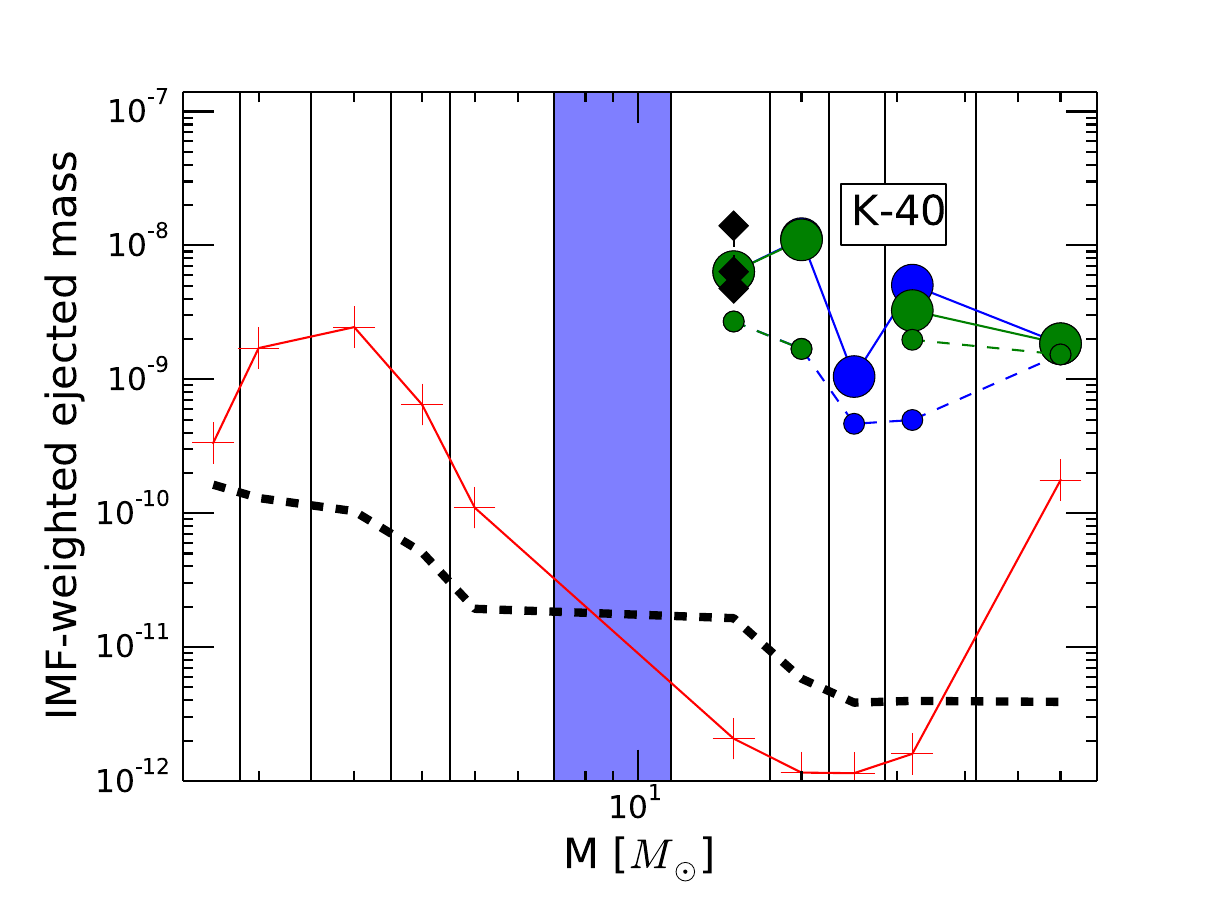}
\caption{Same as \fig{fig:CNONaAl_set1p2} for Si, P, S, Cl, and K and
  some of their stable isotopes.}
\label{fig:p_set1p2}
\end{figure}

Most P is made in massive stars (\fig{fig:p_set1p2}). The amount of
\isotope[31]{P} made by the \sprn\ during the pre-explosive evolution is
further increased during the SN explosion. 
%\citep[see also][]{limongi:00,woosley:02}. 
The dominant contribution is given by
explosive C-burning and explosive He-burning, while this isotope is destroyed by more extreme explosive conditions.

S is mainly composed of \isotope[32]{S}, while \isotope[36]{S} is the least 
abundant stable sulfur isotope (0.01\% in the Solar System). S comes from
massive stars, with the exception of \isotope[36]{S}, which can have a small
contribution from AGB stars (\fig{fig:p_set1p2}). Again, the
contribution from SNIa \citep[e.g.,][]{thielemann:04} are not considered
in our models. S is made during explosive C-burning and O-burning;
%\citep[][]{thielemann:85,woosley:95}, 
while the pre-explosive
production is marginal for \isotope[32]{S} (except for the \setopo\ $15\msun$
\emph{rapid} case), it may be relevant for \isotope[33,34]{S} produced
via neutron captures on \isotope[32]{S}. The neutron-rich isotope \isotope[36]{S} 
is first made by the weak \sprn\ \citep[e.g.,][and references
therein]{woosley:02, mauersberger:04}, mainly via the production channel
\isotope[35]{Cl}(n,$\gamma$)\isotope[36]{Cl}(n,p)\isotope[36]{S}, where the
initial \isotope[35]{Cl} is the main seed \citep[][]{mauersberger:04,pignatari:10}.
 In our models \isotope[36]{S} is mainly produced in explosive C- and He-burning, 
in the latter case also via direct neutron capture on \isotope[34]{S} (\fig{fig:p_set1p2}).

Cl is made in the explosion of massive stars with a small pre-SN
contribution for \isotope[37]{Cl} (\fig{fig:p_set1p2}). \isotope[35]{Cl} may 
also come from neutrino interactions with stellar material that are not considered
here. %\citep{woosley:95}. 
The yields correlate in a non-linear way with
the SN explosion energy. Comparing results for the $25\msun$ model
with those for lower initial masses shows that the yields strongly depend on fallback.
The \sprn\ produces \isotope[37]{Cl} efficiently in massive stars,
\citep[see also][]{woosley:02,rauscher:02} but explosive nucleosynthesis further increases the \isotope[37]{Cl}
yield.

Ar is made in explosive O-burning (\fig{fig:ca_set1p2_sum}). Some
pre-explosive production of \isotope[38]{Ar} is obtained for the $15\msun$
model in the convective O-burning shell; larger masses do not show such a
component because the O shell region is below the fallback coordinate.
The isotope \isotope[40]{Ar}, with a much smaller Solar System abundance, is efficiently
produced by the \sprn\ in all models. An additional contribution
originates in the explosive He-burning shell during the SN explosion due
to the \nprn\ \citep[][]{blake:76,thielemann:79,meyer:00}.

K has 2 stable isotopes, \isotope[39,41]{K}, and a long-lived isotope, 
\isotope[40]{K} ($\tau$$_{1/2}$ = 1.28$\times$10$^9$ years), decaying in part to \isotope[40]{Ca} and in part to \isotope[40]{Ar}.
\isotope[41]{K} and \isotope[39]{K} are efficiently produced in CCSN, 
along with a small \spr\ production of \isotope[41]{K} during the 
pre-explosive phase (\fig{fig:p_set1p2}). A small production of \isotope[41]{K} 
in low-mass AGB stars may be relevant (electronic table, $3\msun$ stellar model,
\setopt). \isotope[40]{K} shows a strong production in AGB and massive
stars. In agreement with the Solar System distribution, \isotope[40]{K}
stellar yields are about 2 orders of magnitude smaller than the total
K yields. In massive stars, \isotope[40]{K} is made by the \sprn\ before
the explosion and during the SN event by explosive He burning.

Most \isotope[40]{Ca} (and therefore most of the calcium) originates in
explosive O-burning, %\citep[e.g.,][]{woosley:95,chieffi:98}, 
with a
minor contribution from models with an $\alpha$-rich freezout component
(\fig{fig:ca_set1p2_sum}). In particular, the large difference between
the $32\msun$ models with \emph{rapid} and \emph{delayed} explosion is due
to the different amount of fallback material. \isotope[44]{Ca} can be 
efficiently produced as \isotope[44]{Ti} in $\alpha-$rich freezout 
conditions \citep[e.g.,][]{magkotsios:10}; a small amount of
\isotope[44]{Ca} may also be produced in more external explosive regions,
mainly as \isotope[44]{Ti} in explosive O- and C-burning or as \isotope[44]Ca and
its neutron rich unstable isobars in explosive He-burning conditions.
\isotope[46]{Ca} is the only Ca isotope with a clear contribution from AGB
stars, in particular from massive AGB stars where high neutron
densities during the convective TP phases allows the \spr\ path to open a branching
at the unstable isotope \isotope[45]{Ca}. In a similar way \isotope[46]{Ca} can
be produced by the \sprn\ in the convective
C-burning shell in massive stars. The explosive contribution is mainly due to the
\nprn\ in the explosive He-burning. \isotope[48]{Ca} originates in the \npr\
in massive stars with a small contribution from the 15
and $20\msun$ stellar models (see full tables online),
but weak compared to the similar \isotope[46]{Ca} production. \isotope[48]{Ca} 
may originate in special conditions in CCSN with a high neutron excess
\citep[][]{hartmann:85}. Alternatively \isotope[48]{Ca} production is
predicted in \ipr-conditions with characteristic neutron densities of
$N_\mathrm{n} \sim 10^{15}\cc$ \citep{herwig:13a}, or by the weak \rpr\ \citep[][]{weissman:12,wanajo:13}.

Mono-isotopic Sc is among the least abundant of the light and intermediate elements in the Solar
System. Because of its low
abundance Sc can be efficiently produced from adjacent Ca 
at high neutron densities obtained in low mass stars \citep[e.g., by the
\iprn,][]{cowan:77,herwig:11}. Besides the pre-explosive
production by the \sprn, in massive stars we find a strong Sc
production mainly in the explosive He-burning (\fig{fig:s_set1p2_sum}).
In the $15\msun$ models with $\alpha$-rich freezout Sc production
is even larger \citep[as previously reported, e.g., by][]{umeda:05}. Sc
production may be also increased if feedback from neutrinos in the deepest SN ejecta is considered
\citep[][]{woosley:95,froehlich:06a,yoshida:08}. 
Sc can receive some contribution from the \spr\ in massive AGB stars \citep[][]{smith:87a,karakas:12}.
In our models we find milder overproduction factors for Sc compared to e.g., \cite{karakas:12} (see \tab{tab:element_prodfac_set1.2_winds}, fully available online). 
In particular, the $4 \msun$ models of \setopt\ show the largest overproduction with 1.316, corresponding to the small production factor 
of 1.047 (see also \fig{fig:s_set1p2_sum}).

\subsection{From Ti to Ni}
\label{iron_elements}
% Ti (Z=22), V (Z=23), Cr (Z=24), Mn (Z=25), Fe(Z=26), Co (Z=27), Ni (Z=28)

The production of most of these elements requires explosive conditions, and therefore in the present set of models they are efficiently made in the CCSN simulations. In \fig{fig:post_versus_pre_set1p2} this is shown for a number of stellar models from \setopt, where the post-explosion yields are compared to the pre-explosive abundances.

\begin{figure}
\centering
\includegraphics[width=0.46\textwidth]{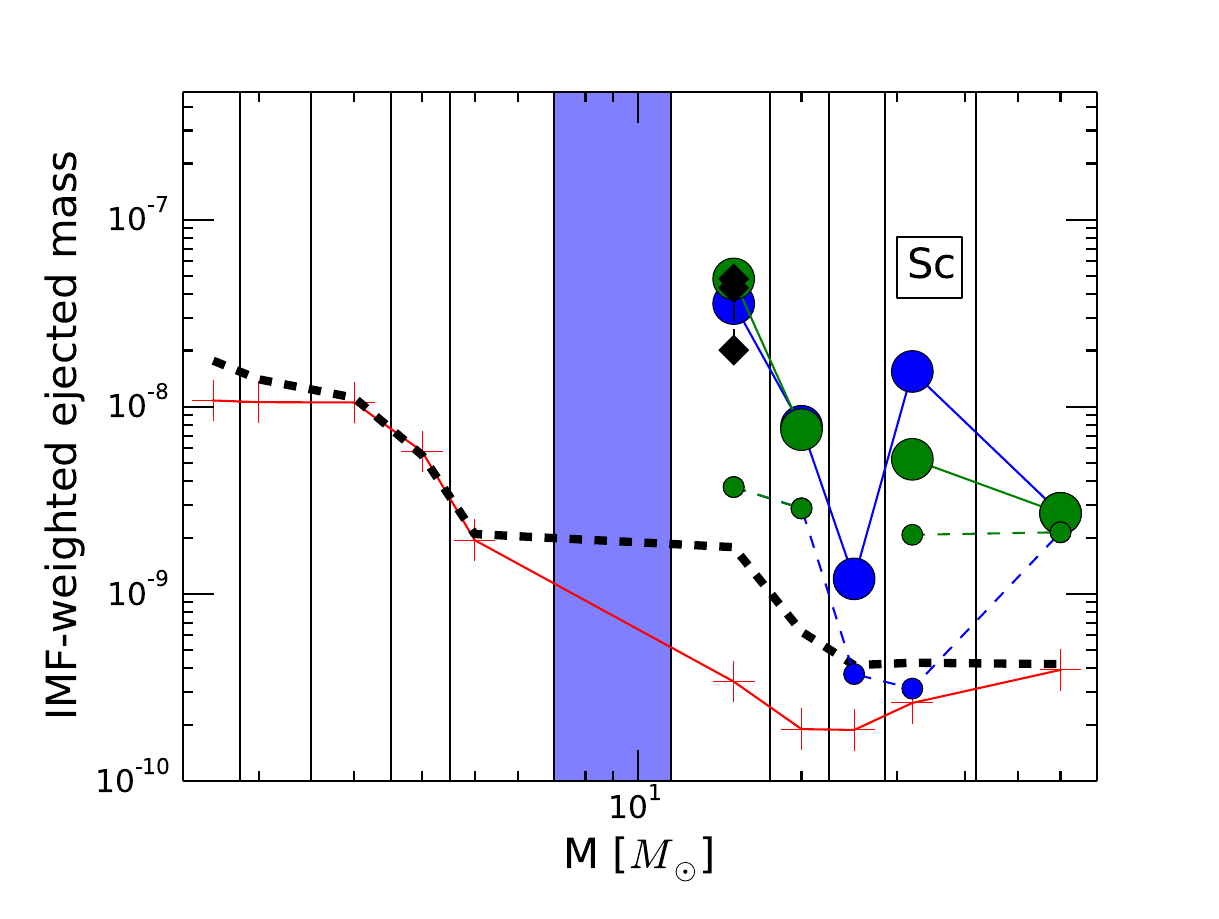}
\includegraphics[width=0.46\textwidth]{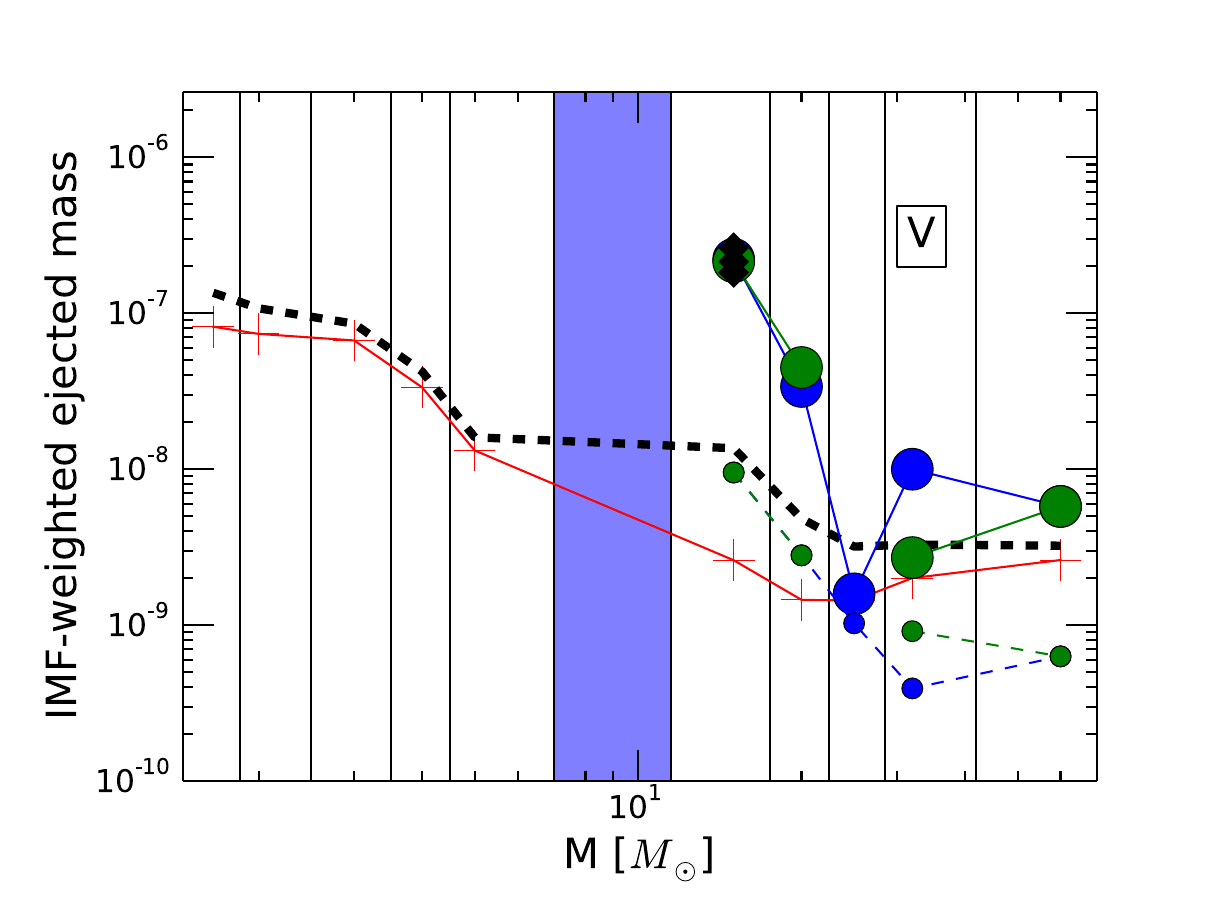}
\includegraphics[width=0.46\textwidth]{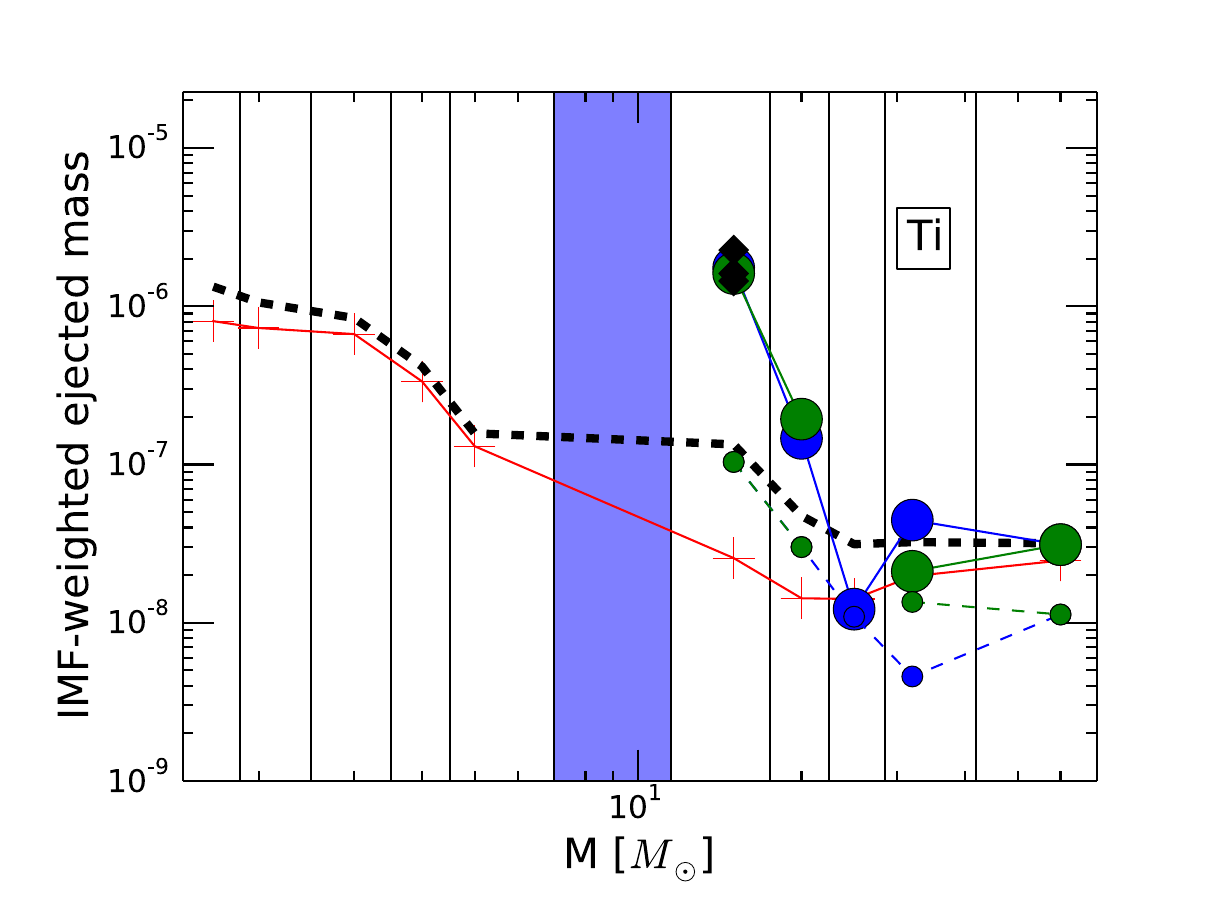}
\includegraphics[width=0.46\textwidth]{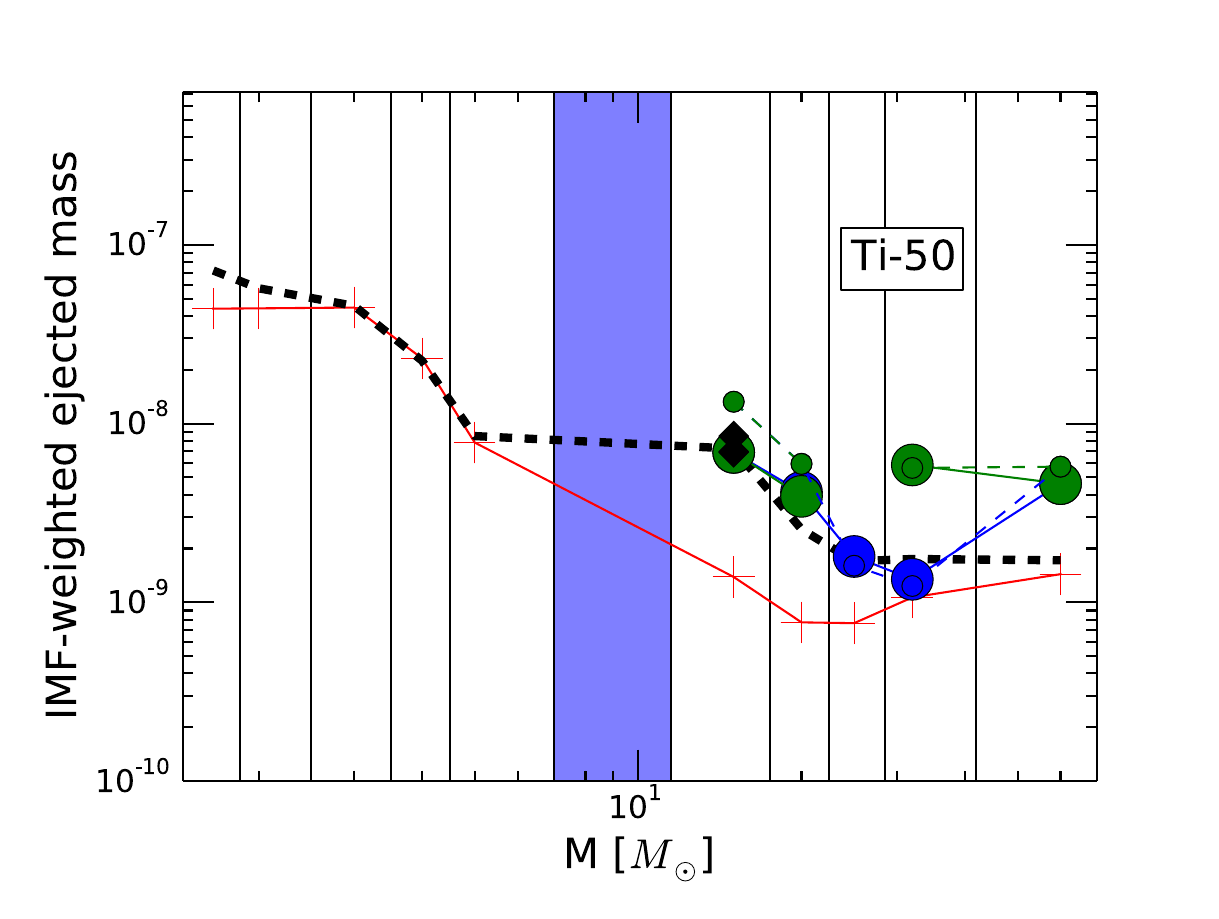}
\includegraphics[width=0.46\textwidth]{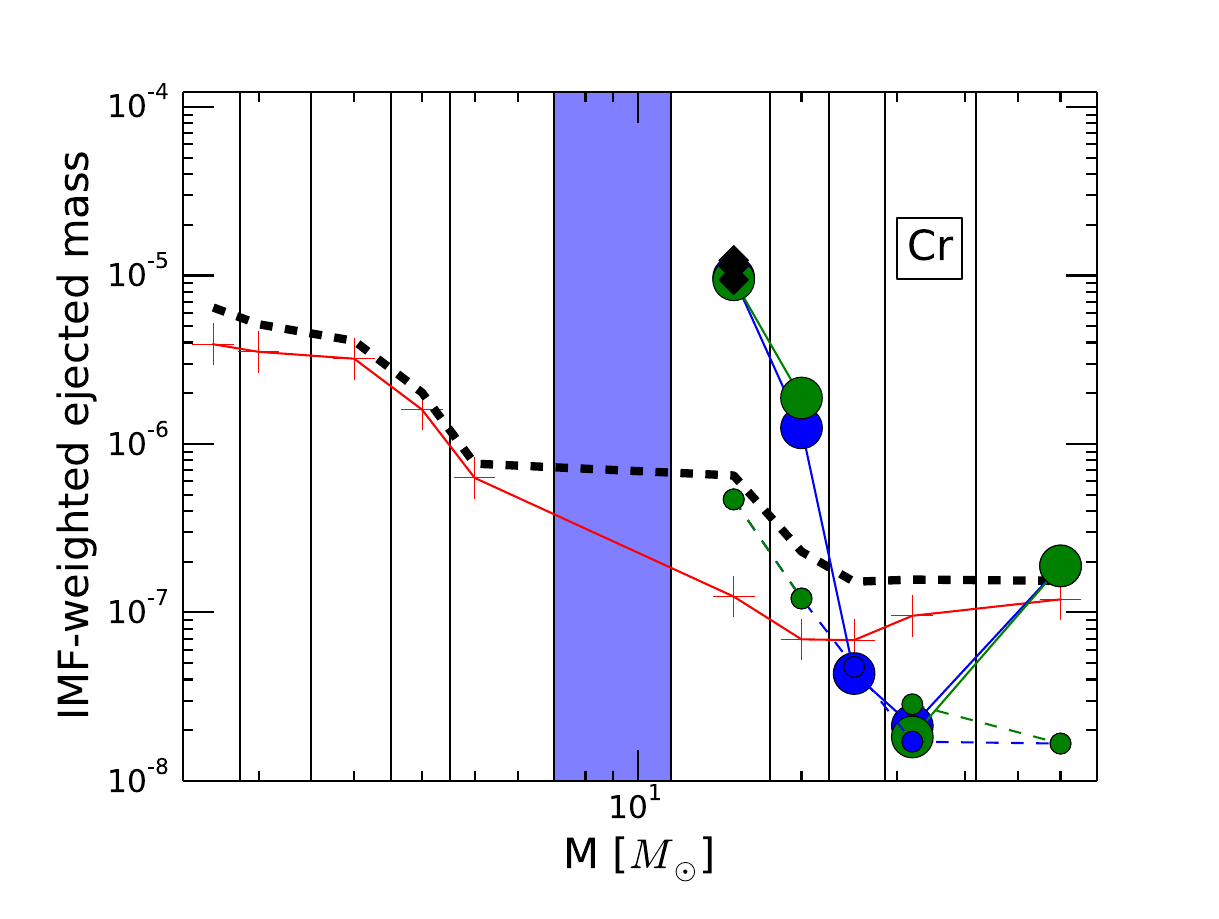}
\includegraphics[width=0.46\textwidth]{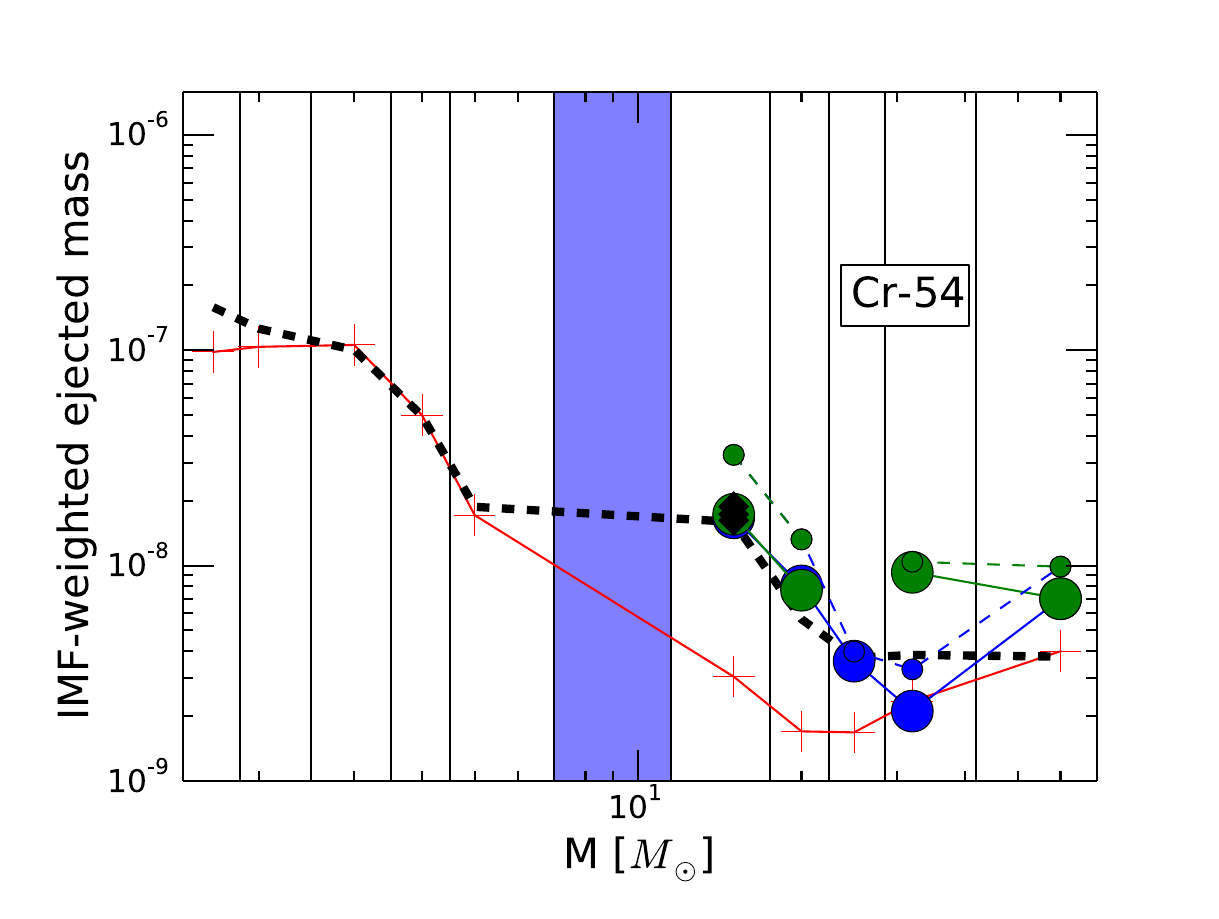}
\includegraphics[width=0.46\textwidth]{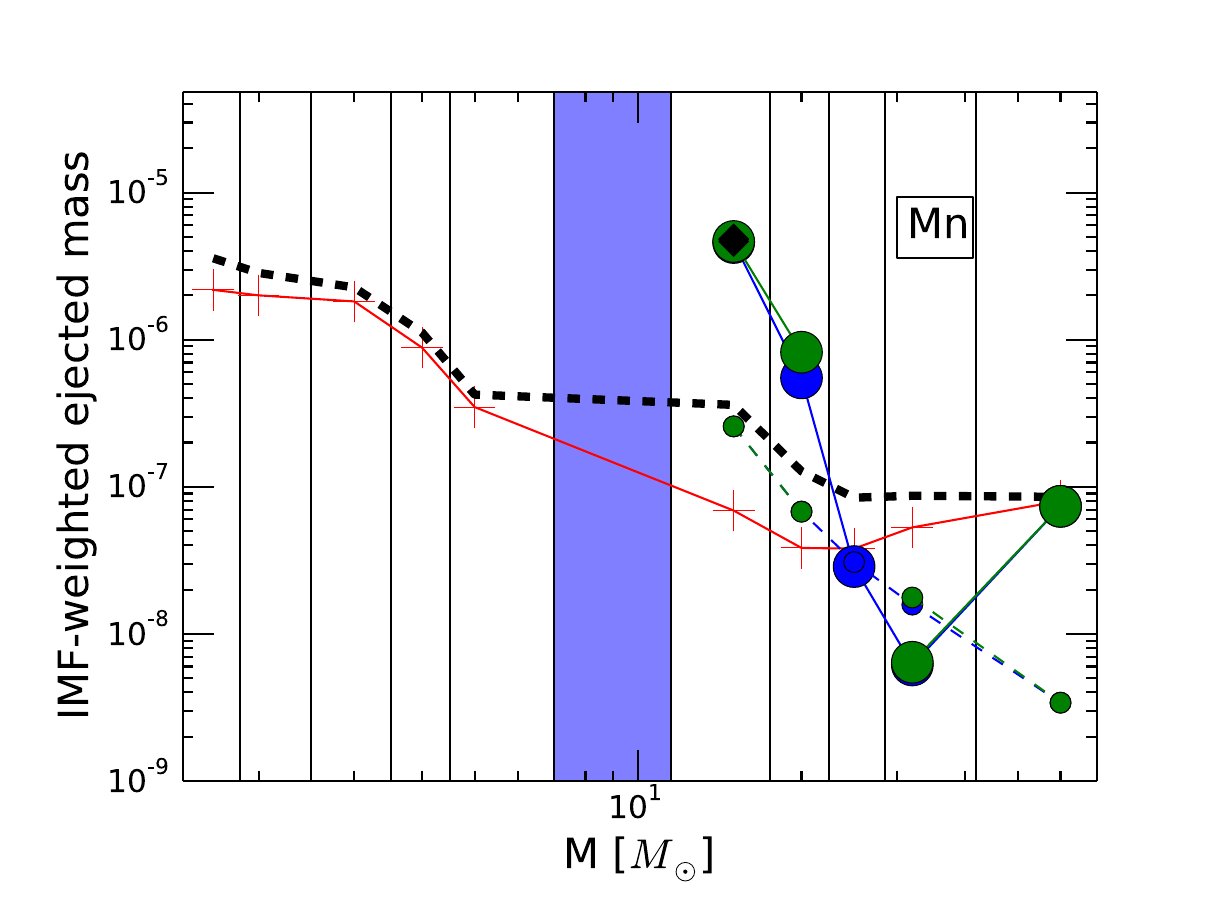}
\includegraphics[width=0.46\textwidth]{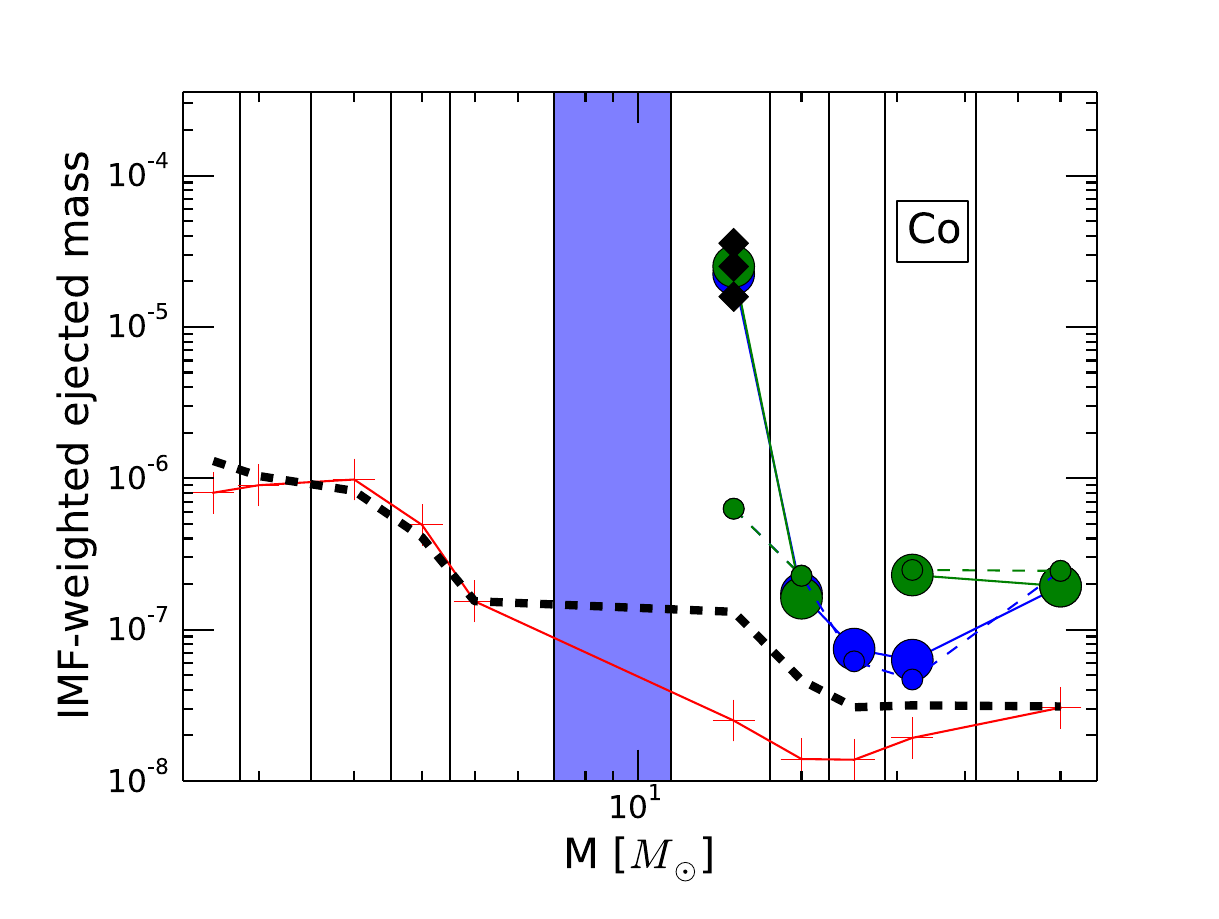}
\caption{Same as \fig{fig:CNONaAl_set1p2} for Sc, V, Ti, Cr, Mn, Ca, and some of their stable isotopes.}
\label{fig:s_set1p2_sum}
\end{figure}

Ti is produced in CCSNe and in SNIa
\citep[e.g.,][]{rauscher:02,seitenzahl:13}. Most production
comes from the mass range 15-$20\msun$ (\fig{fig:s_set1p2_sum}).
For larger masses part of the Ti-rich material falls back, however
looking specifically at the production of individual isotopes of Ti, the
situation is more complex. For example, \isotope[50]{Ti} is underproduced
compared to the other Ti isotopes in several SN models
\citep[e.g.,][]{woosley:95,thielemann:96}. In our calculations, most of the
\isotope[50]{Ti} is made during the pre-explosive evolution by the \sprn\ 
in the convective He-burning core, in the following convective C-burning
shell \citep[e.g.,][]{woosley:02,the:07} and by neutron captures during
explosive He and C burning, which partially compensates for the
destruction of \isotope[50]{Ti} at high temperatures deeper in the star. The
difference in the final \isotope[50]{Ti} yields for the two $32\msun$
explosion cases is due to the larger amount of fallback material in the
$delay$ model. Since recent SNIa models are not producing \isotope[50]{Ti}
efficiently \citep[e.g.,][]{travaglio:11,kusakabe:11}, it is possible
that most of the solar \isotope[50]{Ti} is made in massive stars. In principle,
the final \isotope[50]{Ti} abundance in the SN ejecta would be a good indicator
of the amount of fallback and explosion energy, taking into account
the uncertainties of its \spr\ production.

V is produced in massive stars during the CCSN.
%\citep[e.g.,][]{thielemann:96,woosley:02}. 
The contribution to the V
inventory from SNIa is quite uncertain
\citep[][]{travaglio:11,seitenzahl:13}. \isotope[50]{V} does not receive a
radiogenic contribution and therefore its abundance is a direct
indicator of its production, which is mostly during explosive O-burning
conditions. %\citep[][]{woosley:95}. 
The bulk of the
\isotope[51]{V} is synthesized by the decay of \isotope[51]{Cr} and
\isotope[51]{Mn} during freezout, both of which are produced in
deeper regions and at higher temperature than \isotope[50]{V} in the
explosion. Since most of \isotope[51]{V} is made in extreme conditions, its
total abundance in the ejecta is severely reduced with increasing
fallback. Therefore, V is 
underproduced in the 25 and $32\msun$ models (\fig{fig:s_set1p2_sum}).

Cr is efficiently produced in massive stars \citep[e.g.,][]{woosley:02}
and in SNIa \citep[e.g.,][]{thielemann:04,seitenzahl:13}. The
most abundant stable Cr species (\isotope[52,53]{Cr}) are made mostly as
\isotope[52,53]{Fe}. %\citep{woosley:95}. 
Therefore, Cr is
mostly produced in the 15-$20\msun$ stellar models, whereas for larger initial
masses fallback is limiting the ejection of
Cr-rich material (\fig{fig:s_set1p2_sum}).  \isotope[54]{Cr}
(2.365 \% of solar Cr) originates in the \sprn\,
%\citep[][]{woosley:95,rauscher:02}, 
or via neutron capture in the
explosive He-burning shell, and is destroyed in
explosive O- or Si-burning conditions. 

Mn is produced during CCSNe as \isotope[55]{Co} and \isotope[55]{Fe}.
%\citep[e.g.,][]{woosley:95}. 
\isotope[55]{Mn} is efficiently produced only in
the 15-$20\msun$ stars, whereas it is underproduced in higher mass
models (\fig{fig:s_set1p2_sum}) because the yield strongly decreases
with increasing fallback efficiency. Mn production also shows a significant
dependence on the explosion energy in the $15\msun$ models.

\begin{figure}
\centering

\includegraphics[width=0.46\textwidth]{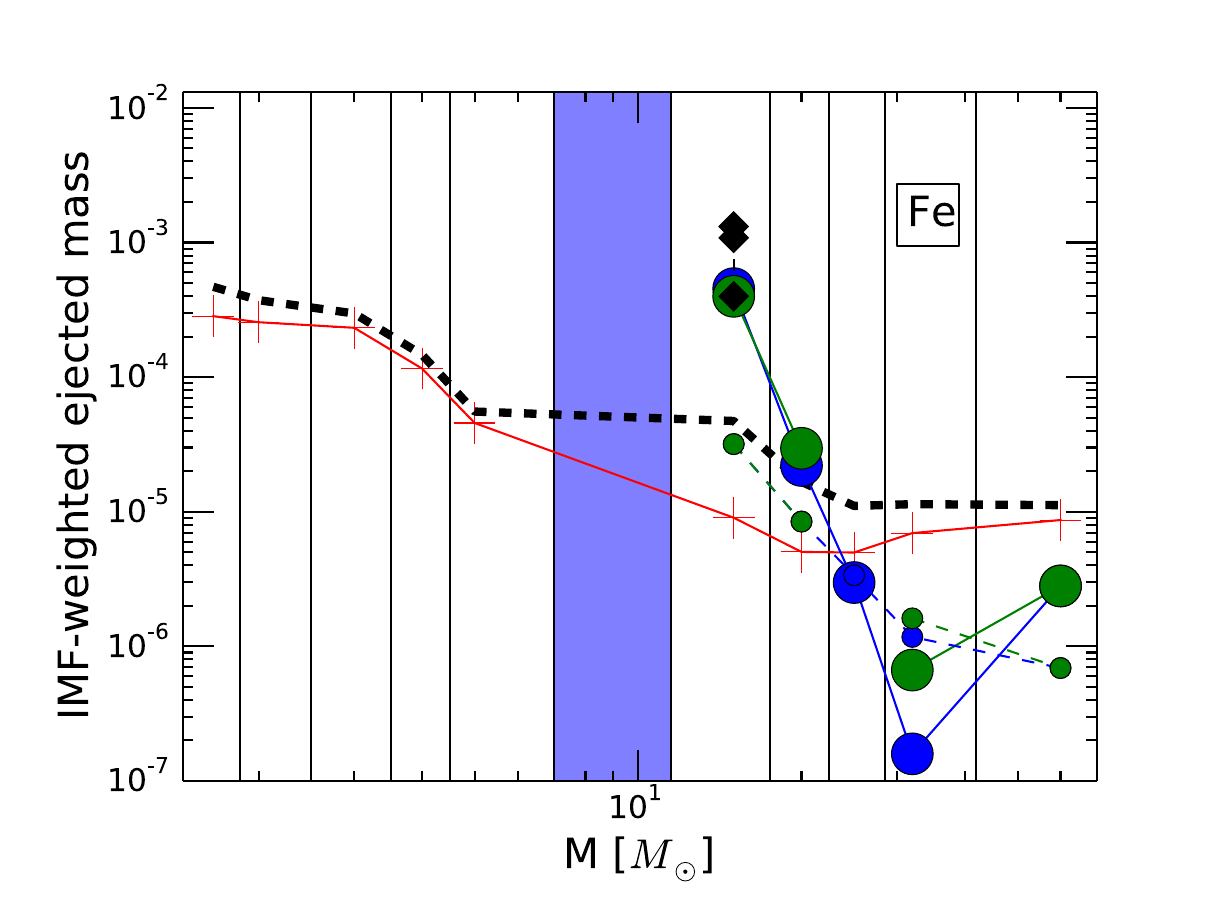}
\includegraphics[width=0.46\textwidth]{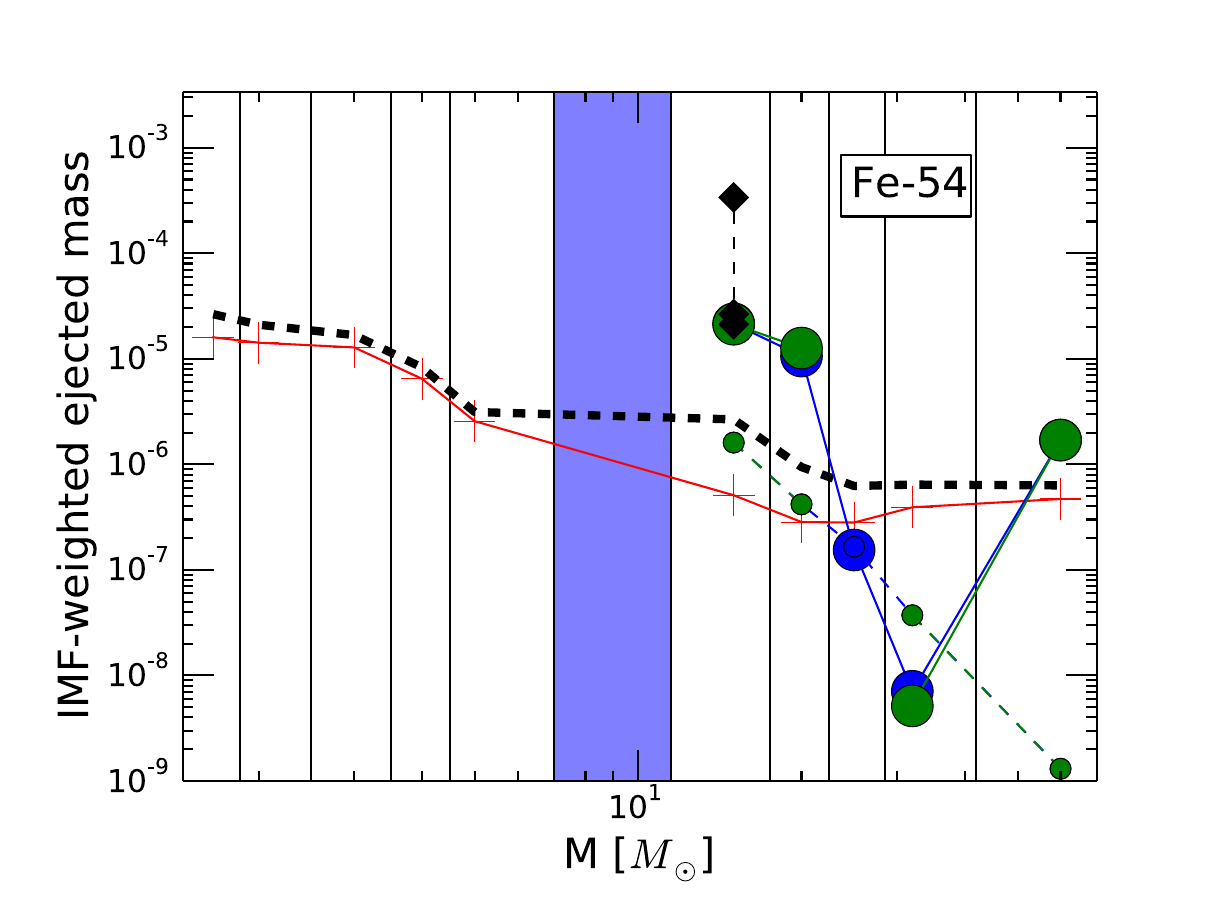}
\includegraphics[width=0.46\textwidth]{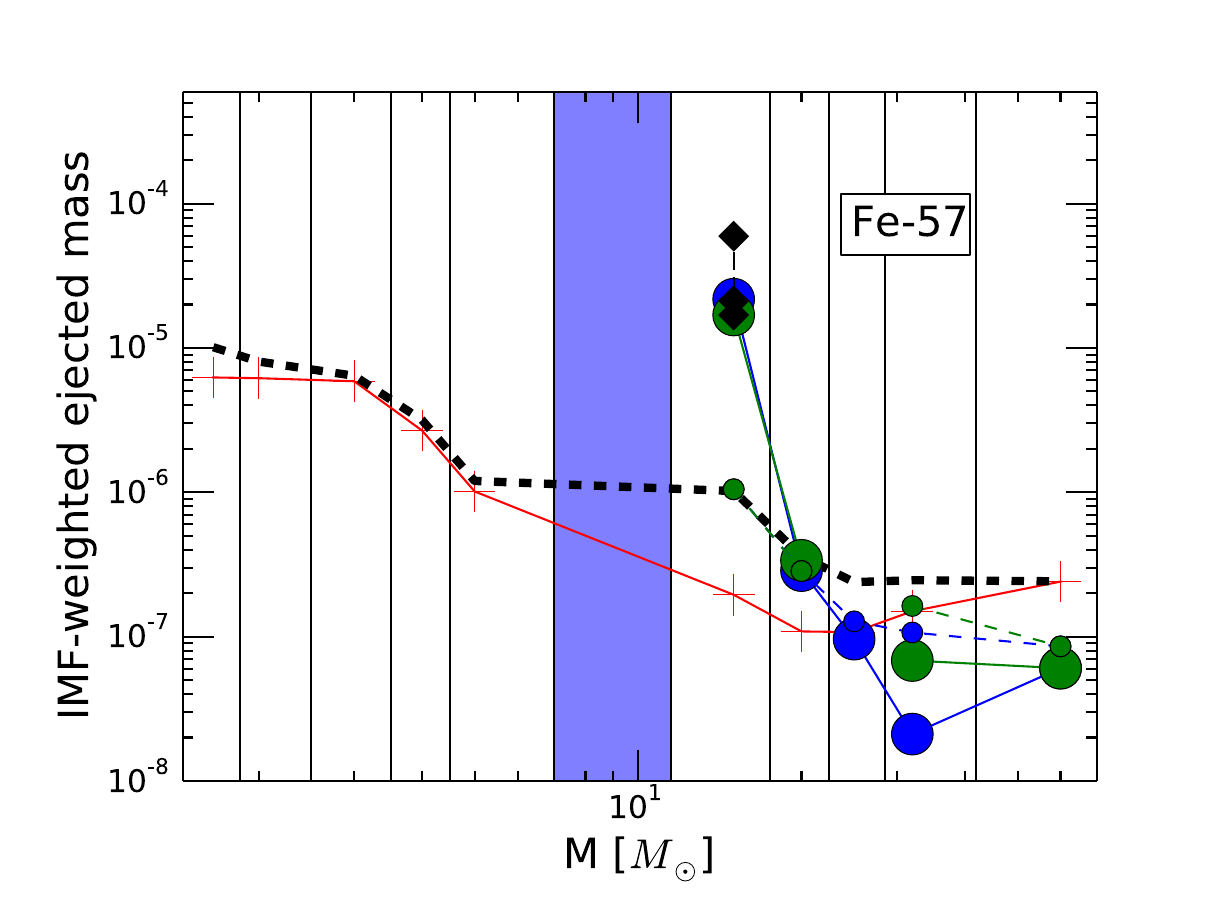}
\includegraphics[width=0.46\textwidth]{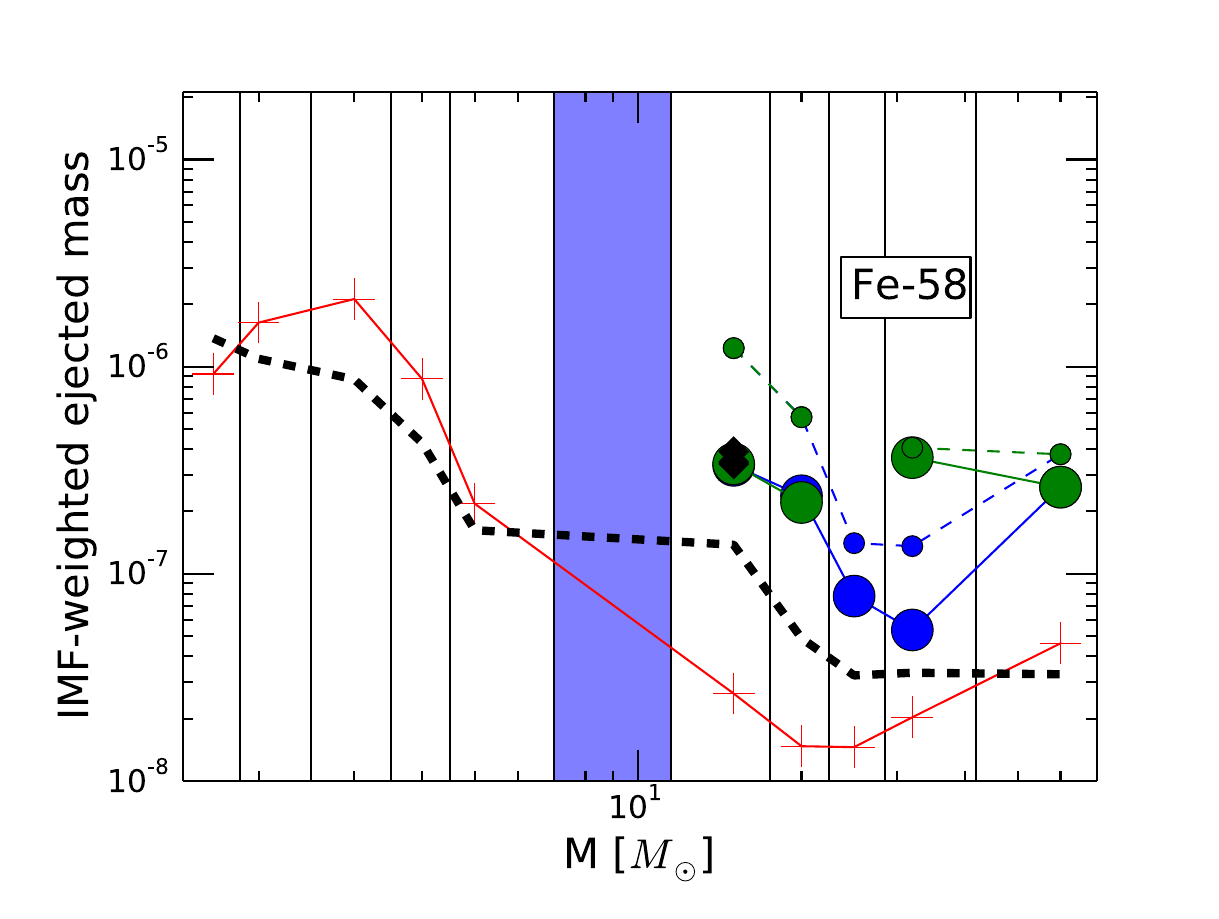}
\includegraphics[width=0.46\textwidth]{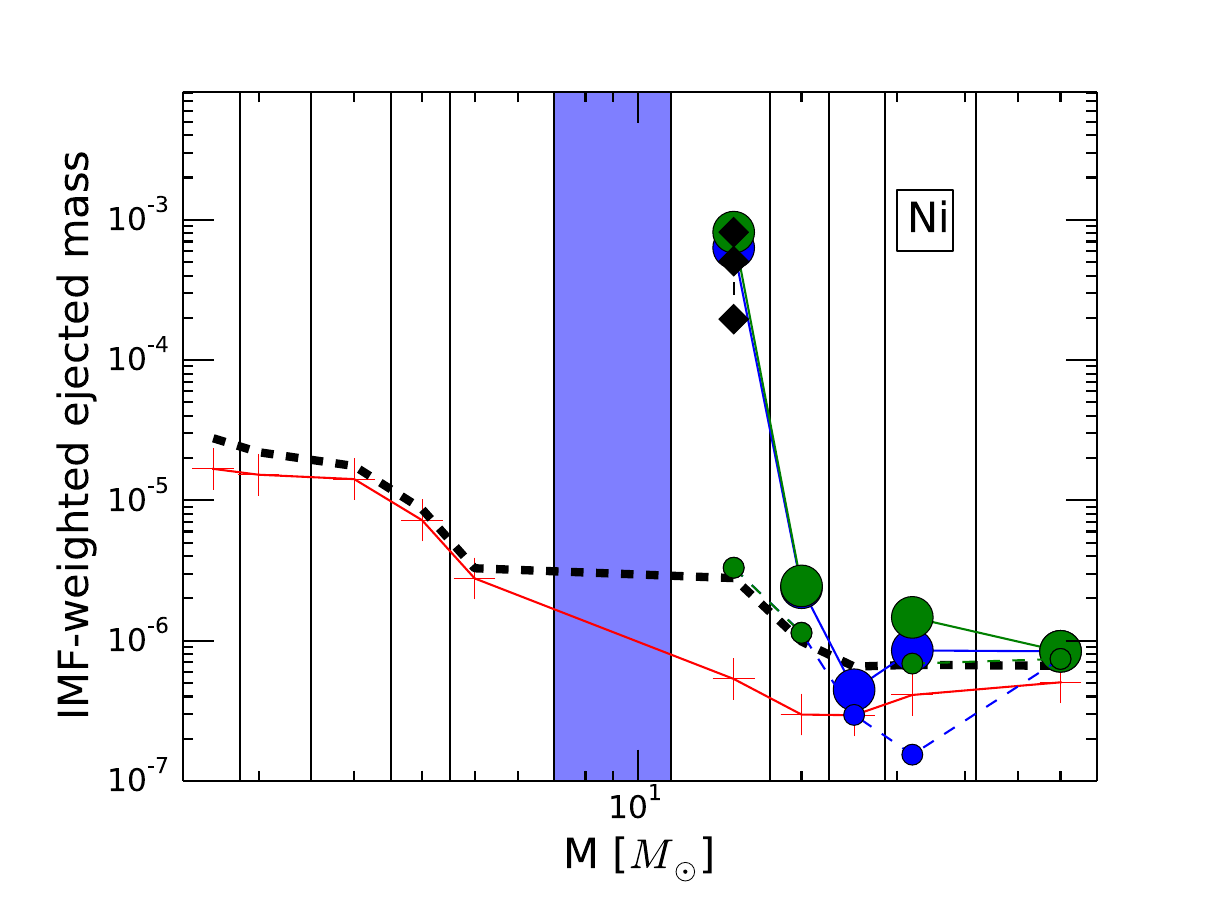}
\includegraphics[width=0.46\textwidth]{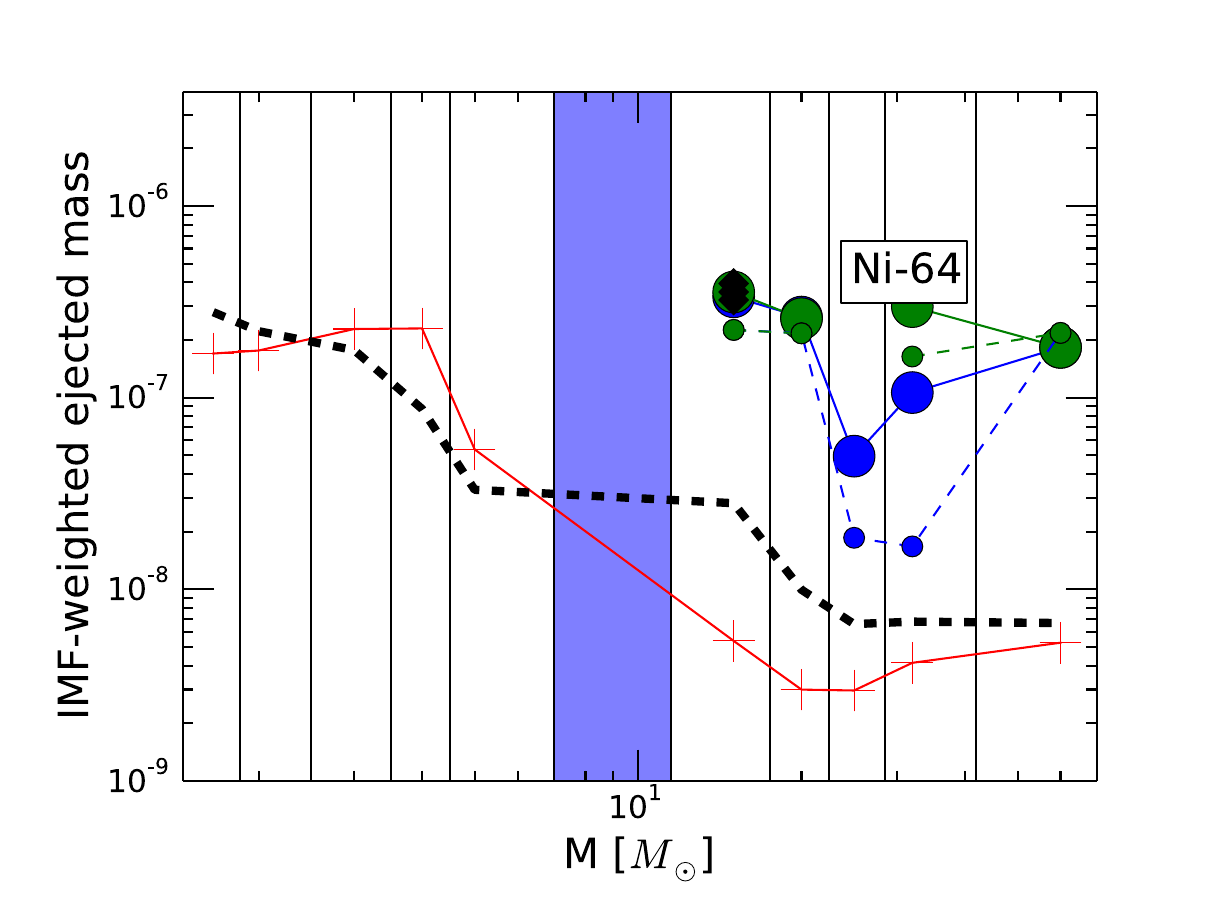}
\caption{Same as \fig{fig:CNONaAl_set1p2} for Fe and Ni and some of their stable isotopes.}
\label{fig:fe_set1p2_sum}
\end{figure}

The dominant Fe isotope \isotope[56]{Fe} is produced in CCSN and in SNIa
as \isotope[56]{Ni} (\fig{fig:fe_set1p2_sum}) Because of fallback, only the
$15\msun$ star efficiently produces \isotope[57]{Fe}, mostly as \isotope[57]{Ni}.
%\citep[see][]{woosley:95}. 
Like \isotope[54,56]{Fe}, \isotope[57]{Fe} also has a strong
dependence on the explosion energy.  
%Contrary to that the \isotope[54]{Fe} is
%increasing with decreasing energy in the $15\msun$ models because
%\isotope[54]{Fe} is mostly made without radiogenic contributions and for larger
%explosion energies \isotope[58]{Ni} is made instead of \isotope[54]{Fe}.
%\citep[][]{woosley:95}. 
The Fe neutron-rich isotope \isotope[58]{Fe} is produced over the whole stellar
range (\fig{fig:fe_set1p2_sum}). In massive stars, it
is mainly produced by the \sprn\ during the pre-explosive phase and
is partially destroyed by the SN explosion. In our models a significant
amount of \isotope[58]{Fe} is also produced in the explosive He-burning shell
by neutron captures. This contribution is particularly important for
the lower-mass CCSNe, such as the $15\msun$ case, where most of the
pre-explosive abundances are strongly affected by the SN explosion. For
the $Z=0.02$ models, the AGB stars provide the largest contribution to the
\isotope[58]{Fe} inventory, via the \sprn\ (\sect{sprocess}).

Besides a small positive contribution to Co from the AGB star \sprn, the
strongest production happens in massive stars (\fig{fig:s_set1p2_sum}). 
The $15\msun$
SN models show a correlation of the Co yields with the explosion energy. In the most
energetic SN models, most of \isotope[59]{Co} is made as \isotope[59]{Cu}, with a
smaller contribution from \isotope[59]{Ni}, \isotope[59]{Co} itself and \isotope[59]{Fe}
from the explosive He-burning shell. At lower explosion energy the \isotope[59]{Cu} and 
\isotope[59]{Ni} production is reduced. In this case \isotope[59]{Co} comes from direct
production and from \isotope[59]{Fe} decay.  This makes Co a possible
nucleosynthesis signature of highly energetic SNe \citep[see
e.g.,][]{nomoto:09}.  For larger masses, where the fallback
contribution in our models is stronger, the explosive
contribution of the radiogenic \isotope[59]{Fe} in the He shell becomes more
relevant for the final Co yields. For weaker fallback 
(e.g., the $25\msun$ SN $rapid$ model or the
$60\msun$ models) most Co originates comes from the \sprn.

The most abundant Ni species, \isotope[58,60]{Ni}, 
%\isotope[58]{Ni} and \isotope[60]{Ni}, 
are produced efficiently 
in CCSNe at high temperatures \citep[for a recent analysis of the Ni production in CCSNe compared to Fe, we refer to][]{jerkstrand:15}, 
with a strong contribution also from SNe Ia
\citep[e.g.,][]{thielemann:04,bravo:10,seitenzahl:13}.  %As already mentioned by
The production in massive stars depends on fallback and explosion
energy. %\citep{woosley:95}. 
For example, the $25\msun$
SN models do not efficiently contribute to the bulk of the Ni inventory
because of the strong fallback (\fig{fig:fe_set1p2_sum}).  
The other stable Ni isotopes, \isotope[61,62,64]{Ni},
%\isotope[61]{Ni}, \isotope[62]{Ni}, and \isotope[64]{Ni},
can have a contribution from AGB stars. The \isotope[64]{Ni} yield from 
the \sprn\ in massive AGB stars is smaller than the massive star yield 
(see e.g., \tabs{tab:isotopic_prodfac_set1.2_winds} and \ref{tab:isotopic_prodfac_set1.2_exp}). On the other hand,
the weighted yields over the Salpeter IMF in \fig{fig:fe_set1p2_sum} become comparable for the two different stellar mass regimes, 
since intermediate-mass stars are more numerous than massive stars.
%favoring yields from the more frequent low-mass and intermediate-mass stars comparing to massive stars.

For models with less fallback and with less energetic explosions, more than 50\% of the 
\isotope[64]{Ni} abundance is produced by the pre-explosive \spr\ contribution. 
The explosive contribution via neutron capture in the explosive
He shell becomes more relevant for models with large fallback and/or
high SN energy, where \isotope[64]{Ni} is produced directly from neutron captures on other Ni isotopes or via the
decay of neutron rich isobars from the lighter iron group elements (e.g., unstable \isotope[64]{Co} and \isotope[64]{Fe}).

\subsection{Trans-iron elements}
\label{sprocess}

%\label{sprocess}

Trans-iron elements are made during the quiescent stellar evolution by
the \sprn\ \citep{meyer:94,busso:99,kaeppeler:11}. Our models contain contributions
from the \sprn\ in AGB and in massive stars. In addition, in massive stars there is a 
relevant contribution from explosive nucleosynthesis during the CCSN
explosion (\sect{subsec:exp}). 
In these conditions, we also follow the activation of the \pprn\ \citep[or $\gamma$ process,][]{arnould:03}, 
the $\alpha$ process \citep[][]{woosley:92},
and the $n$ process in the explosive He shell \citep[e.g.,][]{blake:76}. 
In the following sections we will consider these different processes more in detail, depending on the mass region where their contribution is more relevant.
Our models do not include $\nu$-winds nucleosynthesis components \citep[e.g.,][]{hoffman:96, froehlich:06, kratz:08, qian:08,farouqi:10,magkotsios:10,arcones:11}
or the rapid neutron capture process \citep[\rprn, e.g.,][]{thielemann:11}. 
%Therefore, the stellar yields presented in this work do not include any \rpr\ component.

%A number of nucleosynthesis processes are responsible for the production of heavy elements in different types of stars, showing different time scales in their contribution to chemical evolution. Important constraints are coming from the observation of heavy elements in old stars \citep[e.g.,][]{sneden:08}, the Solar System distribution \citep{asplund:09}, and stellar abundances of field stars in the Milky Way or in other environments \citep[e.g., for Dwarf Spheroidal Galaxies,][and references therein]{tolstoy:09}. Considering the isotopic distribution of nuclides heavier than Fe in the Solar System, most of the abundances are produced by the \sprn\ and by the \rprn. Additionally, between iron and the neutron shell closure at N=50 a number of charged particle reactions triggered by the CCSN event may have contributed to the final stellar yields. These processes include, amongst others, the incomplete $\alpha-$rich freezout in CCSN ejecta and within the neutrino winds produced during the formation of a neutron star \citep[e.g.,][]{woosley:92,hoffman:96, froehlich:06, kratz:08, qian:08,farouqi:10,magkotsios:10,arcones:11}. Finally, a small contribution is provided by the so called \pprn, triggered by photodisintegration reactions during the CCSN \citep[][]{arnould:03}.  

\subsubsection{From Ni to Sr-Y-Zr}
\label{ni-zr}

The abundances between Ni (Z=27) and Zr (Z=40) can be produced by different processes in different types of stars. In this region, the fundamental phenomenological concept of $p-$only, $s-$only and $r-$only isotopes is too uncertain to be used to disentangle the origin of the Solar System abundances, and each case should be considered carefully. 
On the other hand, for a large number of stars spectroscopic observations are available for Cu, Zn, Sr, Y and Zr at different metallicities. A smaller sample of stellar data are available for Ge \citep[e.g.,][]{cowan:05}, As and Se \citep[e.g.,][]{roederer:14} and Rb \citep[][]{abia:01,garcia-hernandez:09,zamora:14}, that will help to better quantify the relative nucleosynthesis contribution of different processes. In this section we discuss our results, and we introduce the relevant nucleosynthesis processes for this mass region: the \sprn\ in massive stars and massive AGB stars, and the \aprn.
The \pprn\ is contributing to \isotope[74]{Se}, \isotope[78]{Kr} and \isotope[84]{Sr}, but most of the $p-$only isotopes are located above Zr. Therefore, the \pprn\ will be discussed in more details 
in the following section.
%Similarly, while the \sprn\ in standard AGB stars produces most of the Sr, Y and Zr in the solar system \citep[e.g.,][]{bisterzo:14}, we will discuss about its main features in the next section. 
We do not consider in this work the neutrino-driven wind ejecta, which may potentially contribute to the nucleosynthesis up to Zr \citep[e.g.,][]{arcones:11}.

\begin{figure}
\centering
\includegraphics[width=0.48\textwidth]{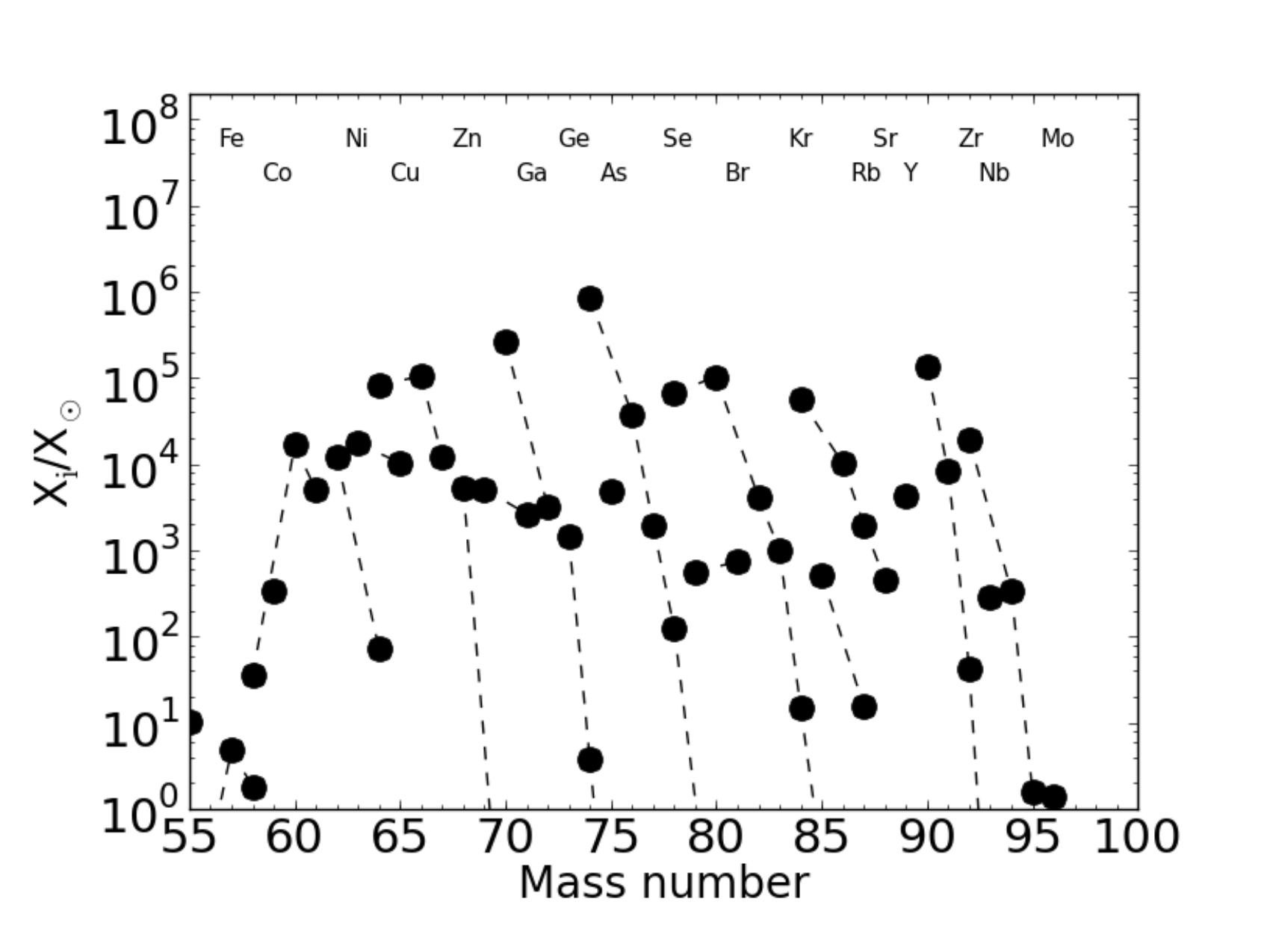}
\caption{Overabundances at mass coordinate $1.849\msun$ in the $15\msun$ $delay$ model from \setopt, due to the $\alpha$ process activation.
}
\label{fig: alpha_process_15_exp_d}
\end{figure}

\begin{figure}
\centering
\resizebox{15cm}{!}{\rotatebox{0}{\includegraphics{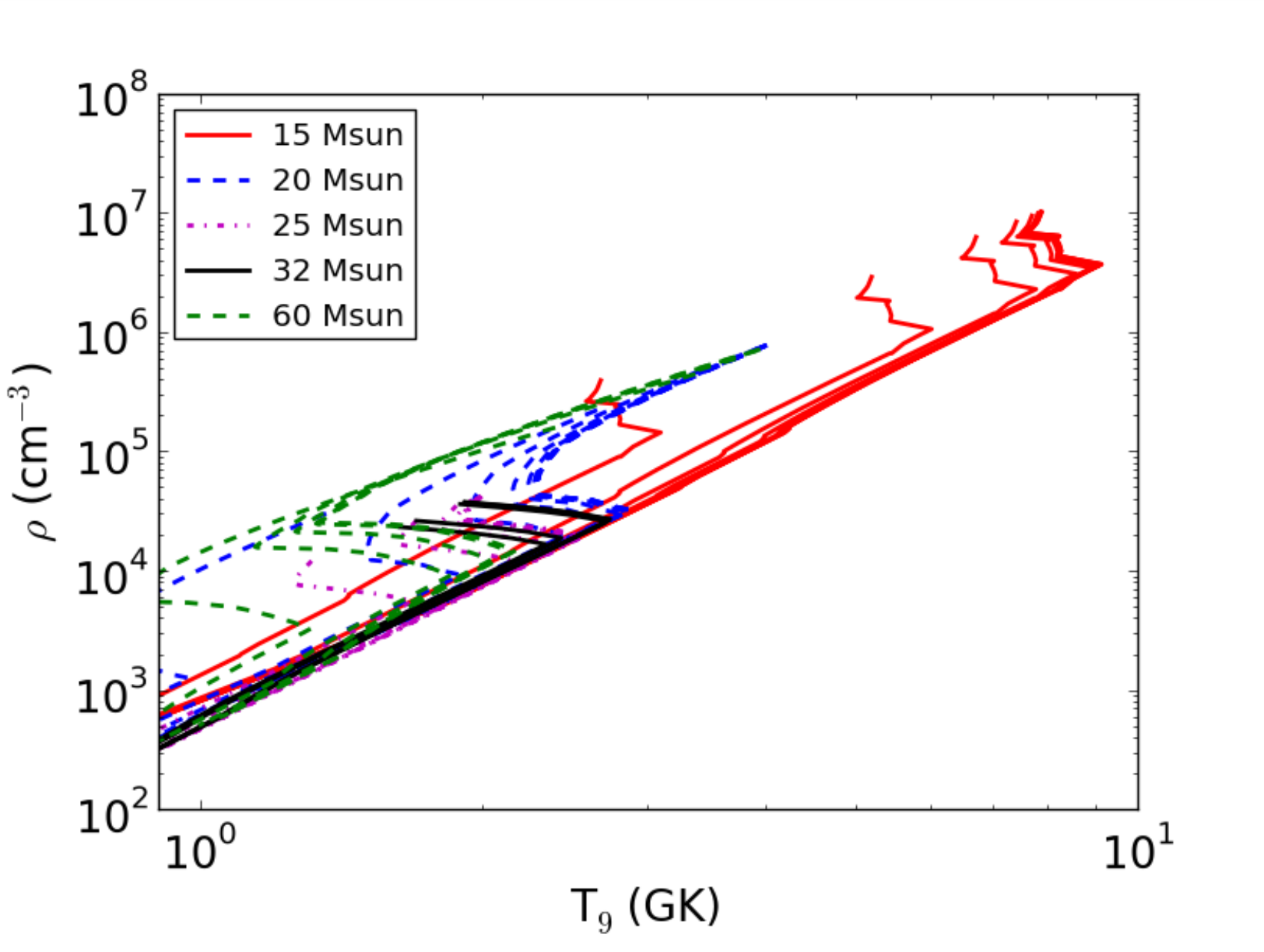}}}
\caption{Evolution of temperature and density profiles during CCSN for massive stars 
from \setopt\ and the SN $delay$ model.
%All the explosive profiles fall in the $\alpha$-rich freezout regime, 
%or in the incomplete burning regime according to the amount of fallback.
For comparison, see the same diagram in \citet{thielemann:11}.
}
\label{fig:rho_t9_exp_diagram_set1p2}
\end{figure}

During the CCSN event, the stellar layers that will be ejected are first exposed to extreme thermodynamics conditions up to the nuclear statistical equilibrium. If the decrease in temperature and density after reaching their peak is fast enough, a large number of $\alpha$ particles are left in the ejecta, eventually leading to the $\alpha$-rich freeze-out nucleosynthesis, or $\alpha$ process \citep[][]{woosley:92}: depending on the thermodynamic conditions and on the initial electron fraction, species heavier than iron can be produced efficiently up to Ag (Z=47). 
\cite{magkotsios:10} renamed the $\alpha$ process as the $\alpha$$n$ process, to distinguish it from the $\alpha$$p$ process.
The present CCSN models are characterized by fast shocks (see also \fig{fig:rho_t9_exp_diagram_set1p2}). Together with the slightly neutron-rich initial conditions (electron fraction $Y_e<0.5$) in the massive star progenitor models, the $\alpha$ process is activated in the $15\msun$ CCSN models of \setone\ in deep stellar layers of few $10^{-2}\msun$.
%: x, y z in \setopt\ and xx, yy and zz in \setopo. [MP: REFER TO YIELDS TABLE]
In \fig{fig: alpha_process_15_exp_d}, we show the abundance distribution obtained at mass coordinate $1.849\msun$ for the $15\msun$ model, $delay$ explosion, from \setopt.
At this location, before the SN explosion the initial electron fraction is $Y_e$ = 0.496. During the CCSN the temperature and density peaks are about 9.3 GK and 3.8$\times$10$^{6}$ g\cc,
%(corresponding to a neutron excess $\eta$=0.008),
and the final \isotope[4]{He} mass fraction will be 0.44 (i.e. 44\% of the material is made of He). The heavy isotope that is most efficiently produced is \isotope[70]{Ge}, with a local overproduction of 8.7$\times$10$^5$, but the production flow is efficient up to the Zr-Mo region. 
%Notice that these results are consistent with the calculations from \citep[][]{woosley:92}, with comparable local overproduction of heavy species. 
The $\alpha$-process is activated only in the simulations for the $15\msun$ models from \setone, affecting their final yields. On the other hand, stars with larger initial mass in the present set of models have a more efficient fallback and do not make any \aprn\ products. Their final abundances in the mass region between Fe and Zr are dominated by the \spr. In the same way, the $15\msun$ stars from \setopt\ with shock velocities reduced by a factor of two and four respectively are not hosting the \apr. The strong sensitivity of the \apr\ to the explosive conditions and to the progenitor mass makes this exotic process more difficult to analyze. On the other hand, its high production efficiency means it has the potential to have an impact on the chemical inventory of the Galaxy, even if associated only to CCSN ejecta with high shock velocities or hypernovae.
At the moment, the only observational confirmation of the \aprn\ activation is given by the observation of the [Zn/Fe] in metal poor stars \citep[][and references therein]{primas:00, bisterzo:05}, where Zn is expected to be produced as \isotope[64]{Zn} and \isotope[66]{Zn} \citep[e.g.,][]{nomoto:13}. While an explosive component for Cu \citep[e.g.,][and references therein]{bisterzo:05,sobeck:08}, Ge \citep[][]{cowan:05}, As and Se \citep[][]{roederer:14} is observed already in old metal poor stars before the \sprn\ contribution becoming relevant, it is not clear at the moment what is the effective relevance of the \apr\ for these elements.

Between iron and strontium (60 $\lesssim$ A $\lesssim$ 90), the \spr\ abundances in the Solar System are mostly 
produced in massive stars \citep[the weak \spr\ component, see for example][and references therein]{kaeppeler:89,beer:92,kaeppeler:11}.
In massive stars, the main neutron source for the
\sprn\ is the \isotope[22]{Ne}($\alpha$,n)\isotope[25]{Mg} reaction
\citep[][]{peters:68,couch:74,lamb:77}. Depending on the initial mass
of the star \citep[e.g.,][]{prantzos:90} and on the \isotope[22]{Ne}+$\alpha$
rates \citep[e.g.,][]{kaeppeler:94}, some \isotope[22]{Ne} is left in the
He-burning ashes, which is activated later in the subsequent C-burning
conditions \citep[e.g.,][]{raiteri:91a}.
%Massive stars are responsible for the weak \spr\ component in the solar system \citep[e.g.,][]{kaeppeler:89}.
The elements produced most efficiently are copper, gallium and
germanium \citep[][and references therein]{pignatari:10}. 

The pre-SN production of the \spr\ elements in our models has been discussed in
the context of an analysis of the \czw+\czw\ nuclear reaction rate
uncertainty by \citet{bennett:12} and \citet{pignatari:13}. 
%The \spr\ elements are partially modified from  the SN explosion.
The SN shock wave partially depletes or changes the original pre-explosive \spr\ abundances
\citep[e.g.,][]{rauscher:02,tur:09}. In this case, the resulting explosive stellar yields of \spr\ elements would still share a similar production efficiency and metallicity dependence with their \spr\ seeds. The relevance
of the feedback of the explosion on the pre-explosive \spr\ signature depends on many details of the SN mechanism.
In a $25\msun$ star, the bulk of the pre-explosive \spr\ abundances lies in the convective C-burning shell and in the
ashes of the He core material located between the C shell and the He shell \citep[e.g.,][and references therein]{the:07,pignatari:10}.
For standard CCSN models, with a SN explosion energy in the order of $10^{51}$ erg and a ``mass cut" located below the bottom of
the convective C shell, most of the \spr-rich material in a $25\msun$ star would be ejected unchanged by the explosion
\citep[e.g.,][]{woosley:95,rauscher:02,limongi:00}.
In the $25\msun$ stellar models discussed here,
most or all of the \spr-rich material falls back forming a BH
\citep[the star in this case ends as a failed SN, see e.g.,][and references therein]{woosley:02}
according to \citet{fryer:12}.
In particular, the central $5.71\msun$ and $6.05\msun$ is not ejected
for the $delay$ explosive calculations of \setopo\ and \setopt, respectively.
For the $rapid$ explosive models, at $Z = 0.02$ no material is
ejected (complete fallback), and for $Z = 0.01$ only the material
external to the mass coordinate $7.91\msun$ (see \tab{tab: coo_fallback}) is
 ejected. 
%[SJ: I would have thought more material was ejected in the rapid models??]
Furthermore, the remaining \spr-rich material will be significantly modified 
by the sudden increase of temperature and density related to the SN shock wave. 
The pre-explosive C shell material could be modified by shell merging 
in the last $\sim$ day before the core collapse starts. This does not happen in our 
simulations, but it has been obtained, for instance, in the 20 and $25\msun$ stars by 
\citet{rauscher:02} and \citet{tur:09}. 
Finally, nuclear uncertainties \citep[e.g.,][]{busso:85,rauscher:02,the:07,pignatari:10,massimi:12,lederer:14,heil:14}
and physics mechanisms not included in our models like rotation \citep[e.g.,][]{frischknecht:12} have a relevant 
impact on \spr\ results.

\begin{figure}
\centering
\includegraphics[width=0.48\textwidth]{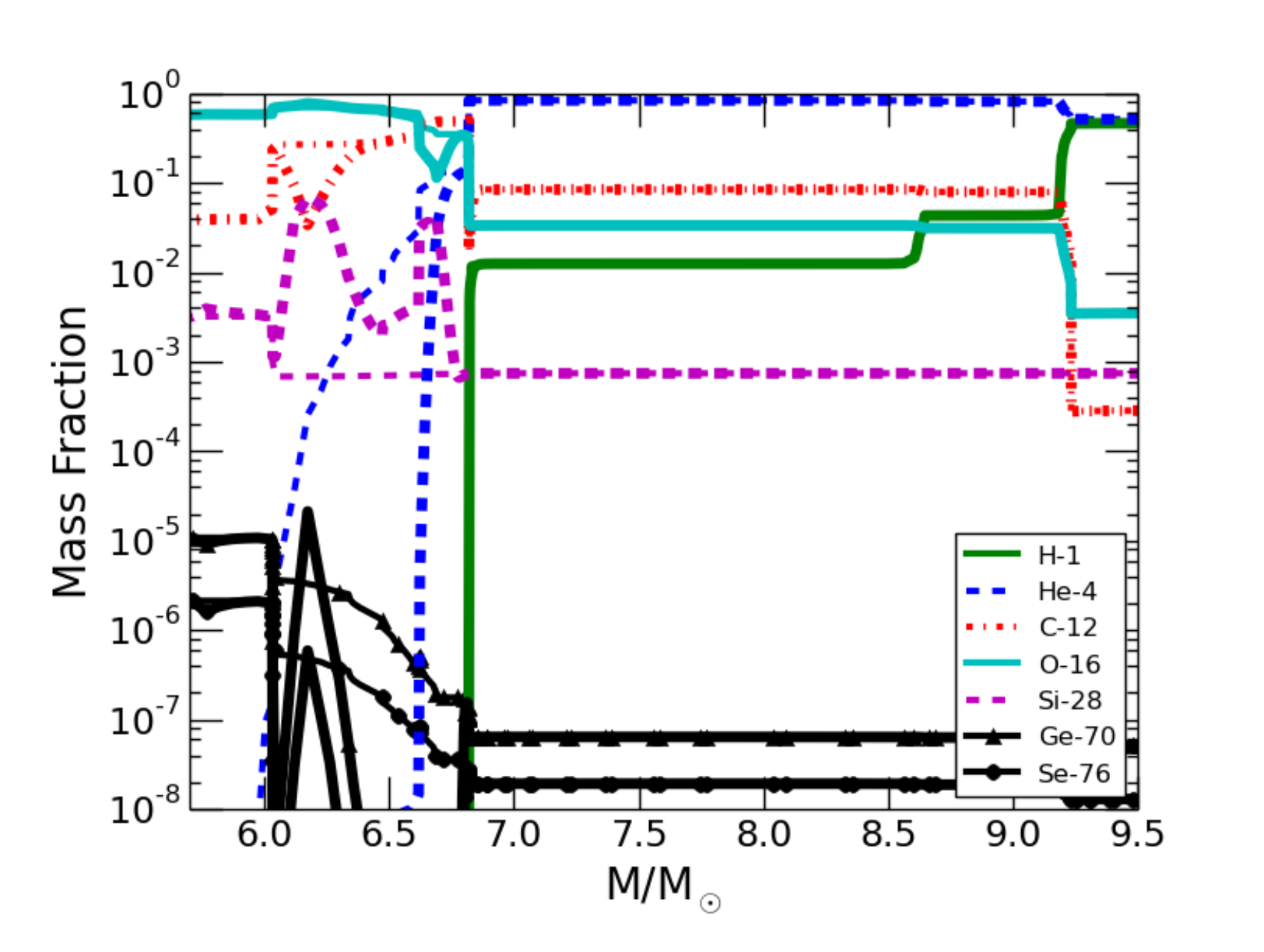}
\includegraphics[width=0.48\textwidth]{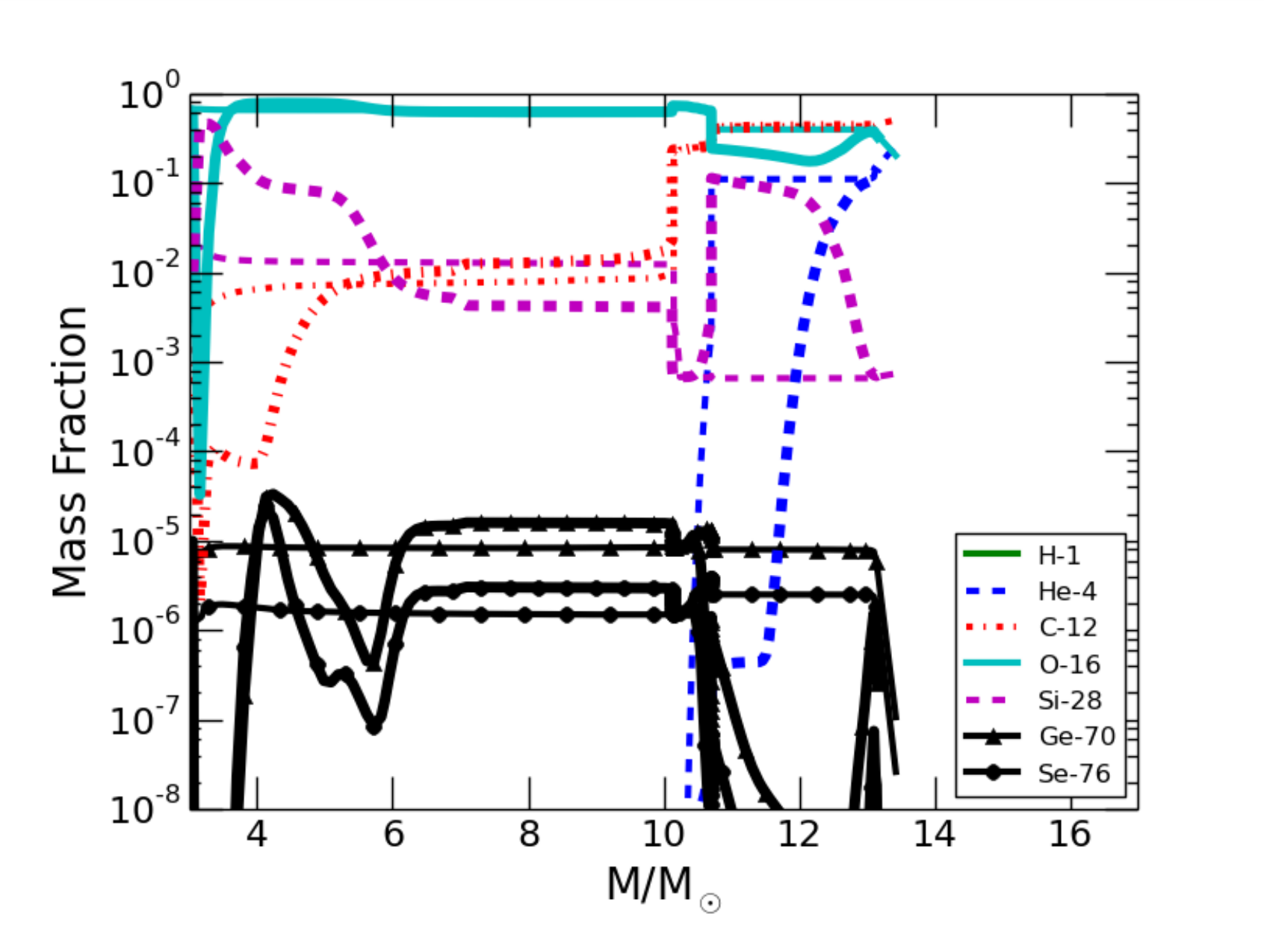}
\caption{Pre-explosive (thin lines) and post-explosive (thick lines) abundances of the $s$-only
species \isotope[70]{Ge} and \isotope[76]{Se} for the $25\msun$ and $60\msun$ models from \setopt\ in the $delay$ model (left and right panel, respectively).
The isotopes \isotope[1]{H}, \isotope[4]{He}, \isotope[12]{C}, \isotope[16]{O}, and \isotope[28]{Si} are reported to identify the different burning zones.
The abundances include the contribution from radiogenic decay.
}
\label{fig: sprocess_massive_pre_post}
\end{figure}

In \fig{fig: sprocess_massive_pre_post}, we show the abundance
profile before and after the SN shock wave for the two $s$-only
species \isotope[70]{Ge} and \isotope[76]{Se}; we compare the $25\msun$ and
$60\msun$ models with $Z = 0.02$ and $delay$.  In the 
$25\msun$ model only about $0.3\msun$ of the \spr-rich
material from the convective C shell is ejected, including small
modifications from the explosion.  The \spr\ abundances are
strongly modified in the He core window and at the bottom of the He
shell by neutron captures, where stellar conditions and fuel are
suitable to trigger explosive He-burning and the efficient neutron
production by the \isotope[22]{Ne}($\alpha$,n)\isotope[25]{Mg} is possible.
In the $60\msun$ model, the amount of fall-back material
is smaller than in the previous case ($3\msun$, see 
\tab{tab: coo_fallback}). However, the pre-explosive \isotope[70]{Ge} and
\isotope[76]{Se} made by the \sprn\ in the regions between about
3 and $6\msun$ and between $10.5\msun$ and the
surface of the star are modified by photodisintegration during
explosive O- and C-burning and by neutron captures due to explosive He
burning respectively.  The external part of the C shell material
(between $6\msun$ and $10.5\msun$) is only modified slightly.

\begin{figure}
\centering
\includegraphics[width=0.48\textwidth]{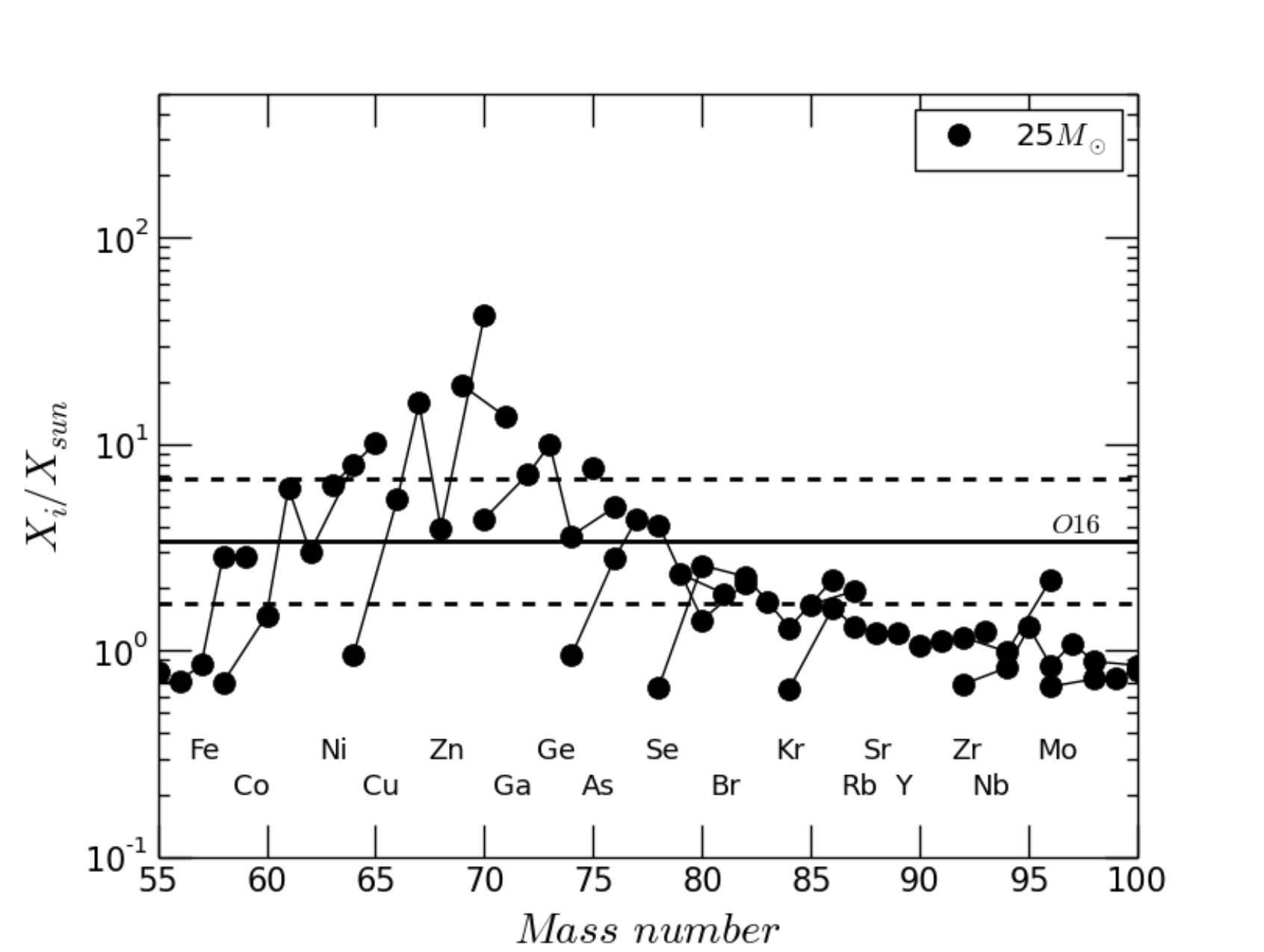}
\includegraphics[width=0.48\textwidth]{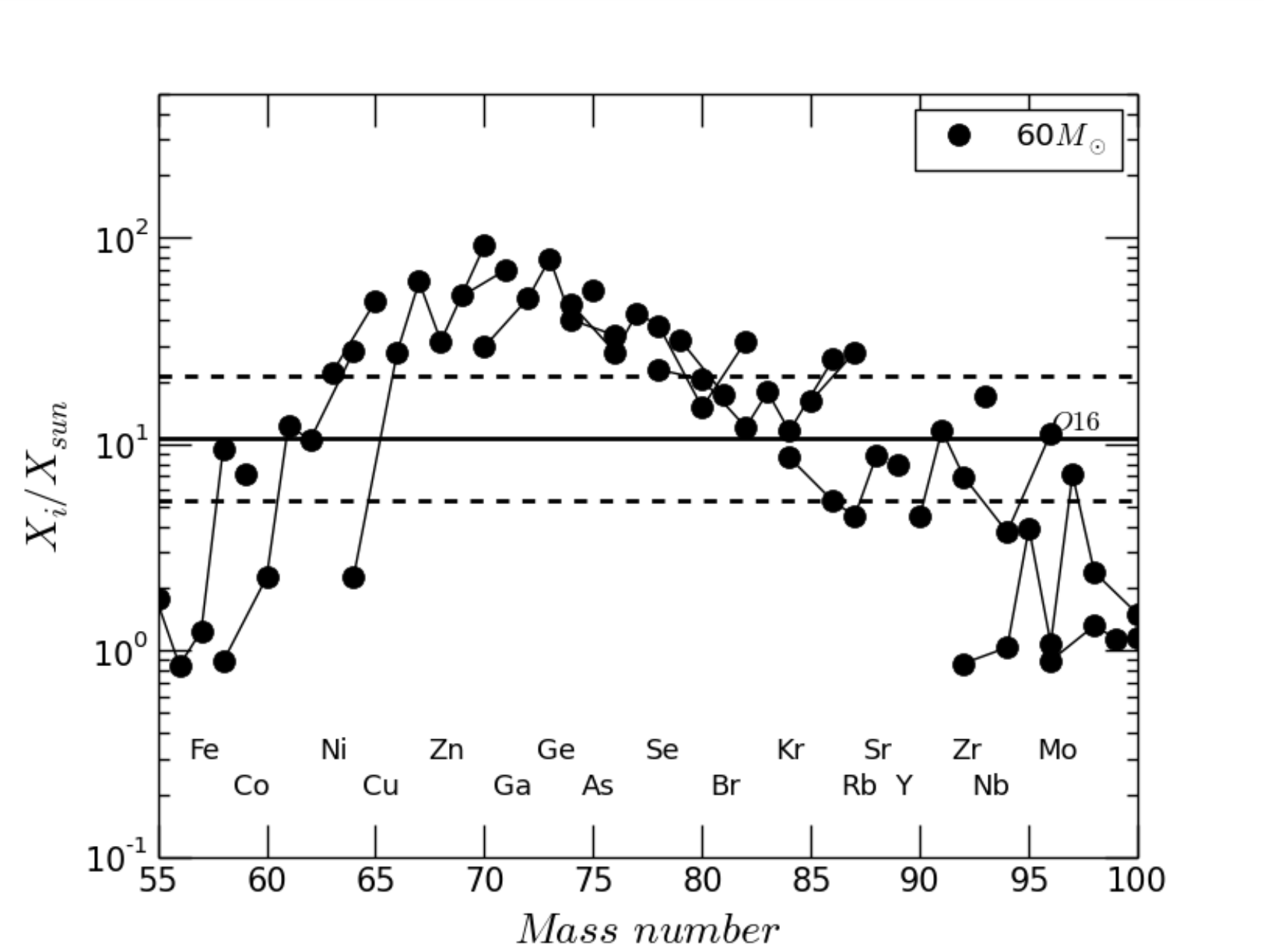}
\caption{Final overproduction factors for isotopes between Fe and Mo 
for the $25\msun$ and $60\msun$ models from \setopt\ 
in the $delay$ model (left and right panel, respectively). 
The production factor of \isotope[16]{O} is reported (continuous line), 
multiplied and divided by a factor of two (dashed lines).
%[FH: CHECK SAME PLOT PRE-SN. SHOULD WE INCLUDE ALSO THAT PLOT? NOT NECESSARY FOR NOW.]
}
\label{fig: sprocess_distribution_25_60}
\end{figure}

In \fig{fig: sprocess_distribution_25_60}, the final isotopic
production factors are given for the same models discussed in
\fig{fig: sprocess_massive_pre_post}.  The abundance distributions
are given compared to the \isotope[16]{O} production factor, which is mainly
produced in massive stars in the same zones where the \spr\
yields are made. However, \isotope[16]{O} is a primary isotope and its yields
therefore do not change with the initial metallicity of the
star. Unlike primary isotopes, \spr\ yields in massive stars (or
more generally any heavy nuclides produced starting from \spr\
seeds) show a direct dependence on the initial stellar metal content,
which is closer to a secondary-like nucleosynthesis %\citep[e.g.,][]{tinsley:80}.  
According to e.g., \citet{tinsley:80},
secondary-like isotopes produced in massive stars are expected to show
an overabundance with a factor of 2 higher than \isotope[16]{O} at solar
metallicity, to be mostly made by the weak \sprn.  Concerning
the $25\msun$ star, fallback reduces the \spr\ and \isotope[16]{O}
yields in a similar way.  Therefore, the tendency to have abundances
lying above the \isotope[16]{O}$\times2$ line in the Cu-As region (see
\fig{fig: sprocess_distribution_25_60}) is conserved, in
agreement with models using different fallback treatment
%\citep[e.g.,][, Fig.\ 6 in that work]{rauscher:02}.  
\citep[e.g.,][]{rauscher:02}.  

The footprint of the \sprn\ in producing different elements of the weak
\spr\ component with different efficiencies is maintained in the
final yields, beside the uncertainties related to the nucleosynthesis
triggered by the SN explosion.  For this reason, the abundances start
decreasing in the Se region, and become marginal above the Sr-Y-Zr
peak, in agreement with the pre-explosive \spr\ distribution.
Concerning the $60\msun$ star, the larger yields between Fe and
Nb compared to the $25\msun$ star are due to a stronger
activation of the \isotope[22]{Ne}($\alpha$,n)\isotope[25]{Mg} in the convective
He-burning core.  Indeed, the central
He-burning temperature tends to increase with the initial mass of the
star, leading to a more efficient \sprn\ in these
conditions \citep[][]{prantzos:90}. Above the Sr neutron magic peak, where the pre-explosive
contribution from the C shell and the He core window material is less
significant, the explosive nucleosynthesis signature in different
parts of the star (including from the explosive He-burning shell)
becomes easier to identify in the total ejecta. For instance, the
isotopic signature of Mo in both masses (but more in the
$60\msun$ star) shows a clear \isotope[95,97]{Mo} enrichment compared to
other Mo isotopes, in agreement with the signature measured in SiC-X
presolar grains \citep[][]{meyer:00} due to the \npr\ activation
\citep[e.g.,][]{blake:76,thielemann:79}.

In general, the present sets of CCSN models may be used to qualitatively
study the impact of fallback and CCSN explosions with high shock velocities
on the weak \spr\ distribution.

\begin{figure}
\centering
\includegraphics[width=0.48\textwidth]{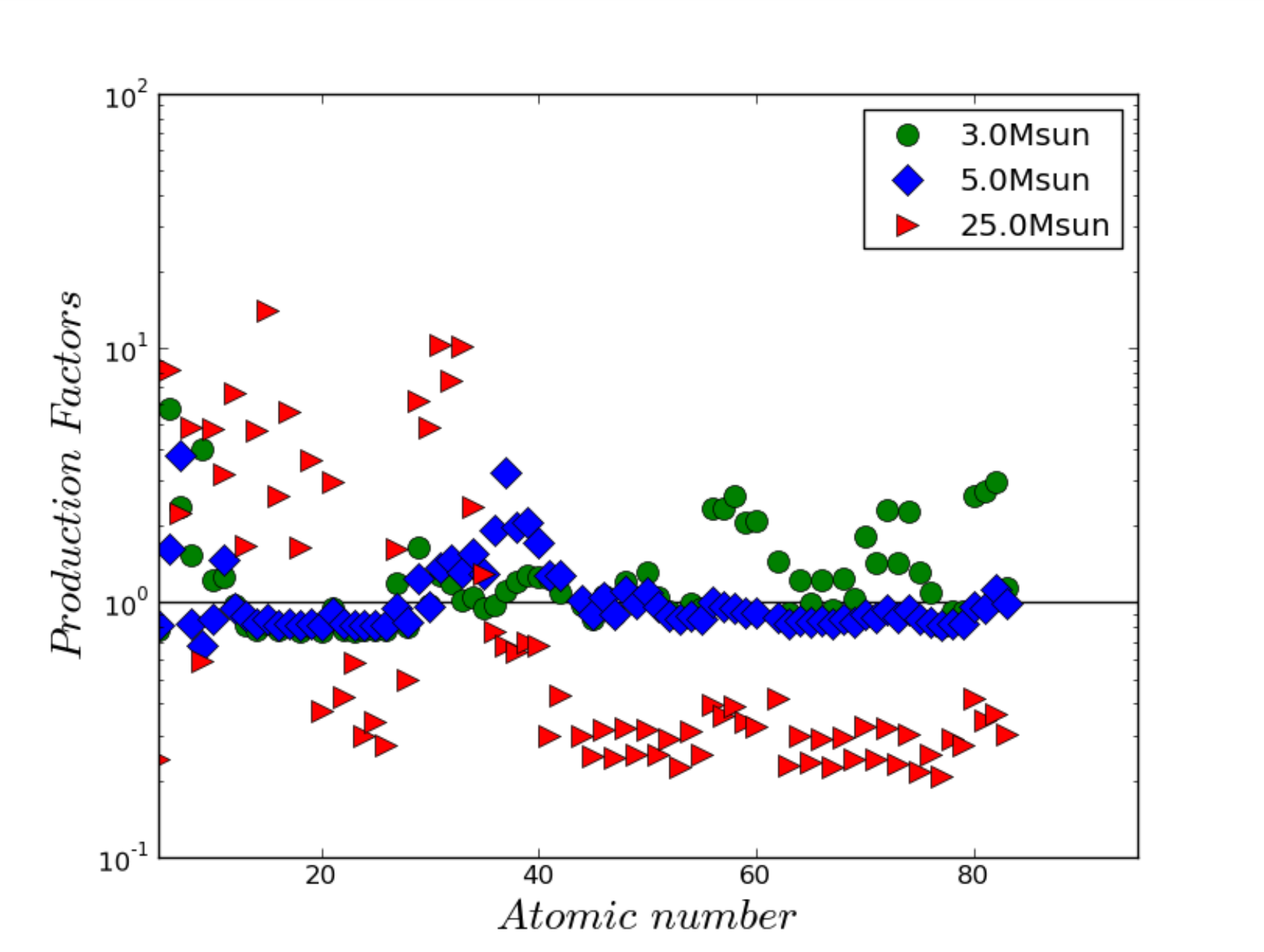}
\caption{Production factors for the $3\msun$, $5\msun$ and $25\msun$ models from \setopo. 
The $delay$ model is shown for the $25\msun$ star.
}
\label{fig: comparison_masses}
\end{figure}

Massive AGB stars (their progenitors are intermediate mass stars massive enough to experience the second dredge-up, represented in our sample by the 4 and 5\msun star models) also contribute to the \spr\ abundances in mass region between Fe and Zr \citep[e.g.,][]{travaglio:04}.
In these stars, \isotope[22]{Ne} is the dominant neutron source in the He-shell flash convection zone during the thermal pulse. 
%The \spr\ in these stars produces elements mostly between iron and the neutron shell closure at N = 50 that are usually assigned to the weak \spr\ component \citep[][]{travaglio:04}. On the other hand, the isotopic distribution is quite different compared to the weak \spr.
In \fig{fig: comparison_masses} the production factors for a $3\msun$ AGB star, a $5\msun$ massive AGB star and a $25\msun$ massive star from \setopo\ are compared. In the weak \spr\ mass region between Fe and Zr, massive stars have a larger production for Cu, Ga and Ge while for heavier elements the production in the $5\msun$ stars is more efficient. Therefore, the \sprn\ isotopic distribution from massive AGB stars is quite different compared to the weak \spr.
%The \spr\ in this $3\msun$ AGB star model becomes the most efficient starting from the Ba peak. We refer to the next section for a detailed discussion of the \spr\ in above the Sr peak.
Recently, the capability of stellar models to reproduce the high [Rb/Zr] ratios
observed in galactic and LMC massive AGB stars was questioned \citep{garcia-hernandez:06,garcia-hernandez:09,vanraai:12}.
However, \cite{zamora:14} showed that this discrepancy between stellar predictions and observations is at least partially 
reconciled thanks to the overestimation of the Rb spectroscopic abundance.

%Most of the Sr, Y and Zr in the solar system are instead made by the \spr\ in AGB stars with initial mass 1.5\msun $\gtrsim$ M $\gtrsim$ 3\msun \citep[e.g.,][]{bisterzo:14}. We will discuss the \spr\ nucleosynthesis in these stars in the next section.

\subsubsection{From Sr-Y-Zr to Pb}
\label{zr-pb}

Beyond the neutron shell closure at N=50, the efficiency of several explosive nucleosynthesis components from SN are rapidly decreasing.
This is the case for the \apr, discussed in the previous section, and for different neutrino-winds components like the weak \rpr\ and the $\nu$$p$ process, that in the most extreme conditions can be efficient up to the Cd-Sn mass region \citep[e.g.,][]{farouqi:10,arcones:11,wanajo:11}.
Therefore, beyond Zr the number of nucleosynthesis processes that efficiently contribute to the chemical inventory of the Galaxy is smaller compared to lighter heavy elements.

The total \spr\ distribution in the Solar System is divided into three
different components. In the previous section we have introduced the weak \spr\ component, between iron and strontium (60 $\lesssim$ A $\lesssim$ 90). 
%We have seen that most of the $s$-abundances are produced in massive stars \citep[e.g.,][and references therein]{kaeppeler:89,beer:92,kaeppeler:11}, with a significant contribution from low-mass and intermediate-mass AGB stars. 
For A $\gtrsim$ 90, AGB stars with initial mass 1.5 $\gtrsim$ M/\msun $\gtrsim$ 3 contribute to most of the \sprn\ abundances \citep[e.g.,][]{arlandini:99,bisterzo:11,bisterzo:14}.
In particular, in the Solar System it is possible to disentangle between the main \spr\ component and the strong \spr\ component, which forms approximately 50\% of the solar \isotope[208]{Pb} and it was produced by low metallicity AGB stars \isotope[208]{Pb} \citep[e.g.,][]{gallino:98}.
According to recent GCE simulations by \cite{bisterzo:14}, beyond Zr the \sprn\ from AGB stars reproduces more than 50\% of the solar Nb, Sn, Ba, La, Ce, Nd, W, Hg, Tl and Pb.  
The element with the smallest \sprn\ contribution is Ir (1.6\% of its solar abundance).

In AGB stars, the neutrons are mainly produced in radiative conditions
in the so-called \isotope[13]{C}-pocket (\fig{fig:c13poc_snuc_form})
via the
\isotope[13]{C}($\alpha$,n)\isotope[16]{O} reaction \citep[][]{straniero:95,gallino:98}. 
%The AGB thermal pulses induce recurrent, interacting mixing episodes including mixing for the formation of the \isotope[13]{C}-pocket and the third dredge-up mixing that brings nuclear processed material from the core to the envelope \citep[][and references therein]{herwig:04c}.
Properties of the \isotope[13]{C}-pocket can be obtained, for example,
from comparison with isotopic information from presolar grains
\citep[][]{lugaro:03a}.  Rotation-induced mixing may have the impact
of prohibiting or lowering the \sprn\ production in AGB stars
\citep{herwig:02a,siess:04,piersanti:13}.  \citet{herwig:02a}
concluded that a convection-induced instability, such as
Kelvin-Helmholz instabilities or internal gravity wave mixing
\citep{denissenkov:03t}, will lead to convective boundary mixing that
generates the \isotope[13]{C}-pocket. As in past work
\citep{herwig:99a} we model this mixing with an exponentially decaying
diffusion coefficient (see \sect{sec:CBM} for details such as the
adopted CBM efficiency).

The \spr\ nucleosynthesis operates in the \isotope[13]{C}-pocket at $T
\sim 10^8 \kelv$ and $\rho \sim 10^3$ g\cc, leading to a low neutron density
of about $10^{6-7}\cc$. These conditions best satisfy the
\spr\ distribution observed in the solar system
\citep[][]{arlandini:99,bisterzo:11}, the study of the \spr\ isotopic
signature in presolar grains \citep[e.g.,][]{lugaro:03b} and the
spectroscopic observations of stars at different metallicities
\citep[][]{lambert:95,busso:01,abia:02,masseron:10,bisterzo:11,lugaro:12,maiorca:12,straniero:14,fishlock:14}
and planetary nebulae \citep[][]{karakas:07a,pignatari:08,karakas:09}.
A smaller contribution to the total neutron exposure comes from the partial activation of the
\isotope[22]{Ne}($\alpha$,n)\isotope[25]{Mg} reaction in He-shell
flash convection zone at $ T\apgeq 2.5\times10^8 \kelv$ and $\rho \sim
10^3$ g\cc with a higher neutron density ($\gtrsim10^{10}\cc$) for up to a
few years. This exposure causes local isotopic shifts in the
\spr\ distribution as are evident in presolar grains
\citep[e.g.,][]{pignatari:06,lugaro:03b,kaeppeler:11,avila:12,liu:14}.
%Beyond the Sr peak, the \spr\ production is dominated by AGB stars, and the contribution from massive stars becomes negligible.

An established methodology to analyze the \spr\ in AGB stars and compare theoretical predictions with observations, is to use the production efficiency at different neutron-magic peaks. 
In particular, the production of the elements at the Sr neutron-magic peak is called ls, and hs is representing the production of the elements at the Ba peak. 
The ratio [hs/ls] is an \spr\ index \citep[][]{luck:91}.

\begin{figure}
\centering
\includegraphics[width=0.48\textwidth]{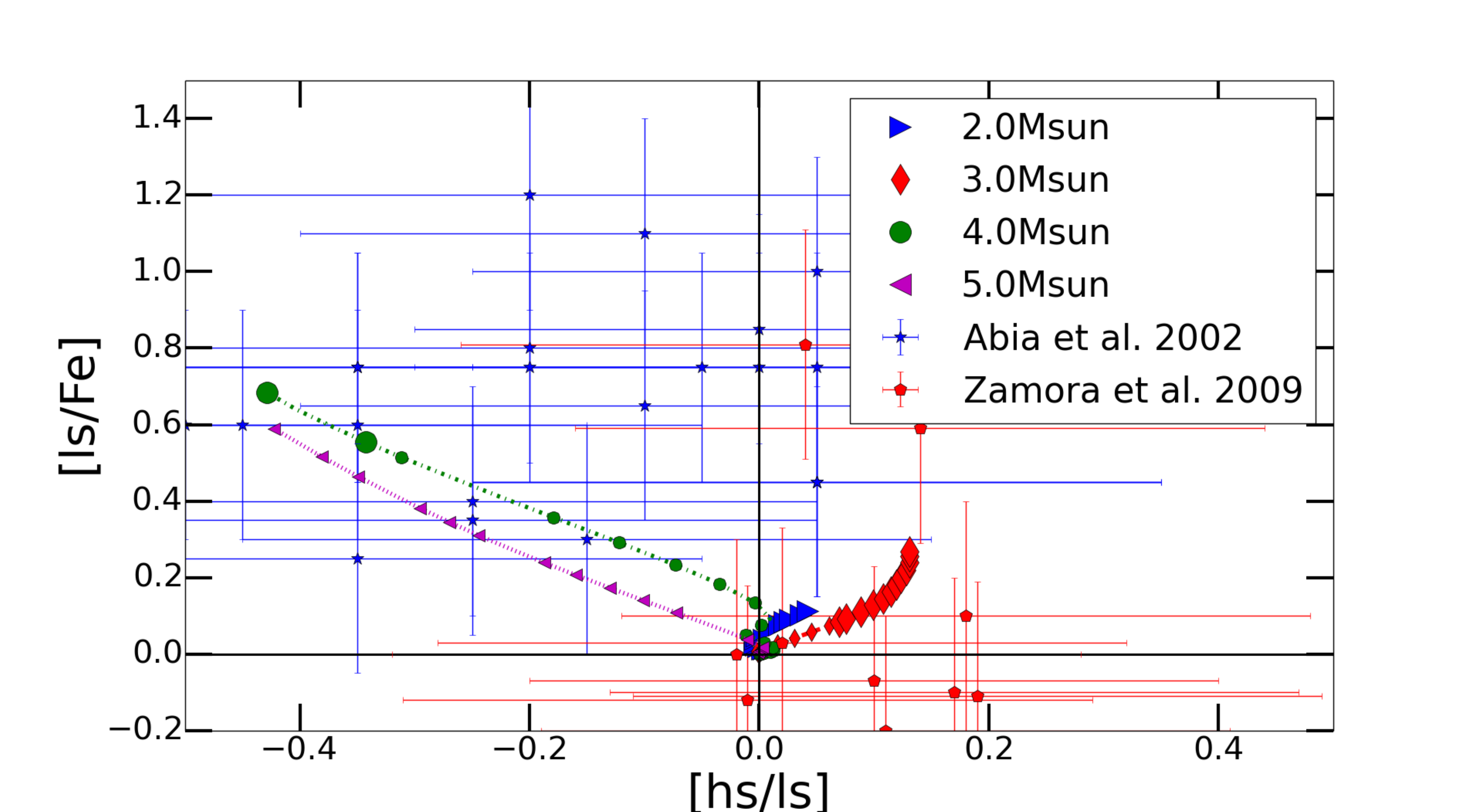}
\includegraphics[width=0.48\textwidth]{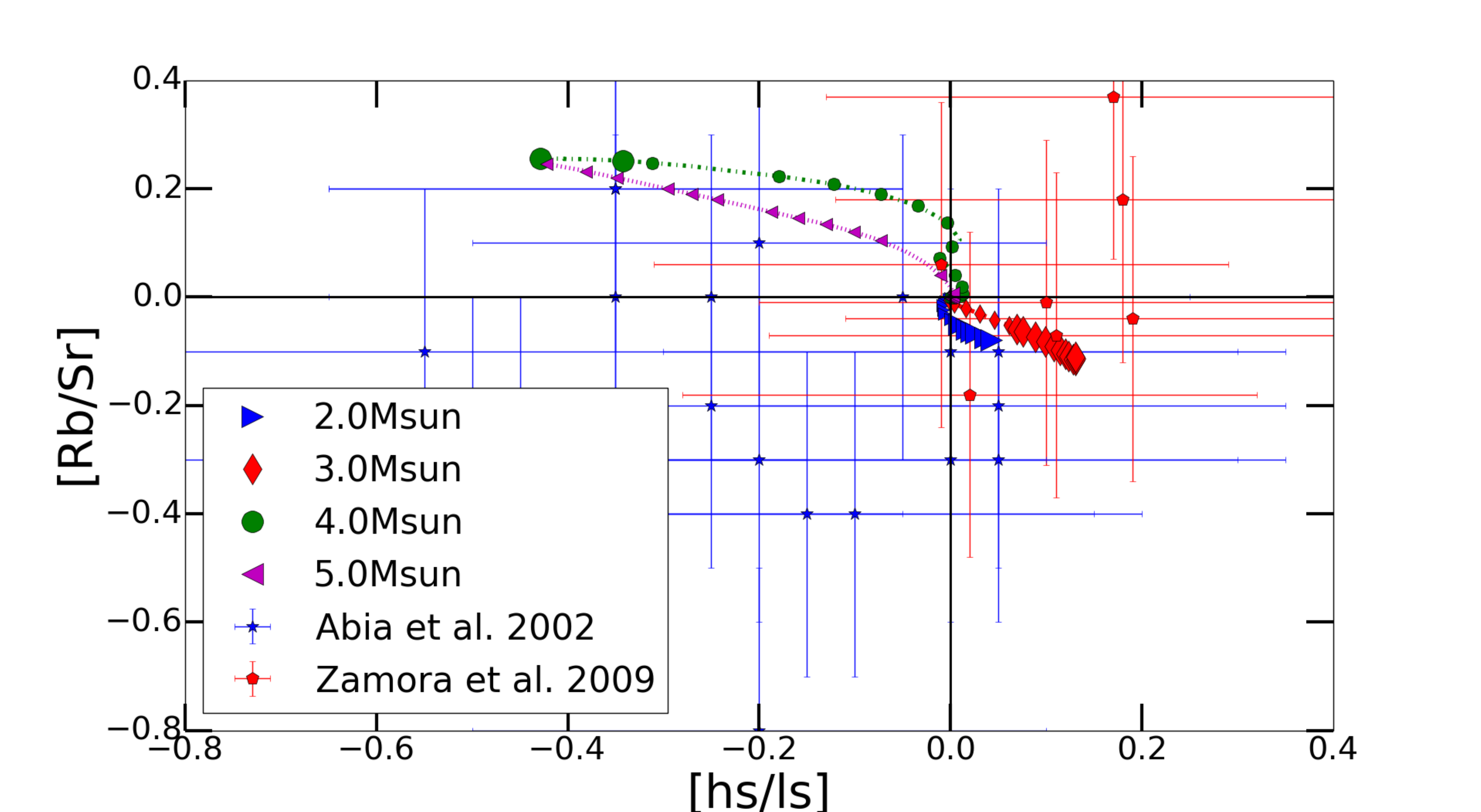}
\caption{$Left$ $Panel$: Evolution of $[ls/Fe]$ at the surface of AGB models
from \setopt\ with respect to the $s$-process index $[hs/ls]$
\citep[][]{luck:91}. The $1.65\msun$ model is not included, since the envelope material is only marginally enriched in \spr\ material. The $ls$ term includes the average of Sr, Y, and Zr production. The $hs$ term includes the average production of the elements Ba, La, Nd and Sm, according to \citet{busso:01}. Observational data from spectroscopy of Carbon stars around solar metallicity from \citet{abia:02} and \citet{zamora:09} are presented as a comparison.
$Right$ $Panel$: The evolution of the $[Rb/Sr]$ with respect to $[hs/ls]$ for the same models in the $Left$ $Panel$.
}
\label{fig:sprocess_index_wind_set1p2}
\end{figure}

In \fig{fig:sprocess_index_wind_set1p2}, we report for the models
of \setopt\ the [ls/Fe] surface evolution, where ls is the average of
Sr, Y, and Zr as a function of the [hs/ls] ratio.  In this case the term hs includes the elements Ba, La, Nd, and Sm. 
%The [ls/Fe] ratio indicates the efficiency in producing elements at the first neutron magic peak. 
The [hs/ls] ratio provides an indication of the average neutron
exposure in the \isotope[13]{C}-pocket.  The $1.65\msun$ model (not
reported in the figure) shows only a negligible \spr\ enrichment in
the envelope.  The $2$ and $3\msun$ stars show an [ls/Fe] lower than
0.4 dex.  This enrichment is lower than the maximum observed in AGB
stars by about 1.0 dex. 
This is shown in \fig{fig:sprocess_index_wind_set1p2}, where the data from observations of C-rich stars 
as reported by \cite{abia:02} and \cite{zamora:09} are also shown for comparison.

In general, the size of a typical \cdr-pocket in the present models is
$2-3 \times 10^{-5}\msun$, similar to the value obtained by
\citet{lugaro:03a}. According to the simple estimate made by
\citet{herwig:03}, the \cdr-pocket should be about 3--4 times larger in
order to reproduce the largest $[ls/Fe]$ $\sim$ 1 observed at
solar-like metallicity in MS-S stars \citep[][]{busso:01}.

In \fig{fig:sprocess_index_wind_set1p2}, the [hs/ls] of low mass AGB models
tends to become positive. We do not reproduce the large spread of observations of AGB stars
at metallicity close to solar, in which a significant fraction of C-rich stars have a negative [hs/ls]. 
This has already been noticed and discussed by \citet{lugaro:03a} and \citet{herwig:03} for models in which CBM is applied
at the bottom of the convective TP, and is even more severe for models at $Z = 0.01$.
Indeed, these models are characterized by a higher \isotope[12]{C} abundance in the 
He intershell, which in turn causes a larger neutron exposure in the
\isotope[13]{C}-pocket.
Therefore, independently from the total \spr\ enrichment in the AGB envelope, 
the larger \isotope[12]{C} concentration in the He intershell causes
the production ratio between the different neutron magic peaks to increase, favoring the $hs$ elements.

In \fig{fig:sprocess_index_wind_set1p2} the [Rb/Sr] ratio is shown with respect to the [hs/ls] ratio for the same models. 
The [Rb/Sr] ratio is affected by the branching point at
\isotope[85]{Kr},  
providing a measure of the relative importance of the
\isotope[22]{Ne}($\alpha$,n)\isotope[25]{Mg} reaction at high neutron density compared to the
\isotope[13]{C}($\alpha$,n)\isotope[16]{O} reaction at low neutron density. Indeed, during the convective TP the nucleosynthesis flow \isotope[84]{Kr}(n,$\gamma$)\isotope[85]{Kr}(n,$\gamma$)\isotope[86]{Kr}(n,$\gamma$)\isotope[87]{Kr}($\beta$$^-$)\isotope[87]{Rb} allows for the production of \isotope[87]{Rb}. Because of the lower neutron capture cross section, \isotope[87]{Rb} is accumulated more efficiently than \isotope[85]{Rb}, increasing the \spr\ production of Rb \citep[e.g.,][]{abia:01}.
%It increases with the neutron density and
%therefore with the relative importance of the
%\isotope[22]{Ne}($\alpha$,n)\isotope[25]{Mg} reaction compared to the
%\isotope[13]{C}($\alpha$,n)\isotope[16]{O} reaction.  
Our AGB models in general show a mildly negative [Rb/Sr] (\fig{fig:sprocess_index_wind_set1p2})
within the range of observations \citep[][]{lambert:95,abia:01,zamora:09}.
In \fig{fig:sprocess_index_wind_set1p2} the abundances from the 4 and $5\msun$ massive AGB star are shown. 
In \sect{sub:st_ev_tr} we mentioned that the present models do not experience hot dredge-up, and form  very small \isotope[13]{C} pockets. Nevertheless, the \spr\ yields beyond iron are 
dominated by the \isotope[22]{Ne}($\alpha$,n)\isotope[25]{Mg} contribution, with a stronger production of the Sr-Y-Zr peak compared to the Ba peak. As expected, the [Rb/Sr] is positive 
due to the high neutron densities during the convective TPs \citep[e.g.,][]{karakas:12}.
Because of the activation of the HBB the $5\msun$ does not become C-rich in these simulations. 
Therefore, the $5\msun$ results cannot be compared with the observations reported in the figure.
The spectroscopic observations of heavy \spr\ elements in massive O-rich AGB stars is still controversial. 
\cite{zamora:14} reported a new estimation of Rb abundances for four of these stars in the galactic disk using new
dynamical atmosphere models, reducing up to 1.6 dex previous measurements by \cite{garcia-hernandez:06}
and confirming the lack of Zr enhancements. In \fig{fig:sprocess_index_wind_set1p2}, the $5\msun$ models of \setopt\ 
show a [Rb/Sr] up to 0.2-0.3 dex, that in first approximation is representative of the [Rb/Zr] ratio. These results is consistent with
\cite{zamora:14}, if we take into account their large observational errors.

\begin{figure}
\centering
\includegraphics[width=0.48\textwidth]{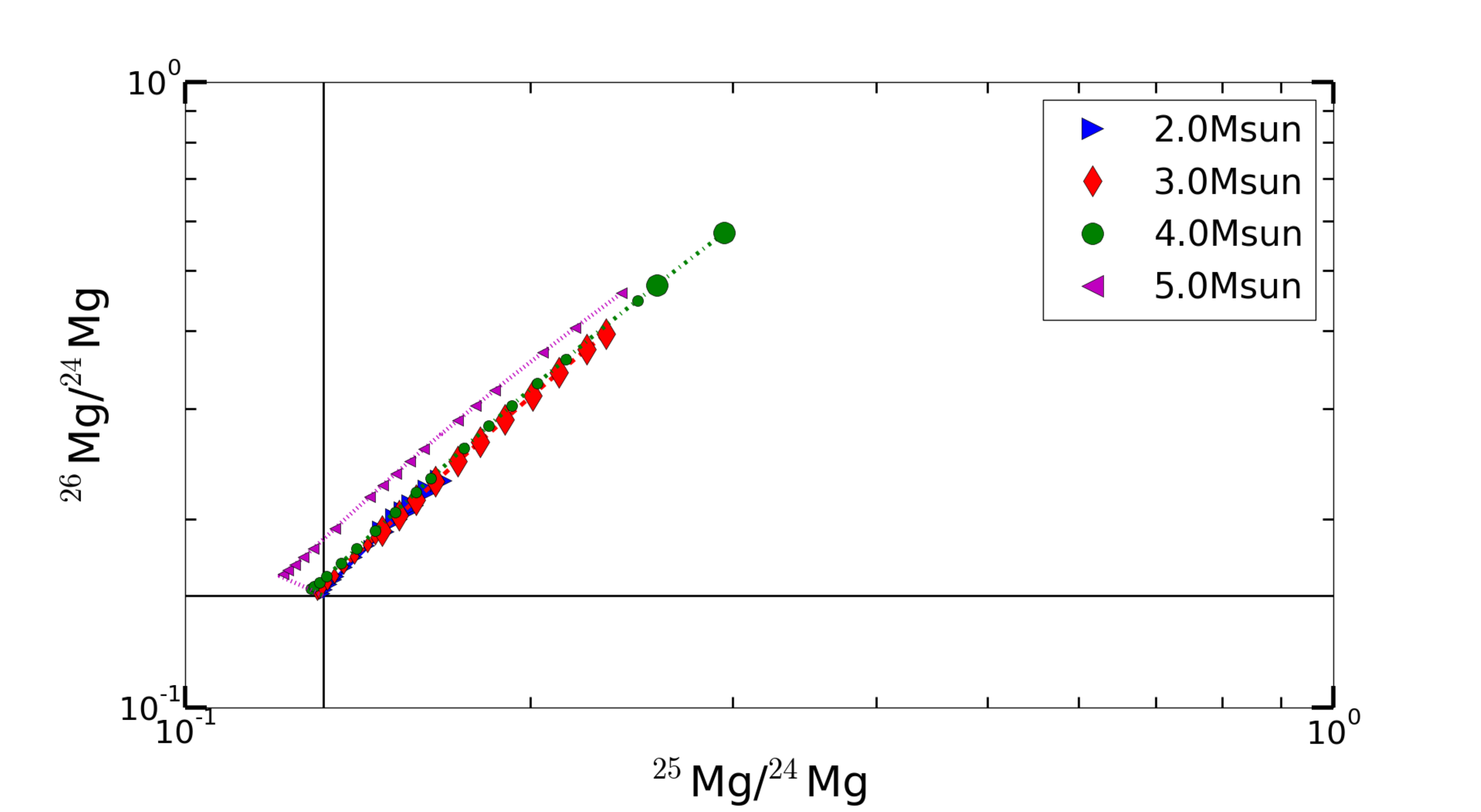}
\includegraphics[width=0.48\textwidth]{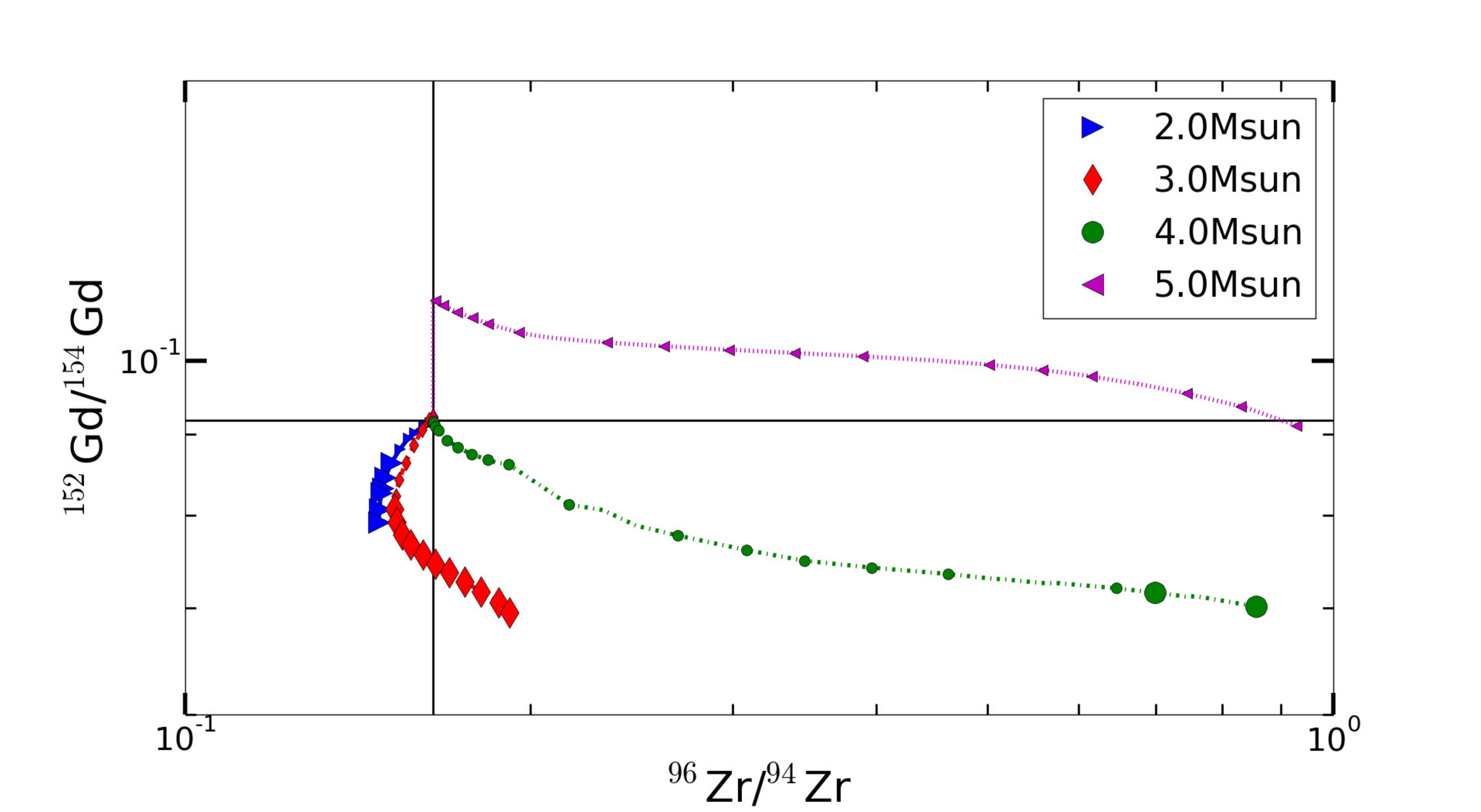}
\caption{$Left$ $Panel$: Evolution of Mg isotopic ratio at the surface of AGB models from \setopt.
$Right$ $Panel$: For the same models in the $Left$ $Panel$, the evolution of the isotopic ratios \isotope[96]{Zr}/\isotope[94]{Zr} and \isotope[152]{Gd}/\isotope[154]{Gd}.
}
\label{fig:isotopic_ratio_wind_set1p2}
\end{figure}

Another signature of AGB models including CBM at the bottom of the
He-shell flash convection zone is the more efficient production of
\isotope[25]{Mg} and \isotope[26]{Mg} compared to models not including CBM, due to a higher temperature at
the bottom of the convective TP and consequently a more efficient
activation of $\alpha$-captures on \isotope[22]{Ne}
\citep{lugaro:03a}.  In our models the isotopic ratios
\isotope[25]{Mg}/\isotope[24]{Mg} and
\isotope[26]{Mg}/\isotope[24]{Mg} increase with the initial mass of
the star (\fig{fig:isotopic_ratio_wind_set1p2}).  For instance,
for the $3\msun$ model they are 0.23 and 0.40 respectively, compared to the solar ratios 0.13 and 0.14.  The
$1.65\msun$ case shows final isotopic ratios of 0.13 and 0.16, also due to the weak pollution of the envelope by the third dredge-up 
(the $1.65\msun$ star of \setopt\ does not become C-rich (\fig{fig:papI_CO-Mstar}).
Notice that AGB stars at solar-like metallicity do not show any significant increase of the Mg isotopic ratios 
within observational uncertainties of about a factor of two \citep[][]{smith:86}. In the future the Mg isotopic ratios might provide a fundamental observational contrain for the CBM to adopt below the convective TP (see \sect{sec:CBM}), once the \isotope[22]{Ne}$+$$\alpha$ rates will be constrained by nuclear experiments with high precision \citep[][]{wiescher:12}. 

%These
%latter values are more in agreement with observations in AGB stars at
%solar-like metallicity, which do not show any significant increase \textbf{within observational uncertainties} \citep[][]{smith:86}.
In \fig{fig:isotopic_ratio_wind_set1p2}, we show the
\isotope[152]{Gd}/\isotope[154]{Gd} isotopic ratio with respect to the
\isotope[96]{Zr}/\isotope[94]{Zr} ratio.  According to Zr measurements
in presolar mainstream SiC grains, \isotope[96]{Zr} is not efficiently
produced in low mass AGB stars \citep[][]{lugaro:03b,
  zinner:03,lugaro:14}, with an observed ratio between solar and 30
times lower than solar.  We obtain a Zr ratio lower than solar for the
$2\msun$ star (\fig{fig:isotopic_ratio_wind_set1p2}), but far to
explain the low Zr isotopic ratio observed in most of mainsteam SiC
grains \citep[][]{barzyk:07,lugaro:14,liu:14b}.  On the other hand,
the $3\msun$ star shows a final \isotope[96]{Zr}/\isotope[94]{Zr}
higher than solar. The same trend was observed by
\citet[][]{lugaro:03a} for models with CBM due to an excessively large
\isotope[22]{Ne}($\alpha$,n)\isotope[25]{Mg} efficiency during the TP
(similar to the Mg isotopic ratios). However, a lower
\isotope[96]{Zr}/\isotope[94]{Zr} ratio is obtained by using the new
Zr neutron-capture cross section rates \citep[][]{lugaro:14}.
Furthermore, for the present models, both the high [hs/ls] and the
weak contribution from the \isotope[13]{C}-pocket to the total neutron
exposure affects the production of \isotope[94]{Zr}, causing a larger
final \isotope[96]{Zr}/\isotope[94]{Zr} in the AGB envelope.  In our
models the \isotope[152]{Gd}/\isotope[154]{Gd} ratio is lower than
solar for low mass AGB models (\fig{fig:isotopic_ratio_wind_set1p2}).
The higher isotopic ratio observed by \citet{lugaro:03a} for models
including CBM is not obtained in the present calculations, and a
plausible explanation is the weaker contribution from the neutron
density freezout, which is caused by the lower
\isotope[22]{Ne}($\alpha$,n)\isotope[25]{Mg} rate adopted in this work
\citep[][]{jaeger:01}.

%In Figs.\ \ref{fig:sprocess_index_wind_set1p2} and
%\ref{fig:isotopic_ratio_wind_set1p2} we provide the abundance
%signature of the 
For the $5\msun$ massive AGB star [hs/ls] is
negative, because of the dominant contribution from the
\isotope[22]{Ne}($\alpha$,n)\isotope[25]{Mg} neutron source at the Sr neutron-magic peak. For the
same reason the intermediate mass model has a positive [Rb/Sr]
\citep[][]{abia:01}. The high neutron density is also responsible for
a large \isotope[96]{Zr}/\isotope[94]{Zr} ratio, whereas the
\isotope[152]{Gd}/\isotope[154]{Gd} ratio, after an initial increase,
tends to decrease to the solar ratio. Similar considerations may be derived from the 4\msun model, not shown in the figure.

Generally, the present AGB stellar models confirm the main
features of AGB models with the CBM prescription described by
\citet{lugaro:03a} and \citet{herwig:03}.  They are able to reproduce
the large C and O abundances observed in H-deficient stars. 
It is known that a range of efficient \isotope[13]{C}-pockets is needed to reproduce the different \spr\ observations in AGB stars \citep[e.g.,][]{busso:01,lugaro:03a}. One physics mechanism that can explain this scatter might be rotation \citep[][]{herwig:03b,piersanti:13}, which is not considered in the present stellar models.

%On the other hand, the size of the \isotope[13]{C}-pocket in \setone\ is too small. 
%%and
%%the \isotope[22]{Ne}+$\alpha$ activation during the TP is too
%%strong. At least 
%Wether this issue could be solved by an alternative
%parameterization of the CBM mixing at the bottom of the convective
%envelope where the partial mixing zone of the \isotope[13]{C}-pocket
%originates, and by the use of an updated set of nuclear reaction rates, it will be discussed in detail elsewhere (Battino et al. 2015, in prep.).

We have seen that neutron capture processes are dominating the nucleosynthesis of heavy elements, at least beyond the neutron shell closure at N=50.
On the other hand, in the solar system distribution there are 35 proton-rich stable nuclides\footnote{\ \isotope[74]{Se}, 
\isotope[78]{Kr}, \isotope[84]{Sr}, \isotope[92,94]{Mo}, \isotope[96,98]{Ru}, 
\isotope[102]{Pd}, \isotope[106,108]{Cd}, \isotope[112,114,115]{Sn}, 
\isotope[113]{In}, \isotope[120]{Te}, \isotope[124,126]{Xe}, \isotope[130,132]{Ba}, 
\isotope[136,138]{Ce}, \isotope[138]{La}, \isotope[144]{Sm}, \isotope[152]{Gd}, 
\isotope[156,158]{Dy}, \isotope[162,164]{Er}, \isotope[168]{Yb}, \isotope[174]{Hf}, 
\isotope[180]{Ta}, \isotope[180]{W}, \isotope[184]{Os}, \isotope[190]{Pt},
and \isotope[196]{Hg}.}. 
A well-established scenario to make most them is the \pprn\ \citep[or $\gamma$ process, e.g.,][]{woosley:78}, during the CCSN explosion of massive stars in the O/Ne-rich layers.
With the relevant exception of \isotope[92,94]{Mo} and \isotope[96,98]{Ru} (14.84, 9.25\%
 and 5.52, 1.88\% of the Mo and Ru solar abundance, respectively),
the abundances of \ppr\ nuclei are 2-3 orders of magnitude lower than other
stable nuclides.  Such isotopes were defined as $p$-only, assuming that they do not receive 
a significant contribution from other processes such as the \sprn\ or the \rprn.
\isotope[152]{Gd} and \isotope[164]{Er} 
receive a dominant \spr\ contribution from low-mass AGB stars and are therefore not associated with \ppr\ \citep[][]{bisterzo:11}.
According to the models presented here \isotope[113]{In} and \isotope[115]{Sn} are not of \ppr\ origin either \citep[][and references therein]{dillmann:08a}.
Therefore, they cannot be indicated as $p$-only nuclides.
Furthermore, \isotope[138]{La} and \isotope[180]{Ta} could not be produced only by the 
\pprn. Indeed, \isotope[138]{La} might receive a significant contribution from neutrino
capture on \isotope[138]{Ba} \citep[][]{goriely:01} and the long-lived
\isotope[180]{Ta} isomer \citep[half-life larger than $1.2\times10^{15}$ yr,][]{cumming:85}
may be efficiently produced by the \sprn\ in low-mass AGB stars
\citep[see for different and controversial predictions][]{arlandini:99,goriely:00,
bisterzo:11} and in massive stars \citep[e.g.,][]{rauscher:02}.

There are three $p-$only isotopes lighter than Zr (\isotope[74]{Se}, 
\isotope[78]{Kr}, \isotope[84]{Sr}). Beside the \ppr\, they may be produced also by the \apr\ (see the previous section, and \fig{fig: alpha_process_15_exp_d}) and in neutrino-wind ejecta \citep[e.g.,][]{froehlich:06,farouqi:10,wanajo:11}. A similar scenario is possible for the first proton-rich species above Zr, \isotope[92,94]{Mo} and \isotope[96,98]{Ru}.
In particular, these isotopes are systematically underproduced by more than an order of magnitude compared to the other 
\ppr\ species in CCSN calculations \citep[][]{arnould:03}, taking into account present nuclear uncertainties
\citep[][and references therein]{rapp:06,rauscher:06}. 
Recently, \citet{pignatari:13} showed that assuming an enhanced  \citep[compared to][]{caughlan:88}
\isotope[12]{C}+\isotope[12]{C}
fusion reaction rate  may lead to a
Mo and Ru $p$-nuclide production up to the level of other
$p$-nuclei ($cp$-component). 

Besides problems in reproducing single isotopes, the average
\ppr\ massive star yields are underproduced by about a factor of three
compared to the amount required to explain the Solar System
distribution \citep[e.g.,][]{rayet:95}, or the secondary nature of the
classical \pprn\ \citep[see][]{pignatari:13}. An alternative
astrophysical source proposed to reproduce, at least in
part, the abundances of \ppr\ nuclides in the Solar System are SNIa \citep[][]{howard:91,howard:93,travaglio:11,kusakabe:11,travaglio:14}.

\begin{figure}
\centering
\resizebox{11cm}{!}{\rotatebox{0}{\includegraphics{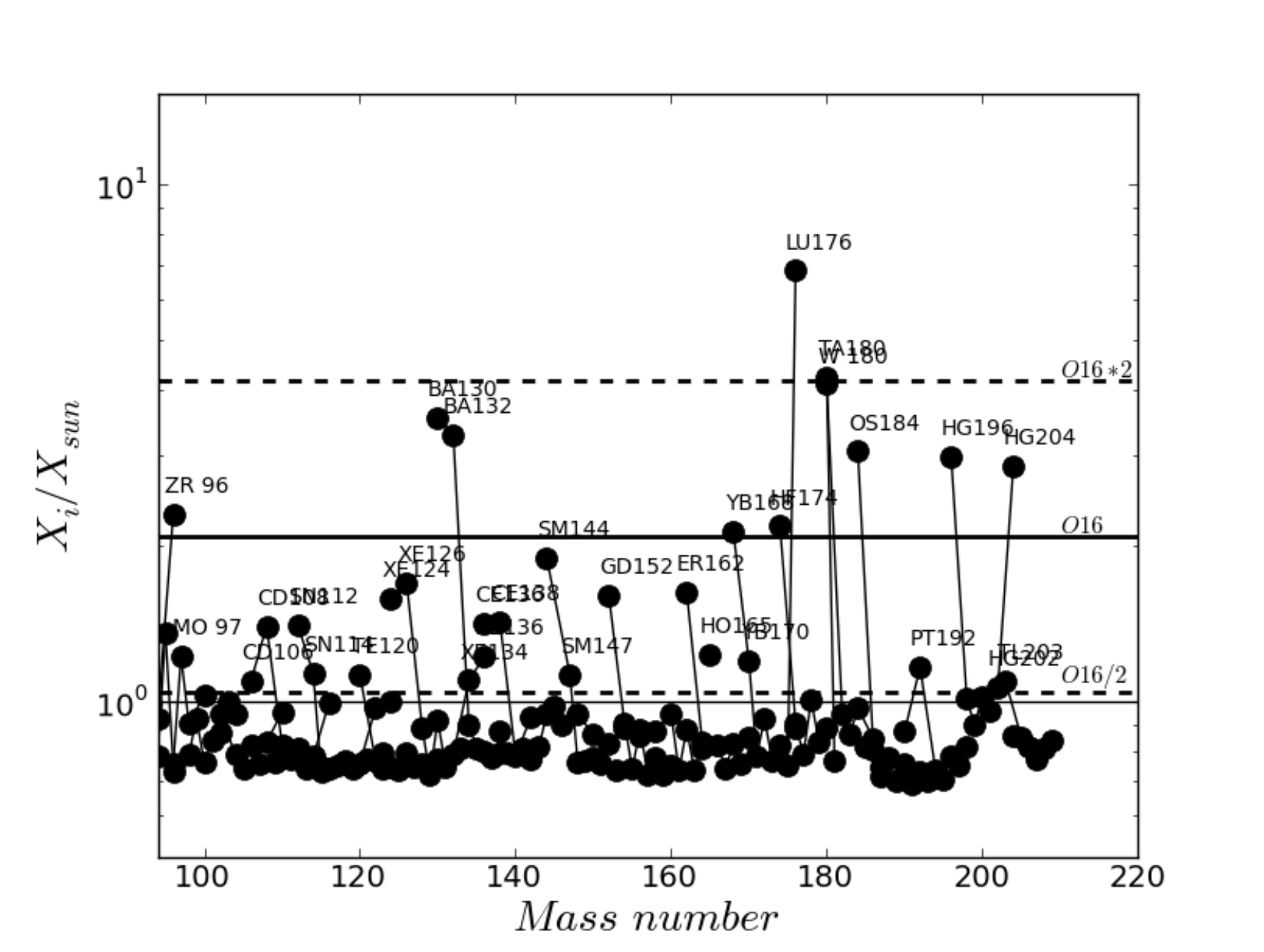}}}
\resizebox{11cm}{!}{\rotatebox{0}{\includegraphics{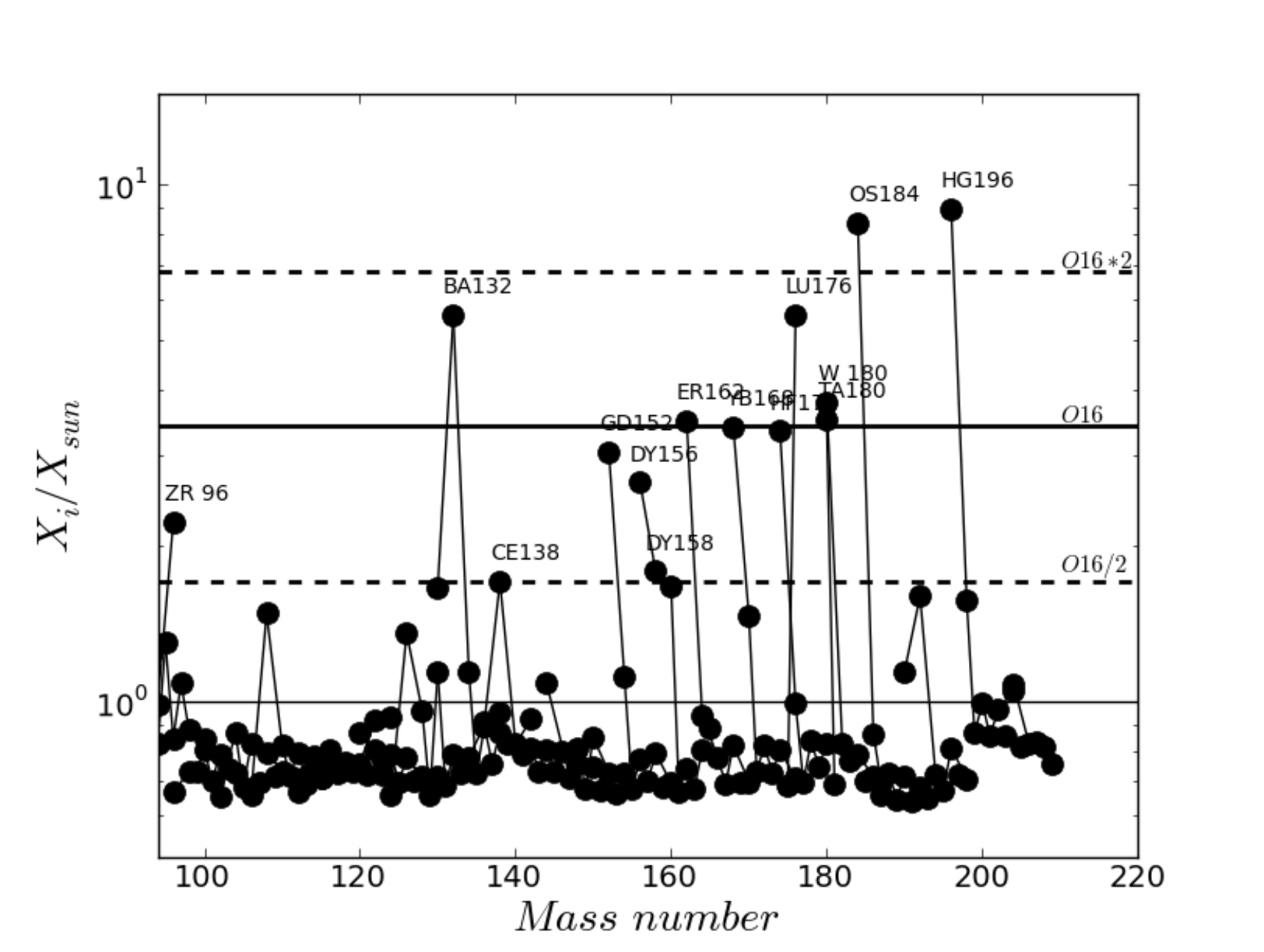}}}
\caption{Final isotopic overproduction factors for the 15 and
$25\msun$ stars from \setopt\ in the mass region A $>$ 95 ($upper panel$ and $lower panel$, respectively). 
The production factor of \isotope[16]{O}, divided and
multiplied by a factor of two are also reported (continuous 
and dashed lines). We label the isotopes with production 
factors larger than \isotope[16]{O} divided by 2.
Among those, different \ppr\ isotopes can be identified.
}
\label{fig:prodfac_15_25_d_set1p2_pprocess}
\end{figure}

Our models represent the \ppr\ contribution from CCSNe with high shock
velocities and including fallback. Similar results are expected for
\ppr\ yields from hypernova or from the high energy component of
asymmetric CCSNe.  
Among all the CCSN models presented in this work, we now focus our
discussion on the \ppr\ distribution of a $15\msun$ star and a $25\msun$
star, $Z = 0.02$ (SN model $delay$, \setopt).  In
\fig{fig:prodfac_15_25_d_set1p2_pprocess}, upper panel, the $15\msun$
star 
%shows the strongest explosive contribution at the lightest
%\ppr\ species \isotope[74]{Se}, \isotope[78]{Kr} and \isotope[84]{Sr},
%with a production factor $\gtrsim10$. No \ppr\ contribution is seen in
does not show a relevant \ppr\ contribution to Ru, in agreement with standard CCSN calculations
\citep[e.g.,][]{rauscher:02}. On the other hand, up to \isotope[92]{Mo} the ejected abundances are dominated by
the $\alpha$-process (see the next section, and e.g., the full \tab{tab:isotopic_prodfac_set1.2_exp} online).
In this specific model, there is no relevant production of \isotope[102]{Pd} either.  The
\ppr\ contribution becomes positive again from \isotope[106,108]{Cd}
to \isotope[196]{Hg} (with a production factor of $\sim1.5$-4), with
the exception of \isotope[156,158]{Dy} and \isotope[190]{Pt}, which
are not efficiently produced. Among those species, the most produced
are \isotope[180]{Ta} and \isotope[180]{W}, with a production factor
of about 4. The high SN explosion energy causes a larger contribution
to the lightest \ppr\ species, in disagreement with the classical flat
\ppr\ distribution. However, the oxygen production factor of this
model is reduced to about 2.1, since the high energy of the explosion
depletes O in a large region of the ejecta.  Therefore, the
\ppr\ production factor in CCSN characterized by high shock velocities
and/or high explosion energies is similar or larger than O, one of the
fundamental requirements in order to reproduce the solar system
\ppr\ abundances.

In \fig{fig:prodfac_15_25_d_set1p2_pprocess}, lower panel,
the $25\msun$ star shows a dominant \ppr\ signature starting from Ba.
Indeed, besides \isotope[108]{Cd} (with a mild production of $\sim 1.5$)
and  \isotope[126]{Te} ($\sim 1.4$), $p$-nuclides lighter than Ba are not ejected.
Above Ba, the production factors range between $\sim 1.1$ (\isotope[144]{Sm}) and
9 (\isotope[196]{Hg}). The reason for this behavior is that the hotter material
carrying the lighter $p$-nuclides falls back onto the forming BH and only
the colder \ppr\ component is ejected. The oxygen production factor is
about 3.4. Compared to standard CCSN models \citep[e.g.,][]{rauscher:02},
the oxygen yields are also smaller. This is due to the large amount of
mass falling back (for this model the central $5.7\msun$ are not ejected).

In summary, for the $15\msun$ model considered the production of the proton-rich heavy isotopes up to the Mo region 
is dominated by the $\alpha$ process, while beyond Mo the standard \ppr\ contribution is becoming the most relevant process.
%the $15\msun$ model with $Z = 0.02$
%shows a stronger \ppr\ contribution at the lightest nuclei,
%but with no production of Mo-Ru $p$-only isotopes.
The $20\msun$ model with the same metallicity and the same explosion energy
shows a more standard \ppr\ distribution
(see \fig{fig:prodfac_15_25_d_set1p2_pprocess}). 
%Therefore, 
%The dependence on the initial stellar mass should be considered before comparing quantitatively
%the \ppr\ yields with the solar system distribution.
On the other hand, a strong fallback (see the $25\msun$ star case discussed here) potentially
favors heavier \ppr\ ejecta. 
%In both of the two cases discussed here, however,
%the oxygen production tends to be reduced compared to standard SN models of the same mass.

\subsection{Comparison with other sets of stellar yields}
\label{comparison}

A number of different sets of stellar yields are available in the literature.
In \tabs{tab:isotopic_yieldscomp_m2z2m2}, \ref{tab:isotopic_yieldscomp_m3z2m2}, \ref{tab:isotopic_yieldscomp_m5z2m2}, \ref{tab:isotopic_yieldscomp_m15z2m2}, \ref{tab:isotopic_yieldscomp_m20z2m2} and \ref{tab:isotopic_yieldscomp_m25z2m2} we show a comparison between the yields presented in this work for different stars in \setopt\ and the yields presented from several works in the literature: for massive stars \cite{thielemann:96,rauscher:02,chieffi:04}, and for intermediate-mass stars \cite{karakas:10a} and \cite{cristallo:11}.
 
The \isotope[16]{O} isotope is the most abundant product of massive stars.
Considering both the CCSN ejecta and the winds (see the yields available online), 
the models 15, 20 and 25 $\msun$ and $Z = 0.02$, delay explosion,
produce 0.30, 1.27 and 0.82 $\msun$ of \isotope[16]{O}.
For instance, for the same masses and metallicity \cite{thielemann:96} provides
0.42, 1.48 and 2.99 $\msun$, \cite{rauscher:02} 0.85, 2.20 and 3.32 $\msun$
(models S15, S20 and S25), and \cite{chieffi:04} 0.52, 1.38 and 2.44 $\msun$. 
For the 15 $\msun$ star, the results change by almost a factor of 
three, and by a factor of 1.7 for the 20 $\msun$ star.
The large fallback included in our simulations causes lower \isotope[16]{O} yield
for the 25 $\msun$ star, which is e.g., about a factor of four smaller than 
\cite{rauscher:02}.
Differences can be even larger if we compare the yields of \isotope[44]{Ti} and
\isotope[56]{Ni}, which critically depend on the explosion parameters applied 
in the simulations \citep[][and references therein]{magkotsios:10}. For these two species, we obtain for the 
same models considered before 1.97$\times$10$^{-4}$ and 0.18 $\msun$, 
1.54$\times$10$^{-5}$ and 0.0087 $\msun$, 1.05$\times$10$^{-7}$ $\msun$ and
no \isotope[56]{Ni} ejected, respectively. In particular, the extended fallback in the
25 $\msun$ model does not allow to eject any relevant amount of 
\isotope[44]{Ti} or \isotope[56]{Ni}.
For the same models and isotopes, \cite{thielemann:96} provides
7.19$\times$10$^{-5}$ and 0.13 $\msun$, 
1.53$\times$10$^{-4}$ and 0.068 $\msun$, 2.11$\times$10$^{-5}$ and 0.052 $\msun$
respectively. \cite{rauscher:02} predicts 
1.39$\times$10$^{-5}$ and 0.11 $\msun$, 4.87$\times$10$^{-5}$ and 0.09 $\msun$,
1.56$\times$10$^{-5}$ and 0.11 $\msun$. Finally, from 
\cite{chieffi:04}, assuming the same amount of \isotope[56]{Ni} ejected equal
to 0.1 $\msun$ for all masses, we obtain for \isotope[44]{Ti}
4.20$\times$10$^{-5}$, 4.03$\times$10$^{-5}$ and 2.19$\times$10$^{-5}$ $\msun$, respectively.

Concerning the impact of neutron capture processes in massive stars, the final yield of the neutron-magic \isotope[88]{Sr} is a good indicator of their total efficiency if alternative processes like the $\alpha$ process are not activated.
%The final yield of the neutron-magic \isotope[88]{Sr} 
%is a good indicator of the neutron capture efficiency in 
%our massive star models. 
In particular, 
the 20 and 25 $\msun$ models with $Z = 0.02$, delay explosion,
produce %6.78$\times$10$^{-7}$, 
2.38$\times$10$^{-6}$ and 1.54$\times$10$^{-6}$ 
$\msun$ of \isotope[88]{Sr}. For comparison, \cite{rauscher:02} predicts
respectively %1.08$\times$10$^{-6}$, 
4.69$\times$10$^{-6}$ and 1.14$\times$10$^{-5}$ 
$\msun$, and \cite{chieffi:04} 
%7.88$\times$10$^{-7}$, 
1.98$\times$10$^{-6}$ and 3.98$\times$10$^{-6}$ $\msun$. Besides the impact of different physics and explosion
choices made in these different models, the differences are also due
to the different nuclear reaction rates used in the simulations,
e.g., for the \isotope[22]{Ne}($\alpha$,n)\isotope[25]{Mg} 
and \isotope[22]{Ne}($\alpha$,$\gamma$)\isotope[26]{Mg} reactions.

%For intermediate-mass stars,
Concerning AGB stars,
the final ejected masses of \isotope[12]{C}, \isotope[14]{N} 
and \isotope[16]{O} for the $2\msun$ AGB model of \setopt\
are 0.0187, 0.0035 and 0.0189 $\msun$, respectively.
For the same isotopes and the same star, \cite{karakas:10a} provides 0.0028, 0.0030 and 0.0130, 
and \cite{cristallo:11} 0.0093, 0.0033 and 0.0118 $\msun$.
For \isotope[12]{C}, we obtain an abundance that is factor of 2.1 and 6.9 higher 
than \cite{cristallo:11} and \cite{karakas:10a}. A higher \isotope[12]{C} enrichment in our models 
is due to the CBM activated at the bottom of convective TPs, while differences between \cite{cristallo:11} and \cite{karakas:10a} might be due to intrinsic differences between the two set of models like dredge-up efficiency and mass loss rates.
%\cite{karakas:10a} show a much weaker production compared to \setone\ and \cite{cristallo:11}.
The \isotope[14]{N} yields are consistent within 20\%.
Concerning \isotope[16]{O}, our models show a larger production, up to 60\%. 
This higher production corresponds to a positive contribution 
to the O inventory of the Galaxy from AGB stars (see \fig{fig:CNONaAl_set1p2} and the data tables).
Galactic chemical evolution simulations are needed to confirm this scenario.
%\cite{karakas:10a} does not provide the abundances for heavy nuclei
%at the three neutron magic peaks.
Concerning the $s$-process nucleosynthesis, for the same model 
of \setone\ the final ejected masses of \isotope[88]{Sr}, \isotope[138]{Ba} 
and \isotope[208]{Pb} are 8.62$\times$10$^{-8}$, 
2.54$\times$10$^{-8}$ and 1.23$\times$10$^{-8}$ $\msun$.
\cite{cristallo:11} predicts a much larger production,
with  7.07$\times$10$^{-7}$, 7.93$\times$10$^{-8}$ 
and 2.52$\times$10$^{-8}$ $\msun$. This is due to the smaller
\isotope[13]{C} pockets obtained in our models compared to \cite{cristallo:11}.

\section{Summary and final remarks}
\label{sec:concl}

In this work we present a set of stellar models and their chemical
yields (\setone). We define 1.65, 2, 3, 4, 5, 15, 20, 25 $\msun$ models;
we also calculated 32 and 60 $\msun$ models at $Z$ = 0.02.  Massive
star models are calculated using the stellar evolution code GENEC and
lower mass models are calculated using MESA.  For low- and
intermediate-mass stars, wind yields are provided in the form of
production factors and absolute yields in solar mass units.  For massive
stars, the yields are given for the stellar wind, pre-explosive and
post-explosive contributions.  Two sets of explosion models are
considered, each with a different fallback prescription. The NuGrid
post-processing code \mppnp\ is used to perform all nucleosynthesis
calculations for AGB stars and for massive stars including their supernova (SN) explosions.

Core collapse SN models are performed in 1D semi-analytic way. The shock
velocity profiles and fallback prescriptions used are motivated by
multi-dimensional hydrodynamic simulations. Due to their simplified nature
they are foremost meant to indicate species that will be affected by
explosive nucleosynthesis in any significant way.
%, rather than provide the basis for a detailed quantitative analysis of explosive yields for complex studies such as galactic chemical evolution. 
The explosive yields therefore provide important insights on the main
features of explosive nucleosynthesis. Furthermore, the \setone\ SN models
represent an example of explosive nucleosynthesis at high
shock velocity (high temperature or high energy), and with a fallback
signature.

For the first time we present a grid of full yields for
$s$-process and $p$-process species for SN models with strong shocks.
%, and with consequent explosion temperatures larger than standard CCSN. 
In particular, models with a large fallback have reduced $s$-process
yields, which are modified significantly in models with higher explosion
energies.  For most cases the $s$-process distribution is affected by
local abundance redistribution. In particular, in the He shell 
the $n$-process from explosive He burning may have a relevant impact.  
The weight of the $n$-process component on the
final yields increases with increasing fallback.  
%The final yield of the neutron-magic \isotope[88]{Sr} 
%is a good indicator of the neutron capture efficiency in 
%our massive star models. In particular, 
%the models 15, 20 and 25 $\msun$ and $Z = 0.02$, delay explosion,
%produce 6.78$\times$10$^{-7}$, 2.38$\times$10$^{-6}$ and 1.54$\times$10$^{-6}$ 
%$\msun$ of \isotope[88]{Sr}. For comparison, \cite{rauscher:02} predicts
%respectively 1.08$\times$10$^{-6}$, 4.69$\times$10$^{-6}$ and 1.14$\times$10$^{-5}$ 
%$\msun$, and \cite{chieffi:04} 7.88$\times$10$^{-7}$, 1.98$\times$10$^{-6}$ 
%and 3.98$\times$10$^{-6}$ $\msun$. Beside the impact of different physics and explosion
%choices made in these different models, the differences in this case are also due
%to the different nuclear reaction rates used in the simulations,
%e.g., for the \isotope[22]{Ne}($\alpha$,n)\isotope[25]{Mg} 
%and \isotope[22]{Ne}($\alpha$,$\gamma$)\isotope[26]{Mg} reactions.
The $15\msun$ models show the activation of the $\alpha$ process in the deeper ejecta up to \isotope[92]{Mo}. Therefore,
for these stars the final yields between Fe and Mo are carrying this nucleosynthesis signature. We showed that 
by reducing the initial shock velocity by only a factor of 2 the $\alpha$ process is suppressed; indeed the $\alpha$ process does not appear in our models with even lower initial shock velocities.
Progenitors with larger masses and/or in general models with larger fallback mass will not eject this component, even assuming the same 
initial shock velocities. Finally, the impact of nuclear uncertainties on the $\alpha$-process yields still need to be explored.

For the $p$-process,
the main effect of a higher energy SN explosion is to move the
$p$-process rich region outwards, without dramatic modification of the
$p$-process efficiency.  On the other hand, the O production tends to
decrease with increasing explosion temperatures (and fallback),
which is used as a reference for $p$-process efficiency.  Furthermore,
different models show local differences in the $p$-process distribution,
but in no case do we obtain a significant $p$-process production
of the $p$-rich isotopes of Mo and Ru. In general, the present yields
could potentially relieve the $p$-process underproduction relative to O.

Low- and intermediate mass stars are evolved to the tip of the AGB,
%until the end of the AGB evolution, 
with the exception of the 4 and $5\msun$ AGB models for which
one-dimensional modeling assumptions are violated before all mass is
lost \citep[e.g.,][]{lau:12}. The remaining envelope mass is assumed to
be ejected without any further processing.  All AGB models include
convective boundary mixing (overshooting) prescriptions. In agreement
with previous work, this causes a larger amount of carbon and oxygen in
the He intershell compared to AGB models without overshooting. The
$s$-process carries the known signature of overshooting applied at the
bottom of the envelope, with large neutron exposures in the
$^{13}$C-pocket. On average, the low-mass AGB models of \setone\ have
$^{13}$C-pockets producing an $s$-process enrichment in the envelope about 3-4 times weaker than the highest abundances observed in AGB stars with metallicity close to solar.
%that are too small, with $s$-process enrichment in the
%envelope about 3-4 times weaker than the abundance observations in AGB stars. 
Despite this, the most efficient producers of the first peak
elements (Y, Sr, Rb, Zr) are the $3-5\msun$ %intermediate mass 
AGB star models.
We are in the process of updating this area of our model
parameterization for the next data release.

%The final ejected masses of \isotope[12]{C}, \isotope[14]{N} 
%and \isotope[16]{O} for the $2\msun$ AGB model of \setopt
%are 0.0187, 0.0035 and 0.0189 $\msun$, respectively.
%For the same isotopes and the same star, for instance
%\cite{karakas:10a} provides 0.0028, 0.0030 and 0.0130, 
%and \cite{cristallo:11} 0.0093, 0.0033 and 0.0118 $\msun$.
%While for \isotope[14]{N} yields are consistent within 20\%
%and for \isotope[16]{O} within 60\%, for \isotope[12]{C}
%\cite{karakas:10a} show a much weaker production compared
%to \setone\ and \cite{cristallo:11}.
%\cite{karakas:10a} does not provide the abundances for heavy nuclei
%at the three neutron magic peaks.
%Concerning the $s$-process nucleosynthesis, for the same model 
%of \setone\ the final ejected masses of \isotope[88]{Sr}, \isotope[138]{Ba} 
%and \isotope[208]{Pb} are 8.62$\times$10$^{-8}$, 
%2.54$\times$10$^{-8}$ and 1.23$\times$10$^{-8}$ $\msun$.
%\cite{cristallo:11} predicts a larger production,
%with  7.07$\times$10$^{-7}$, 7.93$\times$10$^{-8}$ 
%and 2.52$\times$10$^{-8}$ $\msun$.

The present work comprises for the first time 
stellar yields from low mass stars, intermediate mass stars and 
massive stars calculated using the same nuclear reaction network.
%The present work comprises stellar yields from low mass stars \emph{and}
%massive stars.  
We estimate the contribution from different stars to the
nuclides, but note that a more quantitative study would require the use
of a galactic chemical evolution model.  For instance, we show that
although massive stars are generally the dominant source of
$\alpha$-elements beyond carbon, 
%intermediate mass 
AGB stars do show a strong production of oxygen.
In particular, the impact on these results on the galactic chemical evolution of 
oxygen needs to be explored in the future. 
Some preliminary discussion has been presented by \cite{delgado-inglada:15}.

We finally would like to reiterate that our yields have at
this point no contribution for the $r$ process or from SN type Ia, which
again is something we would like to improve upon in the future.

Stellar yields of \setone\ provide stellar abundance data covering both
low mass and massive star models. This data release, however, is based
on simplifications, such as the use of rather basic semi-analytic
explosion assumptions as well as a rather simplistic treatment of mixing
related to convective boundaries, which in fact we assume to be present in
low- and intermediate mass stars at all times at some level, while no
overshooting is assumed during post-He core burning in the massive star
models. We also use two different stellar codes for high-mass and
low-mass stars, which introduces a small amount of inconsistency,
although efforts have been made to minimize these. Our predictions
presently exclude super-AGB stars. Our goal is to remove such
limitations in future data release. In addition we will provide data
sets for lower initial metal content, and such simulations are well
underway.

%\textbf{While the present set of yields can be used as a guidance for qualitative comparisons with stellar observations and limited GCE studies, they need to be used carefully.}

%\newpage
%\include{initalcomposition.tab}

\acknowledgments 
We thanks the anonymous referee for many useful comments and suggestions. We would  also like to thank Sergio Cristallo and Amanda Karakas for the fruitful discussion and sharing information. 
NuGrid acknowledges significant support from NSF grants PHY 02-16783
and PHY 09-22648 (Joint Institute for Nuclear Astrophysics, JINA), NSF grant PHY-1430152 (JINA Center for the Evolution of the Elements) and EU MIRG-CT-2006-046520. 
NuGrid computations are performed at the Arizona State University's Fulton High-performance Computing Center (USA) and the high-performance computer KHAOS at EPSAM Institute at Keele University (UK).
M.~P.\ acknowledges an Ambizione grant of the SNSF, SNF (Switzerland) and support from the "Lendulet-2014" Programme of the Hungarian Academy of Sciences.
M.~P., S.~J., and R.~H.\ thanks the Eurocore project Eurogenesis for support. 
F.~H.\ acknowledges NSERC Discovery Grant funding.
R.~H., and S.~J.\ acknowledge support from the World Premier International Research Center Initiative (WPI Initiative), MEXT, Japan.
M.~G.~B.'s research was carried out under the auspices of the National Nuclear Security Administration of the U.S. Department of Energy at Los Alamos National Laboratory under Contract No. DEAC52-06NA25396.
A.~D.\ acknowledges support from the Australian Research Council under grant FL110100012.
The research leading to these results has received funding from the European Research Council under the European Union's Seventh Framework Programme (FP/2007-2013)/ERC Grant Agreement n. 306901. 
This work used the SE library (LA-CC-08-057) developed at Los Alamos National Laboratory as part of the NuGrid collaboration; SE makes use of the HDF5 library, which was developed by The HDF Group and by the National Center for Supercomputing Applications at the University of Illinois at Urbana-Champaign.
M.~P.\ acknowledges the support to the Milne Astrophysics Center granted by PRACE, through its Distributed Extreme Computing Initiative, for resource allocations on Sisu (CSC, Finland), Archer (EPCC, UK), and Beskow (KTH, Sweden), and by the STFC DiRAC High Performance Computing Facilities. Ongoing resource allocations on the University of Hull’s High Performance Computing Facility - viper - are gratefully acknowledged.
\appendix

\section{NuGrid codes} \label{ap:codes}

The NuGrid nucleosynthesis codes provide a framework for performing both
single-zone (\sppn) and multi-zone parallel (\mppnp) simulations for given
thermodynamic conditions \citep{herwig:08a,Pignatari:2012dw}. Both the
\sppn\ and \mppnp\ drivers use the same solver
(\sect{sub_ap:solver_package}) and physics
(\sect{sub_ap:physics_package}) packages. The single-zone driver is
used, for example, for simplified simulations of trajectories for
reaction rate sensitivity studies. The yields presented in this paper
have been obtained with the multi-zone driver \mppnp.

The stellar structure evolution is calculated with a small network,
just large enough to accurately account for the nuclear energy generation. 
For the \MESA\ AGB simulations the network (\MESA\ agb.net)
contains 14 isotopes, while the GENEC network contains 8 to 15
isotopes.  The stellar structure evolution data for all zones at all
time steps are written to disk using the NuGrid \se\ format,
a data structure based on HDF5\footnote{HDF stands for Hierarchical Data Format; 
\url{http://www.hdfgroup.org/HDF5/‎}}.
All zones at all timesteps are then processed with the \mppnp\ code using 
a dynamic network that includes all relevant reactions automatically.

\begin{figure}
\includegraphics[width=\textwidth]{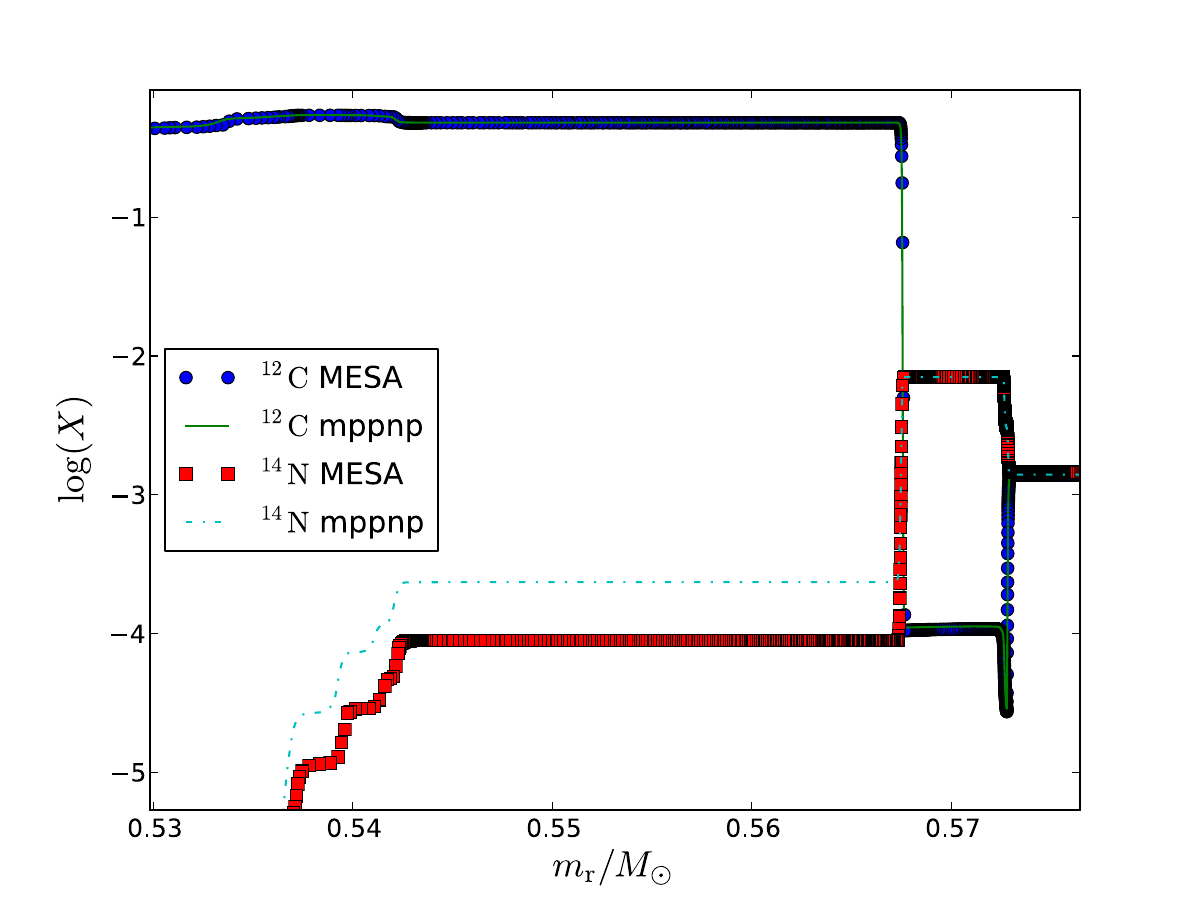}
\caption{ \label{fig:se-mppnp-compare} Comparison of \MESA\ and \mppnp\ 
abundance profiles for the 14th thermal pulse convection zone of the
\setopo\ $2\msun$ AGB sequence. The mass range shown includes
the top of the C/O core, the He-shell flash convection zone, the (now
extinct) H-burning shell, and the bottom of the convective envelope. 
The H-free core mass is $0.572\msun$ for that model.}
\end{figure}

In order for this post-processing approach to work the stellar evolution
code has to include a large enough network to reproduce the energy
generation in the same way the post-processing network would, which
implies that for important reactions like $\nvi(\p,\gamma)\ofu$ and
$\czw(\alpha,\gamma)\ose$ the same nuclear physics has to be adopted in
both cases. The quality of the stellar evolution and post-processing
network consistency is checked by comparing abundance profiles for key
species from both cases, and shows in general good agreement
(\fig{fig:se-mppnp-compare}). The \czw\ abundance agrees well both in
the He-intershell and the H-burning ashes, indicating that both
He-burning and H-burning are treated consistently between the stellar
evolution and post-processing approaches. The \nvi\ abundance agrees for
the two cases in the H-burning ashes. This reflects the consistent
treatment of CNO burning in the stellar evolution and the
post-processing, where \nvi\ is the most important isotope due to its
small $\p$-capture cross section. \nvi\ does not contribute in
significant ways to the energy generation in He-burning, and therefore
the difference between \nvi\ in \MESA\ and \mppnp\ in this isotope in the
He-burning layers (in the mass region $0.540 < m_\mathrm{r}/\msun <
0.567 $)  reflects the more complete nuclear network (including
n-capture reactions) in the post-processing simulation. The latter is
the more realistic solution in that case. 

The advantages of the post-processing approach over a complete inline
network include larger flexibility and 
%better numerical behaviour. 
shorter computing time.
In particular, the higher scale of flexibility is due to the capability to adopt different nuclear reaction rates with no relevance for the energy economy of the stellar structure, without having to calculate again new stellar structures. This means that a large number of different sets of stellar yields can be made for different sets of nuclear reaction rates, but using the same stellar models.
One of the reasons for the superior numerical behavior of the \MESA\ code
during the advanced phases of stellar evolution is the simultaneous
solution of the structure, network and mixing operators. It would be
numerically too time consuming to perform such a joint operator solve
for a full \spr\ network with up to 1000 isotopes.

However, the implementation of a fully coupled 
solver in \MESA\ is a source of inconsistency with the post-processing
approach,
%this difference in the operator split philosophy is also a
%source of inconsistency 
since \mppnp\ solves the mixing and
nucleosynthesis in separate steps. There is little that can be done
about this, except monitor the difference (\fig{fig:se-mppnp-compare})
and, in case they get unacceptably large, force sub-time stepping in
\mppnp. So far this was not necessary.

Further, the post-process approach
allows easy and rapid post-processing of the same stellar evolution
track with modified input nuclear physics, provided the reactions are
not important for energy generation. Realistic sensitivity studies can
be performed in this way for many application. Finally, it was
straight-forward to adopt a distributed parallelized computing model for
the post-processing simulations (\sect{sub_ap:parallel}).

\subsection{Parallel-programming implementation - \mppnp}
\label{sub_ap:parallel}

%As mentioned in \S \ref{sec:Overview}, the post-processing calculation
%can be boiled down to the calculation of Eqn. --- ---, which is a
%system of equations that encompasses changes in abundances due to all
%input reaction rates.  Once all reactions are included in the nuclear
%network, this system of equations can become very large and the typical
%duration of a post-processing run can last approximately $10-12$ months
%with current serial technology. Therefore, the application of parallel
%programming allows the problem to be solved in a reasonable timeframe
%and enables large calculations.
%
The implementation of parallelism in \mppnp\ frame is a simple
master-worker (or Workqueue) routine that assigns a single process
(normally a single processor) to be the `master' with the rest as
`workers', which is coded using the Message-Passing Interface (MPI).  The
main advantage gained by using MPI is the ability to use \mppnp\ over
distributed memory resources, such as cluster networks.  The master
performs all the serial computations, such as initialization,
input/output and simple tasks, and coordinates the assignment of work to
the workers using a first-in first-out (FIFO) scheduler.  The worker
calculates the work and then returns the result to the master.  For
\mppnp, the unit of work is the network calculation for a single
spherical shell (or `zone') at a single time step, which is assigned by
passing a message containing the temperature, density and chemical
composition in the shell to the worker.  We choose this definition of
`work' because network calculations for individual zones do not depend
on each other and therefore no communication is required between
workers.  This allows for an embarrassingly parallel implementation,
which simplifies the parallel implementation and reduces significantly
the communication overhead.  Load balancing in \mppnp\ is simple in that
zones are allocated spatially, in order, from the centre of the star,
through the interior towards the surface.  The reason for this is that
the dynamic network typically assigns larger networks to regions of
higher temperature, so the zones with the most work are allocated first.

The general operation of \mppnp\ can be described as follows.  First,
the initialization is performed by the master, which includes the
loading into memory of reaction rates, input parameters and initial
stellar model data.  The reaction rate data are then passed to all
workers using broadcasts, which provide each processor with a private
copy of the data required to calculate the nuclear reaction network.
Following the broadcasts, the master invokes the scheduler for the first
timestep.  It assigns work to all available workers and then waits for a
reply.  Upon completion of a network calculation, the worker returns the
modified abundances to the master, which it stores in an array.  If
there is more work to be assigned, the master assigns further work to
the worker and waits for further messages.  If no more work is to be
assigned, the worker returns a message indicating that it is to be
terminated.  Once all workers respond with a termination message, all
work has been completed for a single timestep and the master performs
some additional serial tasks, such as a mixing step (in case a specific
zone of the star has mixing coefficient larger than zero, according to
the stellar structure input) and output.  When the next timestep is
calculated, the master invokes the scheduler again and the process is
repeated.

\begin{figure}
\includegraphics[width=\textwidth]{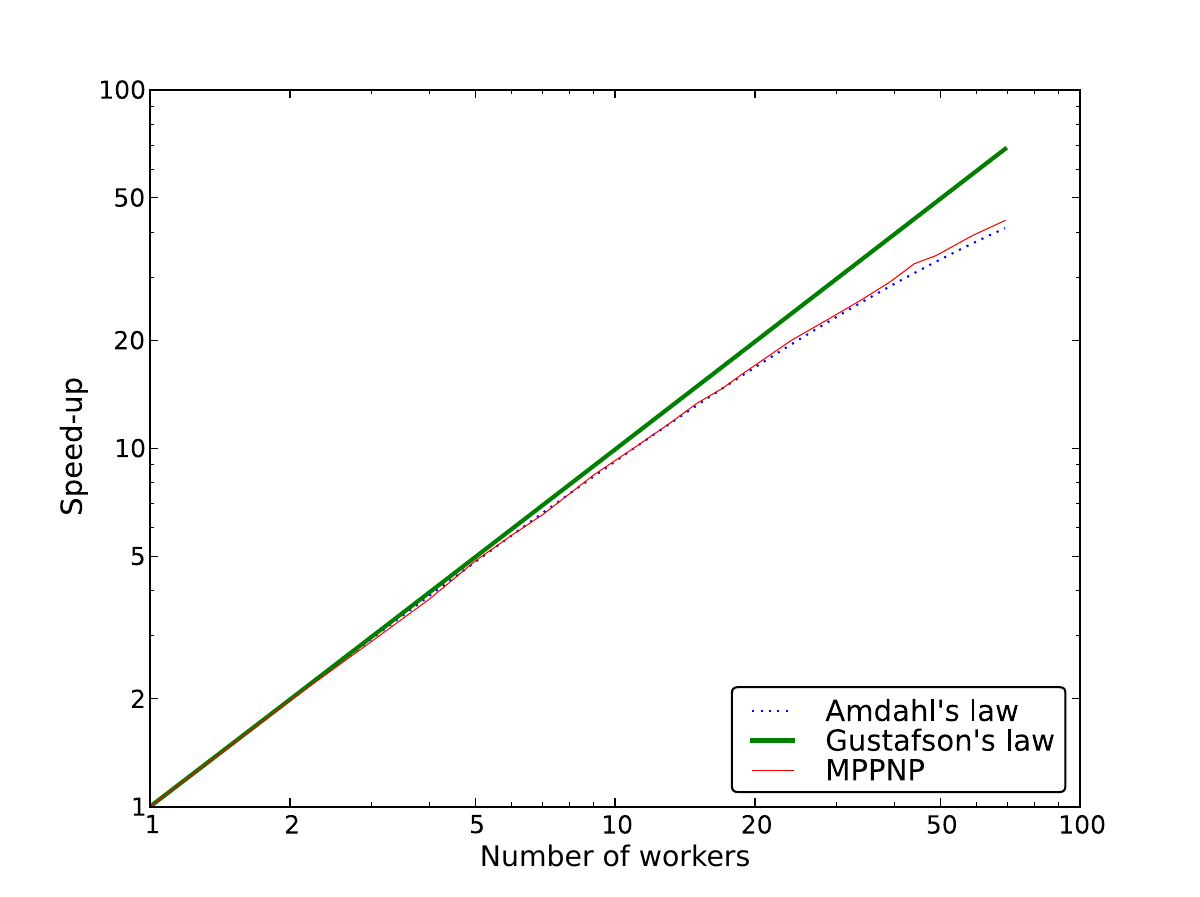}
\caption{Speed-up factor for \mppnp\ with respect to those of 
Gustafson's law and Amdahl's law with a serial fraction of 1\%.}
\label{fig:speedup}
\end{figure}

The parallel performance of the scheduler can be estimated using a
scaling curve, which is a plot of the speed-up factor as a function of
the number of processors.  The scaling of \mppnp\ for a test run with
2500 timesteps of a $15\msun$ massive star model with approximately
250 zones per timestep is shown in \fig{fig:speedup}.
\fig{fig:speedup} also shows the curves for Amdahl's law and Gustafson's
law with a serial fraction of 1\%.  Since the amount of work was fixed
during the test run, it is unsurprising that the 
%scaling curve 
curve in the strong-scaling test
follows
that of Amdahl's law, but it otherwise indicates that the communication
overhead is negligible and that load balancing is reasonably close to
optimal.

\subsection{Physics package} 
\label{sub_ap:physics_package}

The physics package provide to the post-processing code the list of
isotopes and the nuclear reaction network to use in the calculations,
and for every stellar evolution time step and stellar zone the new set
of reaction rates given at the correct temperature, density and electron
fraction $\Ye$.

The species included in the network are defined by a list in a database
file and by two parameters, giving the maximum allowed number of species
($NNN$) and the lower limit of half life of unstable species by
$\beta$-decay ($tbetamin$).  The parameter $tbetamin$ regulates the
width of the network departing from the valley of stability.  In other
words, all the unstable isotopes with an half life shorter than
$tbetamin$ are not included in the network.  For \setone\ the
non-explosive calculations the isotopic list contains 1095 species
($NNN$ = 1095 and $tbetamin$ = 0.5\,s). For explosive simulations the
network is increased up to 5200 species ($NNN$ = 5200 and $tbetamin$ =
$10^{-5}$\,s).  
%Notice that $NNN$ does not take into account long lived
%isomers of few isotopes, that are added as separated species at the end
%of the rest of the network.

The nuclear reaction network is designed as a compilation of different
compilations, with the possibility to select single specific rates from
sandbox. Therefore, for the same reaction is possible to choose
different reaction rates.  Available reaction rate libraries are REACLIB
\citep[available interface for JINA REACLIB, the last tested revision
V1.0, and Basel REACLIB, revision 20090121][respectively]{cyburt:11,
rauscher:00}, KADoNIS \citep[][]{dillmann:06}, NACRE
\citep[][]{angulo:99}, CF88 \citep[][]{caughlan:88},
\citep[][]{iliadis:01}. The available compilations for weak rates are
\citep[][]{fuller:85, oda:94, goriely:99, langanke:00}.  
%A Sandbox (at
%the moment 108 reaction rates) with the option to include single
%reaction rates is available. For instance, with this capability multiple
%options are available for the $^{12}$C($\alpha$,$\gamma$)$^{16}$O
%\citep[][]{caughlan:85,caughlan:88,angulo:99,buchmann:96,kunz:02}.
%Netgen Bruslib interface is also available, to add reactions in this
%format \citep[][]{aikawa:05}.  In case temperature and/or density from
%stellar data are outside the range of Netgen tables, the closest
%provided rate is used. For \setone, no extrapolation is allowed outside
%provided nuclear informations.

For temperatures above $6\times10^{9}$ K, 
network calculations are switched to Nuclear Statistical
Equilibrium (NSE). Temperature-dependent partition function
and mass excess are given by the REACLIB revision used for the
simulations.  Coulomb screening correction is applied according to
\citet{calder:07}. The NSE module is included into a loop where feedback
to the $\Ye$ from weak interactions is checked, and considered for
following NSE steps.

The isomers considered are 
\isotope[26]{Al}$_{\rm m}$, 
\isotope[85]{Kr}$_{\rm m}$,
\isotope[115]{Cd}$_{\rm m}$, 
\isotope[176]{Lu}$_{\rm m}$,
and 
\isotope[180]{Ta}$_{\rm m}$.
Long-lived non-thermalized isomers and ground states 
are considered as separated species.  
For temperatures lower than a given thermalization temperature,
both the ground state and the isomeric state are
produced. In case they are unstable, we use terrestrial $\beta$-decay
rates \citep[e.g.,][]{ward:80}. 
%The parameter $IR$ gives the
%fraction of the nucleosynthesis production channels that goes to the
%isomer, whereas (1.0 - $IR$) is the fraction that goes to the ground
%state.  Note that for \setone\ all the production channels have the same
%$IR$.  This approximation is justified by the fact that the isomers
%considered are mainly produced by one reaction, e.g.,
%$^{25}$Mg(p,$\gamma$)$^{26}$Al for $^{26}$Al and (n,$\gamma$) capture
%for the other cases.  Notice that such approximation may not be correct
%for all isomers in different nucleosynthesis conditions, therefore in
%the future a more consistent treatment of all the production channels
%will be implemented.  
For temperatures higher than the thermalization
temperature, the production channels to the considered isomer are
neglected, and only the thermalized specie is fed. In case the
isotope is unstable, above thermalization temperature 
the stellar $\beta$-decay rate is used.
Such a simple implementation is going to be upgraded in the near future,
to keep into account properly the transition phase to thermalization.

%%%%
\subsection{Solver Package} \label{sub_ap:solver_package}

The solver package used to perform nucleosynthesis post-processing
calculations relies on a Newton-Raphson implicit implementation, which
is controlled on full precision, mass conservation and maximum size of
negative yields. In case convergence criteria are not satisfied,
adaptive sub-time stepping is allowed.  A recursive, dynamic network
generation has been integrated into the solver, i.e., the size of the
network automatically adapts to the conditions given. If, for example, a
neutron source is activated the network will be automatically enlarged
to include all heavy and unstable isotopes as needed according to the
network fluxes. This dynamic network feature ensures that the network
calculation never misses any production/depletion of different species
or reaction chains.

Different numerical solvers based on the fully implicit method are
included, and may be selected according to the architecture of the
machine where the calculations are performed.  At present, the
available solvers are $ludcmp/lubksb$ \citep[][]{press:92}, $leqs$
(solves a linear system of equations $a\,x = b$ via Gauss Jordan
elimination), and standard LAPACK $dgesv$ (double-precision general
solver). The LAPACK solvers are provided from ACML or MKL libraries,
which are optimized respectively for AMD Opteron and Intel processors.
These LAPACK solvers can invert even rather large matrices (650
elements) rapidly ($\sim 0.01 \mathrm{s}$).

\section{NuGrid data products}
\label{database}

Although we have provided the most commonly requested derived data
sets, such as yield tables, the calculations hold much more
information than we can report in this paper. We are therefore making
the entire computed raw data sets available via CADC\footnote{The Canadian
Astronomical Data Center, 
\url{http://www.cadc-ccda.hia-iha.nrc-cnrc.gc.ca/vosui/\#nugrid}} or the 
NuGrid website\footnote{\url{http://data.nugridstars.org}}. The data consists 
of two libraries.
The Stellar Evolution and Explosion (SEE) library contains, for each
time step, profile data needed for nucleosynthesis post-processing as well 
as a few abundance profiles (to check the accuracy of the post-processing) 
for each grid point and some scalar data (like \teff, $L$, etc.). The
Post-Processing Data (PPD) library contains the post-processing
nucleosynthesis data of the SEE library data.  Data is provided in the
\se-flavour of HDF5. These files are normal HDF5 files but
follow a certain structure suitable for the purpose. Software
libraries for writing and reading se files with Fortran, C and Python,
as well as detailed instructions on how to access the data are
available at the NuGrid project website.

The provided data is structured in the following way. NuGrid data
comes in \emph{sets}. Each set corresponds to a model generation,
which is defined by a common (or similar enough) set of modeling
assumptions. The data provided in this paper belong to \emph{\setone},
which are meant to be standard models and which will serve as a
baseline for future, improved sets. In this paper we provide two
subsets, containing models with $Z=0.01$, which are \setopo, and $Z=0.02$,
which are \setopt. This (and the following) structure is reflected in
the directory tree on the CADC data server. In each of the subset
directories (\texttt{set1.1} and \texttt{set1.2}) are four directories.
For both, the SEE and the PPD libraries there are pre-supernova
data (i.e., the stellar evolution output, \texttt{*\_wind}) and the
explosion data for the massive stars (\texttt{*\_exp})
directories. Each of these four directories is populated with one
directory for each of the relevant masses. In the stellar evolution
directories \texttt{see\_wind} output files with the ending
\texttt{.se.h5} can be found.  The directories for low- and
intermediate-mass star directories in \texttt{see\_wind} are the
actual \MESA\ run directories, and the \texttt{se.h5} are found in a
subdirectory. The time evolution of the approximated one-dimensional
explosion profiles (\S \ref{subsec:exp}) are provided in
\texttt{.se.h5} files in the \texttt{see\_exp} directories.

Likewise, the \texttt{ppd\_*} directories contain the \mppnp\ run
directories for each mass with three types of output directories in
each of them. \texttt{H5\_out} contains se-type hdf files with the
ending \texttt{.out.h5}. These contain complete profiles for all
stable and a number of longer-lived unstable (like \isotope[14]{C}) species
for every $20^\mem{th}$ time step. The \texttt{H5\_restart} directory
contains restart files with all species that are considered in these
calculations, every 500 time steps. The \texttt{H5\_surf} directory
contains surface elemental and isotopic, decayed and undecayed
abundance evolutions at each time step in the \texttt{.surf.h5} files.

\begin{figure}
\includegraphics[width=\textwidth]{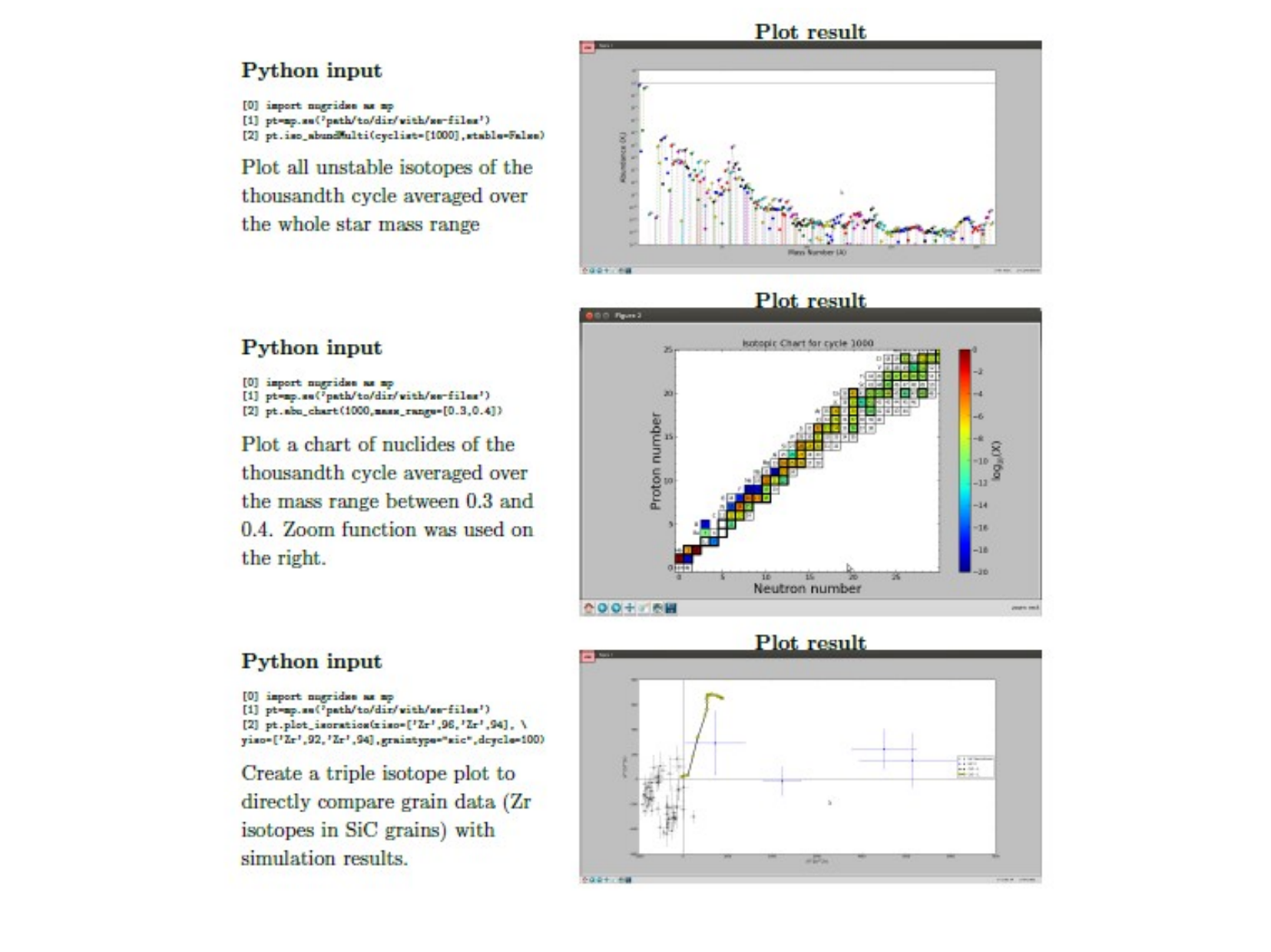}
\caption{Snapshot with examples of possible advanced plots on the right, 
with easy steps to make them and some basic explanation on the left.
Each plot window has a zoom function and the plot can be saved 
in diverse formats.}
\label{fig:python_snapshot}
\end{figure}

\se\ files are ordinary HDF5 files, and any tool that reads HDF5 files
(e.g.\ HDFview) may be used. We are providing the python module
\texttt{nugridse.py} at the project website, which allows
access to \se\ data via Python. \texttt{nugridse.py} also provides
plotting methods for the standard plot types, such as abundance
distribution, table of nuclides, Kippenhahn diagram, as
well as generic plot routines.
The Python environment allows an easy and fast access of the data.
Examples with easy steps to plot more advanced diagrams 
are summarized in \fig{fig:python_snapshot}.
%is shown
%below. Each plot window has a zoom function and 
%the plot can be saved in diverse formats.

%\include{useepp_appendix}
%\include{visualisation}

%\bibliography{astro}
%\bibliography{all}

% list of tables

\include{tables}
%CR: tables temporary added here

\begin{table}
\caption{}
\scalebox{0.7}{
\begin{tabular}{ccccccccccc}
\hline
\multicolumn{11}{c}{\setopt\ Z = 0.02} \\
\hline 
$M_{ini}$ &  $m_c$      & $R_{\ast}$  & $N_{TP}$ & $N_{3DUP}$ & $t_{TPI}$  & $\Delta M_{Dmax}$ & $M_D$              & $t_{ip}$ & $M_{lost}$ & $T_{PDCZ,max}$ \\
{[$M_{\odot}$]}  &  {[$M_{\odot}$]}  & {[$R_{\odot}$]} &         &                &  {[$10^6 yr$]} & {[$10^{-2}M_{\odot}$]} & {[$10^{-2}M_{\odot}$]} & {[$yr$]}     & {[$M_{\odot}$]} &  {[$K$]}\\
\hline 
1.65   & 0.530  & 237&  23    &  6 &  2.270E+03 &  0.3       &  0.751         &  90864  & 0.87       &  8.441  \\
2.00   & 0.510  & 220&  24    &  13 &  1.415E+03 &  0.5       &  4.230         &  116763  & 1.35       &  8.458  \\
3.00   & 0.596  & 309&  23    &  20 &  4.807E+02 &  0.7       &  9.747         &  57700  & 2.34       &  8.473  \\
4.00   & 0.809  & 536&  25    &  24 &  2.148E+02 &  0.4       &  6.522         &  10658  & 3.13       &  8.531  \\
5.00   & 0.865  & 593&  25    &  24 &  1.168E+02 &  0.2       &  3.715         &  5747  & 3.98       &  8.535  \\
\hline
\multicolumn{11}{l}{$M_{ini}$: Initial stellar mass.} \\
\multicolumn{11}{l}{  $m_c$: H-free core mass at the first TP.}\\
%\multicolumn{12}{l}{ $L_{\ast}$: Approximated mean Luminosity.}\\
\multicolumn{11}{l}{ $R_{\ast}$ : Approximated mean radius.}\\
\multicolumn{11}{l}{ $N_{TP}$: NUmber of TPs.}\\
\multicolumn{11}{l}{ $N_{3DUP}$ : Number of TPs with 3DUP.}\\
\multicolumn{11}{l}{ $t_{TPI}$: Time at first TP.} \\
\multicolumn{11}{l}{ $\Delta M_{Dmax}$: Maximum dredged-up mass after a single TP.} \\
\multicolumn{11}{l}{ $M_D$: Total dredged-up mass of all TPs.} \\
\multicolumn{11}{l}{ $t_{ip}$ : Average interpulse duration of TPs.} \\
\multicolumn{11}{l}{$M_{lost}$: Total mass lost during the evolution.} \\
\multicolumn{11}{l}{$T_{PDCZ,max}$: Maximum temperature during the TPAGB phase.}\\
\end{tabular}
}
\label{agbprop_1p2}
\end{table}

\begin{table}
\caption{}
\scalebox{0.7}{
\begin{tabular}{ccccccccccc}
\hline
\multicolumn{11}{c}{\setopo\ Z = 0.01} \\
\hline 
$M_{ini}$ &  $m_c$       & $R_{\ast}$  & $N_{TP}$ & $N_{3DUP}$ & $t_{TPI}$  & $\Delta M_{Dmax}$ & $M_D$              & $t_{ip}$ & $M_{lost}$ & $T_{PDCZ,max}$ \\
{[$M_{\odot}$]}  &  {[$M_{\odot}$]}  & {[$R_{\odot}$]} &         &                &  {[$10^6 yr$]} & {[$10^{-2}M_{\odot}$]} & {[$10^{-2}M_{\odot}$]} & {[$yr$]}     & {[$M_{\odot}$]} &  {[$K$]}\\
\hline

1.65   & 0.533  & 209&  17    &  5 &  1.871E+03 &  0.2       &  0.901         &  109775  & 0.92       &  8.444  \\
2.00   & 0.498  & 173&  25    &  13 &  1.276E+03 &  0.6       &  4.547         &  145367  & 1.37       &  8.456  \\
3.00   & 0.646  & 308&  14    &  13 &  4.123E+02 &  0.8       &  7.425         &  48874  & 2.31       &  8.484  \\
4.00   & 0.831  & 479&  20    &  19 &  1.876E+02 &  0.3       &  3.985         &  9169  & 3.09       &  8.530  \\
5.00   & 0.901  & 559&  22    &  21 &  1.081E+02 &  0.2       &  2.562         &  4362  & 3.92       &  8.539  \\
\hline
\multicolumn{11}{l}{$M_{ini}$: Initial stellar mass.} \\
\multicolumn{11}{l}{  $m_c$: H-free core mass at the first TP.}\\
%\multicolumn{12}{l}{ $L_{\ast}$: Approximated mean Luminosity.}\\
\multicolumn{11}{l}{ $R_{\ast}$ : Approximated mean radius.}\\
\multicolumn{11}{l}{ $N_{TP}$: NUmber of TPs.}\\
\multicolumn{11}{l}{ $N_{3DUP}$ : Number of TPs with 3DUP.}\\
\multicolumn{11}{l}{ $t_{TPI}$: Time at first TP.} \\
\multicolumn{11}{l}{ $\Delta M_{Dmax}$: Maximum dredged-up mass after a single TP.} \\
\multicolumn{11}{l}{ $M_D$: Total dredged-up mass of all TPs.} \\
\multicolumn{11}{l}{ $t_{ip}$ : Average interpulse duration of TPs.} \\
\multicolumn{11}{l}{$M_{lost}$: Total mass lost during the evolution.} \\
\multicolumn{11}{l}{$T_{PDCZ,max}$: Maximum temperature during the TPAGB phase.}\\
\end{tabular}
}
\label{agbprop_1p1}
\end{table}

\begin{center}
\tiny\begin{longtable}{cccccccccc}
\caption{TPAGB evolution properties of set1.2.}\\
\hline
\multicolumn{10}{c}{\setopt\ Z = 0.02} \\
\hline \\
 TP & $t_{TP}$ & $T_{FBOT}$ & $T_{HES}$ & $T_{HS}$ & $T_{CEB}$ & $m_{FBOT}$ & $m_{HTP}$ & $m_{D,max}$ & $M_{\ast}$\\
 & [$yrs$] & [$K$] & [$K$] & [$K$] & [$K$] & [$M_{\odot}$] & [$M_{\odot}$] & [$M_{\odot}$] & [$M_{\odot}$] \\
\\
\hline
\multicolumn{10}{c}{$M=1.65M_{\odot}$} \\
\hline
1  & 0.00E+00   &  8.32 &  8.17 &  7.66     &  6.30  &  0.4948  &  0.5305        &  0.5315          &  1.521 \\
2  & 2.14E+05   &  8.34 &  8.20 &  7.62     &  6.31  &  0.4998  &  0.5361        &  0.5365          &  1.521 \\
3  & 3.31E+05   &  8.29 &  8.14 &  7.71     &  6.37  &  0.5104  &  0.5381        &  0.5401          &  1.520 \\
4  & 4.40E+05   &  8.39 &  8.20 &  7.66     &  6.35  &  0.5078  &  0.5429        &  0.5432          &  1.519 \\
5  & 5.53E+05   &  8.37 &  8.18 &  7.70     &  6.39  &  0.5141  &  0.5465        &  0.5469          &  1.517 \\
6  & 6.65E+05   &  8.39 &  8.20 &  7.70     &  6.39  &  0.5183  &  0.5517        &  0.5519          &  1.514 \\
7  & 7.77E+05   &  8.41 &  8.20 &  7.71     &  6.41  &  0.5250  &  0.5570        &  0.5573          &  1.510 \\
8  & 8.85E+05   &  8.39 &  8.20 &  7.71     &  6.41  &  0.5316  &  0.5627        &  0.5629          &  1.505 \\
9  & 9.88E+05   &  8.40 &  8.20 &  7.72     &  6.41  &  0.5387  &  0.5685        &  0.5686          &  1.498 \\
10 & 1.09E+06   &  8.41 &  8.20 &  7.72     &  6.40  &  0.5459  &  0.5744        &  0.5745          &  1.488 \\
11 & 1.18E+06   &  8.42 &  8.20 &  7.72     &  6.41  &  0.5531  &  0.5803        &  0.5804          &  1.477 \\
12 & 1.27E+06   &  8.42 &  8.20 &  7.73     &  6.41  &  0.5604  &  0.5863        &  0.5863          &  1.462 \\
13 & 1.35E+06   &  8.42 &  8.20 &  7.72     &  6.41  &  0.5677  &  0.5923        &  0.5922          &  1.443 \\
14 & 1.43E+06   &  8.43 &  8.20 &  7.73     &  6.39  &  0.5748  &  0.5983        &  0.5978          &  1.419 \\
15 & 1.51E+06   &  8.41 &  8.20 &  7.74     &  6.40  &  0.5817  &  0.6041        &  0.6030          &  1.387 \\
16 & 1.58E+06   &  8.42 &  8.19 &  7.74     &  6.41  &  0.5883  &  0.6095        &  0.6081          &  1.344 \\
17 & 1.65E+06   &  8.44 &  8.18 &  7.76     &  6.63  &  0.5948  &  0.6148        &  0.6120          &  1.291 \\
18 & 1.72E+06   &  8.44 &  8.19 &  7.75     &  6.62  &  0.6001  &  0.6192        &  0.6179          &  1.251 \\
19 & 1.79E+06   &  8.43 &  8.19 &  7.75     &  6.62  &  0.6074  &  0.6244        &  0.6241          &  1.207 \\
20 & 1.84E+06   &  8.42 &  8.19 &  7.75     &  6.60  &  0.6148  &  0.6302        &  0.6302          &  1.155 \\
21 & 1.90E+06   &  8.41 &  8.19 &  7.75     &  6.57  &  0.6214  &  0.6360        &  0.6361          &  1.090 \\
22 & 1.95E+06   &  8.41 &  8.19 &  7.75     &  6.49  &  0.6277  &  0.6418        &  0.6419          &  1.002 \\
23 & 2.00E+06   &  8.41 &  0.00 &  0.00     &  0.00  &  0.6339  &  0.6475        &  0.0000          &  0.870 \\
 \\
\hline
\multicolumn{10}{c}{$M=2.0M_{\odot}$} \\
\hline
1  & 0.00E+00   &  8.33 &  8.17 &  7.64     &  6.32  &  0.4690  &  0.5103        &  0.5113          &  1.963 \\
2  & 4.24E+05   &  8.36 &  8.19 &  7.65     &  6.37  &  0.4782  &  0.5189        &  0.5194          &  1.963 \\
3  & 7.09E+05   &  8.36 &  8.21 &  7.63     &  6.33  &  0.4877  &  0.5265        &  0.5269          &  1.962 \\
4  & 8.45E+05   &  8.35 &  8.18 &  7.68     &  6.39  &  0.4941  &  0.5294        &  0.5303          &  1.962 \\
5  & 9.67E+05   &  8.39 &  8.20 &  7.67     &  6.40  &  0.4964  &  0.5341        &  0.5345          &  1.961 \\
6  & 1.09E+06   &  8.39 &  8.20 &  7.69     &  6.41  &  0.5021  &  0.5388        &  0.5392          &  1.961 \\
7  & 1.22E+06   &  8.40 &  8.20 &  7.70     &  6.41  &  0.5078  &  0.5441        &  0.5444          &  1.960 \\
8  & 1.34E+06   &  8.40 &  8.21 &  7.70     &  6.42  &  0.5143  &  0.5497        &  0.5500          &  1.958 \\
9  & 1.45E+06   &  8.42 &  8.21 &  7.70     &  6.43  &  0.5212  &  0.5554        &  0.5557          &  1.956 \\
10 & 1.56E+06   &  8.42 &  8.21 &  7.70     &  6.42  &  0.5284  &  0.5613        &  0.5614          &  1.954 \\
11 & 1.66E+06   &  8.42 &  8.21 &  7.71     &  6.44  &  0.5356  &  0.5670        &  0.5670          &  1.951 \\
12 & 1.76E+06   &  8.42 &  8.21 &  7.72     &  6.46  &  0.5428  &  0.5728        &  0.5722          &  1.948 \\
13 & 1.85E+06   &  8.44 &  8.20 &  7.73     &  6.46  &  0.5496  &  0.5782        &  0.5769          &  1.943 \\
14 & 1.94E+06   &  8.44 &  8.20 &  7.73     &  6.48  &  0.5555  &  0.5834        &  0.5815          &  1.937 \\
15 & 2.03E+06   &  8.44 &  8.19 &  7.74     &  6.48  &  0.5613  &  0.5878        &  0.5852          &  1.930 \\
16 & 2.11E+06   &  8.41 &  8.19 &  7.73     &  6.48  &  0.5662  &  0.5921        &  0.5886          &  1.920 \\
17 & 2.20E+06   &  8.44 &  8.19 &  7.56     &  6.37  &  0.5708  &  0.5957        &  0.5920          &  1.908 \\
18 & 2.27E+06   &  8.43 &  8.18 &  7.75     &  6.52  &  0.5751  &  0.5988        &  0.5948          &  1.896 \\
19 & 2.35E+06   &  8.45 &  8.18 &  7.74     &  6.52  &  0.5786  &  0.6023        &  0.5976          &  1.878 \\
20 & 2.42E+06   &  8.46 &  8.18 &  7.75     &  6.51  &  0.5823  &  0.6050        &  0.6002          &  1.859 \\
21 & 2.49E+06   &  8.43 &  8.19 &  7.74     &  6.50  &  0.5854  &  0.6074        &  0.6032          &  1.837 \\
22 & 2.56E+06   &  8.42 &  8.19 &  7.74     &  6.48  &  0.5889  &  0.6101        &  0.6059          &  1.762 \\
23 & 2.62E+06   &  8.44 &  8.19 &  7.74     &  6.46  &  0.5922  &  0.6130        &  0.6093          &  1.623 \\
24 & 2.69E+06   &  8.45 &  0.00 &  0.00     &  0.00  &  0.5966  &  0.6159        &  0.0000          &  1.423 \\
 \\
\hline
\multicolumn{10}{c}{$M=3.0M_{\odot}$} \\
\hline
1  & 0.00E+00   &  8.38 &  8.21 &  7.63     &  6.41  &  0.5708  &  0.5956        &  0.5945          &  2.978 \\
2  & 7.03E+04   &  8.38 &  8.21 &  7.72     &  6.46  &  0.5722  &  0.5969        &  0.5974          &  2.977 \\
3  & 1.28E+05   &  8.41 &  8.21 &  7.73     &  6.49  &  0.5752  &  0.6007        &  0.6008          &  2.977 \\
4  & 1.90E+05   &  8.42 &  8.21 &  7.73     &  6.50  &  0.5799  &  0.6050        &  0.6047          &  2.976 \\
5  & 2.53E+05   &  8.40 &  8.21 &  7.74     &  6.52  &  0.5848  &  0.6094        &  0.6084          &  2.974 \\
6  & 3.15E+05   &  8.42 &  8.21 &  7.75     &  6.55  &  0.5898  &  0.6136        &  0.6118          &  2.972 \\
7  & 3.77E+05   &  8.45 &  8.19 &  7.76     &  6.57  &  0.5943  &  0.6175        &  0.6143          &  2.970 \\
8  & 4.40E+05   &  8.44 &  8.19 &  7.76     &  6.57  &  0.5981  &  0.6207        &  0.6166          &  2.966 \\
9  & 5.02E+05   &  8.43 &  8.19 &  7.76     &  6.59  &  0.6015  &  0.6231        &  0.6186          &  2.962 \\
10 & 5.63E+05   &  8.45 &  8.18 &  7.77     &  6.61  &  0.6043  &  0.6253        &  0.6202          &  2.957 \\
11 & 6.23E+05   &  8.46 &  8.17 &  7.77     &  6.65  &  0.6067  &  0.6271        &  0.6218          &  2.952 \\
12 & 6.83E+05   &  8.44 &  8.17 &  7.77     &  6.63  &  0.6090  &  0.6290        &  0.6230          &  2.946 \\
13 & 7.41E+05   &  8.44 &  8.16 &  7.78     &  6.67  &  0.6106  &  0.6302        &  0.6236          &  2.939 \\
14 & 8.01E+05   &  8.47 &  8.16 &  7.78     &  6.68  &  0.6118  &  0.6314        &  0.6246          &  2.931 \\
15 & 8.58E+05   &  8.46 &  8.17 &  7.77     &  6.66  &  0.6131  &  0.6319        &  0.6255          &  2.869 \\
16 & 9.13E+05   &  8.44 &  8.16 &  7.78     &  6.68  &  0.6142  &  0.6328        &  0.6261          &  2.794 \\
17 & 9.69E+05   &  8.44 &  8.17 &  7.77     &  6.68  &  0.6152  &  0.6338        &  0.6270          &  2.701 \\
18 & 1.02E+06   &  8.46 &  8.16 &  7.78     &  6.68  &  0.6163  &  0.6343        &  0.6279          &  2.598 \\
19 & 1.07E+06   &  8.47 &  8.17 &  7.77     &  6.68  &  0.6174  &  0.6353        &  0.6291          &  2.468 \\
20 & 1.12E+06   &  8.44 &  8.16 &  7.78     &  6.63  &  0.6188  &  0.6362        &  0.6298          &  2.315 \\
21 & 1.18E+06   &  8.46 &  8.18 &  7.77     &  6.44  &  0.6197  &  0.6372        &  0.6317          &  2.086 \\
22 & 1.22E+06   &  8.47 &  8.19 &  7.76     &  6.54  &  0.6218  &  0.6386        &  0.6342          &  1.730 \\
23 & 1.27E+06   &  8.43 &  0.00 &  0.00     &  0.00  &  0.6249  &  0.6407        &  0.0000          &  1.390 \\
 \\
\hline
\multicolumn{10}{c}{$M=4.0M_{\odot}$} \\
\hline
1  & 0.00E+00   &  8.37 &  8.24 &  7.81     &  6.85  &  0.8032  &  0.8093        &  0.8093          &  3.946 \\
2  & 7.40E+03   &  8.40 &  8.24 &  7.83     &  7.00  &  0.8036  &  0.8106        &  0.8098          &  3.943 \\
3  & 1.57E+04   &  8.42 &  8.23 &  7.85     &  7.11  &  0.8043  &  0.8115        &  0.8102          &  3.938 \\
4  & 2.48E+04   &  8.44 &  8.23 &  7.85     &  7.16  &  0.8052  &  0.8123        &  0.8108          &  3.932 \\
5  & 3.47E+04   &  8.46 &  8.23 &  7.85     &  7.22  &  0.8060  &  0.8131        &  0.8114          &  3.925 \\
6  & 4.49E+04   &  8.48 &  8.22 &  7.85     &  7.22  &  0.8070  &  0.8139        &  0.8119          &  3.916 \\
7  & 5.55E+04   &  8.50 &  8.21 &  7.86     &  7.27  &  0.8078  &  0.8146        &  0.8123          &  3.905 \\
8  & 6.62E+04   &  8.46 &  8.20 &  7.87     &  7.33  &  0.8085  &  0.8151        &  0.8125          &  3.894 \\
9  & 7.71E+04   &  8.50 &  8.19 &  7.87     &  7.34  &  0.8090  &  0.8154        &  0.8126          &  3.881 \\
10 & 8.84E+04   &  8.47 &  8.19 &  7.87     &  7.37  &  0.8094  &  0.8157        &  0.8129          &  3.866 \\
11 & 9.95E+04   &  8.50 &  8.17 &  7.88     &  7.44  &  0.8098  &  0.8159        &  0.8128          &  3.851 \\
12 & 1.11E+05   &  8.52 &  8.17 &  7.88     &  7.48  &  0.8098  &  0.8161        &  0.8127          &  3.833 \\
13 & 1.23E+05   &  8.50 &  8.18 &  7.87     &  7.47  &  0.8098  &  0.8160        &  0.8130          &  3.815 \\
14 & 1.34E+05   &  8.51 &  8.16 &  7.88     &  7.47  &  0.8102  &  0.8162        &  0.8128          &  3.795 \\
15 & 1.45E+05   &  8.52 &  8.19 &  7.86     &  7.45  &  0.8101  &  0.8162        &  0.8132          &  3.774 \\
16 & 1.56E+05   &  8.52 &  8.16 &  7.88     &  7.45  &  0.8105  &  0.8164        &  0.8129          &  3.754 \\
17 & 1.68E+05   &  8.50 &  8.15 &  7.88     &  7.52  &  0.8104  &  0.8164        &  0.8126          &  3.731 \\
18 & 1.80E+05   &  8.50 &  8.19 &  7.86     &  7.45  &  0.8161  &  0.8161        &  0.8132          &  3.707 \\
19 & 1.90E+05   &  8.52 &  8.15 &  7.88     &  7.50  &  0.8106  &  0.8162        &  0.8129          &  3.685 \\
20 & 2.02E+05   &  8.52 &  8.18 &  7.87     &  7.47  &  0.8106  &  0.8166        &  0.8133          &  3.657 \\
21 & 2.13E+05   &  8.51 &  8.19 &  7.86     &  7.50  &  0.8109  &  0.8166        &  0.8140          &  3.631 \\
22 & 2.23E+05   &  8.53 &  8.16 &  7.88     &  7.54  &  0.8116  &  0.8171        &  0.8138          &  3.605 \\
23 & 2.34E+05   &  8.52 &  8.19 &  7.87     &  7.48  &  0.8114  &  0.8171        &  0.8145          &  3.578 \\
24 & 2.45E+05   &  8.50 &  8.15 &  7.89     &  7.56  &  0.8122  &  0.8176        &  0.8142          &  3.550 \\
25 & 2.56E+05   &  8.51 &  0.00 &  0.00     &  0.00  &  0.8120  &  0.8178        &  0.0000          &  3.518 \\
 \\
\hline
\multicolumn{10}{c}{$M=5.0M_{\odot}$} \\
\hline
1  & 0.00E+00   &  8.39 &  8.26 &  7.85     &  7.24  &  0.8602  &  0.8648        &  0.8645          &  4.853 \\
2  & 4.33E+03   &  8.42 &  8.25 &  7.86     &  7.33  &  0.8608  &  0.8655        &  0.8650          &  4.840 \\
3  & 9.05E+03   &  8.47 &  8.25 &  7.87     &  7.37  &  0.8613  &  0.8663        &  0.8655          &  4.822 \\
4  & 1.42E+04   &  8.48 &  8.24 &  7.88     &  7.50  &  0.8619  &  0.8670        &  0.8660          &  4.800 \\
5  & 1.96E+04   &  8.49 &  8.24 &  7.88     &  7.55  &  0.8626  &  0.8676        &  0.8663          &  4.772 \\
6  & 2.52E+04   &  8.48 &  8.23 &  7.89     &  7.59  &  0.8631  &  0.8681        &  0.8667          &  4.739 \\
7  & 3.09E+04   &  8.48 &  8.24 &  7.88     &  7.58  &  0.8638  &  0.8685        &  0.8671          &  4.701 \\
8  & 3.67E+04   &  8.50 &  8.23 &  7.89     &  7.64  &  0.8643  &  0.8690        &  0.8675          &  4.658 \\
9  & 4.25E+04   &  8.47 &  8.23 &  7.89     &  7.65  &  0.8649  &  0.8694        &  0.8680          &  4.610 \\
10 & 4.83E+04   &  8.47 &  8.23 &  7.89     &  7.66  &  0.8655  &  0.8700        &  0.8685          &  4.557 \\
11 & 5.41E+04   &  8.51 &  8.23 &  7.89     &  7.69  &  0.8661  &  0.8705        &  0.8690          &  4.498 \\
12 & 5.99E+04   &  8.52 &  8.23 &  7.89     &  7.70  &  0.8667  &  0.8710        &  0.8695          &  4.434 \\
13 & 6.57E+04   &  8.51 &  8.23 &  7.89     &  7.69  &  0.8674  &  0.8715        &  0.8701          &  4.362 \\
14 & 7.15E+04   &  8.53 &  8.22 &  7.89     &  7.70  &  0.8679  &  0.8721        &  0.8706          &  4.283 \\
15 & 7.72E+04   &  8.51 &  8.22 &  7.90     &  7.68  &  0.8685  &  0.8726        &  0.8711          &  4.198 \\
16 & 8.30E+04   &  8.52 &  8.22 &  7.89     &  7.69  &  0.8691  &  0.8731        &  0.8716          &  4.106 \\
17 & 8.88E+04   &  8.53 &  8.22 &  7.89     &  7.67  &  0.8697  &  0.8737        &  0.8721          &  4.003 \\
18 & 9.45E+04   &  8.52 &  8.20 &  7.90     &  7.65  &  0.8703  &  0.8742        &  0.8724          &  3.892 \\
19 & 1.01E+05   &  8.53 &  8.21 &  7.89     &  7.65  &  0.8707  &  0.8746        &  0.8730          &  3.764 \\
20 & 1.06E+05   &  8.53 &  8.20 &  7.89     &  7.56  &  0.8712  &  0.8751        &  0.8733          &  3.623 \\
21 & 1.12E+05   &  8.52 &  8.20 &  7.89     &  7.50  &  0.8716  &  0.8755        &  0.8738          &  3.461 \\
22 & 1.19E+05   &  8.53 &  8.19 &  7.90     &  7.48  &  0.8721  &  0.8760        &  0.8741          &  3.272 \\
23 & 1.25E+05   &  8.50 &  8.17 &  7.90     &  7.33  &  0.8725  &  0.8763        &  0.8742          &  3.042 \\
24 & 1.32E+05   &  8.51 &  8.36 &  7.13     &  6.91  &  0.8727  &  0.8766        &  0.8742          &  2.721 \\
25 & 1.38E+05   &  8.53 &  0.00 &  0.00     &  0.00  &  0.8727  &  0.8766        &  0.0000          &  2.198 \\
\hline
\multicolumn{10}{l}{TP: TP number.} \\
\multicolumn{10}{l}{ $t_{TP}$: Time since first TP.} \\
\multicolumn{10}{l}{$T_{FBOT}$: Largest temperature at the bottom of the flash-convective zone.} \\
\multicolumn{10}{l}{$T_{HES}$: Temperature in the He-burning shell during deepest extend of 3DUP.} \\
\multicolumn{10}{l}{$T_{HS}$: Temperature in the H shell.} \\
\multicolumn{10}{l}{ $T_{CEB}$: Temperature at the bottom of the convective envelope during deepest extend of 3DUP.} \\
\multicolumn{10}{l}{$m_{FBOT}$: Mass coordinate at the bottom of the He-flash convective zone.} \\
\multicolumn{10}{l}{$m_{HTP}$: Mass coordinate of the H-free core at the time of the TP.} \\
\multicolumn{10}{l}{$m_{D,max}$: Lowest mass coordinate at the convective envelope bottom after the TP.} \\
\multicolumn{10}{l}{ $M_{\ast}$: Stellar mass at the TP.} 
\label{agb_model1p2prop2}
\end{longtable}
\end{center}

\begin{center}
\tiny\begin{longtable}{cccccccccc}
\caption{TPAGB evolution properties of set1.1. Same parameter as in Table \ref{agb_model1p2prop2}}\\
\hline
\multicolumn{10}{c}{\setopo\ Z = 0.01} \\
\hline \\
 TP & $t_{TP}$ & $T_{FBOT}$ & $T_{HES}$ & $T_{HS}$ & $T_{CEB}$ & $m_{FBOT}$ & $m_{HTP}$ & $m_{D,max}$ & $M_{\ast}$\\
 & [$yrs$] & [$K$] & [$K$] & [$K$] & [$K$] & [$M_{\odot}$] & [$M_{\odot}$] & [$M_{\odot}$] & [$M_{\odot}$] \\
\\
\hline
\multicolumn{10}{c}{$M=1.65M_{\odot}$} \\
\hline
1  & 0.00E+00   &  8.31 &  8.18 &  7.63     &  6.19  &  0.4974  &  0.5328        &  0.5346          &  1.536 \\
2  & 1.05E+05   &  8.31 &  8.17 &  7.67     &  6.26  &  0.4999  &  0.5343        &  0.5358          &  1.536 \\
3  & 2.23E+05   &  8.36 &  8.18 &  7.68     &  6.30  &  0.5009  &  0.5372        &  0.5379          &  1.536 \\
4  & 3.55E+05   &  8.38 &  8.19 &  7.68     &  6.32  &  0.5051  &  0.5411        &  0.5417          &  1.535 \\
5  & 4.87E+05   &  8.37 &  8.19 &  7.71     &  6.34  &  0.5103  &  0.5453        &  0.5458          &  1.534 \\
6  & 6.21E+05   &  8.40 &  8.20 &  7.70     &  6.35  &  0.5155  &  0.5508        &  0.5512          &  1.532 \\
7  & 7.51E+05   &  8.40 &  8.19 &  7.72     &  6.36  &  0.5227  &  0.5563        &  0.5566          &  1.529 \\
8  & 8.77E+05   &  8.42 &  8.20 &  7.72     &  6.38  &  0.5297  &  0.5624        &  0.5626          &  1.525 \\
9  & 9.96E+05   &  8.42 &  8.20 &  7.73     &  6.38  &  0.5373  &  0.5686        &  0.5687          &  1.520 \\
10 & 1.11E+06   &  8.43 &  8.20 &  7.73     &  6.39  &  0.5451  &  0.5749        &  0.5750          &  1.513 \\
11 & 1.22E+06   &  8.43 &  8.20 &  7.73     &  6.39  &  0.5529  &  0.5812        &  0.5811          &  1.504 \\
12 & 1.32E+06   &  8.43 &  8.19 &  7.74     &  6.40  &  0.5606  &  0.5876        &  0.5866          &  1.493 \\
13 & 1.41E+06   &  8.42 &  8.19 &  7.75     &  6.40  &  0.5676  &  0.5934        &  0.5917          &  1.477 \\
14 & 1.51E+06   &  8.44 &  8.19 &  7.75     &  6.41  &  0.5741  &  0.5987        &  0.5967          &  1.456 \\
15 & 1.59E+06   &  8.44 &  8.19 &  7.74     &  6.41  &  0.5804  &  0.6039        &  0.6017          &  1.429 \\
16 & 1.68E+06   &  8.43 &  8.19 &  7.75     &  6.39  &  0.5866  &  0.6090        &  0.6070          &  1.393 \\
17 & 1.76E+06   &  8.43 &  0.00 &  0.00     &  0.00  &  0.5942  &  0.6140        &  0.0000          &  1.239 \\
 \\
\hline
\multicolumn{10}{c}{$M=2.0M_{\odot}$} \\
\hline
1  & 0.00E+00   &  8.29 &  8.17 &  7.64     &  6.26  &  0.4503  &  0.4979        &  0.4999          &  1.978 \\
2  & 5.25E+05   &  8.36 &  8.18 &  7.64     &  6.28  &  0.4599  &  0.5061        &  0.5069          &  1.978 \\
3  & 8.91E+05   &  8.35 &  8.19 &  7.64     &  6.29  &  0.4693  &  0.5124        &  0.5133          &  1.978 \\
4  & 1.06E+06   &  8.34 &  8.17 &  7.69     &  6.33  &  0.4739  &  0.5152        &  0.5163          &  1.978 \\
5  & 1.22E+06   &  8.38 &  8.19 &  7.68     &  6.33  &  0.4768  &  0.5198        &  0.5205          &  1.978 \\
6  & 1.39E+06   &  8.37 &  8.19 &  7.69     &  6.34  &  0.4824  &  0.5246        &  0.5253          &  1.977 \\
7  & 1.55E+06   &  8.39 &  8.20 &  7.69     &  6.35  &  0.4884  &  0.5302        &  0.5307          &  1.977 \\
8  & 1.72E+06   &  8.38 &  8.20 &  7.70     &  6.37  &  0.4956  &  0.5361        &  0.5365          &  1.976 \\
9  & 1.87E+06   &  8.41 &  8.20 &  7.71     &  6.38  &  0.5033  &  0.5421        &  0.5425          &  1.975 \\
10 & 2.01E+06   &  8.42 &  8.20 &  7.72     &  6.40  &  0.5113  &  0.5484        &  0.5487          &  1.974 \\
11 & 2.15E+06   &  8.42 &  8.20 &  7.72     &  6.41  &  0.5190  &  0.5545        &  0.5547          &  1.973 \\
12 & 2.28E+06   &  8.43 &  8.20 &  7.72     &  6.41  &  0.5270  &  0.5609        &  0.5605          &  1.971 \\
13 & 2.40E+06   &  8.42 &  8.20 &  7.73     &  6.44  &  0.5347  &  0.5668        &  0.5660          &  1.968 \\
14 & 2.51E+06   &  8.42 &  8.19 &  7.74     &  6.44  &  0.5417  &  0.5725        &  0.5707          &  1.965 \\
15 & 2.62E+06   &  8.41 &  8.19 &  7.74     &  6.46  &  0.5479  &  0.5776        &  0.5750          &  1.961 \\
16 & 2.72E+06   &  8.43 &  8.19 &  7.73     &  6.46  &  0.5535  &  0.5821        &  0.5791          &  1.956 \\
17 & 2.82E+06   &  8.42 &  8.19 &  7.74     &  6.47  &  0.5585  &  0.5861        &  0.5826          &  1.950 \\
18 & 2.92E+06   &  8.42 &  8.18 &  7.74     &  6.48  &  0.5634  &  0.5901        &  0.5858          &  1.941 \\
19 & 3.01E+06   &  8.46 &  8.17 &  7.75     &  6.51  &  0.5677  &  0.5936        &  0.5884          &  1.911 \\
20 & 3.10E+06   &  8.46 &  8.18 &  7.75     &  6.50  &  0.5714  &  0.5966        &  0.5912          &  1.857 \\
21 & 3.19E+06   &  8.45 &  8.18 &  7.75     &  6.51  &  0.5749  &  0.5992        &  0.5937          &  1.790 \\
22 & 3.27E+06   &  8.43 &  8.18 &  7.75     &  6.49  &  0.5780  &  0.6013        &  0.5965          &  1.711 \\
23 & 3.35E+06   &  8.45 &  8.19 &  7.74     &  6.48  &  0.5814  &  0.6039        &  0.6001          &  1.608 \\
24 & 3.42E+06   &  8.43 &  8.19 &  7.74     &  6.42  &  0.5860  &  0.6075        &  0.6045          &  1.463 \\
25 & 3.49E+06   &  8.42 &  0.00 &  0.00     &  0.00  &  0.5929  &  0.6117        &  0.0000          &  1.208 \\
 \\
\hline
\multicolumn{10}{c}{$M=3.0M_{\odot}$} \\
\hline
1  & 0.00E+00   &  8.34 &  8.20 &  7.75     &  6.44  &  0.6282  &  0.6461        &  0.6467          &  2.972 \\
2  & 3.91E+04   &  8.40 &  8.22 &  7.74     &  6.47  &  0.6291  &  0.6488        &  0.6485          &  2.971 \\
3  & 8.17E+04   &  8.41 &  8.21 &  7.77     &  6.53  &  0.6324  &  0.6515        &  0.6503          &  2.970 \\
4  & 1.26E+05   &  8.41 &  8.19 &  7.79     &  6.57  &  0.6350  &  0.6544        &  0.6515          &  2.967 \\
5  & 1.74E+05   &  8.43 &  8.19 &  7.79     &  6.60  &  0.6374  &  0.6566        &  0.6525          &  2.964 \\
6  & 2.24E+05   &  8.44 &  8.17 &  7.79     &  6.64  &  0.6394  &  0.6582        &  0.6528          &  2.959 \\
7  & 2.76E+05   &  8.44 &  8.16 &  7.80     &  6.68  &  0.6407  &  0.6594        &  0.6529          &  2.953 \\
8  & 3.29E+05   &  8.47 &  8.16 &  7.80     &  6.69  &  0.6418  &  0.6599        &  0.6530          &  2.903 \\
9  & 3.81E+05   &  8.47 &  8.14 &  7.81     &  6.73  &  0.6425  &  0.6602        &  0.6523          &  2.824 \\
10 & 4.36E+05   &  8.46 &  8.17 &  7.79     &  6.72  &  0.6425  &  0.6602        &  0.6535          &  2.719 \\
11 & 4.84E+05   &  8.47 &  8.15 &  7.80     &  6.72  &  0.6439  &  0.6603        &  0.6533          &  2.613 \\
12 & 5.35E+05   &  8.46 &  8.14 &  7.80     &  6.72  &  0.6441  &  0.6609        &  0.6525          &  2.461 \\
13 & 5.86E+05   &  8.48 &  8.14 &  7.80     &  6.71  &  0.6436  &  0.6604        &  0.6522          &  2.249 \\
14 & 6.35E+05   &  8.47 &  0.00 &  0.00     &  0.00  &  0.6434  &  0.6598        &  0.0000          &  1.932 \\
 \\
\hline
\multicolumn{10}{c}{$M=4.0M_{\odot}$} \\
\hline
1  & 0.00E+00   &  8.41 &  8.24 &  7.84     &  6.96  &  0.8251  &  0.8309        &  0.8306          &  3.930 \\
2  & 6.71E+03   &  8.44 &  8.23 &  7.85     &  7.07  &  0.8256  &  0.8318        &  0.8310          &  3.927 \\
3  & 1.42E+04   &  8.46 &  8.23 &  7.86     &  7.18  &  0.8262  &  0.8326        &  0.8314          &  3.923 \\
4  & 2.25E+04   &  8.45 &  8.22 &  7.87     &  7.27  &  0.8268  &  0.8333        &  0.8318          &  3.917 \\
5  & 3.13E+04   &  8.49 &  8.22 &  7.88     &  7.31  &  0.8276  &  0.8339        &  0.8323          &  3.909 \\
6  & 4.05E+04   &  8.48 &  8.22 &  7.88     &  7.33  &  0.8284  &  0.8345        &  0.8328          &  3.900 \\
7  & 4.98E+04   &  8.50 &  8.22 &  7.88     &  7.42  &  0.8292  &  0.8352        &  0.8332          &  3.890 \\
8  & 5.93E+04   &  8.51 &  8.20 &  7.89     &  7.48  &  0.8299  &  0.8357        &  0.8334          &  3.879 \\
9  & 6.90E+04   &  8.47 &  8.19 &  7.89     &  7.51  &  0.8303  &  0.8360        &  0.8336          &  3.866 \\
10 & 7.87E+04   &  8.49 &  8.18 &  7.89     &  7.49  &  0.8307  &  0.8364        &  0.8336          &  3.852 \\
11 & 8.86E+04   &  8.52 &  8.17 &  7.89     &  7.53  &  0.8308  &  0.8365        &  0.8334          &  3.837 \\
12 & 9.87E+04   &  8.49 &  8.20 &  7.88     &  7.55  &  0.8308  &  0.8365        &  0.8339          &  3.819 \\
13 & 1.08E+05   &  8.48 &  8.20 &  7.88     &  7.59  &  0.8313  &  0.8366        &  0.8345          &  3.803 \\
14 & 1.18E+05   &  8.51 &  8.19 &  7.89     &  7.59  &  0.8318  &  0.8372        &  0.8349          &  3.785 \\
15 & 1.27E+05   &  8.51 &  8.20 &  7.88     &  7.55  &  0.8324  &  0.8377        &  0.8354          &  3.664 \\
16 & 1.37E+05   &  8.53 &  8.19 &  7.89     &  7.60  &  0.8330  &  0.8382        &  0.8357          &  3.524 \\
17 & 1.46E+05   &  8.53 &  8.20 &  7.88     &  7.50  &  0.8334  &  0.8386        &  0.8364          &  3.361 \\
18 & 1.55E+05   &  8.49 &  8.20 &  7.88     &  7.50  &  0.8341  &  0.8391        &  0.8371          &  3.172 \\
19 & 1.65E+05   &  8.49 &  8.21 &  7.87     &  7.35  &  0.8348  &  0.8399        &  0.8380          &  2.936 \\
20 & 1.74E+05   &  8.49 &  0.00 &  0.00     &  0.00  &  0.8358  &  0.8408        &  0.0000          &  2.617 \\
 \\
\hline
\multicolumn{10}{c}{$M=5.0M_{\odot}$} \\
\hline
1  & 0.00E+00   &  8.31 &  8.22 &  7.89     &  7.28  &  0.8989  &  0.9012        &  0.9015          &  4.830 \\
2  & 2.79E+03   &  8.42 &  8.26 &  7.88     &  7.43  &  0.8985  &  0.9019        &  0.9017          &  4.820 \\
3  & 6.06E+03   &  8.42 &  8.26 &  7.89     &  7.49  &  0.8989  &  0.9025        &  0.9021          &  4.805 \\
4  & 9.65E+03   &  8.47 &  8.25 &  7.90     &  7.57  &  0.8993  &  0.9031        &  0.9025          &  4.786 \\
5  & 1.35E+04   &  8.46 &  8.25 &  7.91     &  7.64  &  0.8998  &  0.9037        &  0.9028          &  4.762 \\
6  & 1.76E+04   &  8.50 &  8.24 &  7.91     &  7.70  &  0.9002  &  0.9042        &  0.9031          &  4.731 \\
7  & 2.18E+04   &  8.47 &  8.24 &  7.91     &  7.75  &  0.9007  &  0.9045        &  0.9035          &  4.693 \\
8  & 2.61E+04   &  8.50 &  8.24 &  7.91     &  7.76  &  0.9013  &  0.9050        &  0.9039          &  4.648 \\
9  & 3.04E+04   &  8.48 &  8.23 &  7.92     &  7.79  &  0.9017  &  0.9054        &  0.9042          &  4.594 \\
10 & 3.49E+04   &  8.50 &  8.23 &  7.92     &  7.78  &  0.9022  &  0.9058        &  0.9045          &  4.528 \\
11 & 3.94E+04   &  8.51 &  8.23 &  7.92     &  7.80  &  0.9026  &  0.9061        &  0.9048          &  4.453 \\
12 & 4.40E+04   &  8.51 &  8.22 &  7.92     &  7.79  &  0.9030  &  0.9065        &  0.9051          &  4.365 \\
13 & 4.86E+04   &  8.53 &  8.21 &  7.92     &  7.81  &  0.9034  &  0.9068        &  0.9053          &  4.264 \\
14 & 5.33E+04   &  8.50 &  8.21 &  7.92     &  7.79  &  0.9036  &  0.9071        &  0.9055          &  4.148 \\
15 & 5.81E+04   &  8.50 &  8.22 &  7.92     &  7.78  &  0.9039  &  0.9073        &  0.9058          &  4.020 \\
16 & 6.28E+04   &  8.54 &  8.21 &  7.92     &  7.78  &  0.9042  &  0.9076        &  0.9061          &  3.882 \\
17 & 6.75E+04   &  8.53 &  8.21 &  7.92     &  7.76  &  0.9046  &  0.9079        &  0.9064          &  3.730 \\
18 & 7.23E+04   &  8.50 &  8.21 &  7.92     &  7.74  &  0.9049  &  0.9082        &  0.9067          &  3.563 \\
19 & 7.70E+04   &  8.54 &  8.22 &  7.91     &  7.70  &  0.9053  &  0.9085        &  0.9071          &  3.381 \\
20 & 8.18E+04   &  8.53 &  8.21 &  7.92     &  7.68  &  0.9057  &  0.9089        &  0.9075          &  3.179 \\
21 & 8.66E+04   &  8.51 &  8.21 &  7.91     &  7.47  &  0.9061  &  0.9093        &  0.9079          &  2.939 \\
22 & 9.16E+04   &  8.53 &  0.00 &  0.00     &  0.00  &  0.9066  &  0.9098        &  0.0000          &  2.627 \\
\hline
\hline
\multicolumn{10}{l}{TP: TP number.} \\
\multicolumn{10}{l}{ $t_{TP}$: Time since first TP.} \\
\multicolumn{10}{l}{$T_{FBOT}$: Largest temperature at the bottom of the flash-convective zone.} \\
\multicolumn{10}{l}{$T_{HES}$: Temperature in the He-burning shell during deepest extend of 3DUP.} \\
\multicolumn{10}{l}{$T_{HS}$: Temperature in the H shell.} \\
\multicolumn{10}{l}{ $T_{CEB}$: Temperature at the bottom of the convective envelope during deepest extend of 3DUP.} \\
\multicolumn{10}{l}{$m_{FBOT}$: Mass coordinate at the bottom of the He-flash convective zone.} \\
\multicolumn{10}{l}{$m_{HTP}$: Mass coordinate of the H-free core at the time of the TP.} \\
\multicolumn{10}{l}{$m_{D,max}$: Lowest mass coordinate at the convective envelope bottom after the TP.} \\
\multicolumn{10}{l}{ $M_{\ast}$: Stellar mass at the TP.} 

\label{agb_model1p1prop2}
\end{longtable}
\end{center}

%%%%%%%%%%%%%%%%%%

%%%%%%%%%%%%%%%%%%%%%%%%%%%%%%%%%%%%%%%%%%%%%%
%	TABle isotopic fallback
%%%%%%%%%%%%%%%%%%%%%%%%%%%%%%%%%%%%%%%%%%%%%%

%\addtolength{\voffset}{-0.0cm}
\begin{table}
\begin{center}
%\tabletypesize{\small}
\caption{Final remnant mass coordinates (in solar mass unit)
of massive star models presented in this work. The 25 $\msun$, $Z = 0.02$ stellar model
is directly collapsed into a Black Hole. 
}
\begin{tabular}{ccccc}
\hline
  initial mass &  $Z$ = 0.02    & $Z$ = 0.02    & $Z$ = 0.01    & $Z$ = 0.01	 \\
 ($\msun$) &  $delay$       & $rapid$       & $delay$       & $rapid$        \\
\hline
  15	 & 1.60  & 1.44  & 1.61  & 1.44  \\
  20	 & 2.70  & 2.73  & 2.73  & 1.83   \\
  25	 & 5.71  & 13.8$^{*}$ & 6.05  & 7.91  \\
  32	 & 8.75  & 4.75  & $-$  & $-$  \\
  60     & 3.00  & 3.00  & $-$  & $-$   \\
\hline
 \noalign{\smallskip}
\hline
\end{tabular}
\label{tab: coo_fallback}
\end{center}
\end{table}

%\include{table_kinenergy_ccsn}
%%%%%%%%%%%%%%%%%

%\include{overshoot.tab}
%\include{isotopic_table_set1.2_prodfac_winds_short}
%\include{isotopic_table_set1.2_prodfac_pre_short}
%\include{isotopic_element_table_set1.2_prodfac_exp_short}
%\include{isotopic_table_set1.2_prodfac_exp_short}
%\include{element_table_set1.2_prodfac_winds_short}
%\include{element_table_set1.2_prodfac_pre_short}
%\include{element_table_set1.2_prodfac_exp_short}
%\include{isotopic_table_set1.1_prodfac_winds_short}
%\include{isotopic_table_set1.1_prodfac_pre_short}
%\include{isotopic_table_set1.1_prodfac_exp_short}
%\include{element_table_set1.1_prodfac_winds_short}
%\include{element_table_set1.1_prodfac_pre_short}
%\include{element_table_set1.1_prodfac_exp_short}
%\include{isotopic_table_comparison_other_ref_2msun_set1p2}
%\include{table_snIa_contribution}
%\include{table_fallback_coord}
\clearpage

%%%%%%%%%%%%%%%%%%%%%%%%%%%%%%%%%%%%%%%%%%%%%%
%	TABle isotopic abundance winds
%%%%%%%%%%%%%%%%%%%%%%%%%%%%%%%%%%%%%%%%%%%%%%

%\addtolength{\voffset}{-0.0cm}
\begin{landscape}
\begin{table}
\begin{center}
\tabletypesize{\small}
\caption{The overproduction factors of stable isotopes in stellar winds is given for the stars of \setopt.
The complete table is available online.}
\begin{tabular}{ccccccccccc}
\hline
 isotopes & 1.65 $\msun$ & 2 $\msun$    & 3 $\msun$   & 4 $\msun$  & 5 $\msun$    & 15 $\msun$   & 20 $\msun$   & 25 $\msun$   & 32 $\msun$   & 60 $\msun$    \\
\hline 
%H   1    & 9.703E-01 & 9.430E-01 & 9.258E-01 & 9.408E-01 & 8.993E-01 & 9.784E-01 & 9.598E-01 & 9.043E-01 & 8.139E-01 & 5.501E-01  \\ 
%H   2    & 3.227E-01 & 2.619E-01 & 1.880E-01 & 1.566E-02 & 1.685E-02 & 2.872E-07 & 1.074E-10 & 2.147E-11 & 5.313E-13 & 3.208E-13  \\ 
%HE  3    & 8.067E+00 & 6.184E+00 & 3.075E+00 & 2.032E+00 & 1.328E+00 & 1.093E+00 & 1.197E+00 & 1.097E+00 & 8.841E-01 & 7.326E-01  \\ 
%HE  4    & 1.064E+00 & 1.069E+00 & 1.108E+00 & 1.110E+00 & 1.238E+00 & 1.055E+00 & 1.104E+00 & 1.247E+00 & 1.481E+00 & 1.894E+00  \\ 
%LI  7    & 1.167E+00 & 6.786E-01 & 5.629E-01 & 1.341E+02 & 6.225E+02 & 3.863E-06 & 1.217E-06 & 3.979E-07 & 1.958E-07 & 8.263E-08  \\ 
%B  11    & 9.999E-01 & 9.953E-01 & 9.994E-01 & 1.000E+00 & 9.992E-01 & 1.000E+00 & 1.000E+00 & 1.000E+00 & 1.000E+00 & 1.000E+00  \\ 
 C  12    & 1.359E+00 & 3.960E+00 & 4.869E+00 & 2.812E+00 & 1.605E+00 & 7.313E-01 & 8.061E-01 & 6.971E-01 & 5.245E-01 & 1.831E+01  \\ 
%C  13    & 2.486E+00 & 2.367E+00 & 2.354E+00 & 2.494E+00 & 3.887E+00 & 2.931E+00 & 2.737E+00 & 2.097E+00 & 1.776E+00 & 9.205E-01  \\ 
 N  14    & 2.197E+00 & 2.422E+00 & 2.869E+00 & 3.091E+00 & 3.841E+00 & 2.361E+00 & 2.358E+00 & 3.638E+00 & 5.718E+00 & 6.287E+00  \\ 
%N  15    & 5.200E-01 & 4.810E-01 & 4.477E-01 & 4.178E-01 & 3.009E-01 & 5.219E-01 & 5.833E-01 & 5.275E-01 & 3.963E-01 & 3.301E-01  \\ 
 O  16    & 1.038E+00 & 1.427E+00 & 1.583E+00 & 1.251E+00 & 9.846E-01 & 9.466E-01 & 9.126E-01 & 8.070E-01 & 6.297E-01 & 1.482E+00  \\
\hline
\end{tabular}
\label{tab:isotopic_prodfac_set1.2_winds}
\end{center}
\end{table}
\end{landscape}

%\addtolength{\voffset}{-0.0cm}
\begin{table}
\begin{center}
%\tabletypesize{\small}
\caption{The presupernova overproduction factors of stable isotopes are given for massive stars of \setopt.
The complete table is available online.}
\begin{tabular}{cccccc}
\hline
 isotopes & 15 $\msun$   & 20 $\msun$	& 25 $\msun$   & 32 $\msun$	& 60 $\msun$    \\ 
\hline
%H   1  & 6.054E-01 & 4.869E-01 & 3.801E-01 & 1.297E-11 & 2.560E-15  \\ 
%H   2  & 4.542E-02 & 8.028E-03 & 3.921E-04 & 5.556E-15 & 4.940E-95  \\ 
%HE  3  & 3.959E-01 & 1.936E-01 & 3.412E-02 & 1.744E-25 & 1.491E-95  \\ 
%HE  4  & 1.498E+00 & 1.600E+00 & 1.981E+00 & 2.094E+00 & 1.095E-01  \\ 
%LI  7  & 4.542E-02 & 8.075E-03 & 3.465E-01 & 1.309E-14 & 1.780E-89  \\ 
%B  11  & 1.003E+00 & 1.001E+00 & 1.001E+00 & 1.000E+00 & 1.000E+00  \\ 
 C  12  & 7.485E+00 & 9.933E+00 & 1.734E+01 & 4.957E+01 & 3.926E+01  \\ 
%C  13  & 1.516E+00 & 1.422E+00 & 3.285E+01 & 1.431E-01 & 5.798E-05  \\ 
 N  14  & 3.991E+00 & 4.742E+00 & 6.146E+00 & 1.095E+00 & 3.277E-04  \\ 
%N  15  & 4.718E-01 & 2.553E-01 & 5.664E-02 & 6.119E-01 & 7.094E-04  \\ 
 O  16  & 8.415E+00 & 1.321E+01 & 9.806E+00 & 2.322E+01 & 5.905E+01  \\ 
%O  17  & 1.373E+00 & 1.159E+00 & 1.973E+00 & 1.377E+00 & 2.306E-01  \\ 
%O  18  & 3.540E+01 & 5.511E+00 & 1.187E-01 & 2.744E+00 & 4.353E-04  \\ 
%F  19  & 7.109E-01 & 8.894E-01 & 2.174E+00 & 2.033E+01 & 6.660E-03  \\ 
 NE 20  & 1.292E+01 & 9.048E+00 & 7.880E+00 & 4.376E+00 & 1.089E+02  \\ 
%NE 21  & 5.422E+00 & 6.252E+00 & 6.014E+00 & 2.913E+00 & 4.485E+01  \\ 
%NE 22  & 1.733E+01 & 2.458E+01 & 4.535E+01 & 1.002E+02 & 8.390E+00  \\ 
%NA 23  & 9.258E+00 & 5.741E+00 & 1.517E+01 & 7.987E+00 & 8.901E+01  \\ 
 MG 24  & 1.425E+01 & 2.575E+01 & 3.101E+00 & 2.993E+00 & 5.444E+01  \\ 
%MG 25  & 1.635E+01 & 2.725E+01 & 8.690E+00 & 1.593E+01 & 9.344E+01  \\ 
%MG 26  & 1.583E+01 & 2.165E+01 & 1.602E+01 & 3.983E+01 & 1.589E+02  \\ 
%AL 27  & 5.790E+00 & 9.506E+00 & 3.073E+00 & 2.213E+00 & 3.300E+01  \\ 
 SI 28  & 1.083E+01 & 1.002E+01 & 1.131E+00 & 1.178E+00 & 5.248E+00  \\
\hline
\end{tabular}
\label{tab:isotopic_prodfac_set1.2_pre}
\end{center}
\end{table}

%H this is a test
%H to write strings in the first col
%%%%%%%%%%%%%%%%%%%%%%%%%%%%%%%%%%%%%%%%%%%%%%
%   TABle element abundance exp
%%%%%%%%%%%%%%%%%%%%%%%%%%%%%%%%%%%%%%%%%%%%%%

%\addtolength{\voffset}{-0.0cm}
\begin{landscape}
\begin{table}
%\begin{center}
%\tabletypesize{\small}
\caption{The isotopic overproduction factors in supernova ejecta is given for the stars of \setopt.
The complete table is available online.}
\begin{tabular}{ccccccccc}
\hline
 isotopes & 15 $\msun$ delay & 15 $\msun$ rapid & 20 $\msun$ delay & 20 $\msun$ rapid & 25 $\msun$ delay & 32 $\msun$ delay & 32 $\msun$ rapid & 60 $\msun$ delay  \\
\hline 
% H   1  & 5.752E-01    & 5.758E-01    & 4.566E-01    & 4.570E-01    & 3.462E-01    & 1.974E-12    & 3.590E-09    & 3.783E-14	 \\ 
% H   2  & 4.313E-02    & 4.318E-02    & 7.520E-03    & 7.527E-03    & 3.567E-04    & 2.036E-06    & 7.846E-07    & 2.891E-11	 \\ 
% HE  3  & 3.760E-01    & 3.764E-01    & 1.813E-01    & 1.815E-01    & 3.089E-02    & 6.886E-12    & 1.517E-10    & 4.418E-11	 \\ 
% HE  4  & 1.582E+00    & 1.596E+00    & 1.514E+00    & 1.516E+00    & 1.877E+00    & 1.531E+00    & 8.178E-01    & 2.020E-02	 \\ 
% LI  7  & 4.322E-02    & 4.326E-02    & 8.114E-03    & 8.093E-03    & 4.105E+00    & 7.070E-03    & 2.408E-03    & 2.940E-07	 \\ 
% B  11  & 7.270E-01    & 7.300E-01    & 6.135E-01    & 6.050E-01    & 7.632E-01    & 1.969E-02    & 4.413E-02    & 2.513E-01	 \\ 
  C  12  & 4.683E+00    & 4.749E+00    & 6.769E+00    & 6.911E+00    & 1.545E+01    & 3.082E+01    & 2.195E+01    & 3.488E+01	 \\ 
% C  13  & 1.442E+00    & 1.444E+00    & 1.332E+00    & 1.333E+00    & 3.494E+01    & 9.474E-04    & 2.380E-03    & 1.183E-03	 \\ 
  N  14  & 3.811E+00    & 3.817E+00    & 4.451E+00    & 4.455E+00    & 5.664E+00    & 2.491E-01    & 1.331E-01    & 1.344E-03	 \\ 
% N  15  & 1.066E+00    & 1.100E+00    & 5.592E-01    & 5.451E-01    & 3.315E+00    & 5.823E-01    & 1.423E-01    & 4.596E-03	 \\ 
  O  16  & 2.688E+00    & 2.713E+00    & 1.118E+01    & 1.074E+01    & 9.433E+00    & 1.908E+01    & 4.751E+01    & 5.485E+01	 \\
\hline
\end{tabular}
\label{tab:isotopic_prodfac_set1.2_exp}
%\end{center}
\end{table}
\end{landscape}

%H this is a test
%H to write strings in the first col
%%%%%%%%%%%%%%%%%%%%%%%%%%%%%%%%%%%%%%%%%%%%%%
%	TABle element abundance winds
%%%%%%%%%%%%%%%%%%%%%%%%%%%%%%%%%%%%%%%%%%%%%%

%\addtolength{\voffset}{-0.0cm}
\begin{landscape}
\begin{table}
\begin{center}
%\tabletypesize{\small}
\caption{The element overproduction factors in stellar winds is given for the stars of \setopt.
The complete table is available online.}
\begin{tabular}{ccccccccccc}
\hline
 elements & 1.65 $\msun$ & 2 $\msun$    & 3 $\msun$  & 4 $\msun$  & 5 $\msun$    & 15 $\msun$   & 20 $\msun$   & 25 $\msun$   & 32 $\msun$   & 60 $\msun$    \\
\hline 
%H      & 9.703E-01 & 9.430E-01 & 9.258E-01 & 9.408E-01 & 8.992E-01 & 9.784E-01 & 9.598E-01 & 9.042E-01 & 8.139E-01 & 5.501E-01  \\ 
%He     & 1.065E+00 & 1.069E+00 & 1.108E+00 & 1.110E+00 & 1.238E+00 & 1.055E+00 & 1.104E+00 & 1.247E+00 & 1.481E+00 & 1.894E+00  \\ 
%Li     & 1.167E+00 & 6.786E-01 & 5.629E-01 & 1.341E+02 & 6.225E+02 & 3.863E-06 & 1.217E-06 & 3.979E-07 & 1.958E-07 & 8.263E-08  \\ 
%Be     & 0.000E+00 & 0.000E+00 & 0.000E+00 & 0.000E+00 & 0.000E+00 & 0.000E+00 & 0.000E+00 & 0.000E+00 & 0.000E+00 & 0.000E+00  \\ 
%B      & 9.999E-01 & 9.953E-01 & 9.994E-01 & 1.000E+00 & 9.992E-01 & 1.000E+00 & 1.000E+00 & 1.000E+00 & 1.000E+00 & 1.000E+00  \\ 
C      & 1.372E+00 & 3.941E+00 & 4.839E+00  & 2.808E+00 & 1.632E+00 & 7.577E-01 & 8.292E-01 & 7.138E-01 & 5.395E-01 & 1.810E+01  \\ 
N      & 2.190E+00 & 2.415E+00 & 2.859E+00  & 3.080E+00 & 3.828E+00 & 2.354E+00 & 2.351E+00 & 3.625E+00 & 5.697E+00 & 6.264E+00  \\ 
O      & 1.038E+00 & 1.428E+00 & 1.583E+00  & 1.252E+00 & 9.849E-01 & 9.465E-01 & 9.125E-01 & 8.069E-01 & 6.296E-01 & 1.479E+00  \\ 
F      & 1.356E+00 & 2.906E+00 & 4.718E+00  & 1.189E+00 & 9.000E-01 & 9.391E-01 & 8.975E-01 & 7.619E-01 & 5.591E-01 & 2.033E+00  \\ 
Ne     & 1.068E+00 & 1.390E+00 & 1.630E+00  & 1.179E+00 & 1.053E+00 & 9.949E-01 & 9.924E-01 & 9.830E-01 & 9.665E-01 & 2.870E+00  \\ 
Na     & 1.103E+00 & 1.398E+00 & 1.688E+00  & 1.562E+00 & 1.730E+00 & 1.264E+00 & 1.395E+00 & 1.879E+00 & 2.750E+00 & 4.410E+00  \\ 
Mg     & 1.008E+00 & 1.080E+00 & 1.221E+00  & 1.442E+00 & 1.196E+00 & 1.000E+00 & 1.000E+00 & 1.000E+00 & 1.001E+00 & 1.094E+00  \\
\hline
\end{tabular}
\label{tab:element_prodfac_set1.2_winds}
\end{center}
\end{table}
\end{landscape}

%H this is a test
%H to write strings in the first col
%\addtolength{\voffset}{-0.0cm}
\begin{table}
\begin{center}
%\tabletypesize{\small}
\caption{The presupernova elemental overproduction factors are given for massive stars of \setopt.
The complete table is available online.}
\begin{tabular}{cccccc}
\hline
elements & 15 $\msun$   & 20 $\msun$   & 25 $\msun$   & 32 $\msun$   & 60 $\msun$    \\ 
\hline
%H      & 6.054E-01 & 4.869E-01 & 3.800E-01 & 1.297E-11 & 2.560E-15  \\ 
%He     & 1.498E+00 & 1.600E+00 & 1.980E+00 & 2.094E+00 & 1.095E-01  \\ 
%Li     & 4.542E-02 & 8.075E-03 & 3.465E-01 & 1.309E-14 & 1.780E-89  \\ 
%Be     & 0.000E+00 & 0.000E+00 & 0.000E+00 & 0.000E+00 & 0.000E+00  \\ 
%B      & 1.003E+00 & 1.001E+00 & 1.001E+00 & 1.000E+00 & 1.000E+00  \\ 
C      & 7.413E+00 & 9.831E+00 & 1.753E+01 & 4.897E+01 & 3.879E+01  \\ 
N      & 3.977E+00 & 4.724E+00 & 6.122E+00 & 1.093E+00 & 3.292E-04  \\ 
O      & 8.473E+00 & 1.319E+01 & 9.781E+00 & 2.317E+01 & 5.889E+01  \\ 
F      & 7.109E-01 & 8.894E-01 & 2.174E+00 & 2.033E+01 & 6.660E-03  \\ 
Ne     & 1.323E+01 & 1.020E+01 & 1.067E+01 & 1.153E+01 & 1.013E+02  \\ 
Na     & 9.258E+00 & 5.741E+00 & 1.517E+01 & 7.987E+00 & 8.901E+01  \\ 
Mg     & 1.465E+01 & 2.542E+01 & 5.196E+00 & 8.657E+00 & 7.074E+01  \\
\hline
\end{tabular}
\label{tab:element_prodfac_set1.2_pre}
\end{center}
\end{table}

%H this is a test
%H to write strings in the first col
%%%%%%%%%%%%%%%%%%%%%%%%%%%%%%%%%%%%%%%%%%%%%%
%	TABle element abundance exp
%%%%%%%%%%%%%%%%%%%%%%%%%%%%%%%%%%%%%%%%%%%%%%

%\addtolength{\voffset}{-0.0cm}
\begin{landscape}
\begin{table}
\begin{center}
%\tabletypesize{\small}
\caption{The elemental overproduction factors in supernova ejecta is given for the stars of \setopt.
The complete table is available online.
}
\begin{tabular}{ccccccccc}
\hline
 elements & 15 $\msun$ delay & 15 $\msun$ rapid & 20 $\msun$ delay & 20 $\msun$ rapid & 25 $\msun$ delay & 32 $\msun$ delay & 32 $\msun$ rapid & 60 $\msun$ delay  \\
\hline 
 % H	  & 5.752E-01	  & 5.672E-01	  & 4.566E-01	  & 4.582E-01	  & 3.462E-01	  & 4.148E-11	  & 3.605E-09	  & 3.839E-14	  \\ 
 % He	  & 1.582E+00	  & 1.572E+00	  & 1.514E+00	  & 1.519E+00	  & 1.877E+00	  & 1.531E+00	  & 8.177E-01	  & 2.020E-02	  \\ 
 % Li	  & 4.322E-02	  & 4.262E-02	  & 8.114E-03	  & 8.114E-03	  & 4.105E+00	  & 7.070E-03	  & 2.408E-03	  & 2.940E-07	  \\ 
 % Be	  & 0.000E+00	  & 0.000E+00	  & 0.000E+00	  & 0.000E+00	  & 0.000E+00	  & 0.000E+00	  & 0.000E+00	  & 0.000E+00	  \\ 
 % B	  & 7.270E-01	  & 7.191E-01	  & 6.135E-01	  & 6.066E-01	  & 7.632E-01	  & 1.969E-02	  & 4.413E-02	  & 2.513E-01	  \\ 
   C	  & 4.644E+00	  & 4.639E+00	  & 6.704E+00	  & 6.862E+00	  & 1.569E+01	  & 3.045E+01	  & 2.168E+01	  & 3.446E+01	  \\ 
   N	  & 3.800E+00	  & 3.749E+00	  & 4.436E+00	  & 4.452E+00	  & 5.655E+00	  & 2.504E-01	  & 1.332E-01	  & 1.357E-03	  \\ 
   O	  & 2.729E+00	  & 2.714E+00	  & 1.116E+01	  & 1.075E+01	  & 9.409E+00	  & 1.903E+01	  & 4.739E+01	  & 5.470E+01	  \\ 
   F	  & 2.313E+00	  & 2.219E+00	  & 7.752E-01	  & 7.775E-01	  & 2.111E+00	  & 2.162E+00	  & 1.641E+00	  & 1.333E-02	  \\ 
   Ne	  & 2.077E+00	  & 2.051E+00	  & 4.834E+00	  & 3.869E+00	  & 1.266E+01	  & 1.117E+01	  & 4.042E+01	  & 7.747E+01	  \\ 
   Na	  & 2.739E+00	  & 2.703E+00	  & 4.539E+00	  & 4.191E+00	  & 1.473E+01	  & 6.252E+00	  & 1.513E+01	  & 6.026E+01	  \\ 
   Mg	  & 2.617E+00	  & 2.513E+00	  & 1.740E+01	  & 1.606E+01	  & 8.804E+00	  & 3.146E+01	  & 7.237E+01	  & 9.227E+01	  \\
\hline
\end{tabular}
\label{tab:element_prodfac_set1.2_exp}
\end{center}
\end{table}
\end{landscape}

%H this is a test
%H to write strings in the first col
%%%%%%%%%%%%%%%%%%%%%%%%%%%%%%%%%%%%%%%%%%%%%%
%	TABle isotopic abundance winds
%%%%%%%%%%%%%%%%%%%%%%%%%%%%%%%%%%%%%%%%%%%%%%

%\addtolength{\voffset}{-0.0cm}
\begin{landscape}
\begin{table}
\begin{center}
\tabletypesize{\small}
\caption{The overproduction factors of stable isotopes in stellar winds is given for the stars of \setopo.
The complete table is available online.}
\begin{tabular}{ccccccccc}
\hline
 isotopes & 1.65 $\msun$ & 2 $\msun$    & 3 $\msun$  & 4 $\msun$   & 5 $\msun$    & 15 $\msun$   & 20 $\msun$   & 25 $\msun$    \\ 
\hline
%H   1    & 9.684E-01 & 9.437E-01 & 9.361E-01 & 9.290E-01 & 8.880E-01 & 9.936E-01 & 9.802E-01 & 9.331E-01  \\																							
%H   2    & 3.072E-01 & 2.476E-01 & 1.734E-01 & 1.912E-02 & 1.427E-02 & 4.784E-04 & 1.519E-01 & 2.347E-04   \\ 
%HE  3    & 8.109E+00 & 6.297E+00 & 3.209E+00 & 2.052E+00 & 1.241E+00 & 7.667E-01 & 1.067E+00 & 6.500E-01   \\ 
%HE  4    & 1.070E+00 & 1.086E+00 & 1.106E+00 & 1.169E+00 & 1.298E+00 & 1.018E+00 & 1.055E+00 & 1.187E+00   \\ 
%LI  7    & 3.318E-01 & 2.712E-01 & 1.874E-01 & 2.697E-01 & 4.196E-01 & 4.786E-04 & 1.519E-01 & 2.369E-04   \\ 
%B  11    & 1.000E+00 & 1.000E+00 & 9.987E-01 & 1.000E+00 & 1.000E+00 & 1.000E+00 & 1.000E+00 & 1.000E+00   \\ 
 C  12    & 2.613E+00 & 7.766E+00 & 7.423E+00 & 3.646E+00 & 1.948E+00 & 8.411E-01 & 8.580E-01 & 7.541E-01   \\ 
%C  13    & 2.437E+00 & 2.340E+00 & 2.332E+00 & 2.924E+00 & 4.582E+00 & 3.223E+00 & 2.784E+00 & 2.685E+00   \\ 
 N  14    & 2.287E+00 & 2.623E+00 & 3.077E+00 & 3.493E+00 & 4.617E+00 & 1.650E+00 & 1.837E+00 & 3.046E+00   \\ 
%N  15    & 4.989E-01 & 4.663E-01 & 4.340E-01 & 3.816E-01 & 2.557E-01 & 5.981E-01 & 6.329E-01 & 5.340E-01   \\ 
 O  16    & 1.190E+00 & 2.074E+00 & 1.975E+00 & 1.301E+00 & 1.013E+00 & 9.821E-01 & 9.530E-01 & 8.511E-01   \\ 
%O  17    & 4.385E+00 & 8.867E+00 & 5.857E+00 & 4.218E+00 & 3.541E+00 & 1.167E+00 & 1.065E+00 & 1.028E+00   \\ 
%O  18    & 7.430E-01 & 6.988E-01 & 6.685E-01 & 6.450E-01 & 4.898E-01 & 9.115E-01 & 9.031E-01 & 7.860E-01   \\ 
%F  19    & 1.419E+00 & 4.446E+00 & 5.102E+00 & 1.064E+00 & 8.237E-01 & 9.809E-01 & 9.484E-01 & 8.290E-01   \\ 
 NE 20    & 9.995E-01 & 1.008E+00 & 1.017E+00 & 1.041E+00 & 1.025E+00 & 9.997E-01 & 9.989E-01 & 9.959E-01   \\
\hline
\end{tabular}
\label{tab:isotopic_prodfac_set1.1_winds}
\end{center}
\end{table}
\end{landscape}

%H this is a test
%H to write strings in the first col
%\addtolength{\voffset}{-0.0cm}
\begin{table}
\begin{center}
%\tabletypesize{\small}
\caption{The presupernova overproduction factors of stable isotopes are given for massive stars of \setopo.
The complete table is available online.}
\begin{tabular}{cccc}
\hline
isotopes & 15 $\msun$   & 20 $\msun$   & 25 $\msun$    \\ 
\hline
%H   1  & 6.145E-01 & 4.826E-01 & 3.800E-01  \\ 										 
%H   2  & 6.707E-09 & 1.043E-03 & 6.289E-10  \\   
%HE  3  & 2.862E-01 & 1.863E-01 & 1.490E-02  \\   
%HE  4  & 1.557E+00 & 1.620E+00 & 2.208E+00  \\   
%LI  7  & 9.985E-07 & 1.213E-03 & 8.670E-01  \\   
%B  11  & 1.001E+00 & 1.004E+00 & 1.000E+00  \\   
 C  12  & 1.323E+01 & 2.297E+01 & 3.048E+01  \\   
%C  13  & 1.554E+00 & 1.573E+00 & 5.066E+01  \\   
 N  14  & 4.595E+00 & 4.786E+00 & 7.325E+00  \\   
%N  15  & 2.376E-01 & 1.548E-01 & 4.922E-02  \\   
 O  16  & 1.553E+01 & 2.743E+01 & 1.650E+01  \\   
%O  17  & 1.616E+00 & 1.363E+00 & 1.603E+00  \\   
%O  18  & 1.978E+01 & 6.676E-01 & 4.572E-02  \\   
%F  19  & 5.182E-01 & 1.731E+00 & 1.913E+00  \\   
 NE 20  & 2.194E+01 & 4.006E+01 & 4.366E+00  \\   
%NE 21  & 6.017E+00 & 9.144E+00 & 3.448E+00  \\  
%NE 22  & 1.332E+01 & 2.890E+01 & 4.379E+01  \\   
%NA 23  & 7.803E+00 & 1.865E+01 & 9.769E+00  \\   
 MG 24  & 3.538E+01 & 3.351E+01 & 1.859E+00  \\   
%MG 25  & 2.371E+01 & 2.984E+01 & 6.082E+00  \\   
%MG 26  & 1.879E+01 & 3.209E+01 & 1.460E+01  \\ 
%AL 27  & 9.858E+00 & 9.358E+00 & 1.644E+00  \\ 
 SI 28  & 1.930E+01 & 4.990E+00 & 1.033E+00  \\ 
%SI 29  & 4.711E+01 & 2.646E+01 & 1.301E+00  \\ 
%SI 30  & 1.456E+02 & 9.565E+00 & 1.796E+00  \\ 
%P  31  & 3.296E+01 & 7.382E+00 & 1.595E+00  \\ 
 S  32  & 1.887E+00 & 1.050E+00 & 9.545E-01  \\
\hline
\end{tabular}
\label{tab:isotopic_prodfac_set1.1_pre}
\end{center}
\end{table}

%H this is a test
%H to write strings in the first col
%%%%%%%%%%%%%%%%%%%%%%%%%%%%%%%%%%%%%%%%%%%%%%
%  TABle isotopic abundance exp
%%%%%%%%%%%%%%%%%%%%%%%%%%%%%%%%%%%%%%%%%%%%%%

%\addtolength{\voffset}{-0.0cm}
\begin{table}
\begin{center}
%\tabletypesize{\small}
\caption{The overproduction factors of stable isotopes in supernova ejecta is given for the stars of \setopo.
The complete table is available online.}
\begin{tabular}{ccccccc}
\hline
 isotopes & 15 $\msun$ delay & 15 $\msun$ rapid & 20 $\msun$ delay & 20 $\msun$ rapid & 25 $\msun$ delay  & 25 $\msun$ rapid \\
\hline
% H   1  & 5.837E-01   & 5.832E-01   & 4.843E-01   & 4.836E-01   & 3.468E-01   & 3.443E-01   \\ 
% H   2  & 1.983E-02   & 1.959E-02   & 1.047E-03   & 1.046E-03   & 7.212E-04   & 5.709E-10   \\ 
% HE  3  & 2.717E-01   & 2.715E-01   & 1.870E-01   & 1.867E-01   & 1.350E-02   & 1.347E-02   \\ 
% HE  4  & 1.637E+00   & 1.647E+00   & 1.600E+00   & 1.651E+00   & 2.087E+00   & 1.737E+00   \\ 
% LI  7  & 3.156E-04   & 2.428E-04   & 1.471E-03   & 7.370E+01   & 6.673E+01   & 5.850E-01   \\ 
% B  11  & 6.984E-01   & 6.989E-01   & 6.627E-01   & 6.604E-01   & 7.400E-01   & 6.793E-01   \\ 
  C  12  & 8.389E+00   & 8.588E+00   & 1.843E+01   & 1.945E+01   & 2.458E+01   & 4.680E+00   \\ 
% C  13  & 1.466E+00   & 1.467E+00   & 1.580E+00   & 1.578E+00   & 5.716E+01   & 3.195E+01   \\ 
  N  14  & 4.398E+00   & 4.396E+00   & 4.807E+00   & 4.803E+00   & 6.757E+00   & 6.340E+00   \\ 
% N  15  & 1.184E+00   & 1.197E+00   & 1.405E-01   & 1.529E-01   & 2.672E+01   & 4.318E-02   \\ 
  O  16  & 3.989E+00   & 3.999E+00   & 2.583E+01   & 2.441E+01   & 1.499E+01   & 3.246E-01   \\ 
% O  17  & 1.354E+00   & 1.356E+00   & 1.405E+00   & 1.552E+00   & 1.194E+00   & 7.071E-01   \\ 
% O  18  & 5.750E+00   & 6.309E+00   & 4.429E-01   & 4.914E-01   & 5.877E-02   & 3.994E-02   \\ 
% F  19  & 2.322E+00   & 2.216E+00   & 9.823E-01   & 1.057E+00   & 1.811E+00   & 1.255E+00   \\ 
  NE 20  & 1.900E+00   & 1.918E+00   & 1.094E+01   & 6.742E+00   & 1.304E+01   & 6.958E-01   \\ 
% NE 21  & 2.467E+01   & 2.415E+01   & 9.040E+00   & 8.197E+00   & 5.622E+00   & 1.265E+00   \\ 
% NE 22  & 4.841E+00   & 5.047E+00   & 1.199E+01   & 1.330E+01   & 3.531E+01   & 2.284E+01   \\ 
% NA 23  & 3.215E+00   & 3.186E+00   & 6.223E+00   & 4.977E+00   & 9.820E+00   & 4.308E+00   \\ 
  MG 24  & 7.473E+00   & 7.092E+00   & 3.864E+01   & 3.196E+01   & 2.351E+01   & 7.256E-01   \\
\hline
\end{tabular}
\label{tab:isotopic_prodfac_set1.1_exp}
\end{center}
\end{table}

%H this is a test
%H to write strings in the first col
%%%%%%%%%%%%%%%%%%%%%%%%%%%%%%%%%%%%%%%%%%%%%%
%	TABle element abundance winds
%%%%%%%%%%%%%%%%%%%%%%%%%%%%%%%%%%%%%%%%%%%%%%

%\addtolength{\voffset}{-0.0cm}
\begin{landscape}
\begin{table}
\begin{center}
\tabletypesize{\small}
\caption{The element overproduction factors in stellar winds is given for the stars of \setopo.
The complete table is available online.}
\begin{tabular}{ccccccccc}
\hline
 elements & 1.65 $\msun$ & 2 $\msun$	& 3 $\msun$  & 4 $\msun$  & 5 $\msun$	& 15 $\msun$   & 20 $\msun$	& 25 $\msun$  \\ 
\hline
 %H	& 9.684E-01 & 9.437E-01  & 9.361E-01 & 9.290E-01 & 8.880E-01 & 9.936E-01 & 9.802E-01 & 9.331E-01\\ 
 %He	& 1.071E+00 & 1.087E+00  & 1.106E+00 & 1.169E+00 & 1.298E+00 & 1.018E+00 & 1.055E+00 & 1.187E+00\\ 
 %Li	& 3.318E-01 & 2.712E-01  & 1.874E-01 & 2.697E-01 & 4.196E-01 & 4.786E-04 & 1.519E-01 & 2.369E-04\\ 
 %Be	& 0.000E+00 & 0.000E+00  & 0.000E+00 & 0.000E+00 & 0.000E+00 & 0.000E+00 & 0.000E+00 & 0.000E+00\\ 
 %B	& 1.000E+00 & 1.000E+00  & 9.987E-01 & 1.000E+00 & 1.000E+00 & 1.000E+00 & 1.000E+00 & 1.000E+00\\ 
 C	& 2.611E+00 & 7.701E+00  & 7.362E+00 & 3.637E+00 & 1.979E+00 & 8.697E-01 & 8.811E-01 & 7.772E-01\\ 
 N	& 2.280E+00 & 2.615E+00  & 3.067E+00 & 3.481E+00 & 4.600E+00 & 1.646E+00 & 1.832E+00 & 3.036E+00\\ 
 O	& 1.190E+00 & 2.074E+00  & 1.974E+00 & 1.300E+00 & 1.013E+00 & 9.820E-01 & 9.529E-01 & 8.510E-01\\ 
 F	& 1.419E+00 & 4.446E+00  & 5.102E+00 & 1.064E+00 & 8.237E-01 & 9.809E-01 & 9.484E-01 & 8.290E-01\\ 
 Ne	& 1.088E+00 & 1.591E+00  & 1.566E+00 & 1.131E+00 & 1.046E+00 & 9.983E-01 & 9.957E-01 & 9.854E-01\\ 
 Na	& 1.155E+00 & 1.505E+00  & 1.624E+00 & 1.641E+00 & 1.809E+00 & 1.089E+00 & 1.227E+00 & 1.766E+00\\ 
 Mg	& 1.015E+00 & 1.118E+00  & 1.247E+00 & 1.304E+00 & 1.160E+00 & 1.000E+00 & 1.000E+00 & 1.000E+00\\
\hline
\end{tabular}
\label{tab:element_prodfac_set1.1_winds}
\end{center}
\end{table}
\end{landscape}

%H this is a test
%H to write strings in the first col
\begin{table}
\begin{center}
%\tabletypesize{\small}
\caption{The presupernova overproduction factors are given for massive stars of \setopo.
The complete table is available online.}
\begin{tabular}{cccc}
\hline
 elements & 15 $\msun$   & 20 $\msun$	& 25 $\msun$    \\ 
\hline
 %H	& 6.145E-01 & 4.826E-01  & 3.800E-01  \\ 
 %He	& 1.557E+00 & 1.619E+00  & 2.208E+00  \\ 
 %Li	& 9.985E-07 & 1.213E-03  & 8.670E-01  \\ 
 %Be	& 0.000E+00 & 0.000E+00  & 0.000E+00  \\ 
 %B	& 1.001E+00 & 1.004E+00  & 1.000E+00  \\ 
 C	& 1.309E+01 & 2.271E+01  & 3.073E+01  \\ 
 N	& 4.577E+00 & 4.768E+00  & 7.296E+00  \\ 
 O	& 1.553E+01 & 2.736E+01  & 1.646E+01  \\ 
 F	& 5.182E-01 & 1.731E+00  & 1.913E+00  \\ 
 Ne	& 2.126E+01 & 3.915E+01  & 7.307E+00  \\ 
 Na	& 7.803E+00 & 1.865E+01  & 9.769E+00  \\ 
 Mg	& 3.223E+01 & 3.297E+01  & 3.792E+00  \\
\hline
\end{tabular}
\label{tab:element_prodfac_set1.1_pre}
\end{center}
\end{table}

%H this is a test
%H to write strings in the first col
%%%%%%%%%%%%%%%%%%%%%%%%%%%%%%%%%%%%%%%%%%%%%%
% TABle element abundance exp
%%%%%%%%%%%%%%%%%%%%%%%%%%%%%%%%%%%%%%%%%%%%%%

%\addtolength{\voffset}{-0.0cm}
\begin{table}
\begin{center}
%\tabletypesize{\small}
\caption{The elemental overproduction factors in supernova ejecta is given for the stars of \setopo .
The complete table is available online.}
\begin{tabular}{ccccccc}
\hline
 elements & 15 $\msun$ delay & 15 $\msun$ rapid & 20 $\msun$ delay & 20 $\msun$ rapid & 25 $\msun$ delay & 25 $\msun$ rapid \\ 
\hline
 % H   & 5.837E-01   & 5.741E-01   & 4.843E-01   & 4.481E-01   & 3.468E-01   & 4.456E-01   \\ 
 % He   & 1.637E+00   & 1.621E+00   & 1.600E+00   & 1.530E+00   & 2.087E+00   & 2.248E+00   \\ 
 % Li   & 3.156E-04   & 2.390E-04   & 1.471E-03   & 6.830E+01   & 6.673E+01   & 7.571E-01   \\ 
 % Be   & 0.000E+00   & 0.000E+00   & 0.000E+00   & 0.000E+00   & 0.000E+00   & 0.000E+00   \\ 
 % B   & 6.984E-01   & 6.880E-01   & 6.627E-01   & 6.120E-01   & 7.400E-01   & 8.792E-01   \\ 
   C   & 8.306E+00   & 8.370E+00   & 1.823E+01   & 1.783E+01   & 2.497E+01   & 6.480E+00   \\ 
   N   & 4.386E+00   & 4.316E+00   & 4.789E+00   & 4.434E+00   & 6.835E+00   & 8.174E+00   \\ 
   O   & 3.992E+00   & 3.941E+00   & 2.576E+01   & 2.256E+01   & 1.496E+01   & 4.194E-01   \\ 
   F   & 2.322E+00   & 2.181E+00   & 9.823E-01   & 9.800E-01   & 1.811E+00   & 1.624E+00   \\ 
   Ne   & 2.172E+00   & 2.169E+00   & 1.101E+01   & 6.705E+00   & 1.468E+01   & 3.042E+00   \\ 
   Na   & 3.215E+00   & 3.137E+00   & 6.223E+00   & 4.613E+00   & 9.820E+00   & 5.576E+00   \\ 
   Mg   & 6.772E+00   & 6.359E+00   & 3.285E+01   & 2.507E+01   & 2.020E+01   & 9.486E-01   \\
\hline
\end{tabular}
\label{tab:element_prodfac_set1.1_exp}
\end{center}
\end{table}

\clearpage

%%%%%%%%%%%%%%%%%%%%%%%%%%%%%%%%%%%%%%%%%%%%%%
%	TABle isotopic abundance winds
%%%%%%%%%%%%%%%%%%%%%%%%%%%%%%%%%%%%%%%%%%%%%%

%\addtolength{\voffset}{-0.0cm}
\begin{table}
\begin{center}
%\tabletypesize{\small}
\captionof{table}{Comparison between the present work, \cite{cristallo:11} (Cr11) and 
\cite{karakas:10a} (Ka10) for the $2\msun$ stellar yields, \setopt.}
\begin{tabular}{cccc}
\hline
Isotope & Pi13 & Cr11 & Ka10 \\
\hline 
% H   1  &  9.267E-01 & 9.24950E-01 & 1.2335814E+00 \\
% HE  4  &  4.128E-01 & 3.95700E-01 & 5.6186771E-01 \\
 C  12  &  1.912E-02 & 9.33420E-03 & 2.7675023E-03 \\
 C  13  &  1.365E-04 & 1.31110E-04 & 1.2363438E-04  \\
 N  14  &  3.715E-03 & 3.26650E-03 & 3.0212691E-03 \\
 N  15  &  2.738E-06 & 2.80090E-06 & 3.0777028E-06 \\
 O  16  &  1.998E-02 & 1.18310E-02 & 1.3041234E-02 \\
 O  17  &  4.684E-05 & 4.03320E-05 & 2.4374334E-05 \\
 O  18  &  3.327E-05 & 1.91220E-05 & 2.2738323E-05 \\
\hline
\end{tabular}
\label{tab:isotopic_yieldscomp_m2z2m2}                                                          
\end{center}
\end{table}

%%%%%%%%%%%%%%%%%%%%%%%%%%%%%%%%%%%%%%%%%%%%%%
%	TABle isotopic abundance winds
%%%%%%%%%%%%%%%%%%%%%%%%%%%%%%%%%%%%%%%%%%%%%%

%\addtolength{\voffset}{-0.0cm}
\begin{table}
\begin{center}
%\tabletypesize{\small}
\captionof{table}{Comparison between the present work, \cite{cristallo:11} (Cr11) and 
\cite{karakas:10a} (Ka10) for the $3\msun$ stellar yields, \setopt.}
\begin{tabular}{cccc}
\hline
Isotope & Pi13 & Cr11 & Ka10 \\
\hline 
% H   1 & 1.519E+00 & 1.57880E+00 & 1.5249571E+00 \\ 
% HE  4 & 7.493E-01 & 7.04700E-01 & 7.3096633E-01 \\ 
 C  12 & 4.448E-02 & 1.86110E-02 & 2.0739544E-02 \\ 
 C  13 & 2.252E-04 & 2.20200E-04 & 1.9436399E-04 \\ 
 N  14 & 7.685E-03 & 6.64840E-03 & 5.6565693E-03 \\ 
 N  15 & 4.207E-06 & 4.29400E-06 & 5.0818235E-06 \\ 
 O  16 & 3.828E-02 & 1.94360E-02 & 2.1144016E-02 \\ 
 O  17 & 5.194E-05 & 7.91850E-05 & 5.5763638E-05 \\ 
 O  18 & 3.364E-05 & 3.12110E-05 & 3.6596495E-05 \\
\hline
\end{tabular}
\label{tab:isotopic_yieldscomp_m3z2m2}                                                          
\end{center}
\end{table}

%%%%%%%%%%%%%%%%%%%%%%%%%%%%%%%%%%%%%%%%%%%%%%
%	TABle isotopic abundance winds
%%%%%%%%%%%%%%%%%%%%%%%%%%%%%%%%%%%%%%%%%%%%%%

%\addtolength{\voffset}{-0.0cm}
\begin{table}
\begin{center}
%\tabletypesize{\small}
\captionof{table}{Comparison between the present work and \cite{karakas:10a} (Ka10) for the $5\msun$ stellar yields, \setopt.}
\begin{tabular}{ccc}
\hline
Isotope & Pi13 & Ka10 \\
\hline 
% H   1    & 2.578E+00 & 2.6526232E+00 \\ 
% HE  4    & 1.443E+00 & 1.3753262E+00 \\ 
 C  12    & 2.478E-02 & 1.5313132E-02 \\ 
 C  13    & 5.341E-04 & 2.5495309E-03 \\ 
 N  14    & 1.702E-02 & 1.6426099E-02 \\ 
 N  15    & 5.737E-06 & 3.5156333E-07 \\ 
 O  16    & 4.031E-02 & 3.5938345E-02 \\ 
 O  17    & 6.271E-05 & 5.4988403E-05 \\ 
 O  18    & 5.326E-05 & 8.8449030E-07 \\ 
 F  19    & 2.085E-06 & 3.6023453E-06 \\ 
 NE 20    & 7.695E-03 & 6.6723512E-03 \\ 
 NE 21    & 3.187E-05 & 2.2214339E-05 \\ 
 NE 22    & 9.583E-04 & 1.4162241E-03 \\ 
 NA 23    & 2.921E-04 & 2.2751275E-04 \\ 
 MG 24    & 2.415E-03 & 2.1024081E-03 \\ 
 MG 25    & 5.101E-04 & 3.3637485E-04 \\ 
 MG 26    & 9.401E-04 & 4.3323121E-04 \\ 
 AL 27    & 2.887E-04 & 2.5326770E-04 \\ 
 SI 28    & 3.085E-03 & 2.6992788E-03 \\ 
 SI 29    & 1.650E-04 & 1.4311528E-04 \\ 
 SI 30    & 1.196E-04 & 1.0082583E-04 \\ 
 P  31    & 3.162E-05 & 3.4187142E-05 \\ 
 S  33    & 1.366E-05 & 1.3463593E-05 \\ 
 S  34    & 7.941E-05 & 7.7136338E-04 \\ 
 FE 54    & 3.319E-04 & 2.9173499E-04 \\ 
 FE 56    & 5.418E-03 & 4.8021809E-03 \\ 
 FE 57    & 1.323E-04 & 1.2381890E-04 \\ 
 FE 58    & 3.155E-05 & 2.9148772E-05 \\ 
 CO 59    & 2.109E-05 & 1.7460188E-05 \\ 
 NI 58    & 2.334E-04 & 2.0222510E-04 \\ 
 NI 60    & 9.600E-05 & 8.1889491E-05 \\ 
 NI 61    & 5.639E-06 & 1.9852198E-05 \\ 
 NI 62    & 1.722E-05 & 1.3339656E-08 \\ 
\hline
 \noalign{\smallskip}
\hline
\end{tabular}
\label{tab:isotopic_yieldscomp_m5z2m2}                                                          
\end{center}
\end{table}

%%%%%%%%%%%%%%%%%%%%%%%%%%%%%%%%%%%%%%%%%%%%%%
%	TABle isotopic abundance exp
%%%%%%%%%%%%%%%%%%%%%%%%%%%%%%%%%%%%%%%%%%%%%%

%\addtolength{\voffset}{-0.0cm}
\begin{table}
\begin{center}
%\tabletypesize{\small}
\caption{For stable species the abundance enrichment in supernova ejecta is given in solar mass unit for the stars at solar metallicity. Here we compare results from our $15\msun$ models, \setopt, with \cite{thielemann:96} (Th96), \cite{rauscher:02} (Ra02) and \cite{chieffi:04} (CL04).}                                        
\begin{tabular}{cccccc}
\hline 
 specie & 15 $\msun$ $delay$ & 15 $\msun$ $rapid$ & Th96 & Ra02 & CL04 \\ 
\hline 
% H   1  & 6.262E+00	 & 6.267E+00	 &	     & 7.136E+00 & 7.77E+00\\ 
% H   2  & 6.226E-06	 & 6.233E-06	 &  3.03E-26 & 2.735E-07 & 5.95E-17\\ 
% HE  3  & 3.221E-04	 & 3.223E-04	 &  8.40E-20 & 5.695E-04 & 3.01E-05\\ 
% HE  4  & 5.385E+00	 & 5.425E+00	 &  1.83E+00 & 4.684E+00 & 4.65E+00\\ 
% LI  7  & 3.462E-11	 & 3.466E-11	 &	     & 1.820E-07 & 1.19E-10\\ 
% B  11  & 3.006E-08	 & 3.015E-08	 &  2.03E-15 & 6.905E-07 & 1.55E-15\\ 
  C  12  & 1.761E-01	 & 1.785E-01	 &  8.33E-02 & 1.555E-01 & 1.39E-01\\ 
  C  13  & 9.805E-04	 & 9.813E-04	 &  4.98E-10 & 1.264E-03 & 5.10E-10\\ 
  N  14  & 4.967E-02	 & 4.973E-02	 &  5.37E-03 & 4.662E-02 & 2.95E-07\\ 
  N  15  & 5.308E-05	 & 5.457E-05	 &  1.58E-10 & 1.775E-04 & 9.77E-10\\ 
  O  16  & 2.986E-01	 & 3.011E-01	 &  4.23E-01 & 8.495E-01 & 3.46E-01\\ 
  O  17  & 7.713E-05	 & 7.736E-05	 &  5.08E-09 & 9.941E-05 & 2.14E-08\\ 
  O  18  & 4.882E-03	 & 5.004E-03	 &  1.35E-02 & 3.304E-03 & 8.80E-09\\ 
  F  19  & 1.518E-05	 & 1.483E-05	 &  2.67E-11 & 2.989E-05 & 7.00E-11\\ 
  NE 20  & 3.151E-02	 & 3.141E-02	 &  2.83E-02 & 1.267E-01 & 1.15E-01\\ 
% NE 21  & 1.354E-03	 & 1.326E-03	 &  4.53E-05 & 8.376E-04 & 3.31E-06\\ 
% NE 22  & 1.583E-02	 & 1.607E-02	 &  1.26E-02 & 1.089E-02 & 1.46E-06\\ 
  NA 23  & 1.299E-03	 & 1.301E-03	 &  2.09E-04 & 2.625E-03 & 5.79E-04\\ 
  MG 24  & 1.548E-02	 & 1.505E-02	 &  4.20E-02 & 3.999E-02 & 4.98E-02\\ 
% MG 25  & 2.545E-03	 & 2.510E-03	 &  3.46E-03 & 8.809E-03 & 8.97E-05\\ 
% MG 26  & 4.862E-03	 & 4.809E-03	 &  2.52E-03 & 8.469E-03 & 1.96E-04\\ 
  AL 27  & 1.259E-03	 & 1.264E-03	 &  5.56E-03 & 4.682E-03 & 8.66E-04\\ 
  SI 28  & 9.677E-02	 & 8.910E-02	 &  6.52E-02 & 9.684E-02 & 5.30E-02\\ 
% SI 29  & 3.696E-03	 & 3.633E-03	 &  4.40E-03 & 3.166E-03 & 2.57E-04\\ 
% SI 30  & 5.219E-03	 & 4.963E-03	 &  4.91E-03 & 4.504E-03 & 1.84E-04\\ 
% P  31  & 1.822E-03	 & 1.797E-03	 &  8.67E-04 & 1.096E-03 & 9.72E-05\\ 
  S  32  & 6.575E-02	 & 6.385E-02	 &  2.16E-02 & 4.165E-02 & 2.15E-02\\ 
% S  33  & 5.676E-04	 & 5.286E-04	 &  9.31E-05 & 2.084E-04 & 5.04E-05\\ 
% S  34  & 7.210E-03	 & 4.272E-03	 &  1.03E-03 & 2.211E-03 & 1.94E-04\\ 
% S  36  & 2.926E-05	 & 2.947E-05	 &  5.38E-07 & 4.898E-06 & 2.04E-09\\ 
% CL 35  & 1.875E-03	 & 1.603E-03	 &  5.48E-05 & 1.509E-04 & 2.61E-05\\ 
% CL 37  & 1.419E-04	 & 1.407E-04	 &  5.62E-06 & 6.046E-05 & 2.66E-06\\ 
  AR 36  & 2.651E-02	 & 2.651E-02	 &  3.49E-03 & 7.403E-03 & 4.12E-03\\ 
% AR 38  & 3.386E-03	 & 2.765E-03	 &  3.26E-04 & 1.016E-03 & 4.52E-05\\ 
% AR 40  & 1.011E-05	 & 9.999E-06	 &  4.89E-09 & 3.400E-06 & 5.25E-11\\ 
% K  39  & 7.456E-04	 & 6.952E-04	 &  1.71E-06 & 1.256E-04 & 2.13E-05\\ 
% K  40  & 2.897E-06	 & 2.909E-06	 &	     & 7.592E-07 & 2.58E-09\\ 
% K  41  & 3.320E-05	 & 3.456E-05	 &  1.25E-06 & 8.807E-06 & 7.79E-07\\ 
  CA 40  & 1.971E-02	 & 1.967E-02	 &  3.03E-03 & 6.284E-03 & 3.83E-03\\ 
% CA 42  & 2.751E-04	 & 2.996E-04	 &  6.98E-06 & 3.087E-05 & 1.13E-06\\ 
% CA 43  & 1.926E-05	 & 1.647E-05	 &  1.21E-06 & 2.688E-06 & 3.77E-07\\ 
% CA 44  & 2.132E-04	 & 1.673E-04	 &  7.19E-05 & 3.421E-05 & 5.94E-05\\ 
% CA 46  & 3.026E-06	 & 2.991E-06	 &  5.45E-11 & 5.837E-07 & 5.52E-14\\ 
% CA 48  & 2.946E-06	 & 2.932E-06	 &  6.71E-16 & 1.763E-06 & 4.04E-19\\ 
% SC 45  & 1.644E-05	 & 2.221E-05	 &  4.66E-08 & 2.048E-06 & 5.50E-08\\ 
% TI 46  & 1.231E-04	 & 1.417E-04	 &  2.72E-06 & 1.392E-05 & 1.24E-06\\ 
% TI 47  & 4.996E-05	 & 3.983E-05	 &  4.67E-06 & 5.235E-06 & 9.57E-07\\ 
  TI 48  & 5.717E-04	 & 4.802E-04	 &  1.27E-04 & 1.276E-04 & 1.70E-04\\ 
% TI 49  & 6.317E-05	 & 8.251E-05	 &  4.15E-06 & 9.988E-06 & 2.81E-06\\ 
% TI 50  & 3.785E-06	 & 3.790E-06	 &  4.59E-10 & 4.577E-06 & 1.66E-12\\ 
% V  50  & 1.679E-07	 & 1.972E-07	 &  3.51E-10 & 1.270E-07 & 1.21E-11\\ 
  V  51  & 1.024E-04	 & 9.875E-05	 &  1.01E-05 & 3.146E-05 & 5.32E-06\\ 
% CR 50  & 2.773E-04	 & 3.188E-04	 &  3.85E-05 & 8.179E-05 & 8.49E-0\\ 
  CR 52  & 3.766E-03	 & 3.564E-03	 &  8.24E-04 & 1.597E-03 & 9.22E-04\\ 
% CR 53  & 4.523E-04	 & 4.730E-04	 &  8.92E-05 & 2.007E-04 & 6.52E-05\\ 
% CR 54  & 9.005E-06	 & 9.305E-06	 &  1.55E-08 & 1.520E-05 & 1.23E-10\\ 
  MN 55  & 2.124E-03	 & 2.159E-03	 &  3.39E-04 & 1.271E-03 & 3.02E-04\\ 
% FE 54  & 1.004E-02	 & 9.946E-03	 &  3.51E-03 & 7.118E-03 & 1.06E-03\\ 
  FE 56  & 1.915E-01	 & 1.681E-01	 &  1.30E-01 & 1.261E-01 & 1.00E-01\\ 
% FE 57  & 1.004E-02	 & 7.826E-03	 &  4.64E-03 & 4.211E-03 & 3.43E-03\\ 
% FE 58  & 1.646E-04	 & 1.670E-04	 &  3.62E-09 & 3.939E-04 & 5.51E-11\\ 
  CO 59  & 1.023E-02	 & 1.148E-02	 &  1.36E-04 & 4.542E-04 & 2.14E-04\\ 
  NI 58  & 1.823E-01	 & 1.131E-01	 &  6.64E-03 & 7.326E-03 & 3.03E-03\\ 
% NI 60  & 8.652E-02	 & 2.317E-01	 &  3.13E-03 & 2.539E-03 & 2.87E-03\\ 
% NI 61  & 2.622E-03	 & 3.892E-03	 &  1.46E-04 & 2.148E-04 & 1.53E-04\\ 
% NI 62  & 1.693E-02	 & 2.391E-02	 &  1.00E-03 & 1.405E-03 & 8.48E-04\\ 
% NI 64  & 1.559E-04	 & 1.650E-04	 &  1.73E-15 & 1.169E-04 & 2.16E-16\\ 
% CU 63  & 1.288E-03	 & 3.486E-03	 &  9.56E-15 & 8.870E-05 & 2.17E-06\\ 
% CU 65  & 3.708E-04	 & 1.002E-03	 &  7.69E-07 & 3.667E-05 & 1.79E-06\\ 
% ZN 64  & 2.430E-02	 & 6.339E-02	 &  1.41E-05 & 2.772E-05 & 2.68E-05\\ 
% ZN 66  & 4.254E-03	 & 2.076E-02	 &  1.47E-05 & 5.667E-05 & 2.32E-05\\ 
% ZN 67  & 1.308E-04	 & 3.251E-04	 &  1.94E-08 & 8.346E-06 & 5.47E-08\\ 
% ZN 68  & 9.334E-04	 & 2.259E-03	 &  6.35E-09 & 2.518E-05 & 1.00E-08\\ 
  ZN 70  & 2.820E-06	 & 2.846E-06	 &  3.19E-21 & 2.637E-06 & 1.44E-20\\ 
% GA 69  & 3.396E-05	 & 1.035E-04	 &  6.47E-13 & 4.071E-06 & 1.15E-06\\ 
% GA 71  & 9.149E-06	 & 2.365E-05	 &  2.05E-19 & 2.945E-06 & 1.20E-06\\ 
  GE 70  & 8.356E-04	 & 3.634E-03	 &  5.13E-15 & 3.209E-06 & 6.17E-06\\ 
% GE 72  & 3.334E-05	 & 1.658E-04	 &  5.57E-20 & 3.848E-06 & 2.15E-09\\ 
% GE 73  & 3.145E-06	 & 1.103E-05	 &  7.68E-23 & 1.757E-06 & 1.08E-08\\ 
% GE 74  & 2.467E-06	 & 2.746E-06	 &  1.12E-21 & 4.611E-06 & 2.41E-11\\ 
% GE 76  & 1.000E-06	 & 1.003E-06	 &  4.37E-24 & 1.274E-06 & 3.71E-21\\ 
% AS 75  & 5.123E-06	 & 1.537E-05	 &	     & 1.376E-06 & 1.97E-08\\ 
% SE 74  & 6.317E-05	 & 2.207E-04	 &	     & 1.039E-07 & 1.20E-06\\ 
  SE 76  & 3.013E-05	 & 1.847E-04	 &	     & 8.126E-07 & 7.31E-10\\ 
% SE 77  & 1.797E-06	 & 5.562E-06	 &	     & 8.420E-07 & 2.26E-09\\ 
% SE 78  & 1.287E-06	 & 1.049E-05	 &	     & 1.328E-06 & 1.34E-11\\ 
% SE 80  & 9.905E-07	 & 1.011E-06	 &	     & 2.570E-06 & 1.39E-15\\ 
% SE 82  & 2.836E-07	 & 2.881E-07	 &	     & 8.656E-07 & 1.10E-19\\ 
% BR 79  & 8.562E-07	 & 5.299E-06	 &	     & 7.744E-07 & 1.08E-09\\ 
% BR 81  & 8.827E-07	 & 2.642E-06	 &	     & 6.830E-07 & 1.81E-09\\ 
% KR 78  & 2.331E-06	 & 7.449E-06	 &	     & 9.545E-09 & 1.88E-07\\ 
  KR 80  & 1.555E-05	 & 5.410E-05	 &	     & 1.291E-07 & 5.06E-09\\ 
  KR 82  & 4.350E-06	 & 3.938E-05	 &	     & 4.846E-07 & 1.09E-11\\ 
% KR 83  & 1.086E-06	 & 5.350E-06	 &	     & 5.326E-07 & 1.50E-11\\ 
% KR 84  & 9.514E-07	 & 2.822E-06	 &	     & 1.546E-06 & 9.49E-13\\ 
% KR 86  & 3.492E-07	 & 3.533E-07	 &	     & 1.662E-06 & 5.67E-17\\ 
% RB 85  & 6.672E-07	 & 3.835E-06	 &	     & 7.364E-07 & 4.99E-11\\ 
% RB 87  & 1.035E-07	 & 1.093E-06	 &	     & 2.659E-07 & 3.18E-17\\ 
% SR 84  & 1.291E-06	 & 4.190E-06	 &	     & 5.227E-09 & 8.95E-10\\ 
  SR 86  & 3.747E-06	 & 1.756E-05	 &	     & 1.286E-07 & 3.12E-11\\ 
  SR 87  & 6.093E-07	 & 3.172E-06	 &	     & 6.853E-08 & 1.23E-10\\ 
  SR 88  & 2.648E-06	 & 4.056E-05	 &	     & 1.070E-06 & 1.59E-11\\ 
% Y  89  & 4.141E-06	 & 4.813E-05	 &	     & 2.384E-07 & 1.91E-13\\ 
% ZR 90  & 1.326E-04	 & 8.008E-04	 &	     & 2.460E-07 & 9.60E-11\\ 
% ZR 91  & 2.062E-06	 & 8.317E-06	 &	     & 5.846E-08 & 6.04E-11\\ 
% ZR 92  & 9.813E-08	 & 1.472E-07	 &	     & 8.629E-08 & 9.06E-14\\ 
% ZR 94  & 7.530E-08	 & 7.551E-08	 &	     & 8.407E-08 & 9.74E-14\\ 
% ZR 96  & 3.063E-08	 & 3.048E-08	 &	     & 2.363E-08 & 3.29E-16\\ 
\hline
 \noalign{\smallskip}
\hline
\end{tabular}
\label{tab:isotopic_yieldscomp_m15z2m2}                                                          
\end{center}
\end{table}

%%%%%%%%%%%%%%%%%%%%%%%%%%%%%%%%%%%%%%%%%%%%%%
%	TABle isotopic abundance exp
%%%%%%%%%%%%%%%%%%%%%%%%%%%%%%%%%%%%%%%%%%%%%%

%\addtolength{\voffset}{-0.0cm}
\begin{table}
\begin{center}
%\tabletypesize{\small}
\caption{For stable species the abundance enrichment in supernova ejecta is given in solar mass unit for the stars at solar metallicity. Here we compare results from our $20\msun$ models, \setopt, with \cite{thielemann:96} (Th96), \cite{rauscher:02} (Ra02) and \cite{chieffi:04} (CL04).}                                        
\begin{tabular}{cccccc}
\hline
 specie & 20 $\msun$ $delay$ & 20 $\msun$ $rapid$ & Th96 & Ra02 & CL04 \\ 
\hline 
% H   1  &  7.722E+00	  & 7.725E+00	   &	      & 8.700E+00 & 9.74E+00 \\ 
% H   2  &  1.162E-06	  & 1.163E-06	   & 4.05E-17 & 2.094E-07 & 6.34E-17 \\ 
% HE  3  &  4.203E-04	  & 4.204E-04	   & 3.11E-19 & 6.836E-04 & 2.90E-05 \\ 
% HE  4  &  6.487E+00	  & 6.491E+00	   & 2.10E+00 & 6.229E+00 & 6.26E+00 \\ 
% LI  7  &  6.957E-12	  & 6.937E-12	   &	      & 1.588E-07 & 2.16E-10 \\ 
% B  11  &  3.696E-08	  & 3.669E-08	   & 1.04E-14 & 4.046E-07 & 3.01E-15 \\ 
   C  12  & 2.780E-01	  & 2.825E-01	   & 1.14E-01 & 2.233E-01 & 3.35E-01 \\ 
   C  13  & 1.310E-03	  & 1.310E-03	   & 4.86E-07 & 1.412E-03 & 2.31E-08 \\ 
   N  14  & 6.818E-02	  & 6.822E-02	   & 2.71E-03 & 6.440E-02 & 2.05E-06 \\ 
   N  15  & 4.096E-05	  & 4.028E-05	   & 5.06E-08 & 5.191E-05 & 3.24E-09 \\ 
   O  16  & 1.266E+00	  & 1.211E+00	   & 1.48E+00 & 2.205E+00 & 1.00E+00 \\ 
   O  17  & 7.099E-05	  & 7.114E-05	   & 1.84E-08 & 9.820E-05 & 1.70E-07 \\ 
   O  18  & 5.587E-04	  & 5.624E-04	   & 8.68E-03 & 3.122E-03 & 2.52E-08 \\ 
   F  19  & 7.938E-06	  & 7.938E-06	   & 1.15E-09 & 1.081E-05 & 2.47E-10 \\ 
   NE 20  & 1.034E-01	  & 8.129E-02	   & 2.28E-01 & 6.971E-02 & 2.22E-01 \\ 
% NE 21  &  4.775E-04	  & 4.691E-04	   & 3.11E-04 & 2.965E-04 & 1.46E-05 \\ 
% NE 22  &  1.524E-02	  & 1.532E-02	   & 2.93E-02 & 2.028E-02 & 6.28E-06 \\ 
   NA 23  & 2.383E-03	  & 2.216E-03	   & 1.16E-03 & 2.193E-03 & 1.32E-03 \\ 
   MG 24  & 1.333E-01	  & 1.234E-01	   & 1.46E-01 & 7.260E-02 & 1.02E-01 \\ 
% MG 25  &  7.273E-03	  & 5.784E-03	   & 1.86E-02 & 5.627E-03 & 2.92E-04 \\ 
% MG 26  &  1.151E-02	  & 1.035E-02	   & 1.76E-02 & 8.622E-03 & 4.73E-04 \\ 
   AL 27  & 4.062E-03	  & 3.767E-03	   & 1.59E-02 & 1.205E-02 & 1.50E-03 \\ 
   SI 28  & 3.871E-01	  & 4.039E-01	   & 8.33E-02 & 4.416E-01 & 1.38E-01 \\ 
% SI 29  &  1.981E-02	  & 1.894E-02	   & 1.00E-02 & 1.574E-02 & 7.41E-04 \\ 
% SI 30  &  2.283E-02	  & 2.265E-02	   & 7.76E-03 & 1.672E-02 & 5.05E-04 \\ 
% P  31  &  7.935E-03	  & 7.902E-03	   & 1.18E-03 & 9.892E-03 & 2.30E-04 \\ 
   S  32  & 1.736E-01	  & 1.852E-01	   & 2.40E-02 & 1.922E-01 & 6.13E-02 \\ 
% S  33  &  2.968E-03	  & 2.956E-03	   & 1.21E-04 & 2.453E-03 & 1.29E-04 \\ 
% S  34  &  2.190E-02	  & 2.180E-02	   & 1.26E-03 & 1.578E-02 & 5.49E-04 \\ 
% S  36  &  8.754E-05	  & 8.709E-05	   & 4.78E-07 & 6.185E-05 & 9.52E-09 \\ 
% CL 35  &  6.397E-03	  & 6.327E-03	   & 5.22E-05 & 8.342E-03 & 4.35E-05 \\ 
% CL 37  &  3.220E-04	  & 3.181E-04	   & 6.97E-06 & 7.696E-04 & 7.10E-06 \\ 
   AR 36  & 4.567E-02	  & 5.060E-02	   & 4.14E-03 & 4.493E-02 & 1.18E-02 \\ 
% AR 38  &  7.863E-03	  & 7.856E-03	   & 3.73E-04 & 1.649E-02 & 1.22E-04 \\ 
% AR 40  &  2.272E-05	  & 2.267E-05	   & 4.34E-09 & 2.161E-05 & 2.02E-10 \\ 
% K  39  &  2.515E-03	  & 2.412E-03	   & 2.88E-05 & 1.039E-02 & 1.95E-05 \\ 
% K  40  &  1.935E-05	  & 1.902E-05	   &	      & 2.503E-04 & 6.24E-09 \\ 
% K  41  &  6.229E-05	  & 6.156E-05	   & 2.18E-06 & 4.573E-04 & 1.54E-06 \\ 
   CA 40  & 2.283E-02	  & 2.628E-02	   & 3.72E-03 & 2.391E-02 & 1.07E-02 \\ 
% CA 42  &  2.517E-04	  & 2.481E-04	   & 1.10E-05 & 1.195E-03 & 2.13E-06 \\ 
% CA 43  &  2.973E-05	  & 2.786E-05	   & 2.32E-06 & 1.390E-04 & 1.14E-07 \\ 
% CA 44  &  5.323E-05	  & 5.305E-05	   & 1.53E-04 & 1.802E-04 & 3.00E-05 \\ 
% CA 46  &  6.937E-06	  & 6.987E-06	   & 7.43E-11 & 1.292E-06 & 3.32E-13 \\ 
% CA 48  &  4.297E-06	  & 4.297E-06	   & 8.56E-16 & 2.207E-06 & 4.38E-19 \\ 
% SC 45  &  1.389E-05	  & 1.347E-05	   & 1.39E-07 & 5.518E-05 & 7.93E-08 \\ 
% TI 46  &  8.049E-05	  & 8.211E-05	   & 4.00E-06 & 2.072E-04 & 1.36E-06 \\ 
% TI 47  &  1.813E-05	  & 1.813E-05	   & 5.04E-06 & 7.551E-05 & 3.60E-07 \\ 
   TI 48  & 1.442E-04	  & 2.142E-04	   & 1.99E-04 & 2.390E-04 & 1.85E-04 \\
% TI 49  &  2.542E-05	  & 3.124E-05	   & 4.20E-06 & 3.613E-05 & 5.52E-06 \\ 
% TI 50  &  8.448E-06	  & 8.056E-06	   & 1.39E-10 & 9.085E-06 & 4.40E-12 \\ 
% V  50  &  1.002E-06	  & 9.867E-07	   & 3.41E-10 & 9.219E-06 & 3.07E-11 \\ 
   V  51  & 5.992E-05	  & 7.778E-05	   & 1.22E-05 & 5.739E-05 & 6.70E-06 \\ 
% CR 50  &  6.758E-04	  & 7.359E-04	   & 3.55E-05 & 1.054E-04 & 1.00E-05 \\ 
   CR 52  & 1.334E-03	  & 2.196E-03	   & 9.20E-04 & 1.291E-03 & 2.32E-03 \\ 
% CR 53  &  2.415E-04	  & 3.387E-04	   & 8.83E-05 & 1.580E-04 & 1.24E-04 \\ 
% CR 54  &  1.679E-05	  & 1.613E-05	   & 1.51E-08 & 2.506E-05 & 3.35E-10 \\ 
    MN 55 &  1.014E-03	  & 1.453E-03	   & 3.15E-04 & 9.887E-04 & 3.54E-04 \\ 
% FE 54  &  1.877E-02	  & 2.200E-02	   & 3.29E-03 & 5.634E-03 & 1.03E-03 \\ 
   FE 56  & 2.679E-02	  & 3.560E-02	   & 6.78E-02 & 1.096E-01 & 1.00E-01 \\ 
% FE 57  &  6.813E-04	  & 7.625E-04	   & 3.13E-03 & 5.016E-03 & 1.98E-03 \\ 
% FE 58  &  4.356E-04	  & 4.046E-04	   & 5.43E-09 & 6.953E-04 & 1.24E-10 \\ 
   CO 59  & 3.217E-04	  & 3.007E-04	   & 1.46E-04 & 7.622E-04 & 8.04E-05 \\ 
   NI 58  & 1.386E-03	  & 1.531E-03	   & 9.35E-03 & 7.982E-03 & 1.31E-03 \\ 
% NI 60  &  1.379E-03	  & 1.374E-03	   & 1.99E-03 & 3.642E-03 & 1.47E-03 \\ 
% NI 61  &  3.691E-04	  & 3.602E-04	   & 1.54E-04 & 5.084E-04 & 6.55E-05 \\ 
% NI 62  &  1.025E-03	  & 1.006E-03	   & 2.12E-03 & 2.437E-03 & 3.01E-04 \\ 
% NI 64  &  4.588E-04	  & 4.472E-04	   & 4.77E-13 & 2.152E-04 & 1.40E-18 \\ 
% CU 63  &  3.247E-04	  & 3.142E-04	   & 7.60E-06 & 1.121E-04 & 7.73E-07 \\ 
% CU 65  &  1.638E-04	  & 1.614E-04	   & 2.20E-06 & 9.614E-05 & 6.52E-07 \\ 
% ZN 64  &  4.478E-05	  & 4.391E-05	   & 1.24E-05 & 5.240E-05 & 1.34E-05 \\ 
% ZN 66  &  2.973E-04	  & 2.938E-04	   & 4.80E-05 & 1.502E-04 & 7.78E-06 \\ 
% ZN 67  &  7.696E-05	  & 7.533E-05	   & 3.86E-07 & 1.061E-05 & 1.30E-08 \\ 
% ZN 68  &  1.130E-04	  & 1.075E-04	   & 1.18E-07 & 6.653E-05 & 2.19E-09 \\ 
   ZN 70  & 2.360E-05	  & 2.282E-05	   & 1.58E-25 & 5.420E-07 & 9.50E-24 \\ 
% GA 69  &  4.762E-05	  & 4.767E-05	   & 1.09E-10 & 1.063E-05 & 2.89E-13 \\ 
% GA 71  &  2.064E-05	  & 2.033E-05	   & 3.04E-16 & 9.592E-06 & 3.60E-15 \\ 
   GE 70  & 2.062E-05	  & 1.949E-05	   & 1.04E-12 & 1.553E-05 & 7.85E-17 \\ 
% GE 72  &  3.220E-05	  & 3.167E-05	   & 6.07E-19 & 2.204E-05 & 3.29E-16 \\ 
% GE 73  &  8.284E-06	  & 8.014E-06	   & 6.34E-23 & 1.938E-06 & 7.15E-19 \\ 
% GE 74  &  1.875E-05	  & 1.781E-05	   & 3.65E-24 & 1.115E-05 & 9.14E-19 \\ 
% GE 76  &  6.536E-06	  & 6.190E-06	   &	      & 4.070E-07 & 2.71E-22 \\ 
% AS 75  &  8.472E-06	  & 8.215E-06	   &	      & 2.548E-06 & 8.92E-15 \\ 
% SE 74  &  1.315E-06	  & 1.283E-06	   &	      & 2.788E-07 & 3.90E-16 \\ 
 SE 76  &   3.571E-06	  & 3.329E-06	   &	      & 4.269E-06 & 6.55E-16 \\ 
% SE 77  &  4.858E-06	  & 4.741E-06	   &	      & 2.149E-06 & 5.90E-19 \\ 
% SE 78  &  7.575E-06	  & 7.273E-06	   &	      & 8.425E-06 & 4.76E-17 \\ 
% SE 80  &  5.562E-06	  & 5.342E-06	   &	      & 7.081E-06 & 9.22E-17 \\ 
% SE 82  &  2.278E-06	  & 2.196E-06	   &	      & 2.310E-07 & 1.07E-19 \\ 
% BR 79  &  2.463E-06	  & 2.395E-06	   &	      & 1.132E-06 & 2.52E-15 \\ 
% BR 81  &  1.927E-06	  & 1.850E-06	   &	      & 1.261E-06 & 2.62E-17 \\ 
% KR 78  &  1.467E-07	  & 1.527E-07	   &	      & 3.274E-08 & 4.17E-16 \\ 
 KR 80  &   9.266E-07	  & 9.160E-07	   &	      & 4.501E-07 & 2.08E-15 \\ 
 KR 82  &   8.627E-07	  & 7.899E-07	   &	      & 3.352E-06 & 6.20E-16 \\ 
% KR 83  &  1.571E-06	  & 1.517E-06	   &	      & 1.727E-06 & 1.17E-16 \\ 
% KR 84  &  4.701E-06	  & 4.527E-06	   &	      & 6.501E-06 & 1.15E-15 \\ 
% KR 86  &  4.687E-06	  & 4.513E-06	   &	      & 1.506E-06 & 2.62E-16 \\ 
% RB 85  &  1.266E-06	  & 1.167E-06	   &	      & 1.104E-06 & 1.31E-15 \\ 
% RB 87  &  8.683E-07	  & 8.355E-07	   &	      & 1.121E-06 & 1.64E-16 \\ 
% SR 84  &  5.166E-08	  & 5.184E-08	   &	      & 6.621E-08 & 2.05E-15 \\ 
   SR 86  & 2.194E-07	  & 2.155E-07	   &	      & 1.227E-06 & 1.40E-14 \\ 
   SR 87  & 8.387E-08	  & 7.887E-08	   &	      & 8.588E-07 & 1.27E-14 \\ 
   SR 88  & 2.370E-06	  & 2.277E-06	   &	      & 4.688E-06 & 1.61E-13 \\ 
% Y  89  &  4.657E-07	  & 4.488E-07	   &	      & 8.811E-07 & 1.52E-14 \\ 
% ZR 90  &  6.557E-07	  & 6.503E-07	   &	      & 2.739E-06 & 9.36E-14 \\ 
% ZR 91  &  1.124E-07	  & 1.088E-07	   &	      & 2.137E-07 & 1.27E-14 \\ 
% ZR 92  &  1.993E-07	  & 1.973E-07	   &	      & 2.473E-07 & 2.45E-14 \\ 
% ZR 94  &  1.306E-07	  & 1.264E-07	   &	      & 1.040E-07 & 3.10E-13 \\ 
% ZR 96  &  9.135E-08	  & 8.929E-08	   &	      & 1.905E-08 & 2.24E-16 \\ 
\hline
 \noalign{\smallskip}
\hline
\end{tabular}
\label{tab:isotopic_yieldscomp_m20z2m2}                                                          
\end{center}
\end{table}

1%%%%%%%%%%%%%%%%%%%%%%%%%%%%%%%%%%%%%%%%%%%%%%
% TABle isotopic abundance exp
%%%%%%%%%%%%%%%%%%%%%%%%%%%%%%%%%%%%%%%%%%%%%%

%\addtolength{\voffset}{-0.0cm}
\begin{table}
\begin{center}
%\tabletypesize{\small}
\caption{For stable species the abundance enrichment in supernova ejecta is given in solar mass unit for the stars at solar metallicity. Here we compare results from our $25\msun$ models, \setopt, with \cite{thielemann:96} (Th96), \cite{rauscher:02} (Ra02) and \cite{chieffi:04} (CL04).}                                        
\begin{tabular}{ccccc}
\hline
  specie & 25 $\msun$ $delay$ & Th96 & Ra02 & CL04 \\ 
\hline 
% H   1     & 9.157E+00     &	& 1.012E+01 & 1.16E+01       \\ 
% H   2     & 3.925E-08     & 1.22E-25 & 2.429E-07 & 1.77E-16	   \\ 
% HE  3     & 5.720E-04     & 3.11E-19 & 7.994E-04 & 2.77E-05	\\ 
% HE  4     & 7.964E+00     & 1.95E+00 & 7.597E+00 & 7.78E+00  \\ 
% LI  7     & 2.506E-09     &	& 1.327E-07 & 2.40E-10       \\ 
% B  11     & 4.966E-08     & 5.29E-14 & 1.542E-06 & 3.85E-13	\\ 
    C  12   & 4.518E-01     &  1.48E-01 & 4.093E-01 & 4.01E-01   \\ 
    C  13   & 1.264E-02     &  1.03E-01 & 1.570E-03 & 8.30E-03   \\ 
    N  14   & 9.152E-02     &  9.53E-04 & 8.101E-02 & 5.26E-02   \\ 
    N  15   & 1.358E-04     &  1.04E-08 & 1.391E-04 & 4.32E-06   \\ 
    O  16   & 8.163E-01     &  2.99E+00 & 3.316E+00 & 2.03E+00   \\ 
    O  17   & 1.075E-04     &  7.86E-08 & 1.262E-04 & 1.58E-04   \\ 
    O  18   & 1.964E-04     &  6.69E-03 & 1.205E-03 & 7.55E-05   \\ 
    F  19   & 1.432E-05     &  8.17E-10 & 7.820E-05 & 4.14E-07   \\ 
    NE 20   & 1.754E-01     &  5.94E-01 & 5.356E-01 & 6.73E-01   \\ 
% NE 21     & 4.048E-04     & 3.22E-03 & 1.517E-03 & 6.30E-05	\\ 
% NE 22     & 4.622E-02     & 3.39E-02 & 2.796E-02 & 6.15E-05	\\ 
    NA 23   & 5.575E-03     & 1.81E-02 & 1.281E-02 & 4.00E-03   \\ 
    MG 24   & 4.159E-02     &  1.59E-01 & 1.444E-01 & 1.36E-01   \\ 
% MG 25     & 7.196E-03     & 3.92E-02 & 3.146E-02 & 5.17E-04	\\ 
% MG 26     & 1.284E-02     & 3.17E-02 & 3.901E-02 & 7.40E-04	\\ 
    AL 27   & 3.219E-03     &  1.95E-02 & 2.206E-02 & 2.20E-03   \\ 
    SI 28   & 2.930E-02     &  1.03E-01 & 3.540E-01 & 1.15E-01   \\ 
% SI 29     & 6.340E-03     & 6.97E-03 & 1.042E-02 & 5.13E-04	\\ 
% SI 30     & 4.002E-03     & 6.81E-03 & 1.020E-02 & 2.81E-04	\\ 
% P  31     & 1.291E-03     & 9.02E-04 & 3.993E-03 & 1.46E-04	\\ 
    S  32   & 1.266E-02     & 3.84E-02 & 1.475E-01 & 5.27E-02   \\ 
% S  33     & 3.346E-04     & 2.20E-04 & 9.121E-04 & 9.38E-05	\\ 
% S  34     & 1.720E-03     & 2.77E-03 & 9.039E-03 & 2.62E-04	\\ 
% S  36     & 2.328E-05     & 7.51E-07 & 1.740E-05 & 2.72E-09	    \\ 
% CL 35     & 6.702E-04     & 6.72E-05 & 6.545E-04 & 3.39E-05	\\ 
% CL 37     & 1.019E-04     & 1.32E-05 & 2.830E-04 & 5.91E-06	\\ 
    AR 36   & 1.844E-03     &  6.71E-03 & 2.315E-02 & 1.05E-02   \\ 
% AR 38     & 4.879E-04     & 7.24E-04 & 7.505E-03 & 6.15E-05	\\ 
% AR 40     & 8.987E-06     & 8.91E-09 & 1.569E-05 & 5.16E-11	\\ 
% K  39     & 1.539E-04     & 3.47E-05 & 5.551E-04 & 1.87E-05	\\ 
% K  40     & 3.415E-06     &	& 5.100E-06 & 3.13E-09       \\ 
% K  41     & 1.331E-05     & 2.79E-06 & 4.374E-05 & 1.47E-06	\\ 
    CA 40   & 1.303E-03     &  6.14E-03 & 1.716E-02 & 9.82E-03   \\ 
% CA 42     & 1.731E-05     & 1.77E-05 & 2.225E-04 & 1.34E-06	\\ 
% CA 43     & 4.630E-06     & 2.78E-07 & 8.935E-06 & 9.92E-08	\\ 
% CA 44     & 3.585E-05     & 2.11E-05 & 6.164E-05 & 5.73E-05	    \\ 
% CA 46     & 2.888E-06     & 2.60E-10 & 1.979E-06 & 5.74E-14	\\ 
% CA 48     & 3.312E-06     & 1.70E-14 & 2.895E-06 & 3.70E-16	\\ 
% SC 45     & 4.526E-06     & 8.96E-08 & 8.322E-06 & 8.40E-08	\\ 
% TI 46     & 7.557E-06     & 6.84E-06 & 9.140E-05 & 1.09E-06	\\ 
% TI 47     & 7.903E-06     & 9.11E-07 & 1.505E-05 & 4.52E-07	\\ 
    TI 48   & 5.499E-05     &  8.98E-05 & 2.050E-04 & 2.11E-04   \\ 
% TI 49     & 6.778E-06     & 6.01E-06 & 2.525E-05 & 5.40E-06	\\ 
% TI 50     & 8.361E-06     & 5.90E-10 & 1.803E-05 & 1.02E-12	\\ 
% V  50     & 2.369E-07     & 7.99E-10 & 6.913E-07 & 9.79E-12	\\ 
    V  51   & 9.595E-06     &  9.96E-06 & 6.878E-05 & 9.42E-06   \\ 
% CR 50     & 1.455E-05     & 5.01E-05 & 2.452E-04 & 1.35E-05	\\ 
    CR 52   & 2.977E-04     &  1.31E-03 & 2.947E-03 & 1.89E-03   \\ 
% CR 53     & 3.489E-05     & 1.39E-04 & 3.960E-04 & 1.22E-04	\\ 
% CR 54     & 1.718E-05     & 2.41E-08 & 4.283E-05 & 1.62E-10	\\ 
    MN 55   & 2.170E-04     &  5.02E-04 & 2.321E-03 & 5.11E-04   \\ 
% FE 54     & 1.416E-03     & 4.81E-03 & 1.703E-02 & 1.61E-03	\\ 
    FE 56   & 2.351E-02     &  5.24E-02 & 1.294E-01 & 1.00E-01   \\ 
% FE 57     & 6.627E-04     & 1.16E-03 & 3.414E-03 & 2.41E-03	\\ 
% FE 58     & 3.014E-04     & 8.34E-09 & 1.161E-03 & 8.14E-11	\\ 
    CO 59   & 2.869E-04     &  2.19E-05 & 6.682E-04 & 8.58E-05   \\ 
    NI 58   & 9.998E-04     &  1.33E-03 & 4.840E-03 & 6.56E-04   \\ 
% NI 60     & 8.408E-04     & 6.67E-04 & 2.092E-03 & 2.36E-03	\\ 
% NI 61     & 1.548E-04     & 2.74E-05 & 3.672E-04 & 9.64E-05	\\ 
% NI 62     & 2.445E-04     & 1.70E-04 & 1.744E-03 & 3.31E-04	\\ 
% NI 64     & 1.708E-04     & 6.08E-15 & 7.453E-04 & 4.30E-16	\\ 
% CU 63     & 1.040E-04     & 1.50E-07 & 4.340E-04 & 2.00E-06	\\ 
% CU 65     & 7.663E-05     & 1.41E-07 & 2.013E-04 & 9.84E-07	\\ 
% ZN 64     & 2.717E-05     & 3.10E-06 & 5.525E-05 & 3.49E-05	\\ 
% ZN 66     & 9.065E-05     & 2.58E-06 & 1.966E-04 & 9.06E-06	\\ 
% ZN 67     & 3.943E-05     & 2.94E-09 & 6.660E-05 & 1.28E-08	\\ 
% ZN 68     & 4.532E-05     & 9.29E-10 & 2.521E-04 & 2.52E-09	\\ 
    ZN 70   & 1.665E-05     &  2.44E-18 & 1.996E-05 & 8.71E-20   \\ 
% GA 69     & 2.185E-05     & 8.00E-14 & 4.111E-05 & 2.44E-13	\\ 
% GA 71     & 1.050E-05     & 3.70E-18 & 3.040E-05 & 2.43E-15	\\ 
    GE 70   & 5.622E-06     &  5.15E-16 & 2.856E-05 & 8.86E-15   \\ 
% GE 72     & 1.234E-05     & 4.78E-18 & 3.691E-05 & 4.65E-14	\\ 
% GE 73     & 4.866E-06     & 9.62E-20 & 2.232E-05 & 3.39E-16	\\ 
% GE 74     & 8.309E-06     & 3.22E-19 & 5.030E-05 & 1.09E-15	\\ 
% GE 76     & 2.435E-06     & 2.94E-20 & 1.036E-05 & 2.76E-19	\\ 
% AS 75     & 2.737E-06     &	& 1.408E-05 & 1.32E-14   \\ 
% SE 74     & 2.838E-08     &	& 1.036E-06 & 2.47E-14   \\ 
    SE 76   & 9.150E-07     &    & 8.389E-06 & 8.66E-14   \\ 
% SE 77     & 1.161E-06     &	& 9.806E-06 & 3.02E-16   \\ 
% SE 78     & 3.429E-06     &	& 1.072E-05 & 7.08E-15   \\ 
% SE 80     & 2.541E-06     &	& 2.527E-05 & 5.43E-15   \\ 
% SE 82     & 6.969E-07     &	& 5.419E-06 & 2.36E-18   \\ 
% BR 79     & 8.154E-07     &	& 8.674E-06 & 5.92E-15   \\ 
% BR 81     & 6.543E-07     &	& 6.393E-06 & 2.97E-15   \\ 
% KR 78     & 6.479E-09     &	& 1.382E-07 & 1.10E-14   \\ 
    KR 80   & 1.666E-07     &    & 9.503E-07 & 2.62E-13   \\ 
    KR 82   & 7.514E-07     &    & 5.228E-06 & 7.01E-14   \\ 
% KR 83     & 5.721E-07     &	& 6.170E-06 & 1.74E-14   \\ 
% KR 84     & 2.132E-06     &	& 1.466E-05 & 4.54E-13   \\ 
% KR 86     & 1.127E-06     &	& 1.261E-05 & 9.55E-14   \\ 
% RB 85     & 5.356E-07     &	& 7.717E-06 & 1.55E-13   \\ 
% RB 87     & 2.460E-07     &	& 3.012E-06 & 1.39E-14   \\ 
% SR 84     & 5.285E-09     &	& 1.912E-07 & 2.37E-13   \\ 
    SR 86   & 2.364E-07     &   & 1.917E-06 & 1.52E-12   \\ 
    SR 87   & 1.444E-07     &    & 1.281E-06 & 1.42E-12   \\ 
    SR 88   & 1.537E-06     &    & 1.145E-05 & 2.55E-11   \\ 
% Y  89     & 3.707E-07     &	& 1.991E-06 & 2.46E-12   \\ 
% ZR 90     & 4.095E-07     &	& 1.880E-06 & 1.80E-11   \\ 
% ZR 91     & 9.458E-08     &	& 4.120E-07 & 9.38E-12   \\ 
% ZR 92     & 1.511E-07     &	& 4.426E-07 & 8.03E-12   \\ 
% ZR 94     & 1.335E-07     &	& 3.428E-07 & 1.19E-10   \\ 
% ZR 96     & 4.929E-08     &	& 8.282E-08 & 2.32E-13   \\ 
\hline
 \noalign{\smallskip}
\hline
\end{tabular}
\label{tab:isotopic_yieldscomp_m25z2m2}                                                          
\end{center}
\end{table}

% list of figures

%%%%%%%%%%%%%%%%%%%%%%%%%
%%%%%%%%%%%%%%%%%%%%%%%%%

%%%%%%%%%%%%%%%%%%%%%%%%%

%%%%%%%%%%%%%%%%%%%%%%%%%

%%%%%%%%%%%%%%%%%%%%%%%%%

%%%%%%%%%%%%%%%%%%%%%%%%%

%%%%%%%%%%%%%%%%%%%%%%%%%%%

\clearpage
% These figures were generated using the script TcRhoc.py, option 8: /ngpod1/nugrid/work_keele/mike_python/TcRhoc.py

%%%%%%%%%%%%%%%%%%%%%%%%%
\clearpage
%%%%%%%%%%%%%%%%%%%%%%%%%

%%%%%%%%%%%%%%%%%%%%%%%%%

%%%%%%%%%%%%%%%%%%%%%%%%%

%%%%%%%%%%%%%%%%%%%%%%%%%%
%%%%%%%%%%%%%%%%%%%%%%%%%

%%%%%%%%%%%%%%%%%%%%%%%%%
%%%%%%%%%%%%%%%%%%%%%%%%%
%%%%%%%%%%%%%%%%%%%%%%%%%
%%%%%%%%%%%%%%%%%%%%%%%%%

\clearpage

%%%%%%%%%%%%%%%%%%%%%%%%%

%%%%%%%%%%%%%%%%%%%%%%%%%

%%%%%%%%%%%%%%%%%%%%%%%%%

%%%%%%%%%%%%%%%%%%%%%%%%%

%%%%%%%%%%%%%%%%%%%%%%%%%
%\begin{figure}
%\centering
%\includegraphics[width=0.48\textwidth]{agb_lambda_starmass_set1p1_winds.pdf}
%\includegraphics[width=0.48\textwidth]{agb_lambda_starmass_set1p2_winds.pdf}
%\caption{$Left$ $Panel$: Evolution of dredge-up parameter $\lambda_{DUP}$ at each TP,
%starting with the second pulse. Only AGB models from \setopo\ are shown.
%$Right$ $Panel$: The same diagram as in the $Left$ $Panel$ but for \setopt\ AGB models.}
%\label{fig:agb_lambda_starmass_winds}
%\end{figure}

%%%%%%%%%%%%%%%%%%%%%%%%%

%%%%%%%%%%%%%%%%%%%%%%%%%

%%%%%%%%%%%%%%%%%%%%%%%%%
%\begin{figure}
%\centering
%\resizebox{15cm}{!}{\rotatebox{0}{\includegraphics{15_d_se1p2_prodfac_pprocess.pdf}}}
%\caption{Final isotopic production factors for the 15 $\msun$ star, $Z$ = 0.02,
%in the mass region A $>$ 65. [RH: ADD LABELS FOR P-ONLY ISOTOPES.]
%}
%\label{fig:prodfac_15_d_set1p2_pprocess}
%\end{figure}
%
%%%%%%%%%%%%%%%%%%%%%%%%%
%%%%%%%%%%%%%%%%%%%%%%%%%
\begin{figure}
\centering
\includegraphics[width=\textwidth]{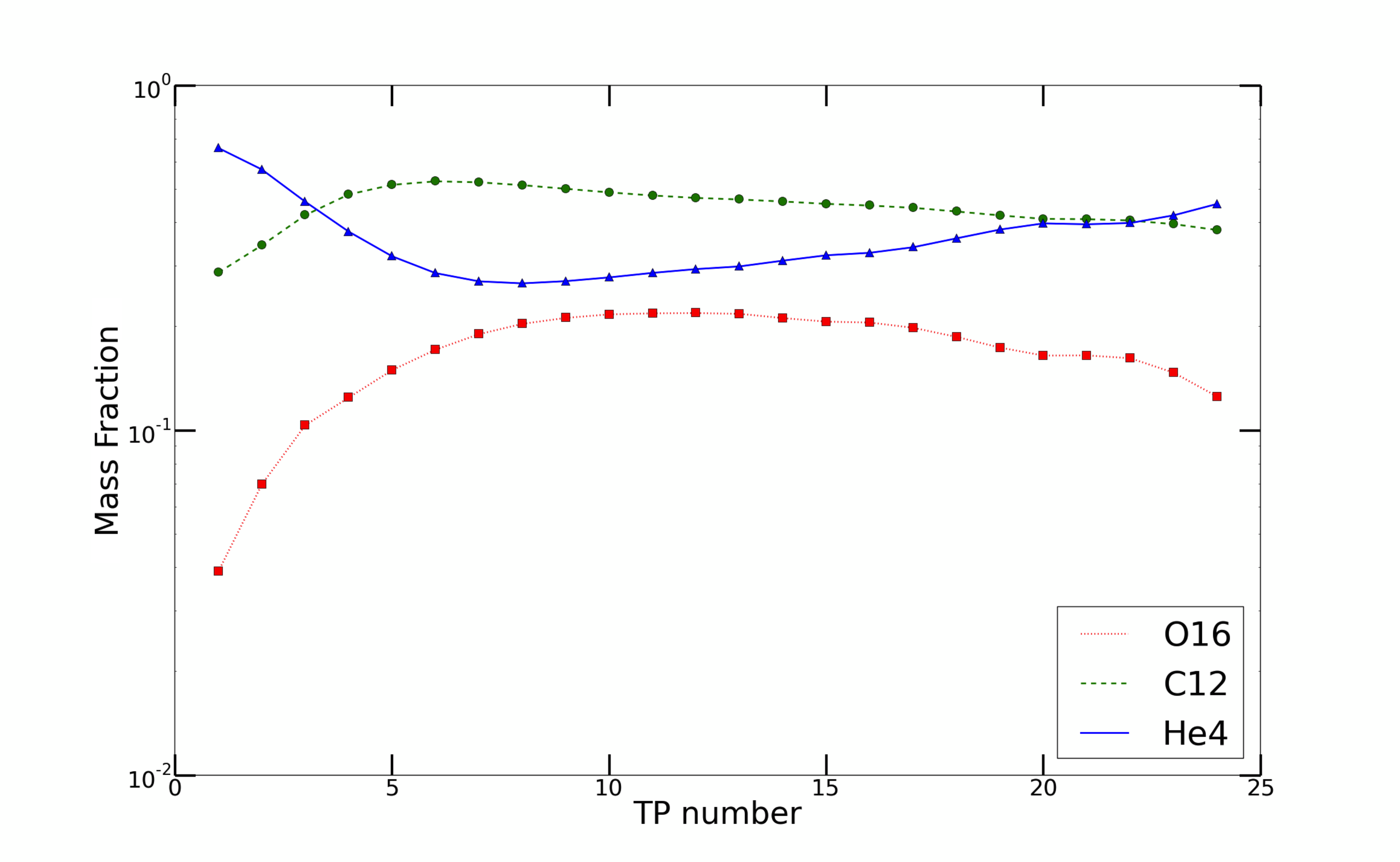}
\caption{Intershell abundances throughout AGB evolution. The horizontal axis refers to the interpulse-period after a given thermal-pulse number. These data refer to the \setopo\ AGB model with $1.65\msun$ as initial mass and metallicity respectively.}
\label{fig:intershell_abu}
\end{figure}

%%%%%%%%%%%%%%%%%%%%%%%%%
%%%%%%%%%%%%%%%%%%%%%%%%%
%\begin{figure}
%\centering
%\resizebox{15cm}{!}{\rotatebox{0}{\includegraphics{15_d_se1p2_prodfac_pprocess.pdf}}}
%\caption{Final isotopic production factors for the 15 $\msun$ star, $Z$ = 0.02,
%in the mass region A $>$ 65. [RH: ADD LABELS FOR P-ONLY ISOTOPES.]
%}
%\label{fig:prodfac_15_d_set1p2_pprocess}
%\end{figure}
%
%%%%%%%%%%%%%%%%%%%%%%%%%

%%%%%%%%%%%%%%%%%%%%%%%%%

%%%%%%%%%%%%%%%%%%%%%%%%%
%\begin{figure}
%\centering
%\resizebox{7.5cm}{!}{\rotatebox{0}{\includegraphics{15_post_pre_comparison_set1p2_rd2.pdf}}}
%\resizebox{7.5cm}{!}{\rotatebox{0}{\includegraphics{15_post_pre_comparison_set1p2_rd4.pdf}}}
%\caption{Final abundance distribution after the explosion are compared to pre-explosive
%abundances for the 15 $\msun$ stellar models of Set 1.2, where the shock velocity
%from the SN explosion $rapid$ is reduced by a factor of 2 and 4 ($rapid/2$ and $rapid/4$, respectively).
%in Fig. \ref{fig:summary_exp_set1.2_15_energytest}.
%}
%\label{fig:post_versus_pre_set1.2_15_energytest}
%\end{figure}
%
%%%%%%%%%%%%%%%%%%%%%%%%%%%%%%%%
%%%%%%%%%%%%%%%%%%%%%%%%%%%%%%%%

\end{document}